\documentclass[a4paper,11pt]{book}
\pdfoutput=1

\usepackage[utf8]{inputenc}
\usepackage[margin=1in]{geometry}
\usepackage[T1]{fontenc}
\usepackage{booktabs,multirow}
\usepackage{subcaption}
\usepackage{amsmath}
\usepackage{amssymb}
\usepackage{mathtools,braket,dsfont}
\usepackage{url}
\usepackage{graphicx}
\usepackage{multicol}
\usepackage{multirow}
\usepackage{dsfont}
\usepackage{float}
\usepackage[dvipsnames]{xcolor}
\usepackage{slashed}
\usepackage[final]{hyperref}
\def\linkcolor{cyan!70!black}
\hypersetup{
	colorlinks=true,
	linkcolor=\linkcolor,
	citecolor=\linkcolor,
	filecolor=\linkcolor,
	urlcolor=\linkcolor
}
\usepackage[numbers, sort&compress]{natbib}
\usepackage[utf8]{inputenc}
\usepackage[T1]{fontenc}
\usepackage{comment}
\usepackage{siunitx}
\usepackage{enumitem}
\usepackage{orcidlink}\usepackage{tocloft}
\usepackage{fancyhdr}
\usepackage{amsthm}
\usepackage{esvect}
\usepackage{wrapfig}
\usepackage[absolute,overlay]{textpos}
\usepackage{cancel}

\newcommand{\beq}{\begin{equation}}
\newcommand{\eeq}{\end{equation}}
\newcommand{\bea}{\begin{eqnarray}}
\newcommand{\eea}{\end{eqnarray}}

\newcommand{\fref}[1]{Figure~\ref{#1}}

\newcommand{\tref}[1]{Table~\ref{#1}}

\def\matlength{\!\!\!\!}

\hypersetup{pdfinfo={
  Title={Towards the Swampland of Flavour Symmetries},
  Author={Full Name},
  Subject={Subject of thesis},
  Keywords={Comma-separated keywords}}
}
\author{Marcel Krause}
\newcommand{\myname}{M.Sc.~Xiyuan Gao}
\newcommand{\mytitle}{Towards the Swampland of\\ Flavour Symmetries}
\newcommand{\myplace}{Qiqihar, China}
\newcommand{\mytimefordefense}{17. Juli 2026}
\newcommand{\mytitleunform}{Towards the Swampland of Flavour Symmetries}

\newcommand{\reviewerone}{Prof. Dr. Ulrich Nierste}
\newcommand{\reviewertwo}{Prof. Dr. Felix Kahlhoefer}

\newcommand{\changefont}[3]{\fontfamily{#1} \fontseries{#2} \fontshape{#3} \selectfont}

\newcommand{\typo}[2]{#2}

\begin{document}

\newcommand{\diameter}{20}
\newcommand{\xone}{-13}
\newcommand{\xtwo}{150}
\newcommand{\yone}{10}
\newcommand{\ytwo}{-246}

\begin{titlepage}
\begin{tikzpicture}[overlay]
\draw[color=gray]  
 		 (\xone mm, \yone mm)
  -- (\xtwo mm, \yone mm)
 arc (90:0:\diameter pt) 
  -- (\xtwo mm + \diameter pt , \ytwo mm) 
	-- (\xone mm + \diameter pt , \ytwo mm)
 arc (270:180:\diameter pt)
	-- (\xone mm, \yone mm);
\end{tikzpicture}
	\begin{textblock}{10}[0,0](1.8,1.4)
		\includegraphics[width=.3\textwidth]{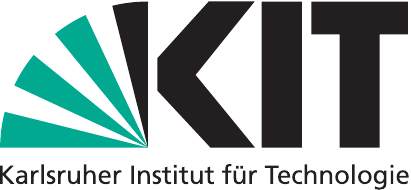}
	\end{textblock}
	\changefont{phv}{m}{n}	
	\vspace*{2.2cm}
	\begin{center}
		\includegraphics[width=0.89\linewidth]{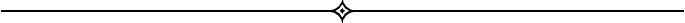} \\
		\vspace*{0.24cm}
		\Huge{\mytitle} \\
		\vspace*{0.24cm}
		\hspace*{0.23cm}\includegraphics[width=0.89\linewidth]{Graphics/TitlepageBorder2.pdf}
		\vspace*{1.0cm}\\
		\Large{Zur Erlangung des akademischen Grades eines}\\
		\vspace*{0.24cm}
		\LARGE{\textsc{Doktors der Naturwissenschaften \\ (Dr. rer. nat.)}}\\
		\vspace*{0.35cm}
		\Large{von der KIT-Fakultät für Physik\\des Karlsruher Instituts für Technologie (KIT)}\\
		\vspace*{0.27cm}
 		\Large{genehmigte}\\
		\vspace*{0.24cm}
		\huge{\textsc{Dissertation}}\\
		\vspace*{0.35cm}
		\Large{von}\\
		\vspace*{0.9cm}
		\huge{\textbf{\myname}}\\
		\vspace*{0.35cm}
		\Large{aus \myplace}\\
		\vspace*{0.5cm}
	\end{center}
	\vspace*{1cm}
\Large{
\begin{center}
\begin{tabular}[ht]{l c l}
  \iflanguage{english}{
Referentin}{Referentin}: & \hfill  & \reviewerone\\
  \iflanguage{english}{
Korreferent}{Korreferent}: & \hfill  & \reviewertwo\\

\end{tabular}
\end{center}
}

	\vspace*{1cm}
\large{
\begin{center}
\begin{tabular}[ht]{l c l}
  Tag der mündlichen Prüfung: & \hfill  & \mytimefordefense
\end{tabular}
\end{center}
}

\begin{textblock}{10}[0,0](1.8,15.1)
\tiny{ 
	\iflanguage{english}
		{KIT –- Die Forschungsuniversität in der Helmholtz-Gemeinschaft}
		{KIT –- Die Forschungsuniversität in der Helmholtz-Gemeinschaft}
}
\end{textblock}

\begin{textblock}{10}[0,0](11.5,15.1)
\large{
	\textbf{www.kit.edu} 
}
\end{textblock}

\end{titlepage}


\thispagestyle{empty}
\null\vfill
\noindent\hbox to \textwidth{\hrulefill}
\begin{figure}[!htb]
	\includegraphics[width=0.07\linewidth]{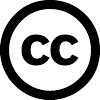}
	\hspace*{0.05cm}
	\includegraphics[width=0.07\linewidth]{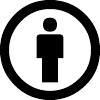}
	\hspace*{0.05cm}
	\includegraphics[width=0.07\linewidth]{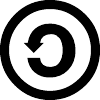}
\end{figure}\\
This document is licensed under a Creative Commons Attribution-ShareAlike 4.0 International License (CC BY-SA 4.0):

\url{https://creativecommons.org/licenses/by-sa/4.0/deed.en}

%
%
%

\cleardoublepage


\thispagestyle{empty}
\null\vfill
\noindent\hbox to \textwidth{\hrulefill} 
Eidesstattliche Versicherung gemäß \S \,13 Absatz 2 Ziffer 3 der Promotionsordnung des Karls\-ruher Instituts für Technologie (KIT) für die KIT-Fakultät für Physik:
\begin{enumerate}
\item Bei der eingereichten Dissertation zu dem Thema
\begin{center}
	\textbf{\mytitleunform}
\end{center}
handelt es sich um meine eigenständig erbrachte Leistung.
\item Ich habe nur die angegebenen Quellen und Hilfsmittel benutzt und mich keiner unzulässigen Hilfe Dritter bedient. Insbesondere habe ich wörtlich oder sinngemäß aus anderen Werken übernommene Inhalte als solche kenntlich gemacht.
\item Die Arbeit oder Teile davon habe ich bislang nicht an einer Hochschule des In- oder Auslands als Bestandteil einer Prüfungs- oder Qualifikationsleistung vorgelegt.
\item Die Richtigkeit der vorstehenden Erklärungen bestätige ich.
\item Die Bedeutung der eidesstattlichen Versicherung und die strafrechtlichen Folgen einer unrichtigen oder unvollständigen eidesstattlichen Versicherung sind mir bekannt.
\end{enumerate}
Ich versichere an Eides statt, dass ich nach bestem Wissen die reine Wahrheit erklärt und nichts verschwiegen habe.
 
\vspace{1.0cm}
\textbf{Karlsruhe, den 09. Juni 2026}
\vspace{1.5cm}
 
\dotfill\hspace*{8.0cm}\\
\hspace*{2.38cm}

%
%
%

\clearpage

\thispagestyle{empty}

{\let\cleardoublepage\clearpage
\chapter*{Declaration of No AI Usage}


\large I hereby declare that no generative artificial intelligence (AI) was used in the writing of this work. 

\vspace{1.5cm}
\textbf{Karlsruhe, den 09. Juni 2026}
\vspace{1.5cm}
 
\dotfill\hspace*{8.0cm}\\
\hspace*{2.38cm}

\thispagestyle{empty}}

\newpage

\pagenumbering{roman}

\chapter*{Zusammenfassung}
\addcontentsline{toc}{chapter}{Zusammenfassung}

Die Replikation von Fermionen --- charakterisiert durch ihren Flavour --- führt im Standardmodell (SM) und in den aus SM-Feldern konstruierten effektiven Feldtheorien zu komplizierten Strukturen. 
Zur Vereinfachung einer Theorie ist es gängige Praxis, Flavour-Sym-metrien aufzuerlegen und weiche Symmetriebrechungsterme zu postulieren. 
Trotz jahrzehntelanger Bemühungen verfügen wir jedoch weiterhin über keinen allgemeinen Rahmen, der zu einem realistischen, symmetriebasierten Modell mit einem oder nur sehr wenigen freien Parametern führt. 
In dieser Arbeit argumentieren wir, dass auch Flavour-Theorien ohne explizite Flavour-Symmetrien, die wir als das „Swampland der Flavour-Symmetrien“ bezeichnen, ernsthaft in Betracht gezogen werden sollten.
Zunächst demonstrieren wir die Überprüfbarkeit solcher Theorien anhand eines einfachen Beispiels, in dem das SM um ein skalares Triplettfeld an der TeV-Skala erweitert wird, das sogenannte Typ-II~Seesaw-Modell. Zusätzlich zu Flavour-Symmetrien, die auf die Leptonfelder wirken, finden wir, dass bestimmte Flavour-Texturen ohne Symmetrieschutz ebenfalls die stark eingeschränkten $\mu\to e$ Übergangsraten unterdrücken können. Dadurch kann das Modell in einer allgemeineren Klasse von Experimenten untersucht werden.
Anschließend zeigen wir zwei Rahmen auf, in denen einige Flavour-Parameter des SM berechenbar sind, selbst wenn die zugrunde liegenden Symmetrien nur implizit vorliegen. 
(i) In der minimalsten SO(10) Theorie kann das $b-\tau$ Massenverhältnis vorhergesagt werden, obwohl die SO(10) Symmetrie bei niedrigen Energieskalen nicht manifest ist. 
Insbesondere wird gefunden, dass, falls die Leptoquarks im skalaren Sektor bei der TeV-Skala liegen, was durch die etablierten B-Anomalien begründet werden kann, der Wert für das $b-\tau$ Massenverhältnis stark vom durch die SO(10) Symmetrie bestimmten Wert abweicht und im Einklang mit Messungen ist.
(ii) Wir untersuchen einen durch Gravitation induzierten nichtperturbativen Effekt, das sogenannte Neutrinokondensat, das im Infrarot-Limes effektive Neutrinomassen erzeugt. Unter Berücksichtigung dieses Effekts identifizieren wir im SM, einschließlich dreier rechtshändiger Neutrinos, eine neue Klasse exakter Symmetrien, die wir als erweiterte $B-L$~Symmetrien bezeichnen und unter denen alle Neutrinos chiral und masselos sind. Unser Aufbau stellt sicher, dass die bekannten Neutrinos, unabhängig davon, ob sie Dirac- oder Majorana-Teilchen sind, ausschließlich durch dynamische Effekte bei niedrigen Energien massiv werden. Bemerkenswerterweise unterscheidet sich die erweiterte $B-L$~Symmetrie im Neutrinosektor von der kanonischen $B-L$~Symmetrie. Wird sie geeicht, kann sie zu charakteristischen Signalen in Neutrinoexperimenten führen.
Des Weiteren kann das Swampland der Flavour-Symmetrien auch leichte neue Physik beinhalten, insbesondere flavour-nicht-diagonal koppelnde Axionen, welche auf natürliche Weise in die oben genannten Modelle eingebettet werden können. Als weitere Studie wird ein vorher übersehener zweischleifen Beitrag zu Axion flavourverletzenden Wechselwirkungen in einem minimalen Axionmodell berechnet und dessen Auswirkung auf einem jüngsten bei Belle II gemessenen Überschuss in $B\to K+E_{\text{miss}}$ analysiert.

{\let\cleardoublepage\clearpage
\chapter*{Abstract}
\addcontentsline{toc}{chapter}{Abstract}}

The replication of fermions --- known as flavour --- gives rise to complicated structures in the Standard Model (SM) and the effective field theories constructed by SM fields. 
To simplify a theory, it is common practice to enforce flavour symmetries and postulate soft symmetry breaking terms. 
However, despite decades of efforts, we still do not have a common framework leading to a realistic symmetry-based model with one or very few free parameters. 
In this thesis, we argue that the flavour theories without explicit flavour symmetries, which we call the `swampland of flavour symmetries', should also be considered seriously.
Firstly, we demonstrate the testability of such theories with a simple example which extends the SM with a TeV-scale triplet scalar field, called the type~II seesaw model. 
In addition to the flavour symmetries acting on the lepton fields, we find certain flavour textures without symmetry protection can also suppress the tightly constrained $\mu\to e$ transition rates, allowing the model to be probed in a more generic category of experiments. 
Next, we show two frameworks in which some SM flavour parameters are calculable even if the underlying symmetries are implicit.
(i) In the most minimal $SO(10)$ theory, the SM $b-\tau$ mass ratio can be predicted although the $SO(10)$ symmetry is not manifest at low energies.
In particular, we find when the leptoquarks contained in the scalar sector lie at TeV scale, motivated by the long-standing $B$ anomalies, the low-energy $b-\tau$ mass ratio largely deviates from the wrong value predicted by $SO(10)$ and stays consistent with the measurements. 
(ii) We revisit a non-perturbative effect induced by gravity, known as neutrino condensate, which generates effective neutrino masses in the infrared limit. 
Considering this, we identify a new class of exact symmetries in the SM (including three right-handed neutrinos), \typo{referred as}{referred to as} enhanced $B-L$, under which all neutrinos are chiral and massless. 
Our setup ensures that the known neutrinos, irrespective of their Dirac or Majorana nature, are massive only via the dynamical effects at low energies. 
Notably, in the neutrino sector, the enhanced $B-L$ symmetry differs from the canonical one. 
If gauged, it can lead to distinctive signals in the neutrino experiments.
Furthermore, we note that the swampland of flavour symmetries could also involve light new physics, in particular the flavourful axions, which can be naturally embedded into the above-mentioned frameworks. 
As a preliminary study, we calculate an overlooked two-loop contribution to the axion flavour violating interactions in a minimal axion model, and analyze its impact on explaining a recent excess at Belle II.


{\let\cleardoublepage\clearpage
\chapter*{Acknowledgments}
\addcontentsline{toc}{chapter}{Acknowledgments}}

\noindent
Firstly, I would like to thank my supervisor, Ulrich Nierste, for his invaluable guidance. 
By working with him, I am able to enjoy the maximal freedom and always get in-time help when I ask. 
To me, Uli is also a very nice friend and he brings a lot of humor to my everyday life. 
I am also grateful to Felix Kahlhöfer for being the second reviewer of my thesis and providing me plenty of advice over the past years.\\

\noindent
Secondly, I want to express my sincere gratitude to 
the supervisors of my bachelor and master thesis, Lorenzo Calibbi and Goran Senjanovic, for their continued support and encouragement after my graduation. \\

\noindent 
Furthermore, I would like to thank Gia Dvali. He kindly gave me the opportunity to visit the Max Planck Institute for Physics and to present my work there. 
I am also thankful to Howard Haber and Nico Gubernari, for inviting me to give talks in the conferences they organized. \\

\noindent
I want to thank the following people for helpful discussions and constructive comments on my research: 
Saiyad Ashanujjaman, Borut Bajc,  Gustavo Branco, Andreas Crivellin,  Gia Dvali, Howard Haber, 
Felix Kahlhöfer, Markus Mosbech, Nudzeim Selimovic, Saurabh Shukla, Ben Stefanek, Di Zhang, Tongxuan Zhang,  Shun Zhou, Robert Ziegler, and Xunwu Zuo.\\

\noindent
I am also grateful to my colleagues and friends for the good times we had together: 
Francesca Acanfora, Benedetta Belfatto, Monika Blanke, Giovani Dalla Valle Garcia, Xiangwen Guan, Syuhei Iguro,  Amir N. Khan, Tim Kretz, Henda Mansour, Jonas Matuszak, Francesco Moretti, Jiaxin Qiao, Lena Rathmann, Maksym Riabokon, Liangliang Su, Benhao Tang, Matteo Tresoldi, Marco Vitti, Yunzhi Wu, Yingxuan Xu, Xiehang Yu, Man Yuan, Chenyang Zhang, Hantian Zhang, Tongxuan Zhang, Robert Ziegler, and Xunwu Zuo. 
My special acknowledgments go to Berkan Demir and Philipp Goller, who translated my English abstract to German manually, and to Dominik Grau for helping me revise and organize the translations.\\


\noindent
My deepest gratitude is to my parents. Most of this thesis was written during my vacation time at home. Without their care for my daily life, it would take me much longer to finish this work. \\

\noindent
The research presented in this thesis was supported by the Deutsche Forschungsgemeinschaft (DFG, German Research Foundation) under grant 396021762 - TRR 257, the BMBF Grant 05H21VKKBA, Theoretische Studien für Belle II und LHCb, and the Doctoral School“Karlsruhe School of Elementary and Astroparticle Physics: Science and Technology.”\\

{\let\cleardoublepage\clearpage
\chapter*{Publications}
\addcontentsline{toc}{chapter}{Publications}}

\noindent
\textbf{This thesis is mostly based on these publications:}\\

\begin{itemize}[label={}, leftmargin=*]
\item X.~Gao and A.~N. Khan, \emph{{Neutrino Masses with Enhanced $B-L$ Symmetry}},  [\href{https://arxiv.org/abs/2601.14376}{{\ttfamily 2601.14376}}].
\item L.~Calibbi, X.~Gao and M.~Yuan, \emph{{Hunting for Neutrino Texture Zeros with Muon and Tau Flavor Violation}}, \href{https://doi.org/10.1007/JHEP05(2026)188}{\emph{JHEP} {\bfseries 05} (2026) 188}, [\href{https://arxiv.org/abs/2511.08679}{{\ttfamily 2511.08679}}].
\item X.~Gao and U.~Nierste, \emph{{TeV-scale scalar leptoquarks motivated by B anomalies improve Yukawa unification in SO(10) GUT}}, \href{https://doi.org/10.1007/JHEP02(2026)082}{\emph{JHEP} {\bfseries 02} (2026) 082}, [\href{https://arxiv.org/abs/2508.11745}{{\ttfamily 2508.11745}}].
\item X.~Gao and U.~Nierste, \emph{{$B\rightarrow K + \text{axion-like particles}$: Effective versus UV-complete models and enhanced two-loop contributions}}, \href{https://doi.org/10.1103/5j2t-2kdf}{\emph{Phys. Rev. D} {\bfseries 112} (2025) 055008}, [\href{https://arxiv.org/abs/2506.14876}{{\ttfamily 2506.14876}}].
\end{itemize}

\vspace{0.5cm}

\noindent
\textbf{The following articles are also relevant to this thesis and are mentioned in the main text:}

\begin{itemize}[label={}, leftmargin=*]
\item X.~Gao, \emph{{Spontaneous CP violation and flavor changing neutral currents in minimal SO(10)}}, \href{https://doi.org/10.1103/PhysRevD.111.055013}{\emph{Phys. Rev. D} {\bfseries 111} (2025) 055013}, [\href{https://arxiv.org/abs/2412.00196}{{\ttfamily 2412.00196}}].
\item L.~Calibbi and X.~Gao, \emph{{Lepton flavor violation in minimal grand unified type II seesaw models}}, \href{https://doi.org/10.1103/PhysRevD.106.095036}{\emph{Phys. Rev. D} {\bfseries 106} (2022) 095036}, [\href{https://arxiv.org/abs/2206.10682}{{\ttfamily 2206.10682}}].
\end{itemize}

\vspace{0.5cm}

\noindent
\textbf{Some of the works shown above are recapped in these conference proceedings:}

\begin{itemize}[label={}, leftmargin=*]
\item X.~Gao, \emph{{$B\to K+$ invisible in a model with axion-like particles}}, \href{https://doi.org/10.22323/1.485.0305}{\emph{PoS} {\bfseries EPS-HEP2025} (2026) 305}, [\href{https://arxiv.org/abs/2511.21811}{{\ttfamily 2511.21811}}].
\item X.~Gao, \emph{{Minimal renormalizable SO(10), spontaneous CP violation, and flavor implications}}, \href{https://doi.org/10.22323/1.481.0038}{\emph{PoS} {\bfseries DISCRETE2024} (2025) 038}, [\href{https://arxiv.org/abs/2503.19028}{{\ttfamily 2503.19028}}].
\end{itemize}

\clearpage

\phantomsection
\tableofcontents
\addcontentsline{toc}{chapter}{Contents}

\clearpage

\phantomsection
\renewcommand{\baselinestretch}{1.5}\normalsize
\listoffigures
\renewcommand{\baselinestretch}{1.0}\normalsize
\addcontentsline{toc}{chapter}{List of Figures}

\clearpage

\phantomsection
\renewcommand{\baselinestretch}{1.5}\normalsize
\listoftables
\renewcommand{\baselinestretch}{1.0}\normalsize
\addcontentsline{toc}{chapter}{List of Tables}

\chapter{Introduction}
\pagenumbering{arabic}

One of the most --- if not the most --- important principles of natural philosophy is that all matter and radiation is composed of fundamental indivisible units~\cite{Feynman:1963uxa}. 
For a long time, atoms are believed to be the elementary building blocks of matter. 
Since the discovery of free electrons in 1897~\cite{Thomson:1897cm}, people have observed a large number of subatomic particles, commonly \typo{referred as}{referred to as} `the particle zoo', which over many years make the physical picture of the elementary particles complicated. 
The modern understanding on the patterns and interactions of elementary particles is based on the electroweak theory~\cite{Glashow:1961tr, Weinberg:1967tq, Salam:1968rm} and quantum chromodynamics (QCD)~\cite{Gross:1973id, Politzer:1973fx}, which altogether constitute the so-called Standard Model (SM) of particle physics. 
Nowadays, the validity of the SM (viewed as an effective field theory) has been verified by the following key experimental observations:

\begin{itemize}

\item \textbf{The discovery of the Higgs boson}:~In 2012, the ATLAS and CMS experiment at the large hadron collider (LHC) at CERN discovered the Higgs boson~\cite{ATLAS:2012yve, CMS:2012qbp}. 
Analysis on the LHC Run 1 and Run 2 dataset proves that the properties of the Higgs boson are in excellent agreement with the SM predictions~\cite{ATLAS:2022vkf, CMS:2022dwd}.

\item \textbf{The confirmation of the Kobayashi-Maskawa mechanism}: The B-factories, including the BaBar experiment at SLAC, the Belle experiment and its upgrade Belle~II at KEK, and the LHC-b experiment at CERN, have confirmed the Kobayashi-Maskawa (KM) mechanism of flavour and CP violation in the quark sector with very high accuracy~\cite{Charles:2004jd, UTfit:2005ras, UTfit:2007eik}. 

\item \textbf{Electroweak precision tests}:~The measurement on the electroweak precision observables (EWPO), in particular the data collected at LEP, SLAC, and Tevatron, shows good agreement with the SM mechanism of spontaneous symmetry breaking, to the level of radiative corrections~\cite{deBlas:2021wap}.

\item \textbf{Measurement of the electron and muon magnetic moments}:~
The magnetic moments of electrons and muons are determined to match the SM predictions to extraordinary precisions of $10^{-12}$ and $10^{-10}$ respectively~\cite{Fan:2022eto, Muong-2:2025xyk}. 

\item \textbf{Tests of conservation laws}:~Despite decades of searches in a wide range of channels, no signals for the violation on the accidental symmetries of the SM Lagrangian (including the baryon number, the lepton number, and the charged leptons flavour symmetries) are found. 

\end{itemize}
As a consequence, if any new particle beyond the SM exists, it must be either much heavier than the known SM particles or weakly interacting with them. 
From this perspective, the SM is a remarkably successful theory and can be viewed as a milestone of modern physics.

However, the SM contains too much arbitrariness so that people rarely view it as the ultimate `theory of everything', even if staying agnostic to gravity. 
It is interesting to note that Weinberg had included a comment in his 1967 paper which established the SM~\cite{Weinberg:1967tq}
\begin{quote}
    \textit{Of course our model has too many arbitrary features for these predictions to be taken very seriously, but it is worth keeping in mind that the standard calculation of the electron-neutrino cross section may well be wrong.}
\end{quote}
Actually, the SM is even not the minimal renormalizable theory to explain the lepton and quark properties known in 1960s.
In 1972, Georgi and Glashow constructed a simpler electroweak model which is based on the gauged $SU(2)$ symmetry only and does not lead to the neutral currents~\cite{Georgi:1972cj}. 
Unfortunately, the Georgi-Glashow $SU(2)$ model got falsified after the experimental discovery of the neutral currents one year later~\cite{GargamelleNeutrino:1973jyy}. 
Introducing a gauged abelian symmetry then turns out to be necessary, but it becomes a weakness of the SM. 
As an abelian group, $U(1)_Y$ can lead to a Landau pole~\cite{Landau:1954nau} at an ultra-high scale and its quantized charges are not guaranteed by any fundamental principle. 
Another shortcoming of the SM is the replication of fermions. 
The SM contains three families --- three copies of each fermion field carrying the same quantum numbers, also known as three  generations. 
As a consequence, the same-type fermions across the three families can have different masses and mix with each other. 
These mass and mixing patterns are all arbitrary within the SM, giving rise to many free model parameters which can only be determined from experiments. 
The measured free SM parameters appear not to be random numbers but rather to follow certain regular interrelationships: for all types of charged fermions, their mass ratios among three families exhibit hierarchical structures. Interestingly, such a pattern is also followed by the quark flavour mixing angles.

In addition to the conceptual puzzles, it is also worth remarking that not all present experimental data are consistent with the SM predictions. 
Many cosmological and astrophysical observations 
indicate the existence of cold non-baryonic components in our universe, which has zero pressure and almost do not interact with the electromagnetic force, commonly \typo{referred as}{referred to as} the `dark matter', see Ref.~\cite{Arbey:2021gdg,Cirelli:2024ssz} for recent reviews. 
The observational evidence for dark matter, assuming they are particles, conflicts with the SM because it contains no dark matter candidates. 
Moreover, the observed matter-antimatter asymmetry \typo{can not}{cannot} be accounted within the SM either, because the SM predicts no large deviation from the thermal equilibrium (no first order phase transitions) in the early universe and incorporates insufficient effects of CP violation. 
Apart from the extra-terrestrial evidences, an increasing number of observations on earth have also shown discrepancies with the SM predictions, some of which, known as `anomalies', are independently reported by differently collaborations with sizable statistical significances, see Ref.~\cite{Crivellin:2023zui} for a recent review.
Over the past decade, many anomalies are even found to be consistently connected to each other. 
An representative example is the so-called `flavour anomalies', which involves the semi-leptonic decays of $B$ mesons including both charged-current and neutral-current transitions~\cite{Alguero:2023jeh, Capdevila:2023yhq, Iguro:2024hyk}, and the CP asymmetries in the hadronic $D$ meson decays~\cite{Iguro:2024uuw}. 
Besides, at the LHC, Higgs-like resonant signals at 95 GeV and 152 GeV and many non-resonant anomalies are reported, see e.g. Ref.~\cite{Crivellin:2023zui, Ashanujjaman:2023etj, Ashanujjaman:2024pky, Crivellin:2024uhc, Crivellin:2026eyk}.
Although we have to admit that many anomalies might be merely fluctuations, it is statistically unlikely that all of them disappear in future with more data collected. 
If (some of) the present anomalies are confirmed in future, it will be evident that the SM is incomplete --- new physics beyond the SM must \typo{exsit}{exist} and manifest as particles not much heavier than $\mathcal{O}(10)$ TeV.

Among the various conceptual puzzles and phenomenological problems introduced above, many of them are closely related to the rich flavour sector of the SM. 
Thus, it is preferable to describe the SM flavour sector within a simplified framework. 
A well-established paradigm is to consider a hypothetical theory with enhanced flavour symmetries and view it as an approximation to the SM. 
For example, the SM can be well approximated with a $U(2)^5$ invariant theory constructed with the same field contents, with the $U(2)$ factors acting on the five types of chiral fermions in the first-two families~\cite{Barbieri:2011ci, Isidori:2012ts, Barbieri:2012uh, Blankenburg:2012nx}. 
The $U(2)^5$ symmetry is chiral and enforces the fermions it acts on to be strictly massless. 
Consequently, in the limit of exact $U(2)^5$ symmetry, only the third-generation charged fermions can be massive and no flavour mixings can arise,
which agrees with the observed flavour pattern in the SM to good precision. 
The flavour symmetries therefore exclude the `would-be' anarchic flavour structure of the SM and impose selection rules consistent with the observed hierarchical patterns, which to some extent explains one of the conceptual puzzles on flavour. 
Furthermore, a number of recent studies found that the new physics models not far above TeV scale, motivated by the present anomalies introduced above, should contain certain flavour symmetries such as $U(2)^5$~\cite{Allwicher:2023shc, Allwicher:2025bub} or $SU(2)_q\times U(1)_X$~\cite{Greljo:2025mwj}.
Without the flavour symmetries, the $1-10$ TeV scale new physics is difficult to reconcile with the absence of deviations from other SM predictions, in particular those related to the stringently constrained flavour changing neutral currents (FCNCs).

Despite the above-mentioned appealing consequences, the origins of the flavour symmetries and the soft flavour symmetry breaking terms in the SM remain puzzling. 
A flavour symmetry can be promoted to a local symmetry so that it manifests as a fundamental principle of nature --- gauge invariance. 
In the presence of gauged flavour symmetries, anomaly cancellation, a condition necessary for ensuring consistency, is not always automatically satisfied and typically requires introducing mirror fermions. 
Since the gauged flavour symmetries must be exact at the Lagrangian level, they can only be spontaneously broken by the vacuum expectation values (VEVs) of the postulated Froggatt-Nielsen type~\cite{Froggatt:1978nt} scalar fields, commonly \typo{referred as}{referred to as} `flavons'~\cite{Dorsner:2002wi, Bazzocchi:2003vh, Antusch:2008gw, Tsumura:2009yf, Bauer:2016rxs}. 
The flavons interact with the SM fermions and Higgs bosons through the higher dimensional effective operators, which manifest as the soft flavour symmetry breaking terms after the flavons develop their VEVs, see Ref.~\cite{Linster:2018avp, Bonnefoy:2019lsn, Fedele:2020fvh, Cornella:2023zme, Calibbi:2025rxn} for recent representative examples. 
These effective operators can be further UV-completed by introducing additional heavy mediators such as the vector-like fermions. 
Comparing with directly gauging the flavour symmetries, an alternative approach called `flavour deconstruction' receives increasing focus over the past few years~\cite{Davighi:2023iks, FernandezNavarro:2023rhv, Davighi:2023evx, Barbieri:2023qpf, Davighi:2023xqn, Fuentes-Martin:2024fpx, Capdevila:2024gki, Davighi:2025cqx, Isidori:2025rci, Masiero:2026qon}. 
In this framework, a factor of the flavour universal gauge group of SM or its left-right symmetric extensions, represented by $G$, is extended into a flavour non-universal gauge group $G^{[1]}\times G^{[2]}\times G^{[3]}$. 
Here, $G^{[i]}$, which is automatically anomaly free, acts \textit{only} on the $i$th generation fermions in the same way as $G$ does, so that $G$ is the diagonal subgroup of the promoted non-universal gauge group. 
Different from the SM where $N_f=3$, the deconstructed theory contains no replication of fermions so that it leads to $N_f=1$. 
Only after the promoted non-universal gauge group breaks down, the flavour symmetry acting horizontally across the three or two generations can arise as a low energy remnant. 
In our view, flavour deconstruction is more economic than many conventional flavour theories. 
To date, concrete minimal models based on this idea have been constructed~\cite{Barbieri:2023qpf, Masiero:2026qon}.

Developments in research on the flavour symmetries encourage people to search for a `standard theory of flavour'. 
Ideally, the SM and new physics needed for addressing the present anomalies (if confirmed in future) should be embedded into a simple framework with very few model parameters. 
Within such a framework, the SM flavour parameters are no longer arbitrary but become predictable. 
Such a theory has been desired for a long time.
The question is: after so many years of efforts, how far we are from it?

Perhaps five or more years ago, the standard answer is that we are even not close to having a common paradigm. 
The popular flavour models often require postulating a large number of new particles and contain numerous new free parameters, so that they in principle \typo{can not}{cannot} be viewed as simpler than the SM.
Meanwhile, many of the flavour models involve high-scale new physics beyond the reach of near future experiments, which makes (fully) probing these models very difficult. 
However, progress over the past few years motives us to re-evaluate this answer. 
Many fairly simple models are constructed, see e.g. Ref~\cite{Fedele:2020fvh, Fuentes-Martin:2022xnb, Antusch:2023shi, Barbieri:2023qpf, Davighi:2025cqx, Arkani-Hamed:2026wwy}. Some of them feature not only explaining the origin of flavour symmetries, but also embedding the particles needed for delivering the symmetry breaking effects. 
On the other hand, recent studies on the effective field theories indicate $\mathcal{O}(1)$ TeV scale new physics with the third-generation specific flavour structures can stay fully compatible with the current experimental bounds~\cite{Allwicher:2023shc, Allwicher:2025bub}. 
In addition, over the past few years, many new flavour structures have been found consistent with new physics not far above TeV scale~\cite{Greljo:2022cah, Greljo:2025mwj}, implying that the flavour structures testable at the collider experiments can be highly diverse.
Interestingly, many of these structures align with the minimal flavour models, which suggests that the standard theory of flavour may lie no higher than a few TeV-scale and can be uncovered by near future experiments. 
Due to these reasons, an optimistic guess is that over the next decade, we will uncover a simple and realistic model at TeV scale, in which all the SM flavour parameters can be correctly predicted. 
Hopefully, in a few more years, certain sum-rules will be identified in one of the realistic flavour deconstruction models. 
The sum-rules can indicate more enhanced symmetries, such as a high-rank group unifying the deconstructed flavour non-universal gauge interactions, under which only very few model parameters are free. 

Nevertheless, it is also worth mentioning that the conservative answer would not be so cheerful. 
There is also a possibility that in the near future, no models constructed with very few UV parameters are found compatible with the known flavour patterns.
The expected sum-rules may not appear at all.  
In this scenario, we need to reflect over the existing paradigms of model building. 
Although the SM features certain approximate flavour symmetries like $U(2)^5$, these symmetries can be \textit{strongly} broken by new physics beyond the SM. 
As a result, it is also possible that no symmetries other than the SM gauge group and the \typo{Lorenz}{Lorentz} group \typo{exsit}{exist} in nature. 
Moreover, some flavour symmetries, despite being exact above certain \textit{ultra-high} energy scales, could be broken by the dynamical effects during the evolution towards infrared. 
At lower energies (for instance, the $1-10$ TeV scale, where a testable flavour theory typically lies), these flavour symmetries become implicit and physically \typo{can not}{cannot} be distinguished with the symmetries broken at all scales. 
Therefore, we argue that the flavour theories without explicit flavour symmetries should also be considered seriously. 
Although similar constructions are partially explored in many extra dimension models~\cite{Arkani-Hamed:1999ylh, Grossman:1999ra, Gherghetta:2000qt, Huber:2000ie, Agashe:2004cp, Blanke:2008zb, Csaki:2008zd, Buchmuller:2017vut, Fuentes-Martin:2022xnb}, to our best knowledge, they have never been investigated thorough from a broader perspective. 
In a recent work of us~\cite{Calibbi:2025ded}, we noticed that predictive models can emerge even if the underlying flavour symmetries remain implicit, and argued that the `swampland of flavour symmetries' could also be relevant to phenomenology. 
Despite this, we find the systematic studies on such a possibility remain overlooked in the present literatures.

In this thesis, we try to pave the way for exploring generic new physics models in which no flavour symmetries are explicit. 
To achieve this goal, we firstly need a concrete definition on the swampland of flavour symmetries --- the concept we informally proposed in Ref.~\cite{Calibbi:2025ded}.  
Identifying such a definition is important because the meaning of flavour is not universal but depends on the effective description to the theory. For example, $N_f$ can take integer values from two to six in QCD, depending on the energy scale. 
After clarifying the definition, we find that for theories in the swampland of flavour symmetries, old problems which would have been solved by the flavour symmetries return and become the major challenges. 
We believe the most important one is about how to suppress the flavour changing processes which are predicted to be rare in the SM and tightly constrained by experiments. 
If this problem \typo{can not}{cannot} be solved, models in the swampland of flavour symmetries must lie at very high scales and are not to be probed in a more generic category of experiments. 
Another challenge is how to promote a model without explicit flavour symmetries to a predictive theory of flavour, because the widely-known paradigm of restricting the model parameters by symmetries can no longer be adopted. 
To avoid this problem, the flavour parameters must appear as dynamical observables which can be directly calculated with certain known methods, or at least are ensured to be calculable. 
The dynamical nature of the flavour parameters implies their low-energy and UV values can significantly deviate from each other, so we refer them as the emerging flavour textures.

The thesis is structured as follows. 
In Chapter~\ref{chapterLandscape}, we will introduce the flavour puzzle, review certain representative symmetry-based solutions, and explain why we need to go beyond the conventional approaches. The concept of the swampland of flavour symmetries will be formally introduced at the end of this chapter. 
In Chapter~\ref{twozerochapter}, we will demonstrate the models without explicit underlying flavour symmetries can also stay predictive and relevant to phenomenology. 
Chapter~\ref{calculating} will show a perturbative model based on $SO(10)$ and a non-perturbative model for gravity. In both cases, some SM flavour parameters emerge dynamically and are calculable, while the restrictions from the underlying flavour symmetries are implicit. 
Some preliminary work generalizing the swampland of flavour symmetries to models containing the flavourful axions will be introduced in Chapter~\ref{flavouredaxions}. 
We will summarize our findings and give a brief outlook in Chapter~\ref{concluchapter}.

\chapter{The Landscape of Flavour Symmetries}
\label{chapterLandscape}

\section{The Flavour Puzzle}
\label{Fpuzzlesection}
To date, a number of fundamental particles with no known internal structures have been discovered. 
Their mass spectrum can be illustrated as:
\begin{itemize}
    \item \textbf{100 GeV scale}: top quark $t$, a Higgs-like boson $h$, weak gauge boson $Z, W$; 
    \item \textbf{GeV scale}: bottom quark $b$, charm quark $c$, tau lepton $\tau$; 
    \item \textbf{100 MeV scale}: strange quark $s$, muon $\mu$;
    \item \textbf{MeV scale}: down quark $d$, up quark $u$, electron $e$; 
    \item \textbf{Sub-eV scale}: three neutrinos $\nu_e, \nu_{\mu}, \mu_{\tau}$; 
    \item \textbf{Massless}: gluons $g$, photon $\gamma$. 
\end{itemize}
The discovered fundamental fermions include six quarks, each of which carries three color indices, and six leptons which are all color singlets. 
The light quarks $u, d, s, c, b$ and the gluons $g$ are confined into bound states and form the hadrons. 
The observation of neutrino oscillation~\cite{Super-Kamiokande:1998kpq, SNO:2001kpb, SNO:2002tuh} implies that at least two neutrinos are massive, while no present experimental evidence can reveal their Dirac or Majorana nature. 
Assuming all three neutrinos are massive Dirac particles, there are at least $6\times 3+6=24$ Dirac fermions discovered. 
Considering the irreducible representation of the Lorentz group, i.e. the Weyl or Majorana spinor, the known fermionic spectrum contains 48 `fundamental particles', corresponding to 96 complex (192 real) degrees of freedom.  
Such a rich fermionic sector is very different from the scalar one, which contains only one physical state $h$.

However, a merely large number of particle does not necessarily leads to any conceptual puzzles.
To explain this, we take the free field theory limit as an example. 
Assuming that all existing interactions are asymptotically free, the fundamental fermions introduced above can be effectively described with a free field theory above some high energy scales. 
The Lagrangian for the 48 non-interacting Weyl fermions read
\begin{equation}
\label{gravity}
\mathcal{L}_{\text{free}}~=~\overline{\psi_{\alpha}}\slashed{\partial}\psi_{\alpha}, \quad \alpha=1,2,...,48.
\end{equation}
Here, all known Weyl fermions are contained in $\psi$
\begin{equation}
\label{48fermions}
    \begin{matrix}
    \psi~=~\matlength &(u_{1L},\matlength & u_{1L},\matlength &u_{3L},\matlength &  d_{1L},\matlength & d_{1L},\matlength & d_{3L},\matlength &  \nu_{eL},\matlength &  e_L,\matlength & u_{1R}^c,\matlength &  u_{2R}^c,\matlength & u_{3R}^c,\matlength & d_{1R}^c,\matlength  & d_{2R}^c,\matlength & d_{3R}^c,\matlength  & \nu_{eR}^c,\matlength &  e_R^c, \\[0.3em]
    &~c_{1L},\matlength &  c_{2L},\matlength &  c_{3L},\matlength &   s_{1L},\matlength & s_{2L},\matlength & s_{3L},\matlength &  \nu_{\mu L},\matlength &  \mu_L,\matlength &  c_{1R}^c,\matlength & c_{2R}^c,\matlength & c_{3R}^c,\matlength &   s_{1R}^c,\matlength  & s_{2R}^c,\matlength  & s_{3R}^c,\matlength  & \nu_{\mu R}^c,\matlength &  \mu_R^c, \\[0.3 em]
     &~t_{1L},\matlength & t_{2L},\matlength &  t_{3L},\matlength &    b_{1L},\matlength & b_{2L},\matlength & b_{3L},\matlength &  \nu_{\tau L},\matlength &  \tau_L,\matlength &  t_{1R}^c,\matlength & t_{2R}^c,\matlength & t_{3R}^c,\matlength &  b_{1R}^c,\matlength & b_{2R}^c,\matlength & b_{3R}^c,\matlength &  \nu_{\tau R}^c,\matlength &  \tau_R^c).
    \end{matrix}
\end{equation}
The charge conjugation operator $^c$ ensures all components in $\psi$ have the same chirality. 
At the Lagrangian level, global $U(48)$ symmetry is enhanced in Eq.~(\ref{gravity}), which corresponds to the unitary transformation in the linear space spanned by $\psi_{\alpha}$
\begin{equation}
    \psi_{\alpha}~\rightarrow~\psi_{\alpha}~=~U_{\alpha\beta}\psi_{\beta}, 
\end{equation}
where $U$ is a unitary matrix and $\alpha, \beta = 1, 2, ..., 48.$
$U(48)$ eliminates all `would-be' mixings among the free fermions and leads to flavour universality. 
In addition, the $U(48)$ symmetry can remain unbroken even if certain interaction terms are included. 
For instance, the free field theory shown in Eq.~\eqref{gravity} can be extended by a scalar field $\phi_s$
\begin{equation}
\label{U48Yukawa}
\mathcal{L}_{\text{int}}~=~\overline{\psi}\slashed{\partial}\psi+\text{Tr}[\partial^{\mu}\phi_s \partial_{\mu} \phi_s^{\dagger}]-y \phi_{s} \overline{\psi}\psi^c.
\end{equation}
Here, the flavor indices are not shown explicitly. $\phi_s$ is a symmetric $48\times48$ complex matrix and transform under $U(48)$ as
\begin{equation}
    \phi_s~\rightarrow~\phi_s~=~U \phi_s U^T. 
\end{equation}
The $U(48)$ invariant Lagrangian Eq.~(\ref{U48Yukawa}) remains flavour universal and contains only one free parameter $y$.
In general, $y$ is a complex number while its phase is unphysical as it can be eliminated by redefining $\psi\to\psi e^{i\theta}$. 
Such a theory leads to no conceptual puzzles related to the large number of degrees of freedom contained in $\psi$.

It is worthy to note that $U(48)$ is only accidentally enhanced in the free field theory Eq.~\eqref{gravity} and could be explicitly broken in extended theories. 
Ideally, $U(48)$ or its $SU(48)$ subgroup~\cite{Pati:1980sb} should be promoted by a fundamental principle, such as a local symmetry. 
However, \textit{fully} gauging the $U(48)$ or $SU(48)$ group leads to the triangle anomalies~\cite{Adler:1969gk, Bell:1969ts}. 
Without extending the minimal theory, only anomaly-free subgroups of $U(48)$ can be gauged, see Ref.~\cite{Allanach:2021bfe} for a systematic study. 
We note a maximal subalgebra decomposition sequence
\begin{equation}
    U(48)~\supset~U(3)\times SU(16) ~ \supset ~ U(3)\times SO(10).
\end{equation}
Here, $U(3)\times SU(16)$ is a special subalgebra of $U(48)$, which \typo{can not}{cannot} be obtained by removing nodes from the Dynkin diagrams\footnote{Special embedding leads to non-trivial anomaly matching conditions, see Ref.~\cite{Hor:2025gxo, Hor:2026dlb} for implications on axion domain walls.}. 
Under the $U(3)\times SU(16)$ group, $\psi$ can be decomposed into three generations
\begin{equation}
\label{16fermions}
    \begin{matrix}
    \psi_{16}^1~=~\matlength &(u_{1L},\matlength & u_{1L},\matlength &u_{3L},\matlength &  d_{1L},\matlength & d_{1L},\matlength & d_{3L},\matlength &  \nu_{eL},\matlength &  e_L,\matlength & u_{1R}^c,\matlength &  u_{2R}^c,\matlength & u_{3R}^c,\matlength & d_{1R}^c,\matlength  & d_{2R}^c,\matlength & d_{3R}^c,\matlength  & \nu_{eR}^c,\matlength &  e_R^c), \\[0.3em]
    \psi_{16}^2~=~\matlength &(c_{1L},\matlength &  c_{2L},\matlength &  c_{3L},\matlength &   s_{1L},\matlength & s_{2L},\matlength & s_{3L},\matlength &  \nu_{\mu L},\matlength &  \mu_L,\matlength &  c_{1R}^c,\matlength & c_{2R}^c,\matlength & c_{3R}^c,\matlength &   s_{1R}^c,\matlength  & s_{2R}^c,\matlength  & s_{3R}^c,\matlength  & \nu_{\mu R}^c,\matlength &  \mu_R^c), \\[0.3 em]
    \psi_{16}^3~=~\matlength &(t_{1L},\matlength & t_{2L},\matlength &  t_{3L},\matlength &    b_{1L},\matlength & b_{2L},\matlength & b_{3L},\matlength &  \nu_{\tau L},\matlength &  \tau_L,\matlength &  t_{1R}^c,\matlength & t_{2R}^c,\matlength & t_{3R}^c,\matlength &  b_{1R}^c,\matlength & b_{2R}^c,\matlength & b_{3R}^c,\matlength &  \nu_{\tau R}^c,\matlength &  \tau_R^c).
    \end{matrix}
\end{equation}
Consequently, the $SU(16)$ group~\cite{Pati:1980sb, Mohapatra:1982yu, Mohapatra:1982aq} acts universally on the each generation of fermions, which corresponds to all possible unitary transformations except for an overall phase rotation
\begin{equation}
\label{U16trans}
    (\psi_{16}^i)_{\alpha'}~\rightarrow~
    (\psi_{16}^i)_{\alpha'}~=~U'_{\alpha'\beta'}(\psi_{16}^i)_{\beta'}.
\end{equation}
Here, $i=1,2,3$ represents the flavour indices and $\alpha',\beta'=1,2,..., 16$. $U'$ is a unitary matrix and $\det U'=1$. 
$SO(10)$ --- a maximal special subgroup of $SU(16)$~\cite{Yamatsu:2015npn, Yamatsu:2017mei} --- is anomaly-free due to its left–right symmetric nature and can be promoted to a gauge symmetry. 
In analogy to the Lorentz group $SO(3,1)$, the generators of $SO(10)$ can be constructed from the commutators of gamma matrices $\Gamma_{\mu}$ of the rank 10 Clifford algebra, about which we discuss more details in subsection~\ref{SO10subsection}, 
so the $SO(10)$ group equips spinorial structures~\cite{Mohapatra:1979nn}. 
The irreducible spinor of $SO(10)$ has dimension $2^{\frac{10}{2}-1}=16$ and can be embedded into the fundamental representation of $SU(16)$. 
As a result, $\psi_{16}^i$ can also stand for the three spinorial representations of $SO(10)$ group. 
Assuming only the $SO(10)$ subgroup of $U(48)$ is exact, due to the principle of gauge invariance, the minimal Yukawa sector of the theory is
\begin{equation}
\label{16Yukawa10}
    -\mathcal{L}_Y^{10}~=~Y_{10}^{ij}\phi^{\mu}_{10}\overline{\psi_{16}^i}\Gamma_{\mu}(\psi^{j}_{16})^c.
\end{equation}
Here, $\mu=1,2,...,10$, and $\phi^{\mu}_{10}$ is a vector representation of $SO(10)$. Without loosing generality, one can choose the basis under which $Y_{10}$ is diagonal and its $(1,1)$ entry is real, so that $Y_{10}$ contains 5 free physical parameters. 
Comparing with the minimal $U(48)$ invariant interacting theory shown in Eq.~(\ref{U48Yukawa}), the minimal $SO(10)$ invariant Yukawa sector contains more free parameters.

The discussions above illustrates how a conceptual puzzle arises with the top-down view. 
The known 48 fundamental fermions \typo{can not}{cannot} be embedded into one single representation of a local symmetry without leading to triangle anomalies, so $\psi$ \typo{can not}{cannot} be promoted as `one particle' by a fundamental principle. Then, even if the scalar sector contains only one representation which can couple to the components in $\psi$, the relevant Yukawa sector always involve more than one free parameters, whose patterns \typo{can not}{cannot} be further explained. 
Such a fact implies the known fermionic spectrum  is inconsistent with any simple ultimate theory of nature, inducing a puzzle which is commonly \typo{referred as}{referred to as} the `flavour puzzle'.
It is crucial to address here that the flavour puzzle is \textit{not} a problem which reflects inconsistencies between experimental observations and theoretical predictions. 
Instead, this is a  conceptual puzzle regarding interpreting the established frameworks.

\begin{table}[t!]
\centering
\renewcommand{\arraystretch}{2}
\begin{tabular}{lcccc}
  \toprule
   & $SU(3)_c$ & $SU(2)_L$ & $U(1)_Y$ & $U(1)_{B-L}$ \\
  \midrule
     $Q_L=\left(\renewcommand{\arraystretch}{1.0}\begin{matrix}u_L\\d_L \\\end{matrix}\right),~
    \left(\renewcommand{\arraystretch}{1.0}\begin{matrix}c_L \\s_L \\\end{matrix}\right),~
    \left(\renewcommand{\arraystretch}{1.0}\begin{matrix}t_L \\b_L \\\end{matrix}\right)$
   & \textbf{3} & \textbf{2} & $+1/6$  & $+1/3$\\
    ~$\ell_L=\left(\renewcommand{\arraystretch}{1.0}\begin{matrix}\nu_{eL} \\ e_L \\\end{matrix}\right),
    \left(\renewcommand{\arraystretch}{1.0}\begin{matrix}\nu_{\mu L} \\\mu_L \\\end{matrix}\right),
    \left(\renewcommand{\arraystretch}{1.0}\begin{matrix}\nu_{\tau L} \\\tau_L \\\end{matrix}\right)$
 & \textbf{1} & \textbf{2} & $-1/2$ & $-1$\\
  $U_R= u_R, c_R, t_R$ & \textbf{3} & \textbf{1} & $+2/3$ & $+1/3$ \\
  $D_R= d_R, s_R, b_R$ & \textbf{3} & \textbf{1} & $-1/3$ & $+1/3$ \\
  $E_R=e_R, \mu_R, \tau_R $ & \textbf{1} & \textbf{1} & $-1$ & $-1$\\
  $~\nu_R=\nu_{eR}, \nu_{\mu R}, \nu_{\tau R}$ & \textbf{1} & \textbf{1} & $0$ & $-1$\\
  \bottomrule
\end{tabular}
\caption[The $G_{\text{SM}}$ representations of the SM fermions.]{The $SU(3)_c, SU(2)_L$ representations and $U(1)_Y, U(1)_{B-L}$ charges of the SM
fermions, including three generations of $\nu_R$.}
\label{SMfermions}
\end{table}

We now start to introduce the flavour puzzle within the SM, while keeping the assumption that all three neutrinos are Dirac particles.
The gauge group of SM is 
\begin{equation}
    G_{\text{SM}}~=~SU(3)_c\times SU(2)_L\times U(1)_Y. 
\end{equation}
This local symmetry connects some of the components in $\psi$, although it does not neatly group them into three copies $\psi_{16}^i$ as what $SO(10)$ does. 
To be specific, the 36 quarks are grouped into 12 $SU(3)_c$ triplets, and six of the 12 triplets form three $SU(2)_L$ doublets, leading to a chiral spectrum
\begin{equation}
    Q_L~=~\left(\begin{matrix}u_L \\d_L \\\end{matrix}\right),
    \left(\begin{matrix}c_L \\s_L \\\end{matrix}\right),
    \left(\begin{matrix}t_L \\b_L \\\end{matrix}\right);\quad 
    U_R~=~ u_R, c_R, t_R; \quad 
    D_R~=~ d_R, s_R, b_R. 
\end{equation}
The same pattern applies for the remaining 12 leptons which are color singlets
\begin{equation}
    \ell_L~=~\left(\begin{matrix}\nu_{eL} \\ e_L \\\end{matrix}\right),
    \left(\begin{matrix}\nu_{\mu L} \\\mu_L \\\end{matrix}\right),
    \left(\begin{matrix}\nu_{\tau L} \\\tau_L \\\end{matrix}\right); \quad 
    E_R~=~ e_R, \mu_R, \tau_R; \quad 
    \nu_R~=~\nu_{eR}, \nu_{\mu R}, \nu_{\tau R}.    
\end{equation}
The 48 Weyl fermions constitute 18 irreducible $G_{\text{SM}}$ representations (18 particles). This number is much larger than those in the $U(48)$ and $SO(10)$ invariant theory, which are one and three, respectively. 
In \tref{SMfermions}, we summarize the $U(1)_Y$ charge of each particles, together with their $G_{\text{SM}}$ representations. 
The charge assignment for $U(1)_{B-L}$, which is also an known exact and anomaly-free in the SM~\cite{Foot:1992ui}, are also included in table \tref{SMfermions}, because the known $U(1)$ gauge symmetry of nature might not be purely $U(1)_Y$ but instead be a combination of $U(1)_Y$ and $U(1)_{B-L}$. This generalized charge set-up leads to de-quantized electric charges which can protect the Dirac nature of neutrinos. 
Moreover, we remark that although the SM admits maximal parity violation, the left-right symmetry when the fermionic spectrum can be restored if $U_R, D_R$ and $\nu_R, E_R$ are grouped into the $SU(2)_R$ pairs~\cite{Mohapatra:1974gc, Senjanovic:1975rk}
\begin{equation}
\label{LRsym}
    Q_L~\leftrightarrow~Q_R=\left(\begin{matrix}U_R \\D_R \\\end{matrix}\right), \quad
    \ell_L~\leftrightarrow~\ell_R=\left(\begin{matrix}\nu_R \\E_R \\\end{matrix}\right).
\end{equation}
To date, there are no experimental hints for the gauged $SU(2)_R$ interactions.

The scalar sector of SM contains one Higgs boson $H$ which is singlet under $SU(3)_c$, doublet under $SU(2)_L$, and carries hypercharge $1/2$. 
The SM Yukawa sector then reads
\begin{equation}
\label{SMYukawa}
-\mathcal{L}_Y~=~Y_u^{ij}\widetilde{H}\overline{Q_L^i}U_R^j+Y_d^{ij}H\overline{Q_L^i}D_R^j+Y_{\nu}^{ij}\widetilde{H}\overline{\ell_L^i}\nu_R^j+Y_e^{ij}H\overline{\ell_L^i}E_R^j+\text{h.c.}
\end{equation}
Here, $\widetilde{H}=i\sigma_2H^*$ and $i,j=1,2,3$ are flavour indices. 
Without loosing generality, one can choose the flavour basis under which $Y_u$ and $Y_{e}$ are diagonal and real. 
However, upon the flavour basis is fixed, there are no freedom redefining the basis of $Q_L^i$ and $\ell_L^i$ to diagonalize $Y_d$ and $Y_{\nu}$. After eliminating the unphysical parameters, the Yukawa coupling matrices in Eq.~(\ref{SMYukawa}) can be represented by
\begin{equation}
\label{diagY}
\begin{aligned}
    Y_u~&=~\left(\begin{matrix}
        y_u & 0 & 0\\
        0 & y_c & 0\\
        0 & 0 & y_t\\
    \end{matrix}\right), \qquad 
    Y_d~=~V_{\text{CKM}}.
        \left(\begin{matrix}
        y_d & 0 & 0\\
        0 & y_s & 0\\
        0 & 0 & y_b\\
    \end{matrix}\right),  \\   
    Y_e~&=~\left(\begin{matrix}
        y_e & 0 & 0\\
        0 & y_{\mu} & 0\\
        0 & 0 & y_{\tau}\\
    \end{matrix}\right), \qquad 
    Y_{\nu}~=~V_{\text{PMNS}}.\left(\begin{matrix}
        y_{\nu_1} & 0 & 0\\
        0 & y_{\nu_2} & 0\\
        0 & 0 & y_{\nu_3}\\
    \end{matrix}\right).
\end{aligned}
\end{equation}
Assuming Dirac neutrinos, $V_{\text{PMNS}}$ is the Pontecorvo-Maki-Nakagawa-Sakata (PMNS) matrix~\cite{Pontecorvo:1957cp, Maki:1962mu}, which can be parameterized by three mixing angles and one CP violating phase
\begin{equation}
\label{PMNSDirac}
    V_{\text{PMNS}} ~=~\left( \begin{matrix} c^{}_{13} c^{}_{12} & c^{}_{13} s^{}_{12} & s^{}_{13} e^{-{\rm i}\delta} \cr -s_{12}^{} c_{23}^{} - c_{12}^{} s_{13}^{} s_{23}^{} e^{{\rm i}\delta}_{} & + c_{12}^{} c_{23}^{} - s_{12}^{} s_{13}^{} s_{23}^{} e^{{\rm i}\delta}_{} & c_{13}^{} s_{23}^{} \cr + s_{12}^{} s_{23}^{} - c_{12}^{} s_{13}^{} c_{23}^{} e^{{\rm i}\delta}_{} & - c_{12}^{} s_{23}^{} - s_{12}^{} s_{13}^{} c_{23}^{} e^{{\rm i}\delta}_{} & c_{13}^{} c_{23}^{} \end{matrix} \right), 
\end{equation}
where $c_{ij}, s_{ij}, \delta$ represents $\cos\theta_{ij},\sin\theta_{ij},\delta$. 
$V_{\text{CKM}}$ is the Cabibbo-Kobayashi-Maskawa (CKM) matrix~\cite{Cabibbo:1963yz, Kobayashi:1973fv}, which can also be represented in a way similar to $V_{\text{PMNS}}$. 
In practice, $V_{\text{CKM}}$ is also commonly described by Wolfenstein parameterization~\cite{Wolfenstein:1983yz}
\begin{equation}
    V_{\text{CKM}}~=~
    \left( \begin{matrix}
    V_{ud}  & V_{us} & V_{ub}\\
    V_{cd} & V_{cs} & V_{cb}\\
    V_{td} & V_{ts} & V_{tb}\\
    \end{matrix}
    \right)~=~
    \left( \begin{matrix}
    1-\lambda^2/2  & \lambda & A\lambda^3(\rho-i\eta)\\
    -\lambda & 1-\lambda^2/2 & A\lambda^2\\
    A\lambda^3(1-\rho-i\eta) & -A\lambda^2 & 1\\
    \end{matrix}
    \right)
    +\mathcal{O}(\lambda^4). 
\end{equation}
The other 12 absolute coupling strengths in Eq.~(\ref{diagY}) is connected to the fermion masses by
\begin{equation}
    \begin{aligned}
        y_f~&=~\frac{\sqrt{2}}{v} m_f, \quad f=u,c,t,b,s,b,e,\mu,\tau, \nu_1, \nu_2,\nu_3.
    \end{aligned}
\end{equation}
Here, $v=246$ GeV is the vacuum expectation value of the Higgs field.
Altogether, the Yukawa sector SM with three Dirac neutrinos contains 20 real physical parameters --- commonly \typo{referred as}{referred to as} `flavour parameters'\footnote{Two Majorana CP violating phases must be included, if all three neutrinos are Majorana type. See Section~\ref{modelsection} for details.}. 
For comparison, the $U(48)$ invariant theory with one $\phi_s$ and the $SO(10)$ theory with only one $\phi_{10}$ contain only one and five flavour parameters, respectively. 
Due to the smaller symmetry group, the flavour puzzle is more serious in the SM.

Moreover, the measured values of these free parameters do not appear to be random and generic. Rather, they exhibit peculiar hierarchical patterns. 
At scale 1 TeV, $y_f$ for quarks are~\cite{Antusch:2025fpm}
\begin{equation}
\label{ynumbers}
\begin{aligned}
    y_u~&=~(6.15\pm 0.14)\times 10^{-6}, \quad y_c~=~(3.11\pm0.05)\times10^{-3}, \quad y_t~=~0.861\pm 0.004,\\
    y_d~&=~(1.35\pm 0.02)\times 10^{-5}, \quad y_s~=~(2.68\pm0.03)\times10^{-4}, \quad y_b~=~0.0140\pm 0.0001.\\
\end{aligned}
\end{equation}
These values vary at different scales, but not by orders of magnitudes. 
The other flavour parameters, encoded in the CKM matrix, are nearly scale invariant in the SM. At 1 TeV scale, $y_f$ for charged leptons  are~\cite{Antusch:2025fpm}
\begin{equation}
\label{ynumbers2}
\begin{aligned}
    y_e~&=~(2.868\pm 0.004)\times 10^{-6}, \quad y_{\mu}~=~(6.042\pm 0.008)\times10^{-3},\\
    y_{\tau}~&=~(1.026\pm 0.001)\times 10^{-2}. 
\end{aligned}
\end{equation}
Three similar hierarchical structures then emerge separately in the sector of up-type quarks, down-type quarks, and charged leptons
\begin{equation}
    y_u~\ll ~y_c~\ll~y_t, \quad y_d~\ll~y_s~\ll~y_b, \quad y_e~\ll~y_{\mu}~\ll~y_{\tau}. 
\end{equation}
These repeating hierarchies constitute another important component of the SM flavour puzzle.
Furthermore, the Wolfenstein parameters for $V_{\text{CKM}}$ are measured to be~\cite{ParticleDataGroup:2024cfk}
\begin{equation}
\begin{aligned}
    \lambda~&=~0.22501\pm 0.00068, \quad &&A~=~0.826_{-0.015}^{+0.016}, \\
    \overline{\rho}~&=~0.1591\pm 0.0094, && \overline{\eta}~=~0.3523_{-0.0071}^{+0.0073},\\
\end{aligned}
\end{equation}
where $\overline{\rho}=\rho(1-\lambda^2/2+\mathcal{O}(\lambda^4))$ and $\overline{\eta}=\eta(1-\lambda^2/2+\mathcal{O}(\lambda^4))$. 
The CKM matrix features small $2-3$, $1-3$ mixing angles, which are of order $\lambda^2\sim 0.04$ and $\lambda^3\sim 0.008$, respectively. 
The $1-2$ mixing angle is sizable and approximately 
\begin{equation}
    \lambda~=~\sqrt{y_d/y_s},
\end{equation}
at all scales.
This relationship was firstly noted in 1977~\cite{Weinberg:1977hb, DeRujula:1977dmn, Wilczek:1977uh, Fritzsch:1977za}. Such a coincidence \typo{can not}{cannot} be dynamically explained within the SM, in which $y_d, y_s,$ and $\lambda$ are all free parameters.

The flavour parameters contained in $Y_{\nu}$ are only partly known. 
The oscillation experiments have measured the two neutrino mass-squared differences~\cite{Esteban:2024eli}
\begin{equation}
\label{Nudiff}
    m_{\nu_2}^2-m_{\nu_1}^2~=~7.537_{-0.10}^{+0.094}\times10^{-5}~\text{eV}^2, \quad |m_{\nu_3}^2-m_{\nu_2}^2|~=~2.521_{-0.018}^{+0.026}\times 10^{-3}~\text{eV}^2.
\end{equation}
On the other hand, the absolute values of the neutrino masses are not measured yet, but upper limits are available:
\begin{equation}
    \label{neutrinobound}
    \begin{aligned}
        &\sum m_{\nu_i}~<~ 0.064~\text{eV}~\text{for~DESI}+\textit{Planck}~\text{\cite{DESI:2025zgx}} \\
        &\sum m_{\nu_i}~<~ 0.061~\text{eV}~\text{for~DESI}+\textit{Planck}+\text{ACT}~\text{\cite{DESI:2025gwf}} \\
        &\sum m_{\nu_i}~<~ 0.055~\text{eV}~\text{for~DESI}+\textit{Planck}+\text{ACT}+\text{SPT}~\text{\cite{SPT-3G:2025bzu}} \\
    \end{aligned}
\end{equation}
at 95\% confidence level. The last bound is incompatible with the inverted neutrino mass ordering (IO, $m_{\nu_3}<m_{\nu_1}<m_{\nu_2}$)
and starts to show mild tension with the normal mass ordering (NO, $m_{\nu_1}<m_{\nu_2}<m_{\nu_3}$), which requires $\sum m_{\nu_i}> 0.060~\text{eV}$.
However, these constraints all come from the cosmological observations, which are indirect and subject to corrections from the non-standard cosmological models and/or new physics feebly interacting with the SM\footnote{The kinematics of beta decay gives a direct and model-independent constrain on the absolute neutrino mass~\cite{KATRIN:2024cdt}
\begin{equation}
    \sqrt{\sum_i m_{\nu_i}^2 |V_{\text{PMNS}}^{1i}|^2}~<~0.45~\text{eV}. 
\end{equation}
However, this bound is looser than the ones from cosmology by about one order of magnitude.}. For instance, light dark sector particles can suppress the neutrino abundance and thereby relax the cosmological constraints, see~\cite{Benso:2024qrg, Das:2025asx} and references therein for further examples.
Nevertheless, if the bound $\sum m_{\nu_i}< 0.064~\text{eV}$ is valid,  the three absolute neutrino masses are must satisfy
\begin{equation}
\label{absNubound}
    0~\leq ~m_{\nu_1}~<~0.004~\text{eV},\quad m_{\nu_2}~\approx~ 0.01~\text{eV}, \quad m_{\nu_3}~\approx~ 0.05~\text{eV}. 
\end{equation}
Here, $m_{\nu_2}$ and $m_{\nu_3}$ are bounded from below due to Eq.~\eqref{Nudiff}. 
After translating Eq.~\eqref{absNubound} into the values for $y_{\nu_i}$, we have
\begin{equation}
     0~\leq~ y_{\nu_1}~<~2\times10^{-14}, \quad y_{\nu_2}~\approx~5\times 10^{-14},\quad y_{\nu_3}~\approx~3\times 10^{-13}. 
\end{equation}
A mild hierarchy emerges among $y_{\nu_i}$. 
Moreover, $y_{\nu_i}$ are much smaller than $y_f$ for quarks and charged leptons by at least $6-13$ orders of magnitude. 
The three neutrino mixing angles in $V_{\text{PMNS}}$ are all well-measured.
In case of NO, the best-fit values and $1 \sigma$ uncertainties for the three neutrino mixing angles in $V_{\text{PMNS}}$ are~\cite{Esteban:2024eli}
\begin{equation}
    \theta_{12}/^{\circ}~=~33.76^{+0.42}_{-0.41}, \quad 
    \theta_{23}/^{\circ}~=~43.27^{+1.0}_{-0.82}, \quad 
    \theta_{13}/^{\circ}~=~8.62^{+0.11}_{-0.11}.
\end{equation}
The values for $\theta_{12}$ and $\theta_{13}$ are almost unchanged under IO. The best-fit IO value for $\theta_{23}$ lies in the upper octet $\theta_{23}/^{\circ}=48.15^{+0.75}_{-0.92}$, while the $3\sigma$ range for $\theta_{23}$ remains nearly identical for both mass orders. 
Similar to the quark mixings in $V_{\text{CKM}}$, there are also approximate relationships:
\begin{equation}
    \sin{\theta_{12}}~=~\frac{1}{\sqrt{3}}, \quad
    \sin{\theta_{23}}~=~\frac{1}{\sqrt{2}}, \quad 
    \sin{\theta_{13}}~=~\frac{\lambda}{\sqrt{2}}.
\end{equation}
This coincidental pattern is called Tri-bimaximal-Cabibbo mixing~\cite{Fritzsch:1995dj, Harrison:2002er, Xing:2002sw, Harrison:2002kp, He:2003rm, Altarelli:2005yx, King:2012vj}.
On the other hand, the CP violating phase $\delta$ is poorly constrained at present. Either CP conservation ($\delta=0,~\pi$) or maximal CP violation ($\delta=\pm \pi/2$) are allowed.

Beyond SM theories can involve many more flavour parameters than the SM. 
For instance, if all types of scalar multiplets which can couple to the SM fermions are included, the extended Yukawa sector can contain a large number of free parameters. 
In total, 15 new Lorentz invariant bi-linear structures containing the different-type Weyl fermions can be constructed:
\begin{equation}
    \begin{aligned}
        &\overline{Q_L}U_R,~\overline{Q_L}D_R, \overline{Q_L}\ell_L^c,~\overline{Q_L}E_R,~\overline{Q_L}\nu_R, \\
        &\overline{U_R^c}D_R,~\overline{U_R^c}\ell_L^c,~\overline{U_R^c}E_R,~\overline{U_R^c}\nu_R,\\
        &\overline{D_R^c}\ell_L^c,~\overline{D_R^c}E_R,~\overline{D_R^c}\nu_R,\\
        & \overline{\ell_L} E_R,~\overline{\ell_L} \nu_R, \\
        &\overline{E_R^c}\nu_R, 
    \end{aligned}
\end{equation}
among which four bi-linear structures are already contained in the SM, as shown in Eq.~(\ref{SMYukawa}).
Besides, six bi-linear spinor structures containing different-type Weyl fermions are also Lorentz invariant:
\begin{equation}
    \overline{Q_L}Q_L^c,~\overline{U_R^c}U_R,~\overline{D_R^c}D_R,~\overline{\ell_L}\ell_L^c,~\overline{E_R^c}E_R,~\overline{\nu_R^c}\nu_R. 
\end{equation}
Each different-type (same-type) bi-linear structures can be combined with a beyond SM scalar particle, whose Yukawa couplings involve generic (symmetric) complex $3\times3$ matrices, see Ref.~\cite{deBlas:2017xtg, Greljo:2023adz} for the explicit expressions. 
As a consequence, if each type of additional scalars only appear once, the extended Yukawa sector can contain at least $6\times 6+11\times 9=135$ independent complex parameters or 270 real flavour parameters which \typo{can not}{cannot} be further eliminated by redefining the fermion basis, in addition to the 20 real SM flavour parameters. 
So, the more general $G_{\text{SM}}$ invariant Yukawa sector contains way more free parameters than that of the SM. 
Yet, if one gives up the requirement of renormalizability, the SM can be systematically extended by the high-dimensional operators, which is commonly \typo{referred as}{referred to as} the Standard Model effective field theory (SMEFT)~\cite{Buchmuller:1985jz, Giudice:2007fh, Grzadkowski:2010es, Henning:2014wua, Brivio:2017vri, Isidori:2023pyp}. 
\tref{EFTnumbers} summarizes the numbers of non-redundant operators up to dimension 10, constructed by one or three generations of SM fermions with $\nu_R$ ($\nu$SMEFT)~\cite{delAguila:2008ir, Aparici:2009fh, Bhattacharya:2015vja, Liao:2016qyd, Li:2021tsq, Talbert:2022unj} and those without $\nu_R$~\cite{Marinissen:2020jmb, Fonseca:2019yya}, as it is commonly assumed $\nu_R$ decouples from the light spectrum.
Every effective operator corresponds to a free parameter.
Therefore, the SMEFT or $\nu$SMEFT contains numerous free parameters, and almost all these parameters originate from the fact that SM contains three generation of same-type Weyl fermions instead of only one. 
The predictivity of beyond SM model building is very limited for flavour.

\begin{table}[t!]
    \centering
    \renewcommand{\arraystretch}{1.5}
    \begin{tabular}{l c c c c c c }
    \toprule
      Dimension   & 5 & 6 & 7 & 8 & 9 & 10\\
      \midrule
      One generation (without $\nu_R$)  & 2 & 84 & 30 & 993 & 560 & 15456\\
      One generation (with one $\nu_R)$ & 4  & 113 & 110 & 1316 & 1918 & 21540\\
      Three generations (without $\nu_R$) &  12 &  3045 & 1542 & 44807 & 90456 & 2092441\\
      Three generations (with three $\nu_R)$ &  30 &  4659 & 5748 & 65207 & 334400 & 3513704\\
    \bottomrule
    \end{tabular}
    \caption{Number of SMEFT and $\nu$SMEFT operators under a given dimension.}
    \label{EFTnumbers}
\end{table}

In summary, the flavour puzzle for SM and beyond include:
\begin{enumerate}[label={(\roman*)}]
    \item The SM contains too many free parameters in the Yukawa sector so that it cannot be viewed as a simple ultimate theory for everything. 
    \item The SM flavour parameters are not merely random numbers, but manifest hierarchical patterns and coincidental relationships which \typo{can not}{cannot} be explained. 
    \item Extending the SM in general allows way more additional couplings, which reduces the predictivity of many beyond SM theories. 
\end{enumerate}
The flavour puzzle leads to two conflicting consequences. On one hand, constructing a theory for everything, even if excluding gravity, requires extending the SM. 
On the other hand, it is difficult extended the SM with simple and predictive models. 

\section{From MFV to \texorpdfstring{$U(2)$}{U(2)}}
\label{EFTsection}

\begin{figure}[t!]
    \centering
    \includegraphics[width=1.0\linewidth]{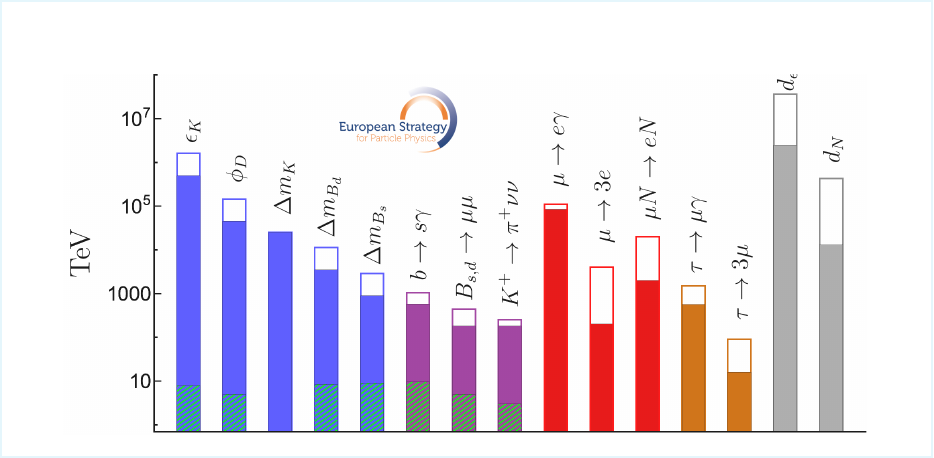}
    \caption[Bounds on the cut-off scales of certain dimension-six operators.]{Bounds on the cut-off scales of the dimension-six operators inducing certain flavour and CP violating observables. The current limits and expected furture improvements are denoted with the colored and empty bars, respectively. The hatched bars indicate current bounds within the MFV framework. This figure is extracted from the input for the 2026 update of the European Strategy for Particle Physics~\cite{deBlas:2025gyz}.}
    \label{EuroStrategy}
\end{figure}

In this section, we review some symmetry based paradigms which reduce the number of `would-be' free flavour parameters in theories beyond the SM. 
These set-ups are related to an important phenomenology-driven motivation, that the flavour structures which can be explored by near-future experiments are very limited. 
The flavour-violating transitions between the light fermions, such as the $\mu\to e\gamma$ decay~\cite{MEGII:2025gzr}, are tightly constrained by experiments. 
As show in Figure~\ref{EuroStrategy}, the cut-off scales of the effective operators inducing certain flavour and CP violating processes must lie at very high scales, typically above $10^2-10^3$ TeV or even higher~\cite{deBlas:2025gyz}. 
By contrast, the flavour transitions involving the third generation fermions, such as $B\to \mu^{\pm}\tau^{\mp}$~\cite{LHCb:2019ujz, Belle:2021rod}, are less constrained by experiments and may only be suppressed by a much lower cut-off scale. 
Consequently, it is very difficult to test the anarchic flavour structures for the beyond SM theories in the near future. 
The TeV-scale new physics must be accompanied with sufficient suppression mechanisms for the $2-1$ flavour changing processes, for instance, through some flavour symmetries.
Ideally, such symmetries should be aligned with the existing ones in the SM. 
In this subsection, we will firstly revisit two extensively studied frameworks based on the $U(3)^5$ and $U(2)^5$ accidental symmetries of the SM gauge sector respectively. 
One is minimal flavour violation (MFV)~\cite{Chivukula:1987py, DAmbrosio:2002vsn, Cirigliano:2005ck, Davidson:2006bd,Isidori:2010kg, Isidori:2012ts}, which assumes that, also in the sector of new physics, the $U(3)^5$ flavour symmetry is only broken by the SM Yukawas coupling matrices, acting as `spurions'. The other assumes only a less restrictive flavour symmetry $U(2)^5$~\cite{Barbieri:2011ci, Isidori:2012ts, Barbieri:2012uh, Blankenburg:2012nx, Faroughy:2020ina, Allwicher:2023shc, Allwicher:2025bub}, which only acts on the two lightest fermion families and is almost exact in all sectors of the SM. 
Then, we will introduce some extended frameworks beyond MFV and $U(2)$. 

\subsection{Minimal Flavour Violation}
In SMEFT, the neutrinos can get the Majorana mass terms via the dimension-5 Weinberg operators~\cite{Weinberg:1979sa}
\begin{equation}
    \label{weinbergoperator}
    \mathcal{L}_{d=5}~=~\frac{\mathcal{C}_{ij}}{\Lambda}\left(\overline{\ell_L^{ic}}\widetilde{H}^*\right)\left(\widetilde{H}^{\dagger}\ell_L^j\right)+\text{h.c.},
\end{equation}
so that the renormalizable Dirac mass terms involving $\nu_R$ are not needed. 
Therefore, we assume that the $\nu_R^j$ (if existing) decouple and consider only the three generations of $Q_L, \ell_L$ and $U_R, D_R, E_R$ as the matter fields of the SM. 
Here, we review the construction of the MFV operators based on Ref.~\cite{DAmbrosio:2002vsn}.

The largest flavour symmetry group that commutes with $G_{\text{SM}}$ is $U(3)^5$, which corresponds to the unitary transformations among three generations of each type of fermions. 
$U(3)^5$ can be decomposed as
\begin{equation}
\begin{aligned}
     U(3)^5~=&~[SU(3)\times U(1)]^5\\
     =&~ SU(3)_{Q_L}\times SU(3)_{U_R}\times SU(3)_{D_R}\times SU(3)_{\ell_L}\times SU(3)_{E_R}\\
     & \times U(1)_{B}\times U(1)_{L}\times U(1)_Y\times U(1)_{\text{PQ}}\times U(1)_{E_R}. 
\end{aligned}
\end{equation}
Under this decomposition, the three generations of each type of fermions transform as $SU(3)$ triplets. 
The baryon number $B$, lepton number $L$, and hyper-charge $Y$ are conserved charges of SM (at the Lagrangian level), among which $Y$ is known gauged. 
$U(1)_{\text{PQ}}$ is the so-called The Peccei-Quinn symmetry~\cite{PQ1, PQ2}, under which the left-handed and right-handed fields are charged oppositely. $U(1)_{E_R}$ is a symmetry under which only $E_R$ rotates. 
The $U(3)^5$ symmetry is respected in the SM gauge sector. In principle, it can be extended to all non-renormalizable operators. 
In such a case, all dimension-5 Weinberg operators must vanish because they violate the lepton number. 
At dimension six, 15 bosonic operators and 32 flavour universal fermionic operators are $U(3)^5$ invariant~\cite{Faroughy:2020ina}, in the so-called Warsaw basis~\cite{Grzadkowski:2010es}.
Although this set-up significantly reduces the number of independent effective operators, it is over-simplified because the flavour violating processes \typo{can not}{cannot} be described.
Besides, the $U(3)^5$ breaking effects of the SM can propagate to the non-renormalizable sector via the loop corrections. The $U(3)^5$ invariant SMEFT requires extensions.

The flavour non-universal effective operators can be constructed via the SM Yukawa couplings matrices. 
The five $SU(3)$ symmetries, the PQ symmetry, and the $U(1)_{E_R}$ symmetry, are all explicitly broken in the SM Yukawa sector. 
Formally, one can recover the broken $SU(3)^5\times U(1)_{\text{PQ}}\times U(1)_{E_R}$ symmetry by promoting the Yukawa couplings $Y_u, Y_d, Y_e$ as (non-dynamical) auxiliary fields, which carry the $U(1)_{\text{PQ}}\times U(1)_{E_R}$ charges and transform non-trivially under $SU(3)^5$ as
\begin{equation}
\label{spurions}
    Y_u~\sim (\textbf{3}, \overline{\textbf{3}}, \textbf{1},\textbf{1},\textbf{1}), \quad Y_d~\sim (\textbf{3},\textbf{1},\overline{\textbf{3}},\textbf{1},\textbf{1}), \quad Y_e\sim (\textbf{1},\textbf{1},\textbf{1},\textbf{3},\overline{\textbf{3}}). 
\end{equation}
These fields are commonly called as `spurions'. 
The criterion of MFV states that a beyond SM theory must be formally invariant under $U(3)^5$, considering the spurions shown in Eq.~(\ref{spurions}). 
In other words, it assumes the new physics contribution breaking the $U(3)^5$ flavour symmetry can only be constructed with $Y_u, Y_d,$ and $Y_e$, which encode the SM fermion mass hierarchies and quark flavour mixings.
In this framework, the flavour symmetry is violated, but only in a `minimal' way.
We note here that MFV does not directly originate from a fundamental principle. Rather, it is an assumption to simplify the renormalizable models and effective field theories beyond the SM.

As we will see, such a setup aligns with the demands of rich phenomenology for experiments. 
Given that the only $\mathcal{O}(1)$ absolute Yukawa coupling in SM is $y_t$, 
the leading order spurion for the FCNC involving the down-type quarks can be constructed by $\epsilon^{ij}=(Y_uY_u^{\dagger})^{ij}$, which transforms as $(\textbf{8}, \textbf{1}, \textbf{1}, \textbf{1}, \textbf{1})$ under $SU(3)^5$. 
If neglecting the small SM Yukawa coupling strengths $y_u, y_d, y_c, y_s$, and take the basis in which $Y_d$ is diagonal, we have
\begin{equation}
\label{DownDiagonal}
        Y_u~=~V_{\text{CKM}}^\dagger.\left(\begin{matrix}
        0 & 0 & 0\\
        0 & 0 & 0\\
        0 & 0 & y_t\\
    \end{matrix}\right), \qquad 
    Y_d~=~ \left(\begin{matrix}
        0 & 0 & 0\\
        0 & 0 & 0\\
        0 & 0 & y_b\\
    \end{matrix}\right),  \\ 
\end{equation}
and the non-diagonal structure for $Y_uY_u^{\dagger}$ is simplified to
\begin{equation}
\label{MFVoctet}
    (Y_uY_u^{\dagger})^{ij}~=~y_t^2 (V_{\text{CKM}}^{ti})^*V_{\text{CKM}}^{tj}, \qquad i\neq j. 
\end{equation}
Using the Wolfenstein parameter, Eq.~(\ref{MFVoctet}) gives
\begin{equation}
\begin{aligned}
(Y_uY_u^{\dagger})^{ij}~\simeq~\left(\begin{matrix}
    -& \lambda^5 & \lambda^3\\
    \lambda^5 & - & \lambda^2\\
    \lambda^3 & \lambda^2 & -\\
\end{matrix}\right)
~\simeq~\left(\begin{matrix}
    -& 3\times 10^{-4} & 0.008\\
    3\times 10^{-4} & - & 0.04\\
    0.008 & 0.04 & -\\
\end{matrix}\right).
\end{aligned}
\end{equation}
As a consequence, the $2\to 1$ FCNC amplitudes are suppressed by a factor of $10^{-4}$ while the $3\to2, 3\to1$ ones only involve $10^{-2}$. 
This is exactly what we needed for observing the third-generation related FCNC processes, with respect to the tight bounds for the $2\to1$ ones.  
On the other hand, FCNC involving up-type quarks are suppressed by small $y_b^2$, which becomes negligible within MFV. 
MFV predicts that the charged lepton flavour violating (CLFV) processes always vanish, because $Y_e$ is diagonal in the basis of charged lepton mass basis.

We now systematically construct the dimension 6 SMEFT operators in the MFV framework.  
At leading order, 25 $SU(3)^5$ invariant fermionic operators can be constructed without spurions. They carry at least two same-type spinors which can contract into flavour singlets, and preserve flavour universality. 
In the limit $y_b=0$ and $y_{\tau}=0$, the $SU(3)^5$ spurions can only be constructed by $Y_u$:
\begin{equation}
    (Y_uY_u^{\dagger})^n, \quad (Y_uY_u^{\dagger})^nY_u 
\end{equation}
Considering $y_t\gg y_c, y_u$, we have $(Y_uY^{\dagger}_u)^n=y_t^{2n-2} (Y_uY_u^{\dagger})$, so the $n\geq2$ spurions do not induce new flavour structures. 
The relevant bilinear structures only include
\begin{equation}
\label{MFVFCNCstructures}
    \overline{Q_L}Y_uY_u^{\dagger}Q_L, \quad  \overline{Q_L}Y_uU_R
\end{equation}
They can be further contracted with the scalars, vector, or tensor currents which are singlets or octets (triplets) under $SU(3)$ ($SU(2)$). 
These currents in general can involve new physics, while only those constructed by the fields within SM generate the SMEFT operators. 
At $\mathcal{O}(Y_u)$, one can construct the following operators
\begin{equation}
\label{mfvnonH}
\begin{aligned}
    &(\overline{Q_L}Y_u\sigma^{\mu\nu}T^AU_R)\widetilde{H}G_{\mu\nu}^A, \quad 
    (\overline{Q_L}Y_u\sigma^{\mu\nu}\tau^IU_R)\widetilde{H}W_{\mu\nu}^I,  \quad 
    (\overline{Q_L}Y_u\sigma^{\mu\nu}U_R)\widetilde{H}B_{\mu\nu}, \\
    &(H^{\dagger}H)\overline{Q_L}Y_u U_R \widetilde{H}.
\end{aligned}
\end{equation}
Here, $T^A=T^1, T^2, ..., T^8$ and $\tau^{I}=\tau^1,\tau^2,\tau^3$ are the $SU(3)$ and $SU(2)$ generators, respectively. $G_{\mu\nu}^A, W_{\mu\nu}^I,$ and $B_{\mu\nu}$ are the field strength tensors of the three gauge groups of $G_{\text{SM}}$. 
These operators do not induce FCNC but are flavour non-universal and can lead to the flavour changing charged currents (FCCC) which are aligned to the CKM matrix in the SM. 
At $\mathcal{O}(Y_u^2)$, there are three operators involving the Higgs fields
\begin{equation}
\begin{aligned}
    &(H^{\dagger}i\overleftrightarrow{D}_{\mu}H)(\overline{Q_L}Y_uY_u^{\dagger}\gamma^{\mu}Q_L), \quad 
    (H^{\dagger}i\overleftrightarrow{D}_{\mu}^IH)(\overline{Q_L}Y_uY_u^{\dagger}\tau^I\gamma^{\mu}Q_L), \\
    &(H^{\dagger}i\overleftrightarrow{D}_{\mu}H)(\overline{U_R}Y_u^{\dagger}Y_u\gamma^{\mu}U_R).
\end{aligned}
\end{equation}
The first two of these operators induce $\Delta F=1$ FCNC for down-type quarks. 
The other $\mathcal{O}(Y_u^2)$ operators are all four-fermion operators, among which 16 contain only quarks, and the other five are mixed quark-lepton operators.
\begin{equation}
    \begin{aligned}
        (LL)(LL):~&~(\overline{Q_L}Y_uY_u^{\dagger}\gamma_{\mu}Q_L)(\overline{Q_L}\gamma^{\mu} Q_L), 
         && (\overline{Q_L}Y_uY_u^{\dagger}\gamma_{\mu}\tau^IQ_L)(\overline{Q_L}\gamma^{\mu}\tau^I Q_L),\\
        ~&~(\overline{Q_L}\gamma_{\mu}Q_L)Y_u^{*}Y_u^T(\overline{Q_L}\gamma^{\mu} Q_L), 
         && (\overline{Q_L}Y_uY_u^{\dagger}\gamma_{\mu}\tau^IQ_L)Y_u^{*}Y_u^T(\overline{Q_L}\gamma^{\mu}\tau^I Q_L),\\
        ~&~(\overline{Q_L}Y_uY_u^{\dagger}\gamma_{\mu}Q_L)(\overline{\ell_L}\gamma^{\mu} \ell_L), 
         && (\overline{Q_L}Y_uY_u^{\dagger}\gamma_{\mu}\tau^IQ_L)(\overline{\ell_L}\gamma^{\mu}\tau^I \ell_L), \\
         (RR)(RR):~&~(\overline{U_R}Y_u^{\dagger}Y_u\gamma_{\mu}U_R)(\overline{U_R}\gamma^{\mu} U_R), 
         && (\overline{U_R}\gamma_{\mu}U_R)Y_u^TY_u^{*}(\overline{U_R}\gamma^{\mu} U_R),\\
        ~&~(\overline{U_R}Y_u^{\dagger}Y_u\gamma_{\mu}U_R)(\overline{D_R}\gamma^{\mu} D_R), 
         && (\overline{U_R}Y_u^{\dagger}Y_u\gamma_{\mu}T^AU_R)(\overline{D_R}\gamma^{\mu}T^A D_R),\\
        ~&~(\overline{U_R}Y_uY_u^{\dagger}\gamma_{\mu}U_R)(\overline{E_R}\gamma^{\mu} E_R),\\
        (LL)(RR):~&~(\overline{Q_L}Y_uY_u^{\dagger}\gamma_{\mu}Q_L)(\overline{U_R}\gamma^{\mu} U_R), 
         && (\overline{Q_L}Y_uY_u^{\dagger}\gamma_{\mu}T^A\tau^IQ_L)(\overline{U_R}\gamma^{\mu}\tau^I T^AU_R),\\
        ~&~(\overline{Q_L}\gamma_{\mu}Q_L)(\overline{U_R}Y_u^{\dagger}Y_u\gamma^{\mu} U_R), 
         && (\overline{Q_L}\gamma_{\mu}T^A\tau^IQ_L)(\overline{U_R}Y_u^{\dagger}Y_u\gamma^{\mu} T^AU_R),\\
        ~&~\text{Tr}[(\overline{Q_L}\gamma_{\mu}Q_L)Y_u^*(\overline{U_R}\gamma^{\mu} U_R)Y_u^T], 
         && \text{Tr}[(\overline{Q_L}\gamma_{\mu}T^A\tau^IQ_L)Y_u^*(\overline{U_R}Y_u^{\dagger}Y_u\gamma^{\mu} T^AU_R)Y_u^T],\\
        ~&~(\overline{Q_L}Y_uY_u^{\dagger}\gamma_{\mu}Q_L)(\overline{D_R}\gamma^{\mu} D_R), 
         && (\overline{Q_L}Y_uY_u^{\dagger}\gamma_{\mu}T^AQ_L)(\overline{D_R}\gamma^{\mu}T^AD_R),\\
        ~&~ (\overline{\ell_L}\gamma_{\mu}\ell_L)(\overline{U_R}Y_u^{\dagger}Y_u\gamma^{\mu} U_R),  &&
        (\overline{Q_L}Y_u^{\dagger}Y_u\gamma_{\mu}Q_L)(\overline{E_R}\gamma^{\mu} E_R).
    \end{aligned}
\end{equation}
The trace sums over the flavour indices, so that $Y_u$ can contract with the spinors in different brackets for operators containing only $Q_L$ and $U_R$.
The $\mathcal{O}(Y_u^2)$ MFV operators predict an approximately third-generation specific flavour pattern, and many of them generate $\Delta F=1$ FCNC.
The $\Delta F=2$ operators arise only at $\mathcal{O}(Y_u^4)$, for instance:
\begin{equation}
    (\overline{Q_L}Y_uY_u^{\dagger}\gamma_{\mu}Q_L)(\overline{Q_L}Y_uY_u^{\dagger}\gamma^{\mu} Q_L),
\end{equation}
which is related to neutral B meson mixing. 
The other SMEFT operators cannot be made $SU(3)^5$ invariant without the suppressed spurion $Y_d$ and/or $Y_e$. 
For example, the operator $(\overline{\ell_L}E_R)(\overline{D_R}Q_L)$ contains charged leptons with opposite chiralities and needs spurion $Y_e$ to flip the lepton chirality.

The discussions above demonstrate the predictive power of the MFV framework, especially on the effective theories. 
There are 3045 dimension-6 SMEFT operators in total.
Within the MFV framework, there are only 45 pure quark operators, 7 pure lepton operators, and 12 mixed quark-lepton operators, up to $\mathcal{O}(Y_uY_u^{\dagger})$~\cite{Greljo:2022cah}. 
The other operators, such as the ones inducing top-quark FCNC, are all suppressed by higher scales, which reflects the model-independent consequence among a wide class of new physics. 
In practice, MFV provides a useful tool analyzing the experimental data. Compared to the generic SMEFT, fitting the experimental observables under the assumption of MFV involves less than a hundred of free parameters. 
Although with the cost of loosing information, this leads to a cleaner physical picture and more understandable outputs.

On the other hand, the MFV framework also contains certain conceptual and practical limits. 
Since the $U(3)^5$ symmetry is explicitly broken in the SM by the large top quark Yukawa coupling $y_t$, it makes more sense to consider new physics which only respects the exact or approximate symmetries of the SM. 
In other words, the effective field theories constructed with only small symmetry-breaking spurions, e.g. those smaller than $\mathcal{O}(10^{-2})$, are more reasonable.
On the practical side, as shown in \fref{EuroStrategy}, the experimental bounds for many cut-offs of the MFV operators remain as high as $\mathcal{O}(10)~\text{TeV}$~\cite{UTfit:2007eik, deBlas:2025gyz}, which are difficult to be lowered by further simple assumptions. 
Two features makes the MFV framework inconsistent with the $\mathcal{O}(1)$ TeV scale cut-offs:
\begin{enumerate}[label={(\roman*)}]
    \item The MFV spurion $Y_uY_u^{\dagger}$ is maximally misaligned with the down-type quark mass basis while remaining diagonal when up-type quarks are in their mass eigenstates. 
    In the MFV framework, it is impossible to transfer the flavour misalignment to the up sector to relax the constraints from FCNC (these bounds are tight only in the down sector). 
    \item The leading order MFV operators are flavour universal, and are constrained by the experimental observables involving the first-two generation fermions, especially the high-$p_T$ data. 
    The $U(3)$ based framework \typo{can not}{cannot} break the flavour universality between the third and first-two generation fermions among the leading order operators.
\end{enumerate}
Assuming a universal cut-off scale above 10 TeV, the MFV operators cannot explain many possible experimental deviations to the SM, in particular on many flavour conserving processes related to high-luminosity LHC and/or FCC-ee. 
As we will see, the $U(2)^5$ framework, with some fair assumptions, goes beyond the limits of MFV~\cite{Allwicher:2023shc, Allwicher:2025bub}.

\subsection{The \texorpdfstring{$U(2)^5$}{U(2)\^~~5} Theory}
\label{U2section}

A subgroup of $U(3)^5$ is $U(2)^5$, under which the first-two generation fermions are $U(2)$ doublets and the third generation fermions are singlets.
The gauge sector of SM is automatically $U(2)^5$ invariant. 
The $U(2)^5$ invariant Yukawa sector reads
\begin{equation}
\label{U2Yukawa}
-\mathcal{L}_Y~=~y_t\widetilde{H}\overline{Q_L^3}t_R+y_bH\overline{Q_L^3}b_R+y_{\tau}H\overline{\ell_L^3}\tau_R+\text{h.c.}
\end{equation}
Eq.~(\ref{U2Yukawa}) induces the massive top quarks, bottom quarks, and charged $\tau$ leptons.
Other quarks and charged leptons are massless (in this limit the $2-1$ quark mixing angle is unphysical) and the $3-2, 3-1$ quark mixing angles vanish. 
Comparing with the SM Yukawa sector shown in Eq.~\eqref{SMYukawa}, the $U(2)^5$ invariant Yukawa interactions account for a good approximation. 
To generating the realistic set of SM flavour parameters, only the spurions softly breaking $U(2)^5$ are needed
\begin{equation}
\label{U2spurions}
\begin{aligned}
    \Delta_u~&\sim~(\textbf{2}, \overline{\textbf{2}}, \textbf{1}, \textbf{1}, \textbf{1}), \quad \Delta_d~\sim ~(\textbf{2}, \textbf{1}, \overline{\textbf{2}}, \textbf{1}, \textbf{1}), \quad
    \Delta_{\ell}~\sim ~(\textbf{1}, \textbf{1}, \textbf{1}, \textbf{2}, \overline{\textbf{2}}), \\
    V_u~&\sim ~(\textbf{2}, \textbf{1}, \textbf{1}, \textbf{1}, \textbf{1}), \quad V_d~\sim ~(\textbf{1}, \textbf{2}, \textbf{1}, \textbf{1}, \textbf{1}), \quad
\end{aligned}
\end{equation}
With these spurions, the SM Yukawas couplings become
\begin{equation}
    Y_u~=~y_t\left(\begin{matrix}
        \Delta_u & V_u\\
        0 & 1 \\
    \end{matrix}\right), \quad 
    Y_d~=~y_b\left(\begin{matrix}
        \Delta_d & V_d\\
        0 & 1 \\
    \end{matrix}\right), \quad 
    Y_e~=~y_{\tau}\left(\begin{matrix}
        \Delta_e & 0\\
        0 & 1 \\
    \end{matrix}\right).
\end{equation}
Comparing this with the SM Yukawa coupling strengths, the $U(2)^5$ spurions shown in Eq.~(\ref{U2spurions}) are all at $\mathcal{O}(10^{-2})$ level. 
The CKM matrix is generated by $V_u$ and $V_d$, see e.g. Ref.~\cite{Fuentes-Martin:2019mun} for the explicit expressions. 
The magnitudes of $V_u$ and $V_d$ depend on the basis defining the $U(2)$ representations and are both non-zero in general. 
In the limit in which the $U(2)$ singlet $Q_L^3$ is aligned with the mass basis of first-two up-type (or down-type) quarks, $V_u$ (or $V_d$) vanishes. 
It is more transparent to encode this information on mixing into the definition of fields instead of couplings. 
In case both $Y_u$ and $Y_d$ are diagonal, $Q_L$ contains combination of multiple quark mass eigenstates.
Taking the approximation $V_{tb}=1$, we can parameterize $Q_L^3$ by
\begin{equation}
\label{U2defbyFields}
    Q_L^3~=~\left(\begin{matrix}
        t_L+(1-\epsilon_F) (V_{cb}^*c_L+V_{ub}^*u_L)\\
        b_L+\epsilon_F (V_{ts}s_L+V_{td}d_L)\\
    \end{matrix}\right).
\end{equation}
Here, $b_L, s_L, d_L$ and $t_L, c_L, u_L$ are all quark mass basis. 
The only free parameter defining $Q_L^3$ is $\epsilon_F$, which quantifies how much the down-type component of $Q_L^3$ is aligned to $b_L$.
Since $Q_L^3$ is defined as $U(2)$ singlet, different $\epsilon_F$ labels distinguished $U(2)$ theories. In particular, the up-aligned limit or down-aligned limit corresponds to $\epsilon_F=1$ or $\epsilon_F=0$, respectively, which reads
\begin{equation}
    Q_L^{3,u}~=~\left(\begin{matrix}
        t_L\\
        b_L+\epsilon_F (V_{ts}s_L+V_{td}d_L)\\
    \end{matrix}\right), \quad
    Q_L^{3,d}~=~\left(\begin{matrix}
        t_L+(V_{cb}^*c_L+V_{ub}^*u_L)\\
        b_L\\
    \end{matrix}\right).    
\end{equation}
Neglecting the small $\Delta_u$ and $\Delta_d$, there are no other source of flavour violating.
The up-aligned $Q_L^{3,u}$ only leads to FCNC for down-type quarks, which is the same as MFV in which $Q_L^3=Q_L^{3,u}$ when $Y_u$ is diagonal. The new feature here is that up-alignment is merely a particular limit for $U(2)$. Given vanishing $\epsilon_F$, $Q_L^3$ is nearly aligned to $b_L$ so that the dangerous FCNC for down-type quarks are suppressed. As we will see, small $\epsilon_F$ is necessary for a consistent $U(2)$ EFT with TeV scale cut-offs.

The $U(2)^5$ invariant dimension six SMEFT operators can be constructed as follows. We neglect $\Delta_u$ and $\Delta_d$, and use the parameterization shown in Eq.~(\ref{U2defbyFields}) so that the $V_u, V_d$ are both eliminated. 
There are then no explicit spurions, and the relevant effective operators are available form simply restricting the $U(2)^5$ symmetry for the generic set of SMEFT operators. 
The non-Hermitian operators involve different fermion species, so $U(2)^5$ only allows those constructed by third-generation fermions, which aligns with structure of SMEFT operators constructed by one single generation of fermions. 
For the Hermitian operators involving Higgs or gauge fields, the fermion bilinears can only be constructed by contracting two third generation fermions or two pairs of first-two generation fermions. For example, 
\begin{equation}
    (H^{\dagger}i\overleftrightarrow{D}_{\mu}H)(\overline{Q_L^3}\gamma^{\mu}Q_L^3), \quad 
    (H^{\dagger}i\overleftrightarrow{D}_{\mu}H)(\overline{Q_L^i}\gamma^{\mu}Q_L^i)
\end{equation}
Here, $i=1,2$ so $Q_L^i$ represents the $U(2)$ doublet. The total number of such operators corresponds to twice of the number of the same-type SMEFT operators with a single generation. 
The four-fermion operators involve more structures. 
There are four patterns for operators involving two different fermion bilinears. For instance, the operators involving the quark and lepton doublets are
\begin{equation}
    (\overline{Q_L^3}\gamma_{\mu}Q_L^3)(\ell_L^3\gamma^{\mu}\ell_L^3), \quad 
    (\overline{Q_L^3}\gamma_{\mu}Q_L^3)(\ell_L^i\gamma^{\mu}\ell_L^i), \quad (\overline{Q_L^i}\gamma_{\mu}Q_L^i)(\ell_L^3\gamma^{\mu}\ell_L^3), \quad (\overline{Q_L^i}\gamma_{\mu}Q_L^i)(\ell_L^i\gamma^{\mu}\ell_L^i).
\end{equation}
For those involving two same bilinears, there are five patterns in total. The structures containing only the quark doublets are
\begin{equation}
    \begin{aligned}
    &(\overline{Q_L^3}\gamma_{\mu}Q_L^3)(Q_L^3\gamma^{\mu}Q_L^3), \quad 
    (\overline{Q_L^i}\gamma_{\mu}Q_L^i)(Q_L^3\gamma^{\mu}Q_L^3), \quad (\overline{Q_L^i}\gamma_{\mu}Q_L^3)(Q_L^3\gamma^{\mu}Q_L^i), \\
    &(\overline{Q_L^i}\gamma_{\mu}Q_L^i)(Q_L^j\gamma^{\mu}Q_L^j), \quad (\overline{Q_L^i}\gamma_{\mu}Q_L^j)(Q_L^j\gamma^{\mu}Q_L^i). \\
    \end{aligned}
\end{equation}
The dimension-6 $U(2)^5$ EFT contains 84 (5) four-fermion and 14 (12) $\psi^2$ type (Non-)Hermitian operators~\cite{Allwicher:2023shc}. In total, there are 124 CP-even and 24 CP-odd independent $\Delta B=0$ dimension-6 operators, after including the pure bosonic ones. 
This number is larger than that of MFV up to $\mathcal{O}(Y_uY_u^{\dagger})$, but still much smaller than that for the generic SMEFT operators.

Comparing with MFV, the $U(2)$ framework does not contain operators universal among all three types of flavour. Rather, it involves operators which are only flavour universal between the first-two generations. 
This feature allows reasonable extensions with additional breaking of flavour universality, which leads to a third-generation specific framework. 
Such setup requires suppressing the operators involving the fields of the first-two generations. To achieve this, one can include factors $(\epsilon_Q)^{n_Q}(\epsilon_L)^{n_L}$ in the corresponding Wilson coefficients, where $n_Q$ and $n_L$ represent the number of first-two generation quarks and leptons in the operators. 
Since the $U(2)^5$ invariant operators always involve even number of the fermions in the first or second generation, these factors are at least squared. 
A factor of $\epsilon_Q\sim0.3$ (or $\epsilon_L\sim0.3$) can then suppress the relevant Wilson coefficient by one order of magnitude. 
Although adding such factors is, strictly speaking, an additional assumption, the physical picture of this third-generation specific framework remains clean. 


As pointed in Ref.~\cite{Allwicher:2023shc, Allwicher:2025bub}, under simple hypotheses for the Wilson coefficients, the $U(2)^5$ invariant EFT with cut-off scale low as 1.5 TeV can be made consistent with all present experimental bounds. 
The construction in Ref.~\cite{Allwicher:2023shc, Allwicher:2025bub} reads as follows.
Firstly, if including factors of $\prod_ig_i/(16\pi^2)$ in the Wilson coefficients of operators including gauge field strength tensors (applicable when operators with field strength tensors are generated beyond the tree level), the maximal current constrains on the $U(2)$ EFT lie at $\mathcal{O}(10)$ TeV. Then, bounds from collider observables, including the High-$p_T$ Drell-Yan tails, LEP-II data for $e^+e^-\to \ell^+\ell^-$, and jet observables typically involve light quarks and/or charged leptons in the first-two generations. These bounds can be relaxed by $\epsilon_Q$ and/or $\epsilon_L$ suppression. 
Next, constraints from electroweak observables, which contain $Z$ and $W$ pole measurements, lepton flavour universality (LFU) tests in $\tau$ decays, and flavour-conserving Higgs decays, can be relaxed by $\epsilon_Q, \epsilon_L$ together with a new factor $\epsilon_H$. 
Here, $\epsilon_H$ is the suppression factor related to the Higgs fields: for operators containing $n_H$ Higgs fields, one should include factor of $(\epsilon_H)^{n_H}$ in the corresponding Wilson coefficients. 
Finally, the restrictions from flavour-changing observables on down-type quarks can be lifted up in the down-alignment limit $\epsilon_F\to 0$. 
If one takes
\begin{equation}
    \epsilon_Q~=~0.16, \quad \epsilon_L~=~0.40, \quad \epsilon_H~=~0.31, \quad \epsilon_F~=~0.15,
\end{equation}
the $U(2)^5$ invariant EFT with cut-off scale satisfying
\begin{equation}
\label{U2bound}
    \Lambda_{U(2)}~\lesssim~1.5~\text{TeV}
\end{equation}
can be made consistent with all current experiments. 
This result can be interpreted as, that new physics with $\mathcal{O}(1)$ coupling to only third-generation fermions can be as light as TeV scale.

\subsection{Beyond MFV and \texorpdfstring{$U(2)^5$}{U(2)\^~~5}}
\label{beyondU2}

Although $U(2)^5$ arises as the approximate symmetry of the SM, other groups can also restrict the generic SM Yukawa couplings to the observed third-generation philic pattern. 
For instance, it is recently noted in Ref.~\cite{Antusch:2023shi} that the SM fermion masses and mixings also follows a minimally broken $U(2)_{q+e}$ flavor symmetry, which is the diagonal subgroup of $U(2)_{Q_L}\times U(2)_{E_R}$ embedded in $U(2)^5$. 
Under this $U(2)_{q+e}$ symmetry, $Q_L^i~(i=1,2)$ transforms as $U(2)$ doublet and so does $E_R^i$, while all other fundamental fermions are singlets. 
For the first-two generations, this symmetry is chiral for both quarks and leptons and forbids the relevant Yukawa couplings, which indicates the $U(2)_{q+e}$ invariant Yukawa interactions thus coincide with ones restricted by $U(2)^5$. 
The light fermion masses and relevant mixings can be generated by including two $U(2)_{q+e}$ breaking spurions. Without loosing generality, one can choose
\begin{equation}
    V_1~=~(0, a),\quad V_2~=~(b,0), \quad b\ll a\ll1. 
\end{equation}
$V_1$ and $V_2$ transform as doublets under $U(2)_{q+e}$. 
The Yukawa couplings matrices are also $U(2)_{q+e}$ and can be constructed by $V_1$ and $V_2$. Without loosing generality, it reads
\begin{equation}
    \label{U2qeYukawa}
    Y_u~\sim~ Y_d~\sim 
        \left(
    \begin{matrix}
        b & b & b\\
        0 & a & a\\
        0 & 0 & 1 \\
    \end{matrix}
    \right), \quad 
    Y_e~\sim~
        \left(
    \begin{matrix}
        b & 0 & 0\\
        b & a & 0\\
        b & a & 1 \\
    \end{matrix}
    \right).    
\end{equation}
Here, the unphysical parameters are eliminated by redefining the $U_R, D_R, \ell_L$ basis. 
It is evident from Eq.~\eqref{U2qeYukawa} that all $Y_u, Y_d,$ and $Y_e$ are third generation specific in the limit $a$ and $b$ vanish. 

Ref.~\cite{Antusch:2023shi} also proposed a variant of $U(2)_{q+e}$, which is \typo{referred as}{referred to as} $U(2)_{q+e^c+u^c}$ and features an attractive realization in SU(5) GUT~\cite{Georgi:1974sy}. 
$U(2)_{q+e^c+u^c}$ can be identified as the diagonal subgroup of $U(2)_{q}\times U(2)_{e^c}\times U(2)_{u^c}$, and commutes with $SU(5)$.
As a consequence, each generation of $Q_L, E_R^c$ and $U_R^c$ can be embedded into an anti-symmetric ten-dimension $SU(5)$ representation, without breaking the $U(2)_{q+e^c+u^c}$ global symmetry of minimal $SU(5)$ GUT. 
Comparing with $U(2)_{q+e}$, the $U(2)_{q+e^c+u^c}$ invariant Yukawa interactions stay philic to the third generation. 
In particular, $Y_d$ and $Y_e$ in Eq.~(\ref{U2qeYukawa}) do not change, and $Y_u$ becomes
\begin{equation}
\label{doublysup}
        Y_u~\sim~ 
        \left(
    \begin{matrix}
        b^2 & ab & b\\
        ab & a^2 & a\\
        b & a & 1 \\
    \end{matrix}
    \right).
\end{equation}
The up and charm quark Yukawas are therefore doubly suppressed, while the other light charged fermions receive only a single suppression. Compared the Yukawa textures restricted by $U(2)^5$ or $U(2)_{q+e}$ symmetry, Eq.~\eqref{doublysup} aligns better with the large mass hierarchies among the up-type quarks.

If viewing the SM as an effective theory enforced by the $U(2)_{q+e}$ or $U(2)_{q+e^c+u^c}$ symmetries, $U(2)^5$ becomes an accidentally enhanced symmetry in the renormalizable sector, while the non-renormalizable operators are not necessarily invariant under $U(2)^5$. 
Notably, there are no selection rules imposed on the dimension-5 Weinberg operator, because it only involves $\ell_L$ and $H$. Therefore, the large mixing angles and small mass hierarchies are allowed in the neutrino sector. 
The $U(2)_{q+e}$ or $U(2)_{q+e^c+u^c}$ invariant dimension-6 operators also contain more generic flavour structures than those in the $U(2)^5$ framework. 
Here, we do not repeat the construction of the corresponding effective theory, but remark that the operators involving flavour violating $\ell_L^i, U_R^i$ or $D_R^i$ currents are suppressed by $U(2)^5$ but can be relevant within the $U(2)_{q+e}$ framework. 
The bounds to these operators remain same as the those to the flavour anarchic ones. 
In particular, operators such as
\begin{equation}
    (\overline{\ell_L^1}\gamma_{\mu}\ell_L^2)(\overline{E_R^i}\gamma^{\mu}E_R^i)\quad \text{and}
    \quad (\overline{\ell_L^1}\gamma_{\mu}\ell_L^2)(\overline{Q_L^i}\gamma^{\mu}Q_L^i)
\end{equation}
are unsuppressed and can induce $\mu\to e$ conversions in nuclei and $\mu\to 3e$, respectively. 
These CLFV processes \typo{can not}{cannot} be described by the $U(2)^5$ EFT with the minimal set of spurions. 
On the other hand, many generic flavour structures remain absent in the $U(2)_{q+e}$ EFT, especially the chirality-flipping structures $\overline{\ell_L}E_R, \overline{Q_L}U_R,$ and $\overline{Q_L}d_R$.
Operators constructed with these structures, for instance, the dipole operator $(\overline{\ell_L^1}\sigma^{\mu\nu}\tau^IE_R)HW_{\mu\nu}^I$, can induce the electron magnetic and dipole moments, but are suppressed in the $U(2)_{q+e}$ framework as in case of $U(2)^5$.

The $U(2)_{q+e}$ and $U(2)_{q+e^c+u^c}$ examples shown above indicate that the possible flavour symmetries predicting the SM-like flavour pattern and restricting SMEFT operators might be are highly diverse. Actually, the flavor structure of dimension-6 SMEFT operators under the following symmetries are systematically explored in Ref.~\cite{Greljo:2022cah}:
\begin{itemize}
    \item \textbf{Quark sector}: $U(2)^3, ~ U(2)^3\times U(1)_b, ~U(2)^2\times U(3)_d, ~ U(3)^3$,
    \item \textbf{Lepton sector}: $U(1)_V^3, ~ U(1)^6, ~ U(2)_V, ~ U(2)^2, ~ U(2)^2\times U(1)^2, ~ U(3)_V, ~ U(3)^2$.
\end{itemize}
Combining the symmetries for quarks and leptons, this set-up leads to $4\times 7=28$ different classes of EFTs, each of them contains $\mathcal{O}(100)$ independent unsuppressed dimension-6 baryon number conserving SMEFT operators. Two of these combinations --- $U(2)^3\times U(2)^2$ and $U(3)^3\times U(3)^2$ --- lead to the $U(2)^5$ EFT and MFV respectively, while the others give rise to EFTs with distinct flavour structures.

\begin{table}[t!]
\centering
\renewcommand{\arraystretch}{1.5} 
\resizebox{0.9\textwidth}{!}{
\begin{tabular}{ccccccccc}
\toprule
\text{Fermions} & $Q_L^i$  &  $d_R,\,u_R$ & $s_R$  & $c_R$ & $b_R$ & $Q_L^3,\,t_R$ & $\ell^i_L$ & $E^i_R$   \\
\midrule
$Q_X$
& 0
& $-X_{V} - 2 X_z$ 
& $X_{V}+X_z$ 
& $-X_W$ 
& $X_z-X_U$ 
& $2 X_z - X_U$ 
& $X_i$  
& $X_i - (4 - i) X_z$ \\
\bottomrule
\end{tabular}}
\caption[The assignment of $U(1)_X$ charges in the MFP framework.]{The assignment of $U(1)_X$ charges to the SM fermions in the MFP framework
introduced in Ref.~\cite{Greljo:2025mwj}.}
\label{U1Xcharges}
\end{table}

Given that many structured EFTs are distinguished from the $U(2)^5$ invariant one, we proposed an interesting question in Ref.~\cite{Calibbi:2025ded}: are there any other flavor textures for the effective theories, that also suppress the light flavor transitions while allow third-generation or Higgs specific signals for near-future experiments as what $U(2)^5$ EFT does?
Recently, Ref.~\cite{Greljo:2025mwj} introduced a minimal flavour symmetry group $SU(2)_q\times U(1)_X$, which aligns with the SM fermion hierarchies and leads to an EFT allowing the TeV-scale cut-offs for every operators. 
The authors of Ref.~\cite{Greljo:2025mwj} refer to this framework as `minimal flavour protection' (MFP). Under $SU(2)_q$, only the two left-handed quarks $Q_L^i~(i=1,2)$ transform as a doublet, and the other fermions are singlets. 
The $U(1)_X$ charges (taken as integers) for SM fermions are summarized in Table~\ref{U1Xcharges}, with the requirement that
\begin{equation}
    |X_V|~\neq~ |X_W|~\neq~ |X_U|,\quad  (X_i-X_j)~\text{mod}~X_z\neq 0, 
\end{equation}
for any $i\neq j$. The SM Higgs is neutral under $U(1)_X$. 
With this assignment, only the top quarks are vector-like fermions under $U(1)_X$, so that $y_t$ can be sizable. All other Yukawa couplings must be small and vanish in the limit that $SU(2)_q\times U(1)_X$ is exact. 
The following spurions are needed to generate the sub-leading Yukawa couplings
\begin{equation}\label{MFPsuprions}
\boldsymbol{z} ~\sim~ (\textbf{1},{X_Z}),\quad 
\boldsymbol{V}
~\sim ~(\textbf{2},{X_V}),
\quad 
\boldsymbol{W} ~\sim ~(\textbf{2},{X_W}),
\quad 
\boldsymbol{U}~\sim~ (\textbf{2},{X_U}).
\end{equation}
The numbers in the brackets indicate the $SU(2)_q$ representations and $U(1)_X$ charges. 
The Yukawa coupling matrices then scale as
\begin{equation}
\label{MFPYukawa}
Y_u ~\sim~
\begin{pmatrix}
\boldsymbol{V}\boldsymbol{z}^2  & \boldsymbol{W} & 0 \\
0 & 0 & 1
\end{pmatrix},
\quad
Y_d \sim~
\begin{pmatrix}
\boldsymbol{V}\boldsymbol{z}^2 & \widetilde{\boldsymbol{V}} \boldsymbol{z}^* & \boldsymbol{U} \boldsymbol{z}^* \\
0 & 0 & \boldsymbol{z}
\end{pmatrix}, 
\quad
Y_e ~\sim~
\begin{pmatrix}
 \boldsymbol{z}^3  & 0 & 0 \\
0 &  \boldsymbol{z}^2 & 0 \\
0 & 0 & \boldsymbol{z}
\end{pmatrix}.
\end{equation}
This generates the SM fermion mass hierarchies and flavour mixings when all components of the $SU(2)_q\times U(1)_X$ breaking spurions take $\mathcal{O}(10^{-2})$ values. 
The $SU(2)_q\times U(1)_X$ invariant effective theory constructed with the spurions given in Eq.~(\ref{MFPsuprions}) features an interesting phenomenological consequence:
For every operator, an effective cut-off scale satisfying
\begin{equation}
    \Lambda_{\text{MFP}}~\lesssim~ 10~ \text{TeV}
\end{equation}
can be made compatible with with all current experimental bounds. 
This result is comparable to that of the third-generation specific $U(2)^5$ EFT shown in Eq.~(\ref{U2bound}).
In addition, the $SU(2)_q\times U(1)_X$ invariant EFT contains 185 CP-even and 15 CP-odd independent dimension-6 baryon-number conserving SMEFT operators, which is not much larger than those of MFV or $U(2)^5$, and  
comparing with 3045 operators of the generic SMEFT, the total number of the operators in the $SU(2)_q\times U(1)_X$ framework remains well under control. 
On the other hand, many new flavour structures which do not exist in the MFV or $U(2)^5$ framework can arise. In particular, when $X_2+X_3=2 X_1$, the operator
\begin{equation}
\label{MFPoperator}
    (\overline{\ell_L^1}\gamma_{\mu}\ell_L^2)(\overline{\ell_L^1}\gamma^{\mu}\ell_L^3)
\end{equation}
is invariant under $SU(2)_q\times U(1)_X$ but not under $U(2)^5$.
This operator induces the flavour violating $\tau^-\to \mu^+e^-e^-$ decays with a special type of charge assignment, that the final state electrons must have the same charges. 


At the end of this subsection, we remark that the landscape of the flavour symmetries introduced above can further be expanded when considering the discrete symmetries. 
The discrete symmetries are motivated by their potential connections to the large neutrino mixing angles. 
Among the various discrete symmetries, the non-abelian ones featuring triplet representations receive particular interest, such as $A_4$~\cite{Feruglio:2008ht, Hagedorn:2009df, Feruglio:2009hu,  delAguila:2010vg,  Ding:2011gt, Altarelli:2012bn, Pascoli:2016wlt, Heinrich:2018nip}, $A_5$~\cite{Feruglio:2011qq, Ding:2011cm, Hernandez:2012ra, Li:2015jxa, DiIura:2015kfa, Ballett:2015wia, Turner:2015mwa, DiIura:2016ols, Cooper:2012bd, Puyam:2023div}, and $S_3$~\cite{Hagedorn:2006ug, Bazzocchi:2008ej, Ishimori:2009ns, Ishimori:2010fs, Ishimori:2011mt, Bazzocchi:2012st, Thapa:2023fxu}. 
Three generations of certain-type SM fermions, especially the left-handed lepton doublets, are commonly promoted to transform as triplets under these discrete symmetry groups. 
Recently, Ref.~\cite{Palavric:2024gvu} systematically studied the role of $A_4, A_5,$ and $S_4$ discrete leptonic flavour symmetries in constraining the UV dynamics. Notably, the one-loop matching contributions to the SMEFT operators are computed in case of $A_4$~\cite{Moreno-Sanchez:2025bzz}. 
An interesting feature of this $A_4$ based leptonic theory for new physics theory is that the tree-level $\mu\to e$ transitions are suppressed in many scenarios. This significantly relaxes the tight constrains of CLFV new physics, and allows discovering multiple CLFV processes in near future, including $\tau$ flavour violating decays.
Such a suppression to $2\to 1$ flavour transitions closely aligns with what implied in the MFV, $U(2)^5$, and MFP frameworks introduced before. 


\section{Deconstructing Flavour}
Although the SM flavour structures can be approximately described by certain patterns restricted by flavour symmetries, the dynamical origin of the hierarchies among the charged fermion masses remains unexplained.
To address this puzzle and remove the small bare Yukawa couplings strengths, many beyond SM theories have been explored, see Ref.~\cite{Altmannshofer:2024hmr} a recent review. 
Over the past few years, the concept of ‘deconstructing flavour’ receives increasing interest, see for instance Ref.~\cite{Davighi:2023iks, FernandezNavarro:2023rhv, Davighi:2023evx, Barbieri:2023qpf, Davighi:2023xqn, Fuentes-Martin:2024fpx, Capdevila:2024gki, Davighi:2025cqx, Isidori:2025rci}.
The word `deconstruction' was rarely used and largely unknown in the daily language~\cite{derrida1988letter}, until being developed by the French philosopher Jacques Derrida in 1967~\cite{Derrida1}. 
As explained in Ref.~\cite{derrida1988letter}, deconstruction should not be interpreted as a negative reduction (such as the Nietzschean `demolition'). Instead, its meaning is closer to the Heideggerian word `destruktion'. To some extent, all sentences of the type ``deconstruction is X'' or ``deconstruction is not X'' are ill-defined, because the defining concepts for deconstruction is in principle also `deconstructible'. 
The concept of deconstruction should only be understood within the certain contexts. 
To our best knowledge, deconstruction is firstly introduced to particle physics by Ref.~\cite{Arkani-Hamed:2001kyx}.
The authors of Ref.~\cite{Arkani-Hamed:2001kyx} pointed out that the extra dimensions (if existing) are not necessarily fundamental. Rather, the action for a five-dimension gauge theory can be built by a class of renormalizable four-dimension theories and dynamically emerge at the low energies. 
The idea of deconstructing flavour is similar to deconstructing dimensions. 
In such a paradigm, it is believed that the existence of flavour (particles behave universally under certain gauge interactions) is not a fundamental feature of nature but manifests as an emerging low-energy phenomena. 
The flavour universal gauge interactions in the SM are viewed as an effective description for the flavour non-universal gauge theories~\cite{Li:1981nk} at higher energies scales (smaller distances). In other words
\begin{equation}
    \text{flavour universal}~G_{\text{SM}}~\xrightarrow{\text{deconstruct}}~\text{flavour non-universal}~G_{F}.
\end{equation}
Here, $G_F$ contains $G_{\text{SM}}$ as a subgroup, and the SM fermions in different families transform differently under $G_F$. 
As a result, the concept of flavour is deconstructed into a wide class of renormalizable theories in which flavour does not exist. 
The puzzle of flavour hierarchies are then automatically explained in the deconstructed flavour models, because the origin of small and sizable Yukawa interactions are different: 
The sizable Yukawa coupling strengths directly arise in the bare Lagrangian, while the small ones are induced by new heavy particles (such as the vector-like fermions) and the flavon VEVs which spontaneously break $G_F$.

\subsection{The Froggatt-Nielsen Mechanism}
\label{FNsection}
The symmetry-based approach to the explanation of the fermion mass hierarchies was firstly proposed in Ref.~\cite{Froggatt:1978nt}, which is commonly \typo{referred as}{referred to as} the Froggatt-Nielsen Mechanism. 
The basic idea is to replace the small Yukawa coupling with $\mathcal{O}(1)$ couplings multiplied by (powers of) the ratio a soft-symmetry breaking term $\mu$ and a cut-off scale $\Lambda$.

We revisit the basic idea of the Froggatt-Nielsen Mechanism in this subsection. In the SM, all entries of the Yukawa coupling matrices $Y_u, Y_d,$ and $Y_{e}$ can vanish in the limit of chiral flavour symmetries. 
One of the simple examples is an $U(1)_f$ symmetry, under which the SM Higgs $H$ is neutral and the fermions $Q_L^i, U_R^i, D_R^i, \ell_L^i, E_R^i$ are charged by $q_Q^i, q_U^i, q_D^i, q_{\ell}^i, q_E^i$ respectively. 
These $U(1)_f$ charges are restricted to be integers and satisfy
\begin{equation}
    q_Q^i~\neq~q_U^j,\quad  q_Q^i~\neq~q_D^j, \quad q_{\ell}^i~\neq~q_E^j, \quad i,j~=~1,2,3. 
\end{equation}
All renormalizable Yukawa interactions are forbidden by the $U(1)_f$ symmetries. To break it, one can include a soft $U(1)_f$ breaking parameter $\mu$ and promoted to a spurion field which breaks $U(1)_f$ by one unit. 
With this set-up, the $U(1)_f$ symmetries remain unbroken at the renormalizable level but can be softly broken by the non-renormalizable operators. The Lagrangian relevant to the charged fermion masses reads
\begin{equation}
\label{FNYukawa}
\begin{aligned}
    \mathcal{L}_{\text{FN}}~=&~C_u^{ij}\left(\frac{\mu}{\Lambda}\right)^{q_Q^i-q_U^j}\widetilde{H}\overline{Q_L^i}U_R^j
    +C_d^{ij}\left(\frac{\mu}{\Lambda}\right)^{q_Q^i-q_D^j}H\overline{Q_L^i}D_R^j\\
    &+C_e^{ij}\left(\frac{\mu}{\Lambda}\right)^{q_{\ell}^i-q_E^j}H\overline{\ell_L^i}E_R^j+\text{h.c.}\\
\end{aligned}
\end{equation}
Then, the Yukawa coupling matrices $Y_u, Y_d,$ and $Y_{e}$ follows the pattern
\begin{equation}
\label{FNYukawa2}
    Y_u~=~C_u^{ij}\left(\frac{\mu}{\Lambda}\right)^{q_Q^i-q_U^j}, \quad Y_d~=~C_d^{ij}\left(\frac{\mu}{\Lambda}\right)^{q_Q^i-q_D^j}, \quad
    Y_e~=~C_e^{ij}\left(\frac{\mu}{\Lambda}\right)^{q_{\ell}^i-q_E^j}.
\end{equation}
By construction, the Eq.~\eqref{FNYukawa} contains no flavour hierarchies, so that all elements in $C_u^{ij}, C_d^{ij},$ and $C_e^{ij}$ take $\mathcal{O}(1)$ values\footnote{We note this requirement must hold under all flavour basis, which indicates $\det C_u^{ij}\sim\det C_d^{ij}\sim\det C_e^{ij}\sim\mathcal{O}(1)$.}. Consequently, the eigenvalues of $Y_u, Y_d,$ and $Y_e$ exhibit the pattern
\begin{equation}
\label{FNyf}
    y_{u_i}~\simeq~\left(\frac{\mu}{\Lambda}\right)^{q_Q^i-q_U^i}, \quad
    y_{d_i}~\simeq~\left(\frac{\mu}{\Lambda}\right)^{q_Q^i-q_D^i}, \quad
    y_{e_i}~\simeq~\left(\frac{\mu}{\Lambda}\right)^{q_{\ell}^i-q_E^i}.
\end{equation}
where $u_i=u,c,t, d_i=d,s,b,$ and $e_i=e, \mu,\tau$.

The magnitudes of the SM Yukawa couplings are then governed by the differences of the $U(1)_R$ charges between the left-handed and right-handed fermion fields. 
Given $\mu\lesssim \Lambda$, the up-type quark mass hierarchies can be generated by
\begin{equation}
    q_Q^1-q_U^1~>~q_Q^2-q_U^2~>~q_Q^3-q_U^3~>~0. 
\end{equation}
The mass hierarchies of the down-type quarks and charged leptons can arise in the similar way. 
It is worthy remarking that the $\Lambda$ is not necessarily far larger than $\mu$. 
As shown in Eq.~(\ref{FNYukawa2}), the tiny Yukawa couplings can arise from the sizable $U(1)_R$ charge differences contained in the exponent. 
For instance, $q_Q^1-q_u^1=8$ and $(\mu/\Lambda)= 0.2$ leads to $y_u\sim0.2^8\approx2.6\times 10^{-6}$, which lies at the same order of the magnitude as the observed value shown in Eq.~(\ref{ynumbers}). 
In the basis that $Y_u$ is diagonal, the quark mixing matrix $V_{\text{CKM}}$ corresponds to the eigenvectors of $Y_dY_d^{\dagger}$.
Up to unphysical phases, the entries of the CKM matrix read
\begin{equation}
\label{FNCKM}
    V_{\text{CKM}}^{ij}~\sim~\left(\frac{\mu}{\Lambda}\right)^{|q_Q^i-q_Q^j|}.
\end{equation}
This relation also explains the hierarchical mixing angles.

The effective operators shown in Eq.~(\ref{FNYukawa}) can be generated by a large class of beyond SM model frameworks. 
As a consequence, certain flavour changing processes will unavoidably deviate from their SM predictions, which indicates $\Lambda$ \typo{can not}{cannot} be arbitrarily low. 
Nevertheless, $\Lambda$ does not necessarily lie at an very high scale, because the flavour transitions break the $U(1)_f$ symmetry and are further suppressed by powers of $\mu$. 
Among the various UV completions, a popular approach is realized by the vector-like fermions. 
In particular, Ref.~\cite{Arkani-Hamed:2026wwy} recently shows that for the class of models with multiple vector-like fermions, the SM fermion mass hierarchies can be generated following the paradigm introduced above with
\begin{equation}
    C_u^{ij}~\sim C_d^{ij}~\sim~C_e^{ij}\sim \left(\frac{\mu}{\Lambda}\right)~\sim~0.1-0.2,
\end{equation}
and the condition
\begin{equation}
    M_{\text{NP}}~\lesssim~ \text{a few TeV},
\end{equation}
does not violate any present experimental constraints. Here, $M_{\text{NP}}$ is the highest mass of the involved new particles.

\begin{figure}[t!]
    \centering
    \includegraphics[width=0.9\linewidth]{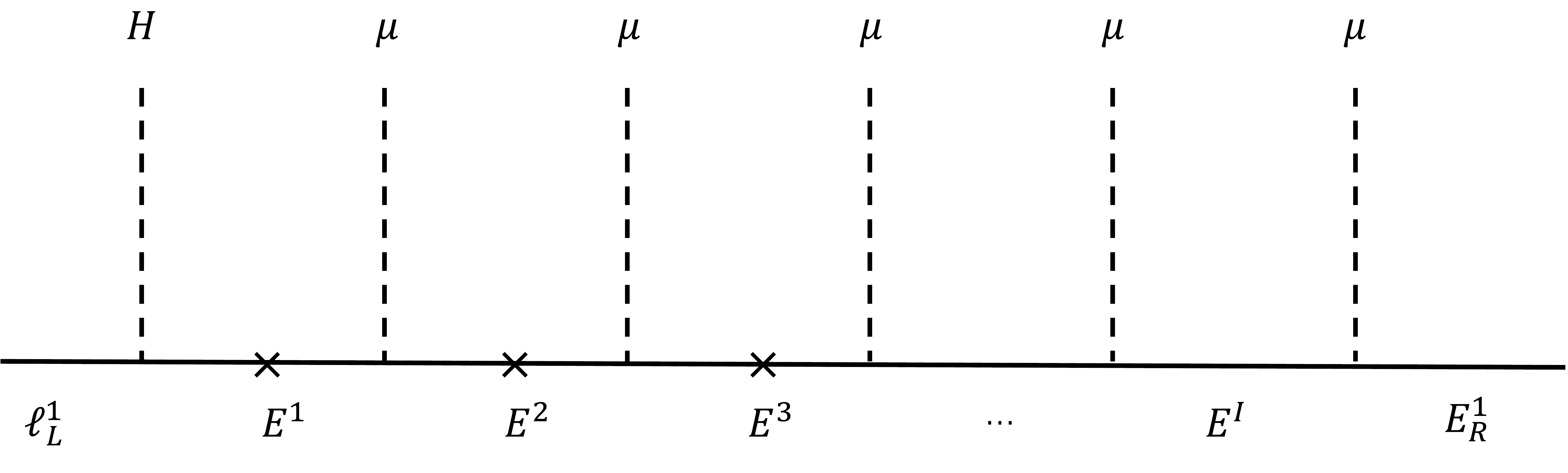}
    \caption[The moose diagram.]{The moose diagram generating the $(1,1)$ entry of the SM charged lepton mass matrix in a model with multiple vector-like fermions.}
    \label{FNdiagram}
\end{figure}

Here, we briefly introduce the model proposed in Ref.~\cite{Arkani-Hamed:2026wwy}, following the notations adopted in this work. 
Taking the charged lepton sector as an example, the authors of Ref.~\cite{Arkani-Hamed:2026wwy} includes multiple the vector-like fermions $E^I$. The left-handed and right-handed spinor component of $E_I$ are the same represents under $G_{\text{SM}}$ and carry the same charge $q_I$ under certain abelian flavour symmetry $U(1)_f$, with $q_I\neq q_J$ if $I\neq J$.
Furthermore, spurions carrying non-zero $U(1)_f$ charges, symbolized by $\mu_{IJ}$ and $\mu_{Ij}$, are also included. 
The $U(1)_f$ invariant renormalizable Lagrangian then takes the form
\begin{equation}
\label{reFNYukawa}
    -\mathcal{L}_{r}~=~y_{iI}H\overline{\ell_L^i} E_I+m^E_I \overline{E^I}E^I+\mu_{IJ}\overline{E^I}E^J+\mu_{Ij} \overline{E^I}e_R^j+\text{h.c.}. 
\end{equation}
Within this framework, the SM Higgs only contribute to the chirality flipping between $\ell_L^i$ and the vector-like fermion $E^I$. 
$\ell_L^i$ is connected to $e_R^j$ via the Feynman Diagrams whose length is controlled by the pattern of $\mu_{IJ}$, commonly \typo{referred as}{referred to as} the `moose diagrams'. 
Here, we show the moose diagram contributing to the $(1,1)$ entry of the charge lepton mass matrix in \fref{FNdiagram}. 
If the vector-like fermion spectrum is (quasi-)degenerate and all non-zero entries of $\mu_{IJ}$ and $\mu_{Ij}$ take value around $\mu$, integrating out the heavy vector-like fermions leads to the effective operators shown in Eq.~(\ref{FNYukawa}). 
Given $y_{iI}\sim\mathcal{O}(1)$, the coefficients $C_u^{ij}, C_d^{ij},$ and $C_e^{ij}$ are all flavour anarchic.

\begin{figure}
    \centering
    \includegraphics[width=0.35\linewidth]{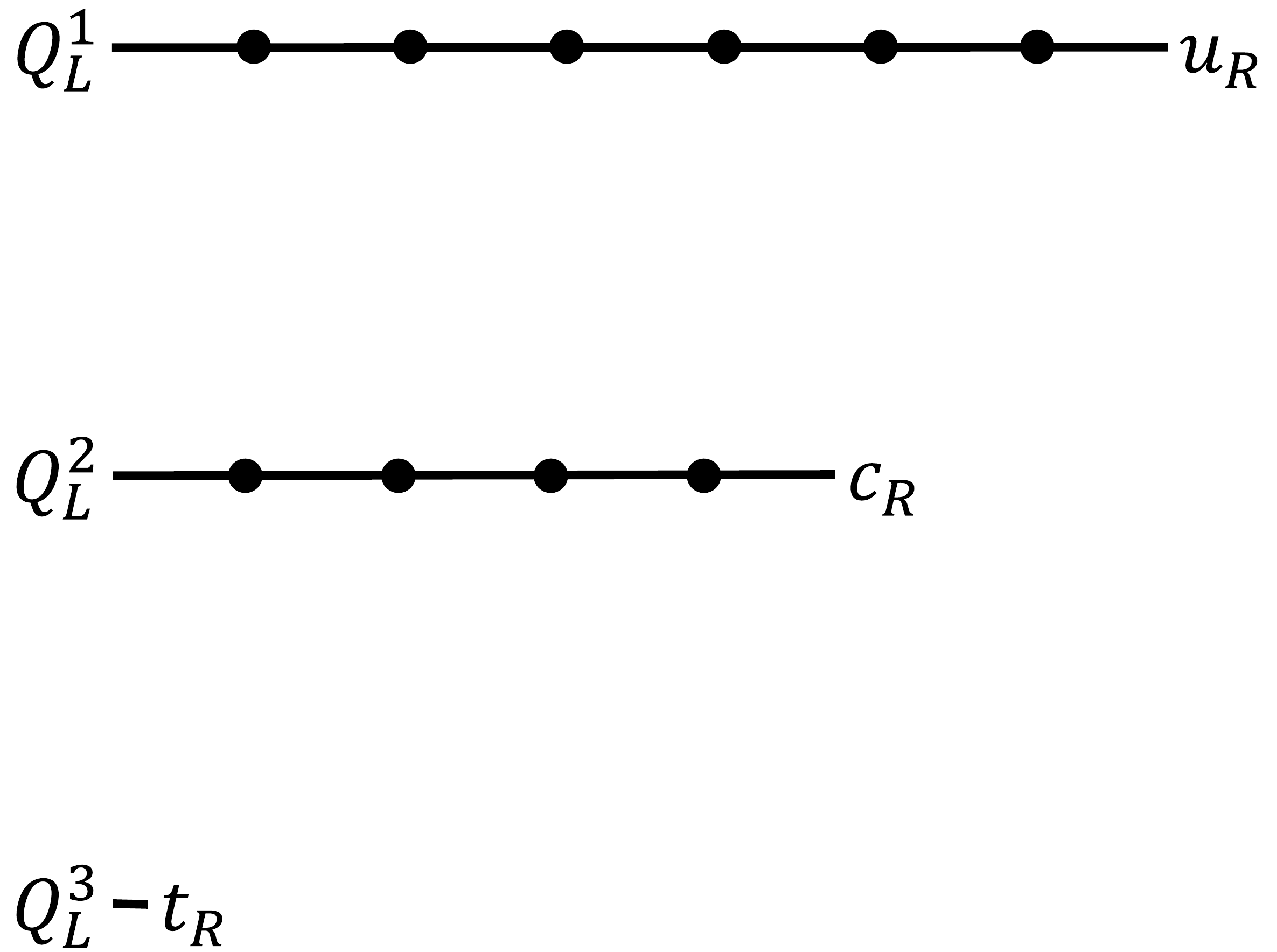}
    \qquad \qquad
    \includegraphics[width=0.35\linewidth]{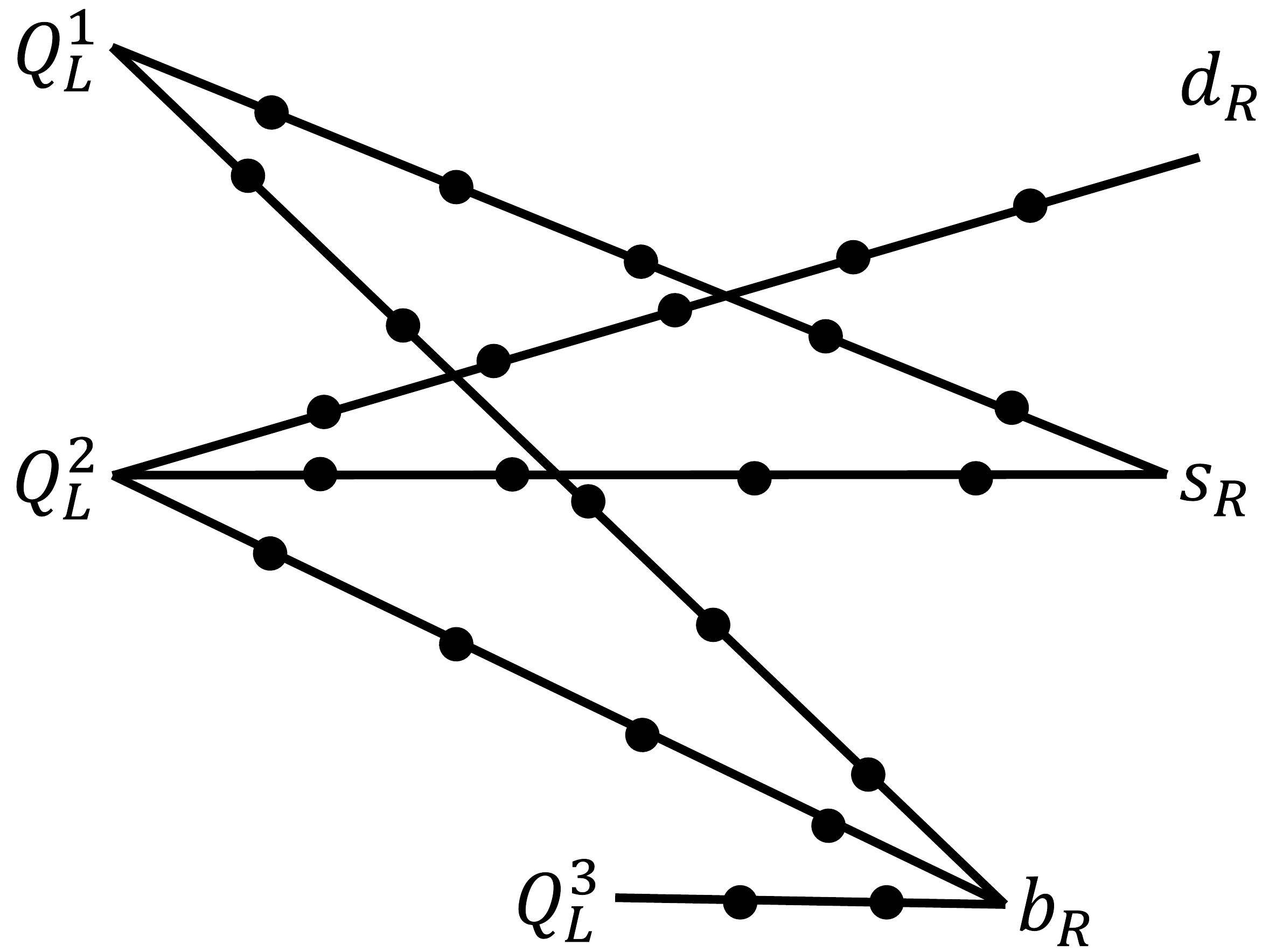}
    \caption[The chain structure of a quark mass model.]{The chain structure of model generating the up-type quark (left panel) and down-type quark mass matrices. The black dots indicate the spurion $\mu$.}  
    \label{TopChain}
\end{figure}

The charged lepton mass model discussed above feature a `chain structure'. 
The information on flavour can then be viewed as encoded in the nodes and edges of the moose diagrams.
Within this framework, the SM flavour pattern corresponds to the following moose diagram structures:
\begin{itemize}
    \item \textbf{Mass hierarchies}: Lengths of different edges;
    \item \textbf{Flavour mixings}: Nodes connecting more than one edges;
    \item \textbf{CP violation}: Closed loops. 
\end{itemize}
In \fref{TopChain}, we show the chain structure for the quark mass matrices introduced in Ref.~\cite{Arkani-Hamed:2026wwy}. 
The edges in \fref{TopChain} represent the structures shown in \fref{FNdiagram}. The number of black dots represent the lengths defined for every edge, which yield the numbers of vector-like fermions pairs (or equivalently the number of $\mu$) contained.
These lengths contain the information on the absolute sizes of the Yukawa couplings in $Y_u$ and $Y_d$. Flavor mixings originate from the nodes connecting more than one edges, which correlate the quarks from different generations. 
\fref{FNdiagram} contains a closed loop
\begin{equation}
    Q_L^1~\to~s_R~\to~Q_L^2~\to~b_R~\to~Q_L^1.
\end{equation}
It represents an invariant constructed by $Y_d^{12}(Y_d^{22})^*Y_d^{23}(Y_d^{13})^*$, whose phase is a physical quantity because it \typo{can not}{cannot} be eliminated by rephasing the quark fields. Hence, $Y_d$ generated in \fref{TopChain} contains a physical phase $\theta$.
Without loosing generality, one can choose every element in $Y_d$ except for $Y_d^{22}$ real, so that $\theta$ is only related to the edge $s_R\to Q_2$.
Then, the Yukawa coupling matrix for $Y_u$ and $Y_d$ reads
\begin{equation}
\renewcommand{\arraystretch}{1.2}
    Y_u~\simeq~\left(\begin{matrix}
        \epsilon_{\mu}^6 & 0 & 0\\
        0 & \epsilon_{\mu}^4 & 0\\
        0 & 0 & 1\\
    \end{matrix}\right), \qquad
    Y_d~\simeq~\left(\begin{matrix}
        0 & \epsilon_{\mu}^5 & \epsilon_{\mu}^5\\
        \epsilon_{\mu}^5 & \epsilon_{\mu}^5e^{i\theta} & \epsilon_{\mu}^4\\
        0 & 0 & \epsilon_{\mu}^2\\
    \end{matrix}\right), \quad \epsilon_{\mu}~=~\frac{\mu}{\Lambda}.
\end{equation}
Given $\epsilon_{\mu}\sim0.1-0.2$, $Y_d$ and $Y_u$ can produces the observed quark mass pattern. 
$Y_u$ is constructed by diagrams without edges connecting quarks from different families so it is diagonal, and all flavour mixing effects are encoded in $Y_d$. 
Due to the physical phase $\theta$, diagonalizing $Y_d$ requires unitary matrices containing complex phases, which can be identified as the physical CP violating phase in the CKM matrix.

\subsection{Non-Universal Gauge Interactions}

The soft symmetry breaking parameter $\mu$ can be promoted as the VEV of a scalar field $\phi$, rendering the restoration of $U(1)_f$ above a certain scale. 
$U(1)_f$ can be further promoted local and manifest as the flavour non-universal gauge interactions. 
The possible flavour symmetries which can forbid the small Yukawa interactions are not limited to $U(1)_f$. Alternatively, one can consider the non-abelian symmetries, in particular the deconstructed flavour non-universal gauge group $G_F$, which contains $G_{\text{SM}}$ as its diagonal subgroup.


Here, we follow the discussion in Ref.~\cite{Davighi:2023iks} to study certain candidates of $G_F$. 
As shown in Eq.~(\ref{LRsym}), each generation of the SM fermion spectrum features the left-right symmetry 
\begin{equation}
    G_{LR}~=~SU(3)_c\times SU(2)_L\times SU(2)_R\times U(1)_{B-L}. 
\end{equation}
$G_{LR}$ can be further extended to products of flavour non-universal groups
\begin{equation}
    G_{LR}~\subset~G_{LR}^{[1]}\times G_{LR}^{[2]}\times G_{LR}^{[3]},
\end{equation}
where the fermion generations are indicated in the square brackets. 
If certain flavour non-universal subgroup of $G_{LR}^{[1]}\times G_{LR}^{[2]}\times G_{LR}^{[3]}$ is gauged, the SM Yukawa coupling matrices are restricted and some of their entries will vanish. 
For instance, if vectorial $U(1)_{B-L}^{[3]}$ group and the diagonal $U(1)_{B-L}^{[12]}$ subgroup\footnote{It acts universally on the first-two generations.} of $U(1)_{B-L}^{[1]}\times U(1)_{B-L}^{[2]}$ is gauged, the SM Yukawa coupling matrices are restricted to be
\begin{equation}
    Y_u~\sim~Y_d~\sim~ Y_e~\sim~
    \left(
    \begin{matrix}
        \times & \times & 0\\
        \times & \times & 0\\
        0 & 0 & \times \\
    \end{matrix}
    \right).
\end{equation}
Here, $\times$ yields the non-zero elements. 
On the other hand, if the chiral $SU(2)_L^{[3]}$ (or $U(1)_{T_{3R}}^{[3]}$) group and the diagonal subgroup $SU(2)_L^{[12]}\subset SU(2)_L^{[1]}\times SU(2)_L^{[2]}$ (or $U(1)_{T_{3R}}^{[12]}\subset U(1)_{T_{3R}}^{[1]}\times U(1)_{T_{3R}}^{[2]}$) is gauged, the SM Yukawa coupling matrices read
\begin{equation}
    Y_u~\sim~Y_d~\sim~ Y_e~\sim~
    \left(
    \begin{matrix}
        0 & 0 & 0\\
        0 & 0 & 0\\
        \times & \times & \times \\
    \end{matrix}
    \right) \quad \text{or} \quad 
    \left(
    \begin{matrix}
        ~0 & 0 & \times \\
        ~0 & 0 & \times \\
        ~0 & 0 & \times \\
    \end{matrix}
\right).
\end{equation}
Here, $U(1)_{T_{3R}}$ is the abelian subgroup generated by the $T_3$ generator of $SU(2)_R$, which is a more minimal choice than gauging $SU(2)_R$ which necessities an additional Higgs doublet. 
The $(3,3)$ elements of the Yukawa couplings are allowed for all cases. 
Gauging any pair among $U(1)_{B-L}^{[12]}\times U(1)_{B-L}^{[3]} , SU(2)_L^{[12]}\times SU(2)_L^{[3]},$ and $U(1)_{T_{3R}}^{[12]}\times U(1)_{T_{3R}}^{[3]}$ leads to Yukawa couplings involving the third-generation fermions only. 
The small flavour mixings and light masses can arise when these deconstructed gauge groups break into the flavour universal $U(1)_Y$ of SM at low energies. 
We do not consider gauging the $SU(3)^{[12]}\times SU(3)^{[3]}$ group here, because it gives no constraint on $Y_e$.

We now recap one of the concrete examples introduced in Ref.~\cite{Barbieri:2023qpf}, which is \typo{referred as}{referred to as} `minimal flavour deconstruction'. 
The $U(1)_Y$ gauge group of SM is deconstructed into products of flavour non-universal groups with two steps
\begin{equation}
\begin{aligned}
    U(1)_Y~&\subset~G^{[12,3]}=U(1)_Y^{[3]}\times U(1)_{B-L}^{[12]}\times U(1)_{T_{3R}}^{[12]}\\
    ~&\subset~G^{[1,2,3]}=U(1)_Y^{[3]}\times U(1)_{B-L}^{[12]}\times U(1)_{T_{3R}}^{[1]}\times U(1)_{T_{3R}}^{[2]}.
\end{aligned}
\end{equation}
The UV gauge group is $G_F=SU(3)_c\times SU(2)_L\times G^{[1,2,3]}$, which acts non-universally on all three generations. It is spontaneously broken to $G^{[12,3]}$ by a high scale VEV $\langle \sigma\rangle$. $G^{[12,3]}$ is flavour universal for the first-two generations but non-universal between the first-two and the third ones. 
At an intermediate scale, $G^{[12,3]}$ is broken to the flavour universal $U(1)_Y$ group by two VEVs $\langle \phi \rangle$ and $\langle \chi^{q,l} \rangle$. 
which break $U(1)_Y^{[3]}\times U(1)_{B-L}^{[12]}$ and  $U(1)_Y^{[3]}\times U(1)_{T_{3R}}^{[12]}$ into their diagonal groups, respectively.
The symmetry breaking chain then reads
\begin{equation}
    G_F~\xrightarrow{\langle \sigma\rangle}~SU(3)_c\times SU(2)_L\times G^{[12,3]} 
    ~\xrightarrow{\langle \phi \rangle, \langle \chi^{q,l} \rangle}~
    SU(3)_c\times SU(2)_L\times U(1)_Y.
\end{equation}
The relevant VEVs are induced by additional scalar fields, which transform under $G^{[1,2,3]}$ as
\begin{equation}
\begin{aligned}
    \phi&\sim(0,0,\frac12,\frac12), ~~\phi_{3}\sim(\frac12,0,-\frac12, 0), ~~
    \chi^{q}\sim(-\frac16, \frac13,0,0), ~~
    \chi^{\ell}\sim (\frac12, -1,0,0). 
\end{aligned}
\end{equation}
The numbers in the brackets show the $U(1)$ charges contained in $G^{[1,2,3]}$. 
The Yukawa coupling matrices then scales as
\begin{equation}
\label{MFDYukawa}
    Y_u~\sim~Y_d~\sim~  \left(
    \renewcommand{\arraystretch}{1.2}
    \begin{matrix}
        \epsilon_{\sigma}\epsilon_{\phi} & \epsilon_{\phi} & \epsilon_{\chi^q}\\
        \epsilon_{\sigma}\epsilon_{\phi} & \epsilon_{\phi} & \epsilon_{\chi^q}\\
        \epsilon_{\sigma}\epsilon_{\phi}\epsilon_{\chi^q}  & \epsilon_{\phi}\epsilon_{\chi^q} & 1 \\
    \end{matrix}
    \right), \quad 
    Y_e~\sim~  \left(
    \renewcommand{\arraystretch}{1.2}
    \begin{matrix}
        \epsilon_{\sigma}\epsilon_{\phi} & \epsilon_{\phi} & \epsilon_{\chi^{\ell}}\\
        \epsilon_{\sigma}\epsilon_{\phi} & \epsilon_{\phi} & \epsilon_{\chi^{\ell}}\\
        \epsilon_{\sigma}\epsilon_{\phi}\epsilon_{\chi^{\ell}}  & \epsilon_{\phi}\epsilon_{\chi^{\ell}} & 1 \\
    \end{matrix}
    \right),
\end{equation}
with
\begin{equation}
    \epsilon_{\sigma}~=~\frac{\langle \sigma\rangle}{\Lambda_{[12]}}, \quad 
    \epsilon_{\phi}~=~\frac{\langle \phi\rangle}{\Lambda_{[23]}}, \quad 
    \epsilon_{\chi^q}~=~\frac{\langle \chi^q\rangle}{\Lambda_{[23]}}, \quad 
    \epsilon_{\phi^{\ell}}~=~\frac{\langle \chi^{\ell}\rangle}{\Lambda_{[23]}}.
\end{equation}
Here, $\Lambda_{[12]}\gg\Lambda_{[23]}$, yielding two cut-off scales.
Given
\begin{equation}
    \langle \phi \rangle~\sim~\langle \chi^{q,l} \rangle~\sim~\mathcal{O}(0.1)\times\Lambda_{[23]}, \quad \langle \sigma \rangle~\sim~\mathcal{O}(0.1)\times \Lambda_{[12]},
\end{equation}
the Yukawa coupling patterns in Eq.~(\ref{MFDYukawa}) generate the correct SM fermion mass hierarchies and mixing angles.

The effective Yukawa couplings can be UV-completed by the following vector-like fermions at $\mathcal{O}(\Lambda_{[23]})$:
\begin{equation}
    U_{L,R}^{i}\sim(\frac12,\frac13,0,0),\quad D_{L,R}^{i}\sim(-\frac{1}{2}, \frac13,0,0),\quad E_{L,R}^i\sim (-\frac12,-1,0,0), \quad i=4,5,
\end{equation}
and at $\mathcal{O}(\Lambda_{[12]})$:
\begin{equation}
    U_{L,R}^{6}\sim(0,\frac13,\frac12,0),\quad D_{L,R}^{6}\sim(0, \frac13,-\frac12,0),\quad E_{L,R}^6\sim (0,-1,-\frac12,0). 
\end{equation}
These fields transform flavour universally under $SU(3)_c\times SU(2)_L$, which are same as those of $U_R, D_R,$ and $E_R$ in the SM. 
The relevant $G_F$ invariant Yukawa interaction for up-type quark masses is ($i=1,2,3$ and $j=4,5$)
\begin{equation}
    \label{DeconstructionYukawa}
    \begin{aligned}
    -\mathcal{L}~=~&\widehat{Y}_u^{33}\widetilde{H}\overline{Q_L^3}U_R^3+ \widehat{Y}_u^{ij} \widetilde{H} \overline{Q_L^i}U_R^j+\widehat{Y}_u^{j3}\chi^q\overline{U_L^j}U_R^3+\widehat{Y}_u^{j2}\phi\overline{U_L^j}U_R^2\\
    &+\widehat{Y}_u^{j6}\phi\overline{U_R^j}U_L^6+\widehat{Y}_u'^{j6}\phi\overline{U_L^j}U_R^6
    +\widehat{Y}_u^{61}\sigma\overline{U_L^6}U_R^1+\text{h.c.}
    \end{aligned}
\end{equation}
Given all Yukawa couplings at $\mathcal{O}(0.1-1)$, Eq.~(\ref{DeconstructionYukawa}) generates $Y_u$ in Eq.~(\ref{MFDYukawa}).
The Yukawa interaction for down-type quarks corresponds to replacing $u$ and $U$ by $d$ and $D$. To generate $y_d\sim\mathcal{O}(1)$, one can further replace $\widetilde{H}$ by an additional Higgs $H'$ whose VEV takes value at $\mathcal{O}(m_b)$, and the $H$ coupling to $\widetilde{H}\overline{Q_L^3}D_R^3$ can be forbidden by a $Z_2$ symmetry. 
The charged-lepton sector is similar, except for additionally replacing $\chi^q$ with $\chi^{\ell}$.

Although the minimally deconstructed flavour model introduced above contains the fewest number of particles, postulating new particles at high scales remains necessary. 
The flavour model building becomes more convincing, if the origin of these postulated new particles can be explained. For instance, they may emerge as resonant states of strongly interacting theories like the mesons and baryons in the confined phase of QCD.
Recently, a number of composite Higgs models with flavour non-universal gauge interactions are discussed in Ref.~\cite{Davighi:2023iks, Covone:2024elw, Davighi:2025cqx}.
Here, we introduce the model constructed in Ref.~\cite{Davighi:2025cqx} here as follows.
The deconstructed flavour non-universal gauge group $G_F$ appears as a subgroup of some global symmetry $G_S$ of a strong sector. 
The SM Higgs field and certain scalars breaking $G_F$ arise naturally as the \typo{pesudo}{pseudo}-Nambu-Goldstone (pNGB) bosons. 
In this model, the UV dynamics is QCD like which only contains fermions and gauge fields, and has global symmetry
\begin{equation}
    G_S~=~Sp(6)\times U(1)_{B-L}^{[12]}\times U(1)_{B-L}^{[3]}.
\end{equation}
At some scale $4\pi f_s$ ($f_s$ can be of TeV), the $Sp(6)$ factor of $G_S$ is supposed to be spontaneously broken into its flavour non-universal subgroup
\begin{equation}
    Sp(6)~\xrightarrow{4\pi f_s}~G_H~=~SU(2)_L\times SU(2)_R^{[12]}\times SU(2)_R^{[3]},
\end{equation}
due to some strong dynamics. This symmetry breaking pattern gives pNGBs which transform under $G_H$ as
\begin{equation}
    H_3\sim (\textbf{2},\textbf{1},\textbf{2}), \quad H_{12}\sim (\textbf{2},\textbf{1}, \textbf{2}), \quad \Sigma\sim (\textbf{1}, \textbf{2},\textbf{2}). 
\end{equation}
SM-like Higgs doublets are contained in $H_3$ and $H_{12}$. 
If $\Sigma$ further takes a VEV $\langle\Sigma\rangle\sim f_s$ from negative quadratic terms of the pNGB potential, $G_H$ is spontaneously broken to its flavour universal subgroup
\begin{equation}
\label{intersection}
    G_H~\xrightarrow{\langle\Sigma\rangle\sim f_s}~SU(2)_L\times SU(2)_R.
\end{equation}
$\Sigma$ also induce the mixing term between $H_{12}$ and $H_3$
\begin{equation}
    \mathcal{L}_{\text{mix}}~=~\mu \langle\Sigma\rangle H_3H_{12}+\text{h.c.}
\end{equation}
Here, $\mu$ is the trilinear coupling strength for the $\Sigma H_3 H_{12}$ interaction, which has the mass dimension 1.

A subgroup of $G_S$ is promoted to be local and acts as the deconstructed flavour non-universal gauge group. 
\begin{equation}
    G_F~=~SU(2)_L\times U(1)_Y^{[12]}\times U(1)_{B-L}^{[3]}\times U(1)_R^{[3]}~\subset~G_S,
\end{equation}
The scalar fields transform under $G_F$ as
\begin{equation}
    H_3^{u,d}~\sim~(\textbf{2},0,\pm1,0), \quad
    H_{12}^{u',d'}~\sim~(\textbf{2},\pm 1,0,0),\quad
    \Sigma^{u,d}_{{u',d'}}~\sim~(\textbf{1},\pm 1,\pm 1,0). 
\end{equation}
Here, $H_3^{u,d}, H_{12}^{u',d'},$ and $\Sigma^{u,d}_{{u',d'}}$ are the $SU(2)_R^{[12]}\times SU(2)_R^{[3]}$ components of $H_3, H_{12},$ and $\Sigma$.
At some high scale $\langle\Omega\rangle\gg\langle\Sigma\rangle\sim f_s$, $U(1)_Y^{[12]}\times U(1)_{B-L}^{[3]}$ spontaneously breaks into its diagonal combination \typo{referred as}{referred to as} $U(1)_X$.
For left-handed fermions, $U(1)_X=U(1)_Y$, and they becomes universal from the gauge point of view; for right-handed fermions, $U(1)_X=U(1)_Y$ for the first-two generations and $U(1)_X=U(1)_{B-L}$ for the third. 
Considering the intersection with the global symmetry breaking, $\langle\Sigma\rangle$ breaks $U(1)_X\times U(1)_R^{[3]}$ into its flavour universal $U(1)_Y$ group. 
The full breaking chain of $G_F$ then reads
\begin{equation}
    G_F~\xrightarrow{\langle\Omega\rangle}~SU(2)_L\times U(1)_X\times U(1)_R^{[3]}~\xrightarrow{\langle\Sigma\rangle\sim f_s}~SU(2)_L\times U(1)_Y. 
\end{equation}
The $G_F$ invariant Lagrangian responsible for the quark masses is
\begin{equation}
\label{YukawaSp6}
    -\mathcal{L}~=~\widetilde{Y}^3\widetilde{H}_3\overline{Q_L^3}Q_R^3
    +\widetilde{Y}^{ij}\widetilde{H}_{12}\overline{Q_L^i}Q_R^j
    +\frac{1}{\Lambda}\widetilde{Y}^{i}\Omega\widetilde{H}_{3}\overline{Q_L^i}Q_R^3
    +\frac{1}{\Lambda}\widetilde{Y}'^{i}\Omega^*\widetilde{H}_{12}\overline{Q_L^3}Q_R^i+\text{h.c.}
\end{equation}
We do not write the $SU(2)_R$ components explicitly here. 
$\Omega$ is the field generating $\langle\Omega\rangle$ and is charged oppositely under $U(1)_Y^{[12]}$ and $U(1)_{B-L}^{[3]}$. 
If all Yukawa couplings are at $\mathcal{O}(1)$, $Y_u$ and $Y_d$ become
\begin{equation}
\label{Sp6Yukawa}
    Y_u~\sim~Y_d~\sim ~
    \left(
    \renewcommand{\arraystretch}{1.2}
    \begin{matrix}
        \epsilon_R & \epsilon_R & \epsilon_{\Omega}\\
        \epsilon_R & \epsilon_R & \epsilon_{\Omega}\\
        \epsilon_R \epsilon_{\Omega} & \epsilon_R \epsilon_{\Omega} & 1 \\
    \end{matrix}
    \right), \qquad \epsilon_{\Omega}~=~\frac{\langle\Omega\rangle}{\Lambda}, \quad \epsilon_R~=~\frac{\langle H_{12}\rangle}{\langle H_3\rangle}.
\end{equation}
In the limit that $H_{12}$ is much heavier than $H_{3}$, $\langle H_{12}\rangle$ is suppressed by its mass $M_{12}$
\begin{equation}
    \langle H_{12} \rangle~=~\frac{\mu\langle\Sigma\rangle}{M_{12}^2}\times \langle H_{3}\rangle.
\end{equation}
Given $\langle\Omega\rangle \ll \Lambda$, and $\mu\sim \Sigma \ll M_{12}$, Eq.~(\ref{Sp6Yukawa}) generated the needed fermion mass hierarchies and mixings. 
The mass splitting between the up-type and down-type quarks originates from the difference between the VEVs of the two $SU(2)_R$ components contained in $H_3$.

\section{Landscape or Swampland}

At first glance, one may expect the flavour non-universal gauge theories to be much simpler than the SM. 
After all, the replication of chiral fermions, commonly viewed as the origin of complexity in the SM, is eliminated within such a framework. 
However, all realistic flavour deconstruction models known today contain far more free model parameters than the SM, which contradict the naive expectation. 
To find a way out, we carefully reflect over the definition of flavour and note that $N_f$ is not a universal constant but depends on the framework  describing the theory. 
Taking the QCD corrections to the SM Higgs boson decay as an example~\cite{Chetyrkin:1997iv}, in the most widely adopted framework, $N_f$ is not three but five, because all light quarks transform equally under the $SU(3)_c$ color group and their Yukawa interactions are negligible. 
Likewise, to compute the evolution of the strong coupling strength $\alpha_s$~\cite{Baikov:2016tgj, Herzog:2017ohr, Luthe:2017ttc}, it is also common practice to stay agnostic to the electroweak interactions and view $N_f$ as an integer input variable ranging from three to six, depending on the quark thresholds $\alpha_s$ have crossed. 
The ambiguous meaning of $N_f$ motivates us to study the set containing \textit{all possible values of $N_f$}. 
The SM involves 48 or 45 chiral fermions assuming all known neutrinos are Dirac or Majorana fermions, respectively, which sets a limit on the the maximal value of $N_f$ and the maximal flavour symmetry group $G_F$. In other words,
\begin{equation}
    N_f^{\text{max}}~=~48, \quad G_F^{\text{max}}~=~U(48). 
\end{equation}
All SM gauge interactions are weak above the TeV scale, so they can be ignored as a rough approximation. Then, the gauge sector of the SM reduces to a free field theory (or strictly speaking a theory with only gravitational interactions) in which $N_f=48$ and $G_F=U(48)$ are self-evident. 
In this content, $N_f$ can take any integer values up to 48 and all subgroups of $U(48)$ can be viewed as flavour symmetries. 

The above-mentioned question on flavour deconstruction can be answered with the generalized definition of flavour. Here, we compare origins of the complexities in the SM, the gauged $U(3)^5$ models, and the flavour deconstruction models together:
\begin{itemize}[label={}, leftmargin=*]
\item \textit{The SM}: The $U(3)^5$ group commutes with the universal gauge group and can be explicitly broken in any directions, giving rise to many free model parameters. 
\item \textit{The gauged $U(3)^5$ models}: The arbitrariness of explicit symmetry breaking is eliminated through promoting $U(3)^5$ to be gauged, but generating the specific soft flavour symmetry breaking terms in the low energy theory requires complicated UV models, which inevitably involves many free model parameters. 
\item \textit{The flavour deconstruction models}: The $U(3)^5$ group does not commute with the non-universal gauge groups so it is automatically eliminated. 
Yet, from a broader perspective, the flavour symmetry persists because the non-universal group $G^{[1]}\times G^{[2]}\times G^{[3]}$ is still a subgroup of $U(48)$. 
Due to its flavourful nature, generating the specific $G^{[1]}\times G^{[2]}\times G^{[3]}$ breaking term at low energies still requires complicated UV model building, just like the case of the gauged $U(3)^5$ models. 
\end{itemize}
Therefore, it is not surprising that the flavour deconstruction models remain much less economical than the SM.

\begin{figure}[t!]
    \centering
    \includegraphics[width=0.8\linewidth]{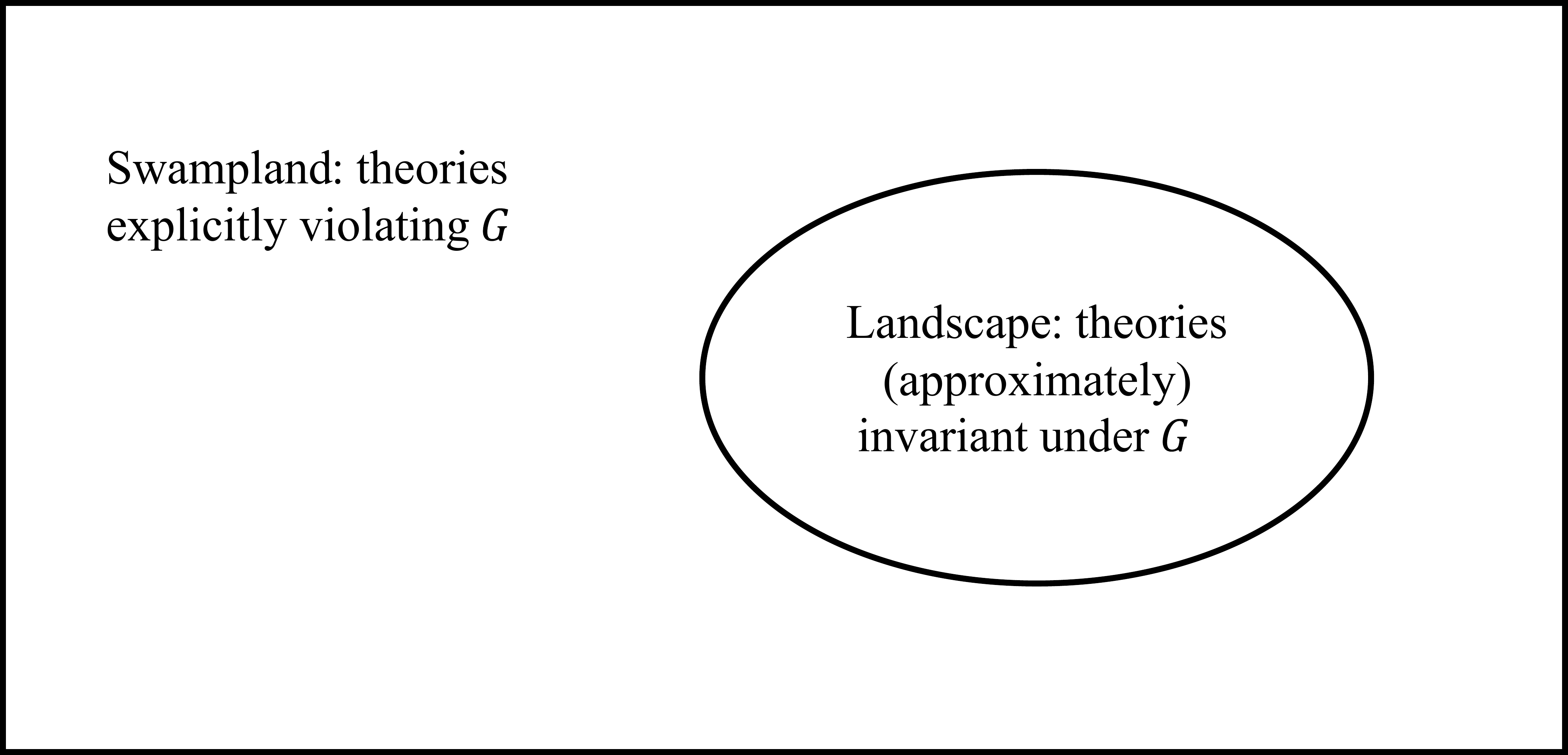}
    \caption[The space of theories in the landscape and swampland.]{A schematic illustration of the space of the theories lying in the landscape and the swampland of flavour symmetry $G$.}
    \label{FigSwampland}
\end{figure}

For this reason, flavour persists in the flavour deconstruction models from a broader perspective. 
We therefore argue for going one further step beyond, that both $G^{[1]}\times G^{[2]}\times G^{[3]}$ and $U(3)^5$, as well as all subgroups of $U(48)$ in addition to $G_{\text{SM}}$, should be eliminated, or at least stay irrelevant to explaining the SM flavour parameters. 
To make this reasoning more concrete, we introduce the following two concepts.
We refer the set of the theories invariant or approximately invariant under a certain symmetry group $G$ as the `landscape of $G$'.
In these theories, the soft $G$ breaking spurions, represented by $\epsilon_G^i$, are all (fairly) much smaller than unity. 
The `swampland of $G$' denotes the complementary set of the landscape of $G$, as illustrated in \fref{FigSwampland}. It contains all theories containing sizable spurion $\epsilon_G$ which strongly break $G$. 
In other words,
\begin{equation}
    \begin{aligned}
        \text{Landscape of $G$:}&\quad \epsilon_G^i~\ll~1, && \forall ~\epsilon_G^i~\in~\mathbb{S}, \\
        \text{Swampland of $G$:}&\quad \epsilon_G^i~\gtrsim~1, &&\exists~\epsilon_G^i~\in~\mathbb{S}. \\    
    \end{aligned}
\end{equation}
Here, $\mathbb{S}$ is the set for all $G$ breaking spurions involved in the theory. 
Taking $U(2)^5$ as an example, the SM and the $U(2)^2$ invariant EFT are in the landscape, while the $U(2)_{q+e}$ invariant EFT stays in the swampland. 
The flavour anarchic EFT lies in the swampland of $U(2)^5$ and of all flavour symmetries. 
It is worthy to remark that, the concepts `swampland' and `landscape' we use here are similar but not identical to those referred in string theory and quantum gravity, see~\cite{Palti:2019pca, vanBeest:2021lhn} for reviews. 
The key difference is that we do not include the UV completion.
For a theory in the swampland of a certain symmetry $G$, it is fully possible that above some scale $\Lambda$, the corresponding UV theory lies in the landscape of $G$. 
In such a case, the unbroken group $G$ exists, but it is hidden by the dynamics at scale $\Lambda$ and stays implicit for at low energies, physically equivalent a strongly broken symmetry.

Supposing a theory rendering all flavour parameters calculable exists, does it necessarily lie in the landscape of flavour symmetries?
To answer this question, we must firstly remark that it is a strong requirement for a theory to explicitly break \textit{all} exact and approximate flavour symmetries, 
because as introduced above, the concept of flavour symmetries can be widely defined. 
In this work, we define all subgroups of the $U(48)$ group acting on the chiral fermions to be flavour symmetries, but exclude $G_{\text{SM}}$ because it is already known to be exact and no realistic theory can live in the swampland of $G_{\text{SM}}$. 
Likewise, it is also reasonable to exclude the baryon number and lepton number, while we do not consider this possibility and still consider them as a flavour symmetries, unless otherwise stated. 
In addition, we note that when consider only the sub-sector of a theory, $N_f^{\text{max}}$ is smaller than 48 and definition of flavour symmetries should be (trivially) restricted to those only acting on the sub-sector. 
With these definitions, the landscape of flavour symmetries contains a wide class of different theories, including the SM itself. 
Consequently, we agree that in practice, it is indeed a reasonable start to assume that the desired standard theory of flavour lies inside the landscape.
The symmetry-based approach remains as one of the most promising known paths solving the flavour puzzle.
However, despite many years of effort, no indications for the expected flavour symmetries are found. 
Considering this, we also argue that one should also take the other possibility seriously. 
The new physics beyond SM, including the long-desired standard theory of flavour, can also lie in the swampland of flavour symmetries. 

%

\chapter{Flavour Textures without Symmetries}
\label{twozerochapter}
This chapter is devoted to discuss perhaps the most important question on the swampland of flavour symmetries: Can we directly probe it in near future experiments? 
Comparing with theories in the landscape, the challenge is how to suppress the flavour transition between the first-two generations.
We take the type~II seesaw model~\cite{Magg:1980ut, Lazarides:1980nt, Schechter:1980gr, Mohapatra:1980yp, Gelmini:1980re} as a minimal example, because it is perhaps the simplest UV-complete model for neutrino masses, which only requires one additional scalar field $\Delta$. 
In particular, we focus on a set of specific patterns for the new Yukawa coupling $Y_\Delta$, which generates vanishing entries in the neutrino mass matrix, called zero textures~\cite{Frampton:2002yf, Xing:2002ta, Kageyama:2002zw, Xing:2002ap, Frigerio:2002fb, Desai:2002sz, Guo:2002ei}. 
Our finding is encouraging, that certain zero textures, despite do not entail any enhanced flavor symmetries for three generations of $\ell_L$, can lead to suppressed $\mu\to e$ transitions and stay consistent with $5-10$~TeV new physics in the leptonic sector.

For the type~II seesaw model, the idea of suppressing $\mu\to e$ transition via the zero entries of $Y_{\Delta}$ has been partly explored in the previous literature. 
Refs.~\cite{Chun:2003ej, Akeroyd:2009nu} are early studies on CLFV within type~II seesaw that also discussed the conditions on $Y_\Delta$ required to suppress the $\mu\to e$ transitions, but the crucial one-loop enhancement for $\mu\to \bar e ee$ was not included. 
In the context of a general EFT analysis, the authors of Ref.~\cite{Ardu:2024bua}
studied the full RG running of the CLFV coefficients (hence capturing all one-loop effects) and mapped them to UV-complete models including type~II seesaw. In this context, 
they considered the one-zero texture with $M^\nu_{e\mu} = 0$ to suppress $\mu\to e$ transitions.
Nevertheless, we find that the predictive patterns of CLFV observables can only be identified within more constraining textures, e.g.~with two vanishing entries. 
The two-zero textures in type~II seesaw were considered in Ref.~\cite{Kitabayashi:2020ajn}, where the one-loop enhanced contribution to $\mu\to \bar e ee$ was again omitted and other loop-induced processes, such as $\mu\to e\gamma$, were not included in the analysis. 
In a footnote of our earlier work~\cite{Calibbi:2022wko}, we remarked that the same zero-entry suppression in a type~II seesaw model embedded in minimal $SU(5)$ GUT, and furthermore commented the potential renormalization group (RG) corrections.

To the best of our knowledge, the first complete analysis on this topic were performed in our recent paper~\cite{Calibbi:2025ded}, which most of the content in this chapter is based on. 
In addition to Ref.~\cite{Calibbi:2025ded}, we add more discussions in this work comparing the two-zero textures with three representative patterns restricted by flavour symmetries, which are $U(2)_{\ell}$, $U(1)_e,$ and $A_4$. 
In particular, we explain how the two-zero textures can suppress $\mu\to e$ transitions by interpreting the theory as in the swampland of $A_4$, which makes the results of Ref.~\cite{Calibbi:2025ded} more concrete and transparent. 
Moreover, we highlight a notable feature on the predicted $\tau$ decay spectrum, that the daughter electrons must be same-sign.
The following sections of this chapter is structured as follows. In Section~\ref{modelsection}, we introduce the minimal type~II seesaw model and the zero textures for the neutrino mass matrix, including analysis on the stability under the renormalization group (RG) running. 
In Section~\ref{texturesection}, we re-examine the low-energy effective operators up to one-loop matching, and then analyze the effective cutoff scale constrained by the current CLFV experiments, to demonstrate the predictivity. 
In Section~\ref{phenosection}, we analyze the correlation among various $\mu$ and $\tau$ CLFV processes and quantify the RG evolution corrections to the predicted ratios of the branching ratios (BRs). We also comment on complementary information that could be obtained at high-energy colliders. 
Finally, our findings and potential generalizations are summarized and discussed in Section~\ref{conclu}.

\section{Zeros in Type~II Seesaw}
\label{modelsection}
If all SM neutrinos are Majorana type, right-handed neutrinos $\nu_R$ are not necessarily contained in the SM fermion mass spectrum. 
Majorana neutrino masses can be generated by introducing a new ${SU}(2)_{L}$ scalar triplet $\Delta$
\begin{equation}
\label{Delta}
    \Delta \equiv 
    \left(\begin{array}{cc}
       \Delta^-/\sqrt{2}  & \Delta^0 \\
        \Delta^{--} & -\Delta^-/\sqrt{2}
    \end{array}\right)\,.
\end{equation}
The SM with additional scalar field $\Delta$ is commonly \typo{referred as}{referred to as} type~II seesaw model for neutrino masses.
The complete Lagrangian reads (see e.g.~\cite{Abada:2007ux, FileviezPerez:2008jbu} for reviews) 
\begin{equation}
\label{minimalTypeII}
\begin{aligned}
    \mathcal{L}~=&~\mathcal{L}_{\text{SM}}
    +\text{Tr}\left[(D_{\mu}\Delta)^{\dagger}(D^{\mu}\Delta)\right]-m_{\Delta}^2\text{Tr}(\Delta\Delta^{\dagger})
    -V^{(4)}(H,\Delta)\\
    &-(Y_{\Delta}\overline{\ell_L}\Delta i\tau_2 \ell_L^c -\mu_\Delta H^T i\tau_2 \Delta H+\text{h.c.})\,,\\
\end{aligned}
\end{equation}
In the above Lagrangian, $Y_{\Delta}$ is a $3\times3$ complex symmetric matrix, whose flavor indices are not shown explicitly, and $\mu$ is the trilinear coupling between the triplet and the SM Higgs doublet. 
The quartic self-interactions involving $\Delta$ and $H$ in $V^{(4)}(H,\Delta)$ have the general form
\begin{equation}
\begin{aligned}
     V^{(4)}(H,\Delta) ~=~& \lambda_1 \left[ \text{Tr}(\Delta^{\dagger}\Delta)\right]^2+\Lambda_{[12]} \text{Det}(\Delta^{\dagger}\Delta) + \Lambda_{[23]} (H^{\dagger}H) \text{Tr}(\Delta^{\dagger}\Delta)\\
     &+\lambda_4 (H^{\dagger}\sigma_iH)\text{Tr}(\Delta^{\dagger}\sigma_i\Delta)\,.
\end{aligned}
\end{equation}
Once the SM Higgs $H$ develops a non-zero VEV $v$ by spontaneous symmetry breaking, the triplet $\Delta$ also gets a non-zero VEV $v_{\Delta}$, from which the Majorana neutrino masses matrix arises
\begin{equation}
\label{massmatrix}
    M^{\nu}~=~ \sqrt{2} Y_{\Delta}v_{\Delta}~=~\mu_\Delta  Y_{\Delta} \frac{v^2}{m_{\Delta}^2}\,. 
\end{equation}
We remark here that $v_{\Delta}$ is a \typo{tadople}{tadpole} VEV induced by a shift $\Delta\to \Delta+v_{\Delta}$ in the scalar potential $V^{(4)}(H,\Delta)$ after the SM Higgs takes its VEV. $v_{\Delta}$ \typo{can not}{cannot} originate from a negative mass term in scalar potential $V^{(4)}(H,\Delta)$, because that leads to an additional source of spontaneous electroweak symmetry breaking with a perturbative unitarity bound for the the triplet mass $m_{\Delta}\lesssim G_F^{-1/2}\sim\mathcal{O}(10^2)$ GeV~\cite{Georgi:1981pg}. Such a non-decoupling theory \typo{can not}{cannot} stay consistent with experiments.

Throughout this chapter, we work in the $\ell_L$ basis under which the charged lepton Yukawa coupling $Y_e$ in Eq.~(\ref{SMYukawa}) is in its diagonal form. $M^{\nu}$ takes the form of a generic symmetric $3\times3$ matrix and is connected to the absolute neutrino masses by a unitary transformation~\cite{Pontecorvo:1957cp, Maki:1962mu}:
\begin{equation}
M^{\nu} ~=~ V'_{\text{PMNS}} \left( \begin{matrix} m_{\nu_1} & 0 & 0 \cr 0 & m_{\nu_2} & 0 \cr 0 & 0 & m_{\nu_3} \end{matrix} \right) V_{\text{PMNS}}'^{\rm T}\,,
\label{eq:mnu}
\end{equation}
where the mixing matrix $V'_{\text{PMNS}}$ relates Majorana neutrino mass and flavour eigenstates. It can be parametrized by multiplying a diagonal matrix containing two CP violation phases to the Dirac neutrino mixing matrix $V_{\text{PMNS}}$ shown in Eq.~(\ref{PMNSDirac})
\begin{eqnarray}
V'_{\text{PMNS}} ~=~ V_{\text{PMNS}}.
\left( \begin{matrix} e^{i\rho} & 0 & 0 \cr 0 & e^{i\sigma} & 0 \cr 0 & 0 & 1 \end{matrix} \right),
\label{eq:pmns}
\end{eqnarray}
$\rho$ and $\sigma$ are two new physical free parameters when neutrinos are Majorana fermions. 
In total, $M^{\nu}$ contains 9 real physical parameters, as shown by Eqs.~\eqref{eq:mnu} and~\eqref{eq:pmns}: two mass differences $\Delta m_{12}$, $\Delta m_{13}$, three mixing angles $\theta_{12}$, $\theta_{13}$, $\theta_{23}$, one Dirac CPV phase $\delta$, two Majorana CPV phases $\rho$ and $\sigma$, and the lightest mass $m_{0}$, which is equal to $m_{\nu_1}$ in case of NO or $m_{\nu_3}$ in case of IO. As introduced in Section~\ref{Fpuzzlesection}, the oscillation experiments can precisely measure the first five parameters, while they constrain the Dirac CPV phase $\delta$ only poorly at present and they are not sensitive to the absolute mass and the Majorana phases. Latest cosmological observations~\cite{DESI:2025zgx} require that $m_{0}$ must take small values and hint for a mildly hierarchical neutrino mass pattern, while such a bound is indirect and can be (easily) modified by new physics. 
$\beta$-decay experiments such as KATRIN~\cite{KATRIN:2024cdt} can also set upper limits but \typo{can not}{cannot} decide whether neutrino masses are hierarchical or quasi-degenerate. 
As a consequence, the absolute neutrino masses $m_{\nu_1}, m_{\nu_2},$ and $m_{\nu_3}$ are only constrained but not yet measured. The flavor texture of the neutrino mass matrix $M^\nu$ and the directly aligned Yukawa coupling matrix $Y_{\Delta}$ cannot be fully inferred from the current data.


The SM features an accidental lepton number symmetry, under which the Majorana neutrino mass terms vanish. 
The fermion bi-linear $\overline{\ell_L}\ell_L^c$ in Eq.~(\ref{minimalTypeII}) explicitly breaks the lepton number,
but in the limit $\mu_\Delta=0$, the neutrino masses remain vanishing and the Lagrangian in Eq.~\eqref{minimalTypeII} is invariant under
\begin{equation}
    \Delta\to\Delta e^{-2i\theta_L}, \quad \ell_L\to \ell_L e^{i\theta_L},  \quad H\to H\,.
\end{equation}
This transformation can be identified as the (generalized) lepton number symmetry $U(1)_L$. 
The magnitudes of Majorana neutrino masses are thus controlled by $\mu_\Delta$, which explicitly breaks $U(1)_L$. 
As a consequence, large Yukawa couplings, $Y_{\Delta}\sim\mathcal{O}(1)$, and/or a relatively light triplet mass, $m_{\Delta}\sim \mathcal{O}(\text{TeV})$, are not in conflict with the observed smallness of neutrino masses.
The tiny neutrino masses can thus be suppressed by a small value of $\mu_\Delta$ --- a situation that can be regarded as ``natural'' since it leads to the restoration of the lepton number symmetry of SM. 
This opens up the possibility of testing the theory searching for $L$-conserving CLFV processes, see e.g.~\cite{Abada:2007ux,Dinh:2012bp}.


Given sizable $Y_{\Delta}$, the approximate $SU(3)_{\ell}$ flavour symmetry in the SM Yukawa sector must be violated. The bi-linear $\overline{\ell_L}\ell_L^c$ transforms as $\textbf{3}\times\textbf{3}=\textbf{6}+\overline{\textbf{3}}$ under $SU(3)_{\ell}$, instead of a singlet. 
Nevertheless, certain other flavour symmetries can arise in the Lagrangian of type~II seesaw model. For instance, if $Y_{\Delta}$ takes the pattern 
\begin{equation}
\label{YdeltaSymTexture}
    Y_{\Delta}^{SU(2)_{\ell}}~=~\left(
    \begin{matrix}
        0 & 0 & 0\\
        0 & 0 & 0\\
        0 & 0 & \times\\
    \end{matrix}
    \right), \quad 
    Y_{\Delta}^{U(1)_{e}}~=~\left(
    \begin{matrix}
        0 & 0 & 0\\
        0 & \times & \times\\
        0 & \times & \times\\
    \end{matrix}
    \right), \quad \text{or} \quad 
    Y_{\Delta}^{A_4}~=~\left(
    \begin{matrix}
        \times & 0 & 0\\
        0 & 0 & \times\\
        0 & \times & 0\\
    \end{matrix}
    \right),
\end{equation}
the approximate $SU(2)_{\ell}, U(1)_e$, or $A_4$ symmetries of the SM remains unbroken. 
$SU(2)_{\ell}$ forbids all elements in $Y_{\Delta}$ except for $Y_{\Delta\tau\tau}$, because the $SU(2)_L$ singlet constructed by its two fundamental representations is anti-symmetric while $Y_{\Delta}$ is symmetric. 
The electron number symmetry $U(1)_e$ requires vanishing $Y_{\Delta e e}, Y_{\Delta e \mu},$ and $Y_{\Delta e \tau}$.
The discrete symmetry $A_4$, under which the three generations of $\ell_L$ transform as a triplet, allows only $Y_{\Delta ee}$ and $Y_{\Delta \mu\tau}$ to be non-zero. 
However, the patterns shown in Eq.~(\ref{YdeltaSymTexture}) significantly deviate from $M^{\nu}$, even though $M^{\nu}$ is not yet fully determined from the current data.
The disagreements are mainly due to the large neutrino mixing angles and the not-so-hierarchal mass spectrum.

In addition to the symmetry protected textures shown in Eq.~(\ref{YdeltaSymTexture}), the zero textures for $M^{\nu}$ are widely discussed in the literatures~\cite{Frampton:2002yf, Xing:2002ta, Kageyama:2002zw, Xing:2002ap, Frigerio:2002fb, Desai:2002sz, Guo:2002ei}.
The current experimental measurements of neutrino oscillations allow up to two vanishing elements in $M^\nu$, and some of such one- and two-zero textures can be tested in the near future --- see, for instance, Refs.~\cite{Fritzsch:2011qv, Meloni:2014yea, Zhou:2015qua, Alcaide:2018vni,Singh:2019baq, Denton:2023hkx, Chauhan:2023faf, Treesukrat:2025dhd} for analyses published after the measurement of the reactor mixing angle.
Interestingly, even without knowledge of the absolute neutrino masses, future oscillation experiments such as JUNO~\cite{JUNO:2015zny} and DUNE~\cite{DUNE:2015lol} can shed light on certain two-zero textures, once the Dirac CPV phase is measured or constrained. 
A notable feature of such textures is that no new symmetries are enhanced comparing with the flavour anarchic pattern. 
If requiring the zero textures are protected by (generalized) flavour symmetries, the minimal seesaw model needs extensions in general --- see, for instance, Refs.~\cite{Grimus:2004hf, Grimus:2004az, Hirsch:2007kh, Dev:2011jc, Fritzsch:2011qv, Araki:2012ip, Dev:2014dla, GonzalezFelipe:2014zjk, Lamprea:2016egz, Linster:2018avp, Borgohain:2018lro, Verma:2018lro, Zhang:2019ngf, Bjorkeroth:2019rat, Linster:2020fww, Barreiros:2022aqu, Ding:2022nzn, Ding:2022aoe,  Ding:2023htn, Rocha:2024twm, Fang:2024qtx, Rocha:2025ade, Jiang:2025psz}. 
More new particles except for $\Delta$ is needed. In case these new particles are much heavier than $\Delta$, the flavour symmetries leading to the zero textures become implicit at the scale of $m_{\Delta}$.

In the charged-lepton mass basis, all six possible one-zero textures for the neutrino mass matrix $M^{\nu}$ are consistent with the neutrino mixing angles and mass splittings measured in neutrino oscillation experiments~\cite{Xing:2003ic, Xing:2004ik, Merle:2006du, Lashin:2011dn, Deepthi:2011sk, Bora:2016ygl, Kitabayashi:2020ajn, Chauhan:2023faf}
\begin{equation}
\label{onezero}
\begin{aligned}
&~ \mathcal{A^{}}: \left( \begin{matrix} 0 & \times & \times \cr \times & \times & \times \cr \times & \times & \times \end{matrix}\right) \; , \quad \mathcal{B^{}}: \left( \begin{matrix} \times & 0 & \times \cr 0 & \times & \times \cr \times & \times & \times \end{matrix}\right) \; , \quad \mathcal{C^{}}: \left( \begin{matrix} \times & \times & 0 \cr \times & \times & \times \cr 0 & \times & \times \end{matrix}\right) \;  ,\\
&~  \mathcal{D^{}}: \left( \begin{matrix} \times & \times & \times \cr \times & 0 & \times \cr \times & \times & \times \end{matrix}\right) \; , \quad { \mathcal{E}^{}}: \left( \begin{matrix} \times & \times & \times \cr \times & \times & 0 \cr \times & 0 & \times \end{matrix}\right) \; , \quad { \mathcal{F}^{}}: \left( \begin{matrix} \times & \times & \times \cr \times & \times & \times \cr \times & \times & 0 \end{matrix}\right) \; .
\end{aligned}
\end{equation}
This is not the case for two-zero textures. The only patterns that, at present, are compatible with neutrino oscillation data are~\cite{Zhou:2015qua, Alcaide:2018vni,Singh:2019baq, Denton:2023hkx, Chauhan:2023faf,Treesukrat:2025dhd}:
\begin{equation}
\label{twozero}
\begin{aligned}
&~ \mathbf{A_1}: \left( \begin{matrix} 0 & 0 & \times \cr 0 & \times & \times \cr \times & \times & \times \end{matrix}\right) \; , \quad \mathbf{A_2}: \left( \begin{matrix} 0 & \times & 0 \cr \times & \times & \times \cr 0 & \times & \times \end{matrix}\right) \; , \quad \mathbf{B_1}: \left( \begin{matrix} \times & \times & 0 \cr \times & 0 & \times \cr 0 & \times & \times \end{matrix}\right) \; , \quad \mathbf{B_2}: \left( \begin{matrix} \times & 0 & \times \cr 0 & \times & \times \cr \times & \times & 0 \end{matrix}\right) \; ,\\
&~ \mathbf{B_3}: \left( \begin{matrix} \times & 0 & \times \cr 0 & 0 & \times \cr \times & \times & \times \end{matrix}\right) \; , \quad \mathbf{B_4}: \left( \begin{matrix} \times & \times & 0 \cr \times & \times & \times \cr 0 & \times & 0 \end{matrix}\right) \; , \quad \mathbf{C^{~}_{~}}: \left( \begin{matrix} \times & \times & \times \cr \times & 0 & \times \cr \times & \times & 0 \end{matrix}\right) \; .
\end{aligned}
\end{equation}
The above two-zero textures are also consistent with
beta decay and neutrinoless double beta decay ($0\nu\beta\beta$) data~\cite{Treesukrat:2025dhd}.\footnote{Only textures $\mathbf{A_1}$ and $\mathbf{A_2}$ are compatible with the stringent cosmological bound on the sum of neutrino masses $\Sigma m_{\nu} < 0.09$ eV~\cite{Denton:2023hkx}, while all seven textures remain viable under more conservative cosmological limits~\cite{Treesukrat:2025dhd}.} They serve as the maximal possible hierarchal structure for $M^{\nu}$ and, in the context of type~II seesaw, $Y_{\Delta}$. Textures with three or more zeros are excluded by oscillation experiments~\cite{Xing:2004ik, Fritzsch:2011qv}.

In presence of a two-zero texture, $M^{\nu}$ only contains five independent real physical parameters, which we can choose to be the precisely measured ones: the three mixing angles $\theta_{12}$, $\theta_{13}$, $\theta_{23}$ and the two mass splittings $\Delta m_{12}$, $\Delta m_{13}$. 
The CP violating phase $\delta$ is determined by the three mixing angles and the dimensionless ratio $\Delta m^2_{12}/|\Delta m^2_{13}|$, whose explicit expressions are reported in Ref.~\cite{Fritzsch:2011qv}. 
Then, up to an overall scale, $M^{\nu}$ or $Y_{\Delta}$ is fully fixed by four parameters 
$\theta_{13},\, \theta_{12},\, \theta_{12},$ and $\delta$~\cite{Kitabayashi:2015jdj}. 
The other quantities become the theory predictions~\cite{Meloni:2014yea,Zhou:2015qua, Treesukrat:2025dhd}, including the neutrino mass ordering. 
The $\textbf{A}$ textures always lead to the NO mass spectrum.
For $\textbf{B}$ textures, the mass ordering depends on the octant of $\theta_{23}$: 
$\theta_{23}<45^{\circ}$ yields the NO spectrum for $\mathbf{B_1}$, $\mathbf{B_3}$ and IO spectrum $\mathbf{B_2}, \mathbf{B_4}$; 
Conversely, $\theta_{23}>45^{\circ}$ generates the dual solution, IO spectrum for $\mathbf{B_1}$, $\mathbf{B_3}$ and NO spectrum $\mathbf{B_2}, \mathbf{B_4}$.
The texture $\textbf{C}$ produces the NO spectrum only for $\theta_{23}=45^{\circ}$; otherwise, it predicts the IO spectrum.
The one-zero textures are less restrictive and only poorly constrained by data at present. Although they contains the same number of free parameters and experimental inputs, the non-zero elements can only be inferred with large uncertainties, unless $\delta$ and $m_0$ are measured or constrained at a better precision level. Typically, in order to narrow down the allowed ranges for the parameters in Eq.~\eqref{onezero}, the sensitivity for $m_0$ should reach the $\mathcal{O}(0.05)$~eV level so that the uncertainty of the dimensionless variable $(m_0^2/\Delta m_{13}^2)$ becomes smaller than $\mathcal{O}(1)$,
a result that is supported by the $m_0$ dependence of the $M^{\nu}$ elements~\cite{FileviezPerez:2008jbu}.

We generate $10^4$ points of the parameter sets $\{\theta_{13},\,\theta_{12},\, \theta_{12},\,\Delta m^2_{12},\,\Delta m^2_{13}\}$ that satisfy the distributions resulting from the fit to neutrino oscillation data reported in~\cite{Esteban:2024eli}. 
Using these experimental inputs, we calculated calculate the $10^4$ possible two-zero textures for $Y_{\Delta}$ for further analysis. 
To illustrate how these textures look like, we fix $Y_{\Delta\mu\tau}=1$ and provide here one numerical benchmark of the flavor structure of $|Y_{\Delta ij}|$ for each texture
\begin{equation}
\label{twozerobenchmark}
\begin{aligned}
\mathbf{A_1} &: \left(
\begin{array}{ccc}
 0 & 0 & 0.43 \\
 0 & 1.07 & 1\\
 0.43 & 1& 1.12 \\
\end{array}
\right),~  \mathbf{A_2}: \left(
\begin{array}{ccc}
 0 & 0.45 & 0 \\
 0.45 & 1.15 & 1\\
 0 & 1& 1.09 \\
\end{array}
\right),
~\mathbf{B_1}: \left(
\begin{array}{ccc}
 0.83 & 0.10 & 0 \\
 0.10 & 0 & 1\\
 0 & 1& 0.42 \\
\end{array}
\right),\\ \mathbf{B_2} &: \left(
\begin{array}{ccc}
 0.87 & 0 & 0.07 \\
 0 & 0.31 & 1\\
 0.07 & 1& 0 \\
\end{array}
\right),
~ \mathbf{B_3} : \left(
\begin{array}{ccc}
 0.84 & ~~0~~ & 0.07 \\
 0 & 0 & 1\\
 0.07 & 1& 0.38 \\
\end{array}
\right),
~\mathbf{B_4} :\left(
\begin{array}{ccc}
 0.87 & 0.06 & 0 \\
 0.06 & 0.29 & 1\\
 0 & 1& ~~0~~ \\
\end{array}
\right),\\
\mathbf{C^{~}_{~}}&: \left(
\begin{array}{ccc}
 0.74 & 0.06 & 0.40 \\
 0.06 & 0 & 1\\
 0.40 & 1& 0 \\
\end{array}
\right).
\end{aligned}
\end{equation}
Comparing with Eq.~(\ref{YdeltaSymTexture}), the above two-zero textures do not obey the patterns originated from $SU(2)_{\ell}$, $U(1)_{e}$, and $A_4$ flavor symmetries, even approximately.  
In other words, $SU(2)_{\ell}$, $U(1)_{e}$, and $A_4$ are all broken explicitly. 
To our best knowledge, no other explicit symmetries for $\ell_L$ for the seven two-zero textures can arise without extending the type~II seesaw. Therefore, we conclude that given any flavour symmetry $G$ under which three generations of $\ell_L$ transform non-trivially, the type~II seesaw model with two zero entries contained $Y_{\Delta}$ lies in the swampland of $G$. 
Here, we do not consider the symmetry groups acting only on quarks and/or right-handed electrons.
In addition, we note that the lepton number $L$ can be decomposed as the $U(1)_{E_R}$ the known gauged $U(1)_Y$ and, so it should not be regarded as a flavour symmetry. 
Comparing with the two zero textures, perhaps the most closed symmetry protected patterns is those restricted by $A_4$. If ignoring the $\mathcal{O}(30\%\sim 40\%)$ explicit symmetry breaking entries, the $\textbf{B}$ and $\textbf{C}$ textures align with $Y_{\Delta}^{A_4}$ in Eq.~\eqref{YdeltaSymTexture}. Consequently, although $\textbf{B}$ and $\textbf{C}$ textures are in the swampland of $A_4$, they are relatively close to the boundary of its landscape.

Due to the absence of explicit symmetry protection, the texture zeros are typically not stable under the RG running above $m_{\Delta}$. Strictly speaking, the considered textures above can only be consistently defined at a certain high-energy scale $\Lambda_\text{UV}$ where they originate (e.g.~from spontaneous breaking of a flavor symmetry). Radiative effects will in general generate non-zero entries.
The relevant RG equations above the scale $m_\Delta$ are~\cite{Chao:2006ye, Schmidt:2007nq} 
\begin{equation}
\label{rges}
    \begin{aligned}
        16\pi^2\frac{d Y_e}{d\log{\mu}}~=&~ 3Y_eY_{\Delta}^{\dagger}Y_{\Delta}+Y_e\left(3y_t^2-\frac{9}{4}g_1^2-\frac{9}{4}g_2^2\right)\,,\\
        16\pi^2\frac{d Y_{\Delta}}{d\log{\mu}}~=&~6Y_{\Delta}Y_{\Delta}^{\dagger}Y_{\Delta} 
        +Y_{\Delta}\left(2\text{tr}\left[Y_{\Delta}^{\dagger}Y_{\Delta}\right]-\frac{9}{10}g_1^2-\frac{9}{2}g_2^2\right)\,,
    \end{aligned}
\end{equation}
where we neglect all the couplings much smaller than $\mathcal{O}(1)$, including $Y_e^{\dagger}Y_e\sim y_{\tau}^2$. 
If $\Lambda_\text{UV} > m_{\Delta}$, running effects from $\Lambda_\text{UV}$ down to $m_{\Delta}$ can change the $Y_\Delta$ flavor structure. The resulting matrices at $m_{\Delta}$ are:
\begin{equation}
\label{sol1}
\begin{aligned}
    Y_e(m_{\Delta})~=&~I_{tg}  Y_e(\Lambda_\text{UV})\cdot \left(\frac{m_{\Delta}}{\Lambda_\text{UV}}\right)^{\frac{3}{16\pi^2}\left(Y_{\Delta}^{\dagger}Y_{\Delta}\right)_{\text{eff}}}\,, \\
    Y_{\Delta}(m_{\Delta})~=& ~I_{\Delta g}\left(\frac{m_{\Delta}}{\Lambda_\text{UV}}\right)^{\frac{3}{16\pi^2}\left(Y_{\Delta}^{\dagger}Y_{\Delta}\right)^T_{\text{eff}}}\cdot  Y_{\Delta}(\Lambda_\text{UV})\cdot \left(\frac{m_{\Delta}}{\Lambda_\text{UV}}\right)^{\frac{3}{16\pi^2}\left(Y_{\Delta}^{\dagger}Y_{\Delta}\right)_{\text{eff}}}\,.
\end{aligned}
\end{equation}
Here, $I_{tg}$ and $I_{\Delta g}$ are overall factors accounting for the flavor-conserving contribution and are given by
\begin{equation}
\label{Idef}
\begin{aligned}
    I_{tg}~&=~ \left(\frac{\Lambda_\text{UV}}{m_{\Delta}}\right)^{\frac{1}{16\pi^2}\left(3y_t^2- \frac{9}{4}g_2^2- \frac{9}{4}g_1^2 \right)_{\text{eff}}}, \quad
    I_{\Delta g}~=~ \left(\frac{\Lambda_\text{UV}}{m_{\Delta}}\right)^{\frac{1}
    {16\pi^2}\left( \text{Tr} \left[Y_{\Delta}^{\dagger}Y_{\Delta}\right]-\frac{9}{10}g_1^2-\frac{9}{2}g_2^2 \right)_{\text{eff}}}.\\
\end{aligned}
\end{equation}
The effective matrix $\left(Y_{\Delta}^{\dagger}Y_{\Delta}\right)_{\text{eff}}$ can be interpreted as the average of $\left(Y_{\Delta}^{\dagger}Y_{\Delta}\right)$ between the scales $m_{\Delta}$ and $\Lambda_\text{UV}$. Its explicit definition is
\begin{equation}
\label{Ydef}
\left(Y_{\Delta}^{\dagger}Y_{\Delta}\right)_{\text{eff}}~=~\frac{\int^{\log{m_{\Delta}}}_{\log{\Lambda_\text{UV}}} \left(Y_{\Delta}^{\dagger}(\mu')Y_{\Delta}(\mu')\right) d\log{\mu'}}{\log{\left(m_{\Delta}/\Lambda_\text{UV}\right)}}, 
\end{equation}
and other effective couplings are defined analogously. Replacing the effective couplings with their values at $m_{\Delta}$ is a good approximation, since these couplings vary slowly and always appear together with the loop factor $1/(16\pi^2)$. 
The matrix functions in Eq.~\eqref{sol1} are defined as: 
\begin{equation}
\label{matrixpower}
     \left(\frac{m_{\Delta}}{\Lambda_\text{UV}}\right)^{\frac{3}{16\pi^2}\left(Y_{\Delta}^{\dagger}Y_{\Delta}\right)_{\text{eff}}}~=~\exp{\left[\frac{3}{16\pi^2}\log{(m_{\Delta}/\Lambda_{\text{UV}})}\left(Y_{\Delta}^{\dagger}Y_{\Delta}\right)_{\text{eff}}\right] }.
\end{equation}
Here, $\exp{M}\equiv\mathds{1}+M+\frac{1}{2!}M\cdot M+...$ can be evaluated as a matrix power expansion when $M$ is a square matrix. 
In Eq.~\eqref{sol1}, $Y_e(m_{\Delta})$ is no longer diagonal. To quantify how radiative corrections destabilize the zero textures defined in the charged lepton basis, one must further diagonalize $Y_e$ by rotating the $\ell_{L}$ and $E_{R}$ basis, which also modifies $Y_{\Delta}(m_{\Delta})$. Assuming $Y_{\Delta}$ to vary slowly with respect to $\mu$, we derive the leading-log expansion for $Y_{\Delta}(m_{\Delta})$ in the charged lepton basis
\begin{equation}
\label{Ydiag}
    Y_{\Delta}(m_{\Delta})=I_{\Delta g} \left(Y_{\Delta}(\Lambda_\text{UV}) +\frac{3}{16\pi^2}\log\left(\frac{m_{\Delta}}{\Lambda_\text{UV}}\right) \left(2Y_{\Delta} Y_{\Delta}^{\dagger}Y_{\Delta}-(Y_{\Delta}^{\dagger}Y_{\Delta})_{ij}\left[Y_{\Delta}, \mathcal{O}_{\{ij\}}\right]  \right)(\Lambda_\text{UV}) \right). 
\end{equation}
To obtain this expression, we take into account the hierarchy $m_{\tau}\gg m_{\mu}\gg m_e$, which implies that $Y_e(m_{\Delta})$ in Eq.~\eqref{sol1} can be diagonalized by rotating the $\ell_{L}$ basis only. Here, $\mathcal{O}_{\{ij\}}$ are matrices defined in terms of the Gell-Mann matrices~\cite{Gell-Mann:1962yej} as $\mathcal{O}_{\{12\}}=i\Lambda_{[12]}$, $\mathcal{O}_{\{13\}}=i\lambda_5$, $\mathcal{O}_{\{23\}}=i\lambda_7$, while the others are zero. 
Below $m_{\Delta}$, $Y_{\Delta}$ is matched to $M^{\nu}$, and the texture zeros are absolutely stable~\cite{Hagedorn:2004ba, Fritzsch:2011qv}.

We proceed assuming that $Y_{\Delta}(\Lambda_\text{UV})$ displays one of the zero textures in Eq.~(\ref{onezero}) and Eq.~(\ref{twozero}), plug it into the solution Eq.~(\ref{Ydiag}), and check whether such texture is preserved for $Y_{\Delta}(m_{\Delta})$. 
As expected, we find most of other zero textures we considered are not RG stable. The elements that are zero at $\Lambda_\text{UV}$ receive additive corrections at one loop. These radiative contributions must be included in the phenomenological analysis and, as we will see, introduce some dependence of the results on the cutoff scale $\Lambda_\text{UV}$. 
Interestingly, we find that the zero textures $\mathcal{A^{}}$ and $\mathbf{A_1}$ remain stable at any scale below $\Lambda_\text{UV}$. 
Starting with vanishing $Y_{\Delta ee}(\Lambda_\text{UV})$ and $Y_{\Delta\mu e}(\Lambda_\text{UV})$, we obtain $Y_{\Delta ee}(m_{\Delta})=Y_{\Delta \mu e}(m_{\Delta})=0$, hence the RG running does not give rise to these vanishing entries. In addition, if requiring only $Y_{\Delta ee}(\Lambda_\text{UV}) =0$, the conclusion $Y_{\Delta ee}(m_{\Delta})=0$ remains. 
Although we cannot identify a simple symmetry responsible for this result, we note that one-loop RG stability is not necessarily enforced by symmetries. In some cases, special flavor structures can also forbid loop corrections up to certain orders. Known examples include the rank of the neutrino mass matrix~\cite{Zhang:2024weq} and $CP$ violation in extended Higgs sectors~\cite{Fontes:2021znm, deLima:2024lfc}. 

\section{The Low Energy Predictivity}
\label{texturesection}

\begin{figure}[!t]
    \centering
    \includegraphics[width=0.9\linewidth]{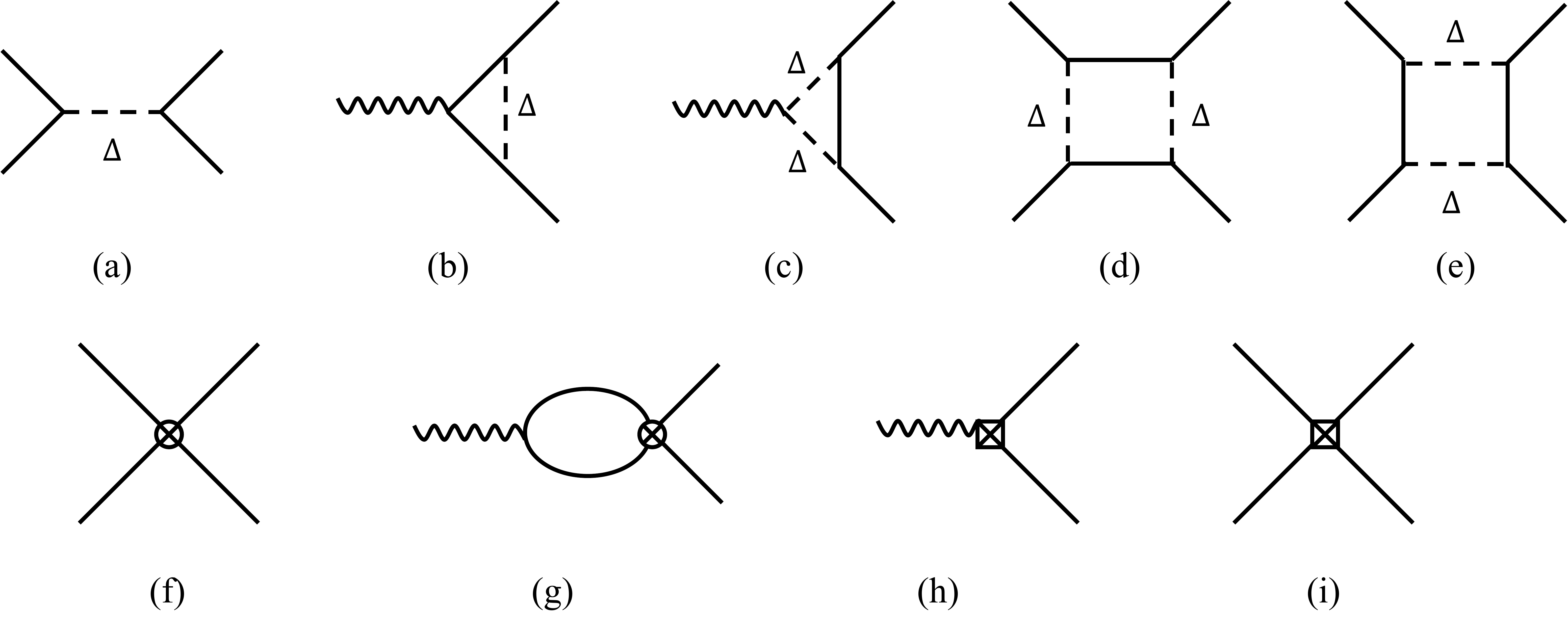}
    \caption[The CLFV Feynman diagrams.]{(a)-(e): Feynman diagrams contributing to CLFV processes within the full type~II seesaw model. (f)-(i): Corresponding Feynman diagrams in the effective theory at low-energy scales, $q^2\ll m_{W}^2$. We do not apply the on-shell condition so a pair of
    quarks or charged leptons can be attached to the external gauge boson lines. Self-energy diagrams are not included for simplicity.}
    \label{Feynmandiagrams}
\end{figure}

We demonstrate the predictive power of theories in the swampland of flavour symmetries, using the type~II seesaw model with two zero entries in $Y_{\Delta}$. 
We take the $L$-conserving limit through this section, so that the $\Delta$ is required relatively light and some entries of $Y_{\Delta}$ are sizable.
The doubly-charged state of the triplet directly contributes to processes such as $\mu\to \bar e ee$ via the diagram (a) of Figure~\ref{Feynmandiagrams}. 
Upon integrating out $\Delta$, this amplitude matches to the following low-energy CLFV operators:
\begin{equation}
\label{eftTree}
\delta\mathcal{L}^{\text{d=6}}_{\text{tree}}~=~ \frac{1}{2m_{\Delta}^2} Y_{\Delta jl}Y_{\Delta ik}^{*}(\overline{\ell_{i}}\gamma_{\mu}P_L\ell_{j}) (\overline{\ell_{k}}\gamma^{\mu} P_L\ell_{l})\,,
\end{equation}
where $\ell_i=e_{L\,i}+E_{R\,i}$ with $e_{L}$ ($E_{R}$) denoting the left-handed (right-handed) charged leptons and $i,j,k,l=e,\mu, \tau$ being flavor indices. 
For two-zero textures $\mathbf{B_1}$, $\mathbf{B_4}$, and $\mathbf{C}$, $Y_{\Delta\,e\mu}$ and $Y_{\Delta\,ee}$ are both non-vanishing so $\mu\to \bar e ee$ decay occur without any suppression. 
This is not the case for the other textures, that $\mu\to \bar e ee$ vanishes at the tree level. 
Except for the ones shown in Eq.~\eqref{eftTree}, all other Wilson coefficients of the dimension-six operators are proportional to $\mu_\Delta$ at tree level and should be ignored in our $L$-conserving limit~\cite{Li:2022ipc}. 
In other words, for $\mathbf{A_1}$, $\mathbf{A_2}, \mathbf{B_2}$, and $\mathbf{B_4}$, there are no unsuppressed operators inducing the $\mu\to e$ transition.
This interesting situation is in stark contrast to the standard expectation for type~II seesaw with a anarchic flavor structure of $Y_\Delta$, where the strong limit set by $\mu\to \bar eee$ typically prevents the rates of other CLFV modes (especially for $\tau$ leptons) from resulting in the observable range; see~e.g.~\cite{Abada:2007ux,Dinh:2012bp} for detailed discussions based on the generic flavor structure.

The zero-textures of $Y_{\Delta}$ motivates us to carefully re-examine the operators beyond tree-level matching. 
At one-loop level, a set of new CLFV operators arise and can be sizable. Comparing with the tree-level operator shown in Eq.~(\ref{eftTree}), these one-loop operators are also at leading order because of the different flavor structures. 
At low energy scales $q^2\sim m_{\mu}^2$ or $q^2\sim m_{\tau}^2$, the EFT relevant for CLFV is
\begin{equation}
\label{eft}
\begin{aligned}
    \delta\mathcal{L}^{\text{d=6}}_{\text{FV}}~=&~\mathcal{C}_{\text{dipole}}^{ik} \left(\overline{\ell_{i}}\sigma^{\mu\nu} P_R \ell_{k}\right) F_{\mu\nu}+\mathcal{C}_{\text{penguin}}^{ik}(\overline{\ell_{i}}\gamma_{\mu}P_L\ell_{k}) \left(\overline{\ell_{l}}\gamma^{\mu} \ell_{l}+\frac{2}{3}\overline{u_{l}}\gamma^{\mu}  u_{l}-\frac{1}{3}\overline{d_{{l}}}\gamma^{\mu} d_{l}\right)\\
    ~&+~ \mathcal{C}_{\text{box}}^{ijkl}(\overline{\ell_{i}}\gamma_{\mu}P_L\ell_{j}) (\overline{\ell_{k}}\gamma^{\mu} P_L\ell_{l})\,, 
\end{aligned}
\end{equation}
where 
\begin{equation}
\label{eftWilson}
    \begin{aligned}
        \mathcal{C}_{\text{penguin}}^{ik}~=&~\frac{e^2}{12\pi^2 m_{\Delta}^2} Y_{\Delta\,ij}Y_{\Delta\,kj}^{*}\left(\frac{1}{12}+f\left(\frac{q^2}{m_{\Delta}^2}\,,\frac{m_j^2}{m_{\Delta}^2}\right)\right),\\
        \mathcal{C}_{\text{dipole}}^{ik}~=&~\frac{3e }{64 \pi^2 m_{\Delta}^2} (Y_{\Delta}Y_{\Delta}^{\dagger})_{ik} m_k\,, \\
        \mathcal{C}_{\text{box}}^{ijkl}~=&~ -\frac{1}{32\pi^2m_{\Delta}^2}(Y_{\Delta}Y_{\Delta}^{\dagger})_{ij}(Y_{\Delta}Y_{\Delta}^{\dagger})_{kl}-\frac{1}{8\pi^2m_{\Delta}^2}(Y_{\Delta}Y_{\Delta}^{\dagger})_{il}(Y_{\Delta}Y_{\Delta}^{\dagger})_{kj}\,.\\
    \end{aligned}
\end{equation}
These one-loop matching is also discussed in~\cite{Dinh:2012bp, Li:2022ipc, Ardu:2023yyw}. The corresponding Feynman diagrams of the full renormalizable theory are illustrated in the panels (b)-(e) of Figure~\ref{Feynmandiagrams}. Notice that we do not use the standard operator basis (see e.g.~\cite{Jenkins:2017jig}) of the low-energy effective field theory (LEFT), in order to keep a clear correspondence between operators and Feynman diagrams.
%
Both diagram (b) and (c) contribute to $\mathcal{C}_{\text{penguin}}$ and complete form of the loop function $f$ appearing in Eq.~\eqref{eftWilson} reads~\cite{Raidal:1997hq}
\begin{equation}
\label{loopfunctionfull}
\begin{aligned}
    f\left(\frac{-q^2}{m_{\Delta}^2},\frac{m_j^2}{m_{\Delta}^2}\right)~=&~\log\left(\frac{m_j^2}{m_{\Delta}^2}\right)-\frac{4 m_j^2}{q^2}\\
    &+\left(\left(1+\frac{2 m_j^2}{q^2}\right)\sqrt{1-\frac{4 m_j^2}{q^2}} \log\left( \frac{\sqrt{1-\frac{4 m_j^2}{q^2}}+1}{\sqrt{1-\frac{4 m_j^2}{q^2}}-1}\right)\right)\,.
\end{aligned}
\end{equation}
In the limit $q^2 \lesssim m_j^2 \ll m_{\Delta}^2$ (corresponding to a $\mu$ decay through $\mu$ or $\tau$ in the loop), the loop function $f$ is reduced to~\cite{Raidal:1997hq}:
\begin{equation}
\label{loopf}
    f\left(\frac{-q^2}{m_{\Delta}^2},\frac{m_j^2}{m_{\Delta}^2}\right)~=~\log\left(\frac{m_j^2}{m_{\Delta}^2}\right)+\frac{5}{3}\,, \quad m_j=m_{\mu}~\text{or}~m_{\tau}\,.
\end{equation}
The threshold correction $5/3$ is absent if $-q^2\gg m_j^2$, which corresponds to the electron loop contributing to the $\mu\rightarrow \bar e  ee $ decay.
The large mass hierarchy ensures that varying $m_{\Delta}$ from 1 to 10 TeV does not significantly change the numerical value of Eq.~\eqref{loopf}, so for simplicity it is a good approximation to take $m_{\Delta}=3$ TeV when evaluating this $\log$ function. In the view of the effective theory, the $\log$ function contained in $\mathcal{C}_{\text{penguin}}$ encodes the RG running of the tree-level WCs in diagram (f). This running evolves from $m_{\Delta}$ to $m_{\mu}$ or $m_{\tau}$, and its anomalous dimension can be calculated from diagram (g) of Figure~\ref{Feynmandiagrams}. Diagrams (h) and (i) cancel the divergent part of diagram (g) as counter terms. Their finite parts are matched from diagram (b)-(e) and lead to $\mathcal{C}_{\text{dipole}}$ and $\mathcal{C}_{\text{box}}$ in Eq.~\eqref{eftWilson}.

\begin{table}[t!]
\renewcommand\arraystretch{2.5}
\resizebox{\textwidth}{!}{
    \begin{tabular}{l c  c  c }
    \toprule
       Observable  & Tree-level ($\times m_{\Delta}^4G_F^2 $) & One-loop ($\times 10^5 \,m_{\Delta}^4G_F^2$)  & 90~\% CL UL \\
    \midrule
      $\text{BR}(\mu\xrightarrow{}\overline{e}ee)$   & $0.25\left|Y_{\Delta \mu e}Y_{\Delta ee}\right|^2$ & $\left(25\left|(Y_{\Delta}Y_{\Delta}^{\dagger})_{ee}  \right|^2+19\, \text{Re}\left[(Y_{\Delta}Y_{\Delta}^{\dagger})_{ee} \right]+5.8\right)  \left|(Y_{\Delta}Y_{\Delta}^{\dagger})_{e\mu}  \right|^2 $ & $1.0\times 10^{-12}$~\cite{SINDRUM:1987nra}\\
      $\text{BR}(\tau\xrightarrow{}\overline{e}ee)$   & $0.085\left|Y_{\Delta \tau e}Y_{\Delta ee}\right|^2$ & $\left(25\left|(Y_{\Delta}Y_{\Delta}^{\dagger})_{ee}  \right|^2+19\, \text{Re}\left[(Y_{\Delta}Y_{\Delta}^{\dagger})_{ee} \right]+6.3 \right)  \left|(Y_{\Delta}Y_{\Delta}^{\dagger})_{e\tau}  \right|^2 $ & $2.5\times 10^{-8}$~\cite{Belle-II:2025urb}\\
      $\text{BR}(\tau\xrightarrow{}\overline{e}e\mu)$   & $0.043\left|Y_{\Delta \tau e}Y_{\Delta e \mu}\right|^2$ & $\left(0.085\left|(\mathcal{Y}_{\Delta}\mathcal{Y}_{\Delta}^{\dagger})_{ee}  \right|^2+0.32\,\text{Re}\left[{(\mathcal{Y}_{\Delta}\mathcal{Y}_{\Delta}^{\dagger})}_{ee} \right]+0.76\right)  \left|(Y_{\Delta}Y_{\Delta}^{\dagger})_{\mu\tau}  \right|^2 $ & $1.6\times 10^{-8}$~\cite{Belle-II:2025urb} \\
      $\text{BR}(\tau\xrightarrow{}\overline{\mu}ee)$   & $0.043\left|Y_{\Delta \tau\mu}Y_{\Delta ee}\right|^2$ &  $25\left|(Y_{\Delta}Y_{\Delta}^{\dagger})_{e\mu}\right|\left|(Y_{\Delta}Y_{\Delta}^{\dagger})_{e\tau}\right|^2$ & $1.5\times 10^{-8}$~\cite{Hayasaka:2010np}\\
      $\text{BR}(\tau\xrightarrow{}\overline{e}\mu\mu)$   & $0.043\left|Y_{\Delta \tau e}Y_{\Delta \mu\mu}\right|^2$ & 25$\left|(Y_{\Delta}Y_{\Delta}^{\dagger})_{\mu e}\right|\left|(Y_{\Delta}Y_{\Delta}^{\dagger})_{\mu\tau}\right|^2$ & $1.3\times 10^{-8}$~\cite{Belle-II:2025urb}\\
      $\text{BR}(\tau\xrightarrow{}\overline{\mu}\mu e)$   & $0.085\left|Y_{\Delta \tau\mu}Y_{\Delta \mu e}\right|^2$ & $\left(0.085\left|(\mathcal{Y}_{\Delta}\mathcal{Y}_{\Delta}^{\dagger})_{\mu\mu}  \right|^2+0.32\,\text{Re}\left[(\mathcal{Y}_{\Delta}\mathcal{Y}_{\Delta}^{\dagger})_{\mu\mu} \right]+0.63\right) \left|(Y_{\Delta}Y_{\Delta}^{\dagger})_{e\tau}  \right|^2 $ & $2.4\times 10^{-8}$~\cite{Belle-II:2025urb}\\
      $\text{BR}(\tau\xrightarrow{}\overline{\mu}\mu\mu)$   & $0.043\left|Y_{\Delta \tau\mu}Y_{\Delta \mu\mu}\right|^2$ & $\left(4.3\left|(Y_{\Delta}Y_{\Delta}^{\dagger})_{\mu\mu}  \right|^2+3.2 \,\text{Re}\left[(Y_{\Delta}Y_{\Delta}^{\dagger})_{\mu\mu} \right]+0.93\right)  \left|(Y_{\Delta}Y_{\Delta}^{\dagger})_{\mu\tau}  \right|^2 $ & $1.9\times 10^{-8}$~\cite{Belle-II:2024sce}\\
      \midrule
    $\text{BR}(\mu\xrightarrow[]{}e\gamma)$ & 0 &$98 \left|(Y_{\Delta}Y_{\Delta}^{\dagger})_{e\mu}\right|^2$  & $1.5\times 10^{-13}$~\cite{MEGII:2025gzr}\\
    $\text{BR}(\tau\xrightarrow[]{}e\gamma)$ & 0 & $17\left| (Y_{\Delta}Y_{\Delta}^{\dagger})_{e\tau}\right|^2 $ & $3.3\times 10^{-8}$~\cite{BaBar:2009hkt}\\
    $\text{BR}(\tau\xrightarrow[]{}\mu\gamma)$ & 0 & $17 \left|(Y_{\Delta}Y_{\Delta}^{\dagger})_{\mu\tau}\right|^2$ & $4.2\times 10^{-8}$~\cite{Belle:2021ysv}\\
    \midrule
    $\text{CR}(\mu\text{Au}\xrightarrow[]{} e \text{Au})$ & 0 &  $71 \left|(Y_{\Delta}Y_{\Delta}^{\dagger})_{e\mu}\right|^2$ & $7.0\times 10^{-13}$~\cite{SINDRUMII:2006dvw}\\
    $\text{CR}(\mu\text{Al}\xrightarrow[]{} e \text{Al})$ & 0 &  $32 \left|(Y_{\Delta}Y_{\Delta}^{\dagger})_{e\mu}\right|^2$ & ---\\
    \bottomrule
    \end{tabular}}
    \caption[Branching ratios of various CLFV processes in the type~II seesaw model.]{Branching ratios of various CLFV processes in minimal type~II seesaw. 
    The last column displays the current experimental 90~\% confidence level upper limits. The notation $(\mathcal{Y}_{\Delta}\mathcal{Y}_{\Delta}^{\dagger})_{ee/\mu\mu}$ is defined in Eq.~\eqref{calY}.}
    \label{3body}
\end{table}

As a result, although $\mu\to \bar e ee$ still arises in texture $\mathbf{A_1}$, $\mathbf{A_2}, \mathbf{B_2}$, and $\mathbf{B_4}$, the branching ratios (BRs) are suppressed by a squared loop factor $1/(4\pi)^4=4\times10^{-5}$. 
$\mu\to \bar e ee$ is no longer the leading probe for type~II seesaw. 
Due to the dipole operator, $\mu\rightarrow e\gamma$ becomes also phenomenologically relevant. 
For all textures, its branching ratio is proportional to $(Y_{\Delta}Y_{\Delta}^{\dagger})_{\mu e}$ and thus directly connected to the neutrino oscillation parameters by the following expression~\cite{Ardu:2023yyw}
\begin{equation}
\label{universal}
    \text{BR}(\mu\to e\gamma)~\propto~\left|(Y_{\Delta}Y_{\Delta}^{\dagger})_{\mu e}\right|^2~\propto~ \left(\Delta m_{13}^2 \cos{\theta_{13}}\sin{\theta_{13}}\sin{\theta_{23}} \right)^2\,.
\end{equation}
This relation is a good approximation in the limit $\Delta m_{12}^2 \ll \Delta m_{13}^2$. 
Eq.~\eqref{universal} does not contain the absolute neutrino masses and CP violation phases, and is valid for both normal and inverted ordering. 
Consequently, $\text{BR}(\mu\to e\gamma)$ only weakly depends on the unknown parameters in $M^{\nu}$ and is non-vanishing for all textures. 
In principle, simultaneously observing a multiple CLFV processes, including $\mu\to \bar e ee$, $\mu\to e\gamma$, and, in some cases, tree-level $\tau$ CLFV decays, becomes possible at current and upcoming experiments under these two-zero textures. 
We calculate the BRs of these CLFV processes by inserting the Wilson coefficients in Eq.~\eqref{eftTree} and Eq.~\eqref{eft} into the expressions provided in Ref~\cite{Kuno:1999jp} and show the results in Table~\ref{3body}. The one-loop BRs are calculated in the limit that the tree-level amplitudes are zero (as the former would be subdominant otherwise), so interference terms are not displayed. However, in the following numerical analysis, these contributions are included when they are relevant. For simplicity, we take $m_{j}=m_{\tau}$ to evaluate Eq.~\eqref{loopf}, which is a good approximation for all one-loop $\tau$ decay BRs.\footnote{Instead, for what concerns $\text{BR}(\mu\to \bar e ee)$, the expression of the table should be modified by adding the term $\left(25\left|(Y_{\Delta}Y_{\Delta}^{\dagger})_{ee} \right|^2+26\,\text{Re}\left[(Y_{\Delta}Y_{\Delta}^{\dagger})_{ee} \right]+10 \right) \left|(Y_{\Delta}Y_{\Delta}^{\dagger})_{e\mu}  \right|^2$ to account for $\mu$ running in the loop, which enhances the log term by a factor $\approx 1.8$.} 
Moreover, we define:
\begin{equation}
\label{calY}
    \begin{aligned}
    (\mathcal{Y}_{\Delta}\mathcal{Y}_{\Delta}^{\dagger})_{\mu\mu}~=~&(Y_{\Delta}Y_{\Delta}^{\dagger})_{\mu\mu}-4\frac{(Y_{\Delta}Y_{\Delta}^{\dagger})_{e\mu}(Y_{\Delta}Y_{\Delta}^{\dagger})_{\tau\mu}^*}{(Y_{\Delta}Y_{\Delta}^{\dagger})_{e\tau}}\,, \\
    (\mathcal{Y}_{\Delta}\mathcal{Y}_{\Delta}^{\dagger})_{ee}~=~& (Y_{\Delta}Y_{\Delta}^{\dagger})_{ee}-4\frac{(Y_{\Delta}Y_{\Delta}^{\dagger})_{\mu e}(Y_{\Delta}Y_{\Delta}^{\dagger})_{\tau e}^*}{(Y_{\Delta}Y_{\Delta}^{\dagger})_{\mu \tau}}\,,\\
    \end{aligned}
\end{equation}
in order to simplify the expressions of box diagram contributions.

As an $SU(2)_L$ triplet, $\Delta$ also couple to the $W$, $Z$, and $H$ bosons, thus giving rise to the following SMEFT operators 
\begin{equation}
\label{EFTZpole}
\begin{aligned}
    \delta\mathcal{L}^{\text{d=6}}_{\text{Z,H-FV}}~=&~
    \mathcal{C}_{eW}(\overline{\ell_L}\sigma^{\mu\nu}e_R)\tau^IHW_{\mu\nu}^I+
    \mathcal{C}_{eB}(\overline{\ell_L}\sigma^{\mu\nu}e_R)HB_{\mu\nu}+
    \mathcal{C}_{eH}(H^{\dagger}H)(\overline{\ell_L}e_RH)\\
    &~+\mathcal{C}^{(3)}_{H\ell}(H^{\dagger}\tau^Ii\overset{\leftrightarrow}{D}_{\mu} H)(\overline{\ell_L}\tau^I\gamma^{\mu}\ell_L)+
    \mathcal{C}^{(1)}_{H\ell}(H^{\dagger}i\overset{\leftrightarrow}{D}_{\mu} H)(\overline{\ell_L}\gamma^{\mu}\ell_L)\,.
\end{aligned}
\end{equation}
At the scale below $m_W$, these operators are matched to the LEFT operators and their contributions are already included in the coefficients of dipole and four-fermion operators shown in Eq.~(\ref{eftTree}) and Eq.~(\ref{eft})\footnote{It is worth remarking that they are the LEFT operators instead of the SMEFT ones.}. 
Around the $m_W$ scale, the operators in Eq.~\eqref{EFTZpole} can induce flavor-violating $Z$ and $H$ decays.
Compared to the CLFV decays of $\mu$ and $\tau$ leptons, the $Z$-pole and Higgs observables such as $Z \to \bar \ell_i \ell_j$ currently provide weaker constraints on the set of coefficients \{$\mathcal{C}_{eW},\mathcal{C}_{eB},\mathcal{C}_{eH},\mathcal{C}^{(3)}_{H\ell}, \mathcal{C}^{(1)}_{H\ell}$\}~\cite{Calibbi:2021pyh}, so we do not include them in the following analysis. 
In the limit $\mu_\Delta\to 0$, quark flavor-violating operators do not arise at tree and one-loop level~\cite{Li:2022ipc}. As a consequence, constraints from lepton flavor universality (LFU) tests in flavor-changing hadronic decays, such as the bounds from $R(K^{(*)})\equiv\Gamma(B\to K^{(*)}\mu^+\mu^-)/\Gamma(B\to K^{(*)}e^+e^-)$~\cite{Covone:2025lee}, are automatically evaded.


All other operators are flavor conserving and the resulting constraints are in general weaker than those from CLFV processes.
In addition, almost all flavor conserving operators are suppressed by $\mu_\Delta$, loop factors, and/or chirality flipping terms~\cite{Li:2022ipc}.
We do not discuss these phenomenologically irrelevant observables here.
The only exception is provided by the lepton flavor conserving but LFU violating four-fermion operators in Eq.~\eqref{eftTree}, such as $(\overline{\mu}\gamma^{\mu}P_L \tau)(\overline{\tau}\gamma_{\mu}P_L \mu)$. The LFU tests in $\tau$ decays can constrain the cut-off scales of such kind of operators to be above $4.5$~TeV~\cite{Allwicher:2023shc}. As we will see, this bound is still weaker than those we obtain from the CLFV $\tau$ decays.

\begin{table}[t!]
\renewcommand\arraystretch{2.}
    \centering
$\begin{array}{lccccccc}
\toprule
 ~& \bf{A_1}& \bf{A_2}& \bf{B_1} & \bf{B_2}& \bf{B_3}& \bf{B_4} & \bf{C} \\
 \midrule
 \mu \to \bar{e}ee & 8.3^{+ 0.6}_{-0.5} & 10.1^{+ 0.8}_{-0.9} & \textcolor{red}{22^{+ 7}_{-9}} & 3.3^{+ 0.7}_{-0.8} & 3.1^{+ 0.5}_{-0.6} & \textcolor{red}{20^{+ 5}_{-9}} &
   \textcolor{red}{75^{+ 49}_{-58}} \\
 \mu \to e\gamma  & \textcolor{red}{28.0^{+ 2.3}_{-1.9}} & \textcolor{red}{29.4^{+ 2.9}_{-4.0}} & 9.3^{+ 2.8}_{-4.0} & \textcolor{red}{9.8^{+ 3.5}_{-5.0}} & \textcolor{red}{8.9^{+ 2.6}_{-4.0}} & 9.5^{+ 3.3}_{-5.0} & 38^{+ 18}_{-35} \\
 \tau \to \bar{\mu }\mu \mu  & 6.1^{+ 0.7}_{-0.7} & 6.0^{+ 0.4}_{-0.4} & 0.5^{+0.2}_{-0.2} & 2.9^{+ 0.9}_{-1.3} & 0.2^{+0.2}_{-0.2} & 2.7^{+ 0.8}_{-1.3} & 0.4^{+0.5}_{-0.3} \\
 \tau \to \bar{\mu }ee & 0.64^{+0.04}_{-0.06} & 0.63^{+0.05}_{-0.06} & 5.7^{+ 0.2}_{-0.2} & 5.7^{+ 0.2}_{-0.2} & 5.7^{+ 0.2}_{-0.2} & 5.7^{+ 0.2}_{-0.2} &
   6.4^{+ 1.5}_{-1.7} \\
   \bottomrule
\end{array} $
    \caption[The lower limit on the effective cut-off scale of the type~II seesaw model.]{Central values and $3\sigma$ uncertainties for the lower limit on the type~II seesaw scale $\Lambda_{\Delta}\equiv m_{\Delta}/(2|Y_{\Delta\mu\tau}|)$ (in TeV) set by various CLFV processes for the two-zero textures defined in Eq.~\eqref{twozero}. The uncertainties originate from the neutrino oscillation parameters. For each texture, the strongest constraint is marked in red. }
    \label{cutoff}
\end{table}

The discussions above implies that in type~II seesaw model, the induced processes rare or forbidden in the SM can all be suppressed by the zero textures. 
We therefore expect the theory, although not explicitly protected by symmetries, can lie at relatively low scales, ideally $\mathcal{O}(\text{TeV})$. 
To quantify it, we define the type~II seesaw scale as
\begin{equation}
\label{cut}
    \Lambda_{\Delta}~\equiv~\frac{m_{\Delta}}{2|Y_{\Delta \mu\tau}|}\,,
\end{equation}
where the factor 2 accounts for the fact that $Y_{\Delta\mu\tau}=Y_{\Delta\tau\mu}$ and both entries lead to the same interaction term. 
Across all seven allowed two-zero textures, $Y_{\Delta\mu\tau}$ is always non-vanishing, and the 
other non-zero Yukawa couplings are be connected to $Y_{\Delta \mu\tau}$ by neutrino oscillation observables. 
In other words, all information on flavour are contained in the zero textures, as in the symmetry protected patterns, so $\Lambda_{\Delta}$ can serve as the over-all cut-off scale for the low-energy theory, and $\Lambda_{\Delta}\to \infty$ restores the SM. 
For every CLFV process, its experimental constraint can then be interpreted as a lower limit on $\Lambda_{\Delta}$. 
In Table~\ref{cutoff}, we summarize the $\Lambda_{\Delta}$ lower limits within the seven two-zero textures, derived from $\mu\to\bar e ee, \mu\to e\gamma, \tau\to\bar\mu\mu\mu,$ and $\tau\to\bar\mu ee$. 
To generate this table, we firstly fix $m_{\Delta}=3$ TeV, since varying $m_{\Delta}$ within the range $1-10$~TeV does not significantly impact on the numerical value of Eq.~\eqref{loopf}.
Then, we use the expressions shown in Table~\ref{3body} and obtain the various CLFV BRs as a function of $Y_{\Delta\mu\tau}$. Here, we apply the one-loop expressions only when the tree-level contribution vanishes. 
Finally, we compare the calculated CLFV BRs with the experimental limits collected in the last column of Table~\ref{3body} and get constraints for $Y_{\Delta\mu\tau}$, which, after combined with $m_{\Delta}=3$ TeV, are translated into the lower limits on $\Lambda_{\Delta}$.
We further quantify the uncertainties for these lower limits.
For this purpose, we utilize the generated $10^4$ sets of seven-types of $Y_{\Delta}$, as explained in Section~\ref{modelsection}. 
For each texture and process, we get $10^4$ lower bounds on $\Lambda_\Delta$ as outputs. 
The uncertainty displayed for each entry of the table represents the range that 99.7~\% of the output numbers lie within ($3\sigma$ range). 
As we can see, the variance for texture $\mathbf{C}$ is especially sizable because, within this texture, the entry $Y_{\Delta\mu e}$ can be tuned to be small when the CPV phase $\delta$ takes specific values. 
In Table~\ref{cutoff}, we highlight in red the strongest bound for each texture, in all cases coming from $\mu\to e$ transitions. 
In particular, Table~\ref{cutoff} also shows that $\mu\to e\gamma$ provides tight constraints for all the textures, although up to large uncertainties. 
These results inform us that the textures $\mathbf{A_1}$, $\mathbf{A_2}$, $\mathbf{B_1}$, $\mathbf{B_4}$, and $\mathbf{C}$ are inconsistent with an effective scale $\Lambda_{\Delta}\lesssim 10$~TeV.
On the other hand, $\mathbf{B_2}$ and $\mathbf{B_3}$ are more effective in suppressing $\mu\to e$ transitions, so that they allow $m_\Delta$ to be as light as of $5-6$~TeV even for $Y_{\Delta\mu\tau}\approx 0.5$.
As we can see for $\mathbf{B_2}$ and $\mathbf{B_3}$, the scales probed by $\tau \to\bar \mu ee$ and $\mu\to e\gamma$ are comparable within uncertainties. This result implies observing both $\tau$ and $\mu$ CLFV decay simultaneously is possible in near future. 

\begin{figure}[t!]
    \centering
    \includegraphics[width=0.8\linewidth]{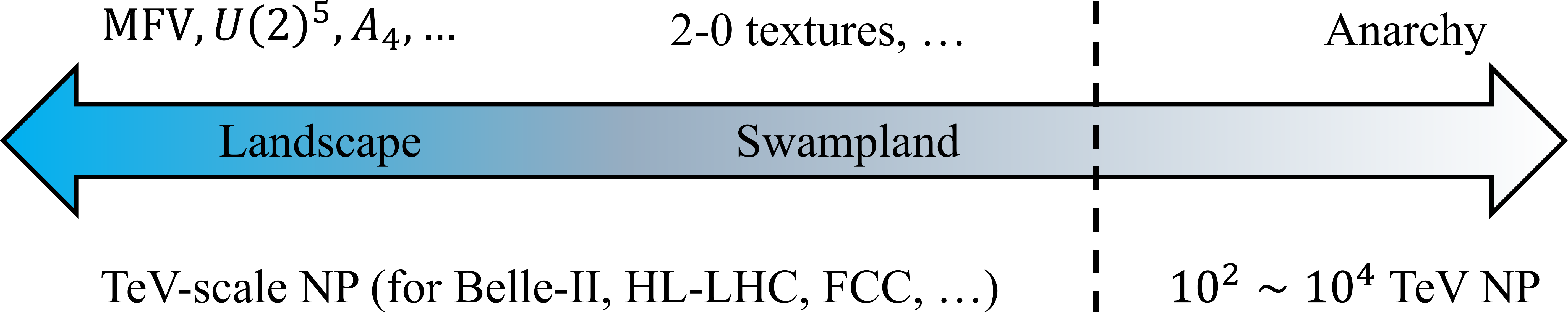}
    \caption[Representative textures for the theories in the landscape and the swampland.]{Representative textures for the theories lying in the landscape and the swampland of the flavour symmetries, and the lowest energy scale they can stay consistent with.}
    \label{TeVswampland}
\end{figure}

The $5\sim 6$ TeV cut-off scale is in stark contrast to what one expects from the anarchic flavour structure. 
In the type~II seesaw model, if all entries $Y_{\Delta}$ are of the same order of magnitude, 
the experimental constraints require $\Lambda_{\Delta}$ to be larger than $\mathcal{O}(100)$ TeV~\cite{Abada:2007ux,Dinh:2012bp}. 
Rather, the results under certain zero textures are more similar to those from the $U(2)^5$ EFT, which we introduced in Section~\ref{U2section}. 
The $U(2)$ symmetry forbids all CLFV transitions, unless the next-to minimal symmetry breaking spurions are included, see Ref.~\cite{Covone:2025lee} for a recent analysis. 
The strongest bound for the cutoff scale in the lepton sector is about 5~TeV, mainly following from LFU tests in $\tau$ decays~\cite{Allwicher:2023shc, Allwicher:2025bub}.
This bound can be further relaxed to values even closer to the TeV scale if the NP couplings to $e$ and $\mu$ are mildly suppressed~\cite{Allwicher:2023shc}. 
Here, we want to highlight that, for textures $\mathbf{B_2}$ and $\mathbf{B_3}$, constraints on the NP scale are comparable to those within the $U(2)^5$ framework. 
We regard these examples as critical evidence supporting the predictivity of the theories in the swampland of flavour symmetries. 
This result is illustrated in \fref{TeVswampland}.
The blue region represents the landscape of flavour symmetries, including the $U(2)^5$ invariant theories. The gray and white regions yield the swampland, in which the two-zero textures and flavour anarchic theories lie. 
The blue (landscape) and part of the gray (swampland) regions corresponds the TeV-scale new physics, relevant for present or future experiments such as Belle II, LHC, and FCC. 
The theories requiring $10^2\sim 10^4$ TeV cut-off scales are separated a the dashed line.

It is worthy to remark here that simply the zero entries in $Y_{\Delta}$ are insufficient to explain why $\Lambda_{\Delta}$ can become so low. 
The one-loop induced $\mu\to e\gamma$ transition is also tightly constrained by experiments, which does not vanish under the two-zero textures. 
Instead, we need the full pattern of $Y_{\Delta}$ to understand \tref{cutoff}. 
The coefficient for the dipole operator inducing $\mu\to e\gamma$ is proportional to $(Y_{\Delta}Y_{\Delta}^{\dagger})_{\mu e}$, which reduces to $Y_{\Delta\tau e}$ in texture $\mathbf{B_2}, \mathbf{B_3}$. 
These entries are smaller than $Y_{\Delta\mu\tau}$ by about a factor of 0.1, and thus suppress the $\mu\to e\gamma$ transition, allowing $\Lambda_\Delta \lesssim 10$ TeV. 
The explanation above will become more transparent if we interpret the two-zero textures as theories in the swampland of $A_4$ symmetry. 
All $\mu\to e$ transitions are forbidden by the $Z_3$ subgroup of $A_4$~\cite{Bigaran:2022giz, Lichtenstein:2023iut, Lichtenstein:2023vza, Palavric:2024gvu, Moreno-Sanchez:2025bzz}
\begin{equation}
    \label{Z3A4}
    \Delta\to \Delta, \quad e\to e, \quad \mu \to e^{i\frac{2\pi}{3}}\mu, \quad  \tau\to e^{i\frac{4\pi}{3}}\tau.
\end{equation}
With texture $\mathbf{B_2}$ and $\mathbf{B_3}$, 
the approximate $A_4$ symmetry is broken by $Y_{\Delta\mu\mu}$ and $Y_{\Delta\tau\tau}$, respectively, but such symmetry-breaking couplings are irrelevant for $\mu\to e\gamma$ at one-loop.
Therefore, the $\mathbf{B_2}$ and $\mathbf{B_3}$ textures inherit the $A_4$ suppression to flavour transition, although they correspond to more general theories. 
One can further argue that these textures as theories lying in the swampland of $U(1)_e$ symmetry, which, although explicitly broken by $Y_{\Delta ee}$, restricts $Y_{\Delta \mu\tau}$ to be small. 
We believe these examples represent an important feature, that the flavour symmetries do not only restrict the theories in its landscape, but also those in its swampland.

In addition, the tree-level $\mu\to \bar e e e$ amplitude in texture $\mathbf{B_1}, \mathbf{B_4},$ and $\mathbf{C}$ are suppressed by small $Y_{\Delta\mu e}$, which can similarly be regarded as originated from the \textit{explicitly broken} $A_4$ or $U(1)_e$ symmetries. 
The swampland argument then also explains that, although texture $\mathbf{B_1}, \mathbf{B_4},$ and $\mathbf{C}$ do not forbid tree-level $\mu\to \bar e e e$, 
their cut-off scales $\Lambda_{\Delta}$ can still stay at a relative low scale of $\mathcal{O}(10\sim20)$ TeV, which is nevertheless much lower than what one expects from the flavour anarchic patterns.

\section{Patterns of CLFV observables}
\label{phenosection}

\begin{figure}[t!]
  \centering
  \includegraphics[width=1.0\textwidth]{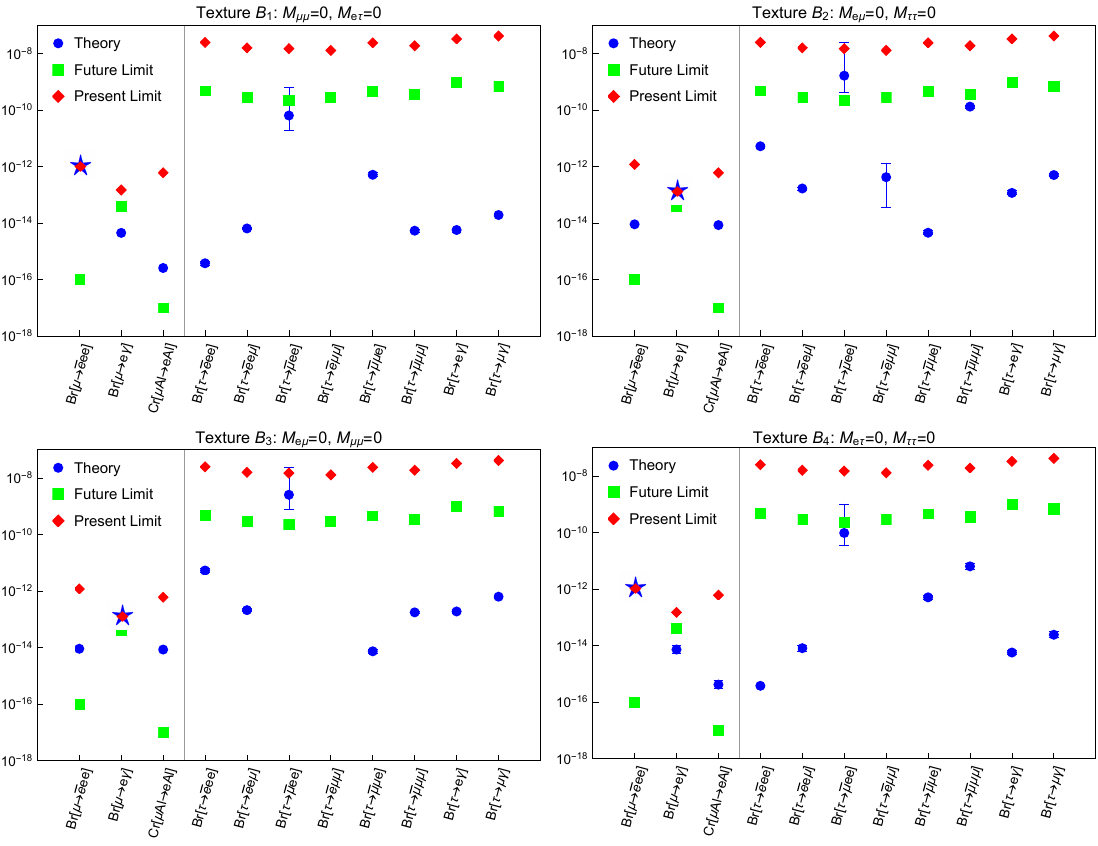}
      \caption[The CLFV branching ratios under textures $\mathbf{B_1}$-$\mathbf{B_4}$.]{Predicted branching ratios of various CLFV processes for textures $\mathbf{B_1}$-$\mathbf{B_4}$. The strongest constraint, $\text{BR}(\mu\to e\gamma)$ or $\text{BR}(\mu\to \bar e ee)$, is fixed to the value that saturates the current experimental limit (blue star). The predicted BRs for the other processes are denoted by blue dots with error bars corresponding to $3\sigma$ uncertainties from neutrino oscillation data. Red diamonds and green squares respectively indicate current and future experimental sensitivities. See the main text for details. }
   \label{B1B4}
\end{figure}

\begin{figure}[t!]
  \centering
  \includegraphics[width=1\textwidth]{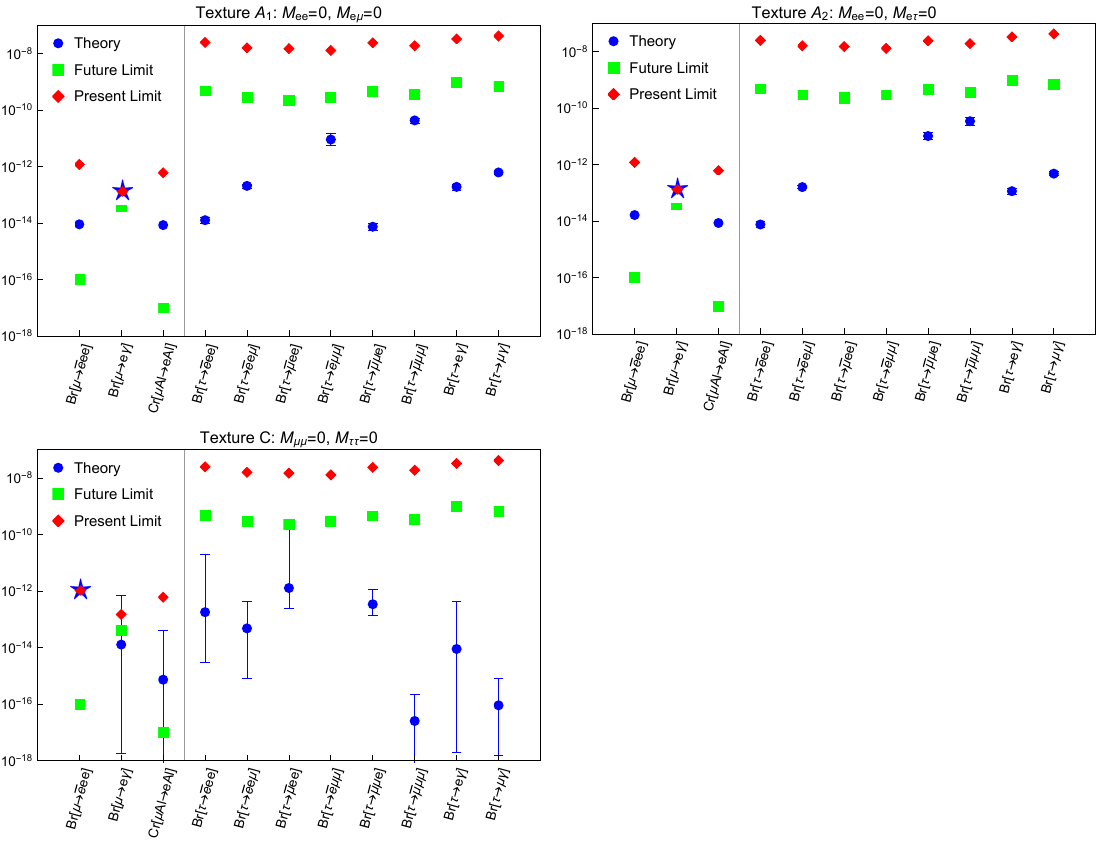}
    \caption[The CLFV branching ratios under textures $\mathbf{A_1}, \mathbf{A_2},$ and $\mathbf{C}$.]{Same as Figure~\ref{B1B4} for textures $\mathbf{A_1}, \mathbf{A_2}$ and $\mathbf{C}$.}
   \label{A1A2C}
\end{figure}

In this section, we study with more detail on distinctive patterns that CLFV observables can display in the type~II seesaw model, when $Y_\Delta$ features the two-zero textures. 
Our main results are shown in Figures~\ref{B1B4} and~\ref{A1A2C}, which illustrate the correlations among the relevant CLFV processes for the seven possible two-zero textures in Eq.~\eqref{twozero}, together with current and future experimental sensitivities. 
\tref{cutoff} indicates that $\text{BR}(\mu\to \bar e ee)$ provides the strongest constraints to textures $\mathbf{B_1}$, $\mathbf{B_4}$ and $\mathbf{C}$. 
In Figures~\ref{B1B4} and~\ref{A1A2C}, we set the prediction of this observable to its upper limit $1.2\times10^{-12}$, and mark it as a blue star. This number comes from the SINDRUM experiment in 1988~\cite{SINDRUM:1987nra} and is still the best. 
For the other textures, the strongest constraints come from $\text{BR}(\mu\to \bar e ee)$, which we set the theory prediction as $1.2\times10^{-12}$, same as the recent MEG~II limit~\cite{MEGII:2025gzr}. 
In order to illustrate correlations with other processes independent of the absolute size of $|Y_\Delta|$, we assume $|Y_{\Delta\mu\tau}|$ to be small enough that the terms proportional to $(Y_{\Delta}Y_{\Delta}^{\dagger})_{ee}$ contained in the expression for $\text{BR}(\mu\to \bar e ee)$ shown in Table~\ref{3body} do not dominant.
This is a loose condition and does not restore the $SU(3)_{\ell}$ symmetry: 
For $\mathbf{B_2}$, $\mathbf{B_3}$, it only requires $|Y_{\Delta\mu\tau}|\lesssim 0.7$. This number is relaxed to about $1.5$ for $\mathbf{A_1}, \mathbf{A_2}$.
With this assumption, all relevant CLFV BRs, no matter if they are dominated by tree-level or one-loop amplitudes, scale as $|Y_{\Delta\mu\tau}|^4$, and the ratios between different BRs are independent of the absolute size of the Yukawa couplings. 
In other words, all CLFV BRs within the same zero texture are connected by neutrino oscillation data only. In our plots, the error bars illustrate the $3\sigma$ uncertainties arising from oscillation parameters, which we calculate using the $10^4$ sets of $Y_{\Delta}$ introduced in Section~\ref{texturesection}. The red diamonds represent the current experimental limits for each process, as listed in Table~\ref{3body}. For $\mu \to e$ conversion in nuclei, we use the expressions in Table~\ref{3body} to rescale the bound on  $\text{CR}(\mu\,\text{Au}\to e \,\text{Au})$ translating it into a limit on $\text{CR}(\mu\,\text{Al}\to e \,\text{Al})$. Expected future sensitivities~\cite{MEGII:2018kmf,Blondel:2013ia,Kuno:2013mha,Mu2e:2014fns,Banerjee:2022xuw} are represented as green squares.

To explain the patterns shown in Figures~\ref{B1B4} and~\ref{A1A2C}, we firstly discuss about the results on $\mu\to e$ transitions. 
For texture $\mathbf{B_1}$ and $\mathbf{B_4}$, observing $\mu\to e\gamma$ within the sensitivity of the MEG~II experiment~\cite{MEGII:2018kmf} would then directly exclude these textures. 
Due to the large theoretical uncertainties, this is not the case for texture $\mathbf{C}$, and simultaneous observation of $\mu\to \bar e ee$ and $\mu\to e\gamma$ becomes possible. 
For the remaining textures, observing $\mu\to \bar e ee$ at the Mu3e experiment~\cite{Blondel:2013ia} which is projected to significantly improve sensitivity comparing with the SINDRUM 1988 result~\cite{SINDRUM:1987nra}, is not in conflict with the tight bound on $\mu\to e\gamma$. 
For all textures, as long as either $\mu\to e\gamma$ or $\mu\to \bar e ee$ are discovered, observing $\mu\to e$ conversion in nuclei would be ensured at the upcoming COMET~\cite{Kuno:2013mha} and Mu2e~\cite{Mu2e:2014fns} experiments.

Results in $\tau$ sector are particularly interesting. 
All $\mathbf B$ textures (and marginally the $\mathbf C$ texture) allow the possibility of detecting $\tau$ flavour violating decay in near future, even under the strict constraints from $\mu\to e$ transitions. 
Notably, textures $\mathbf{B_2}$ and $\mathbf{B_3}$ ensure the detection of $\tau$ flavour violating decay given that $\mu\to e \gamma$ is observed, 
and it is even possible to discover the CLFV decay of $\tau$ before that of $\mu$. 
A particular interesting pattern is that the detectable $\tau$ CLFV decay spectrum only contains a single processes, $\tau\to \bar \mu ee$, which features two same-sign electrons in the final state.
Observing any other $\tau$ CLFV decay process (for instance $\tau\to \bar e e \mu$, which contains two opposite-sign electrons) would immediately exclude all the considered textures.
This pattern is to be compared with the $U(2)^5$ invariant theories. 
The minimal $U(2)^5$ invariant EFT \typo{can not}{cannot} describe CLFV. 
Next-to minimal spurions must be added. 
Recently, Ref.~\cite{Covone:2025lee} point out that given the tightness of the bounds on processes involving electrons flavour changing, the $U(2)_{\ell}\times U(2)_e$ breaking spurions, if sizable, must be restricted to those couple to the third and second generation leptons only. 
Consequently, the $U(2)^5$ EFT allows relatively sizable $\tau\to \mu$ transition ($\tau\to \bar\mu \mu\mu$ and $\tau\to \bar e e\mu$), which can lie within the limit of future experiments despite the tight constraints for $\mu\to e$,
but restricts that $\tau\to \bar \mu e e $ must be small. 
Operators with more generic flavour textures are needed to describe $\tau\to \bar \mu e e $, for instance, the $SU(2)_q\times U(1)_X$ invariant operator $(\overline{\ell_L^1}\gamma_{\mu}\ell_L^2)(\overline{\ell_L^1}\gamma^{\mu}\ell_L^3)$ introduced in Eq.~\eqref{MFPoperator} of Section~\ref{beyondU2}. 
Therefore, the zero textures do not only allow sizable $\tau$ CLFV decay as what the third-generation specific EFT do, but can also describe processes which arise in more general frameworks.


The discussions above provides more theoretical motivations to search for $\tau$ CLFV in high-intensity frontier experiments, as a probe complementary to the currently running MEG II and the upcoming Mu3e, COMET, and Mu2e experiments.
We remark that such a task is tailor-made for Belle II~\cite{Banerjee:2022xuw,Belle-II:2018jsg}. 
As shown in the upper panel of \tref{tauexp}, up to $4.5\times 10^{10}$ $\tau^+\tau^-$ pairs are expected at Belle II, which are two and one orders of magnitudes more than those from BESIII and STCF, respectively. 
Future $e^+e^-$ collider such as CEPC can produce $\mathcal{O}(10^{11})$ tau pairs at $Z$-pole, which is a little more than Belle~II, but not by orders of magnitude.
Consequently, as demonstrated in the lower panel, Belle II can improve the current limit on BR($\tau \to \bar \mu\mu\mu$) by nearly two orders of magnitudes, almost as good as the projected sensitivity for CEPC. Significant improvements over Belle can be achieved only on FCC, but merely by around one order of magnitude. 
According to Ref.~\cite{Ai:2024nmn}, dedicated studies on $\tau\to\bar \mu ee$ are still missing, the sensitivities are expected similar to those for $\tau \to \bar \mu\mu\mu$.

\begin{table}[t!]
    \centering
    \resizebox{0.9\textwidth}{!}{\begin{tabular}[t]{ccccc}
    \toprule[1pt]
        Particle
        & BESIII & STCF (1 ab$^{-1}$) & Belle II (50 ab$^{-1}$)    & CEPC (TDR) \\
    \midrule
        $\tau^{+}\tau^{-}$
        & $3.6 \times 10^{8}$ & $3.6 \times 10^{9}$ & $4.5 \times 10^{10}$ &   $1.2 \times 10^{11}$ \\
    \bottomrule[1pt]
    & & & & \\
    \toprule[1pt]
    Measurement & Current & Belle II & CEPC prelim. & FCC \\
    \midrule
    BR($\tau \to \bar \mu\mu\mu$) & $< 2.1\times 10^{-8}$ &  $3.6\times 10^{-10}$  &  $10^{-10}$ & $1.4\times 10^{-11}$  \\
    \bottomrule[1pt]
    \end{tabular}}
    \caption[Expected yields of $\tau$ leptons and sensitivities for $\tau \to \bar \mu\mu\mu$.]{Up: Expected yields of $\tau$ leptons at BESIII, STCF, Belle II, and CEPC ($4\times 10^{12}$ $Z$ bosons). Down: Current and projected sensitivities for $\tau \to \bar \mu\mu\mu$ at Belle~II, CEPC, and FCC-$ee$. This table is extracted from the CEPC white paper~\cite{Ai:2024nmn}.}
    \label{tauexp}
\end{table}

\begin{figure}[t!]
  \centering
  \includegraphics[width=0.47\textwidth]{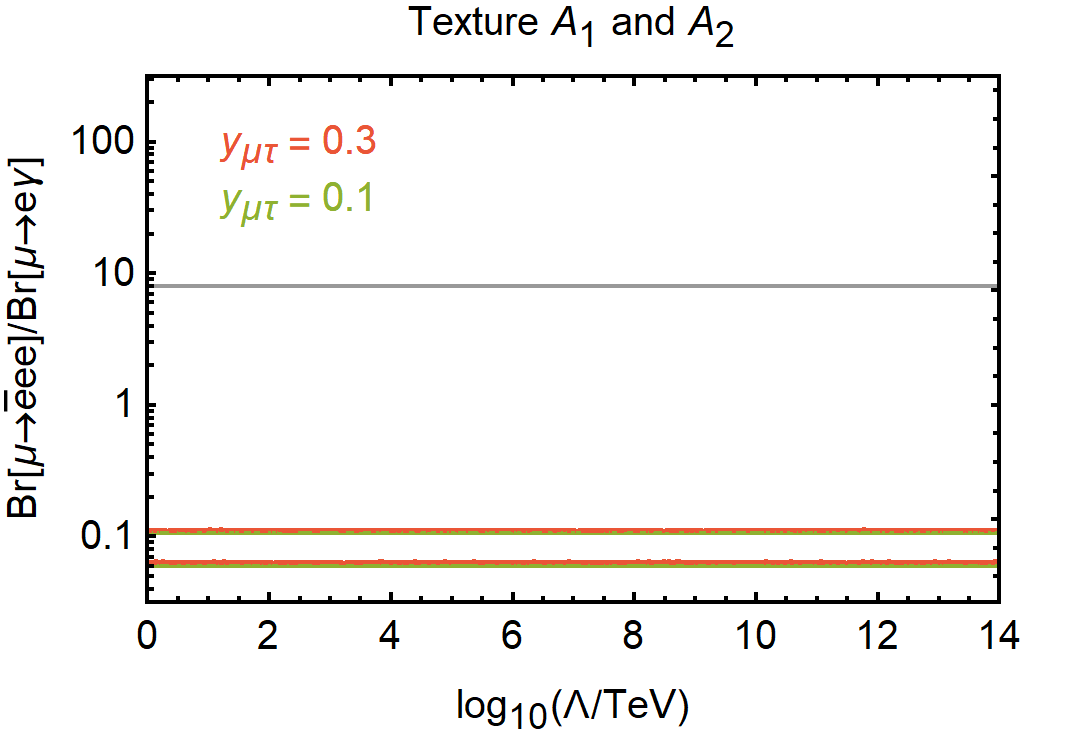} \quad
  \includegraphics[width=0.47\textwidth]{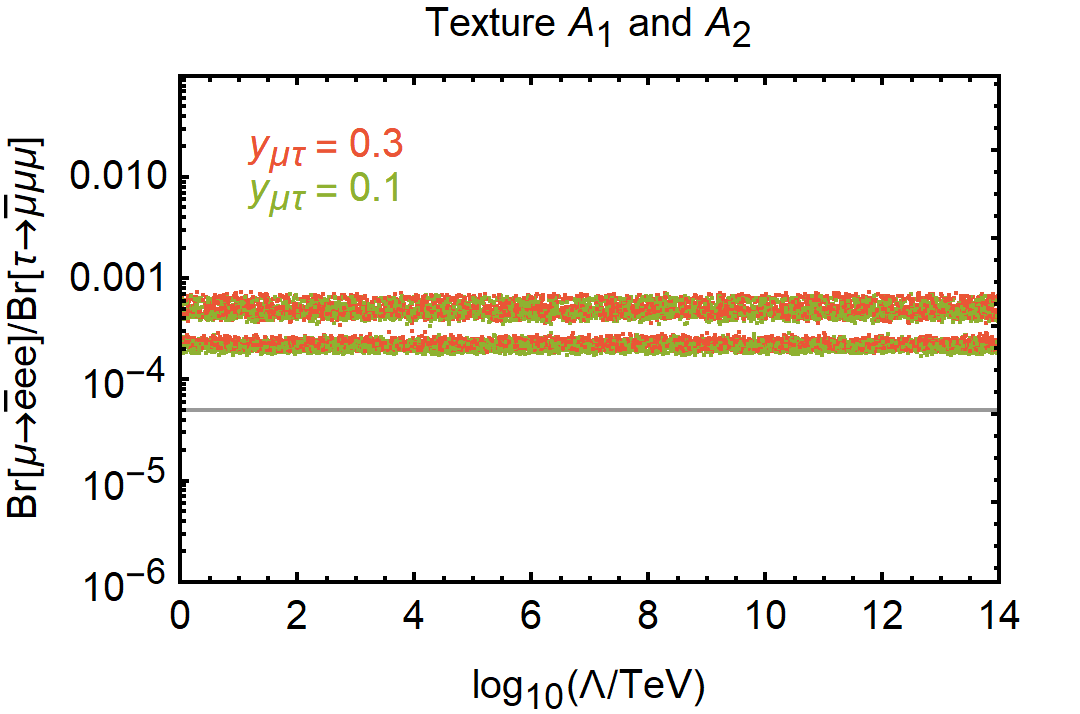}\\
  \vspace{10pt}
  \includegraphics[width=0.47\textwidth]{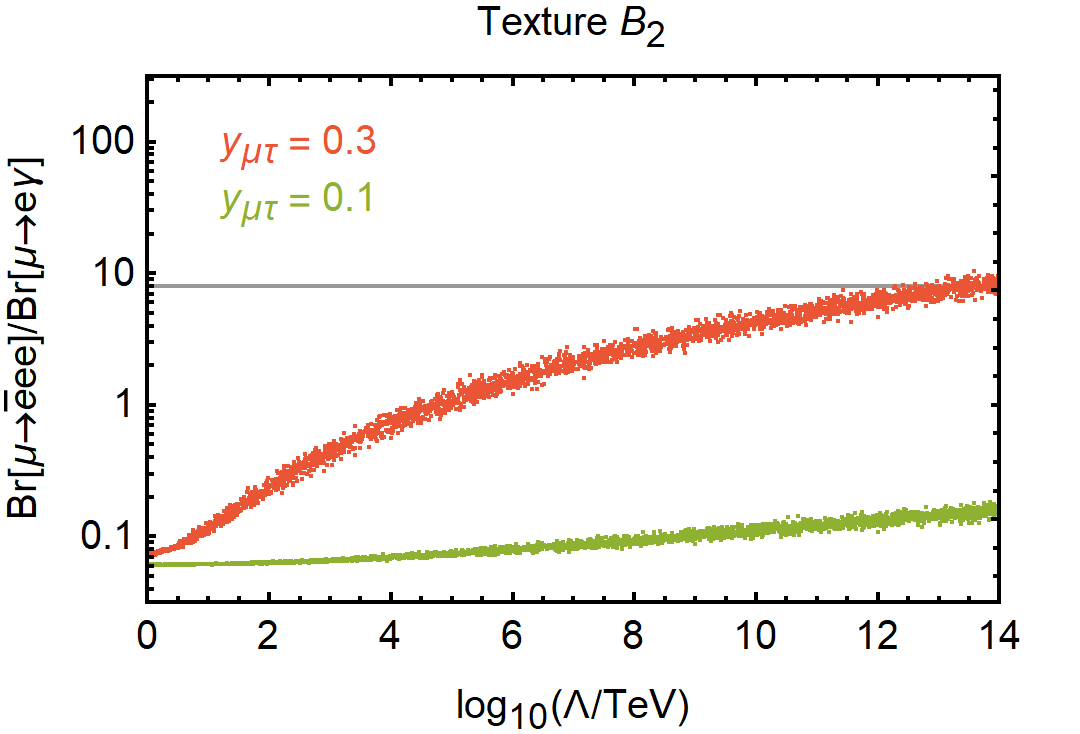}\quad
  \includegraphics[width=0.47\textwidth]{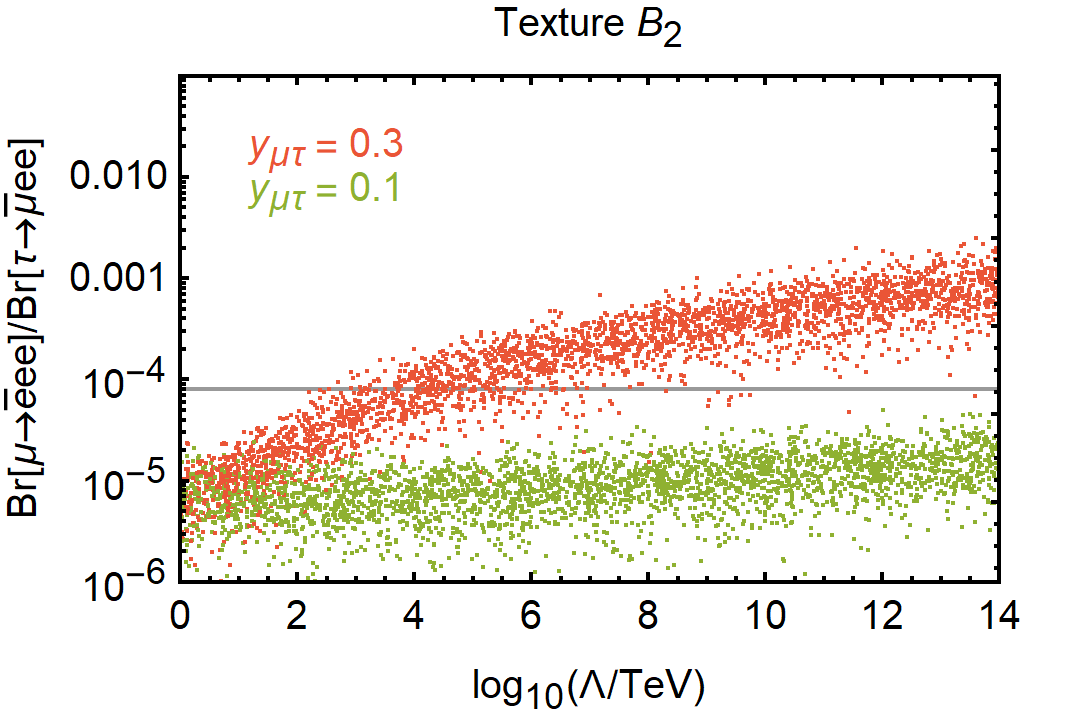}\\
  \vspace{10pt}
  \includegraphics[width=0.47\textwidth]{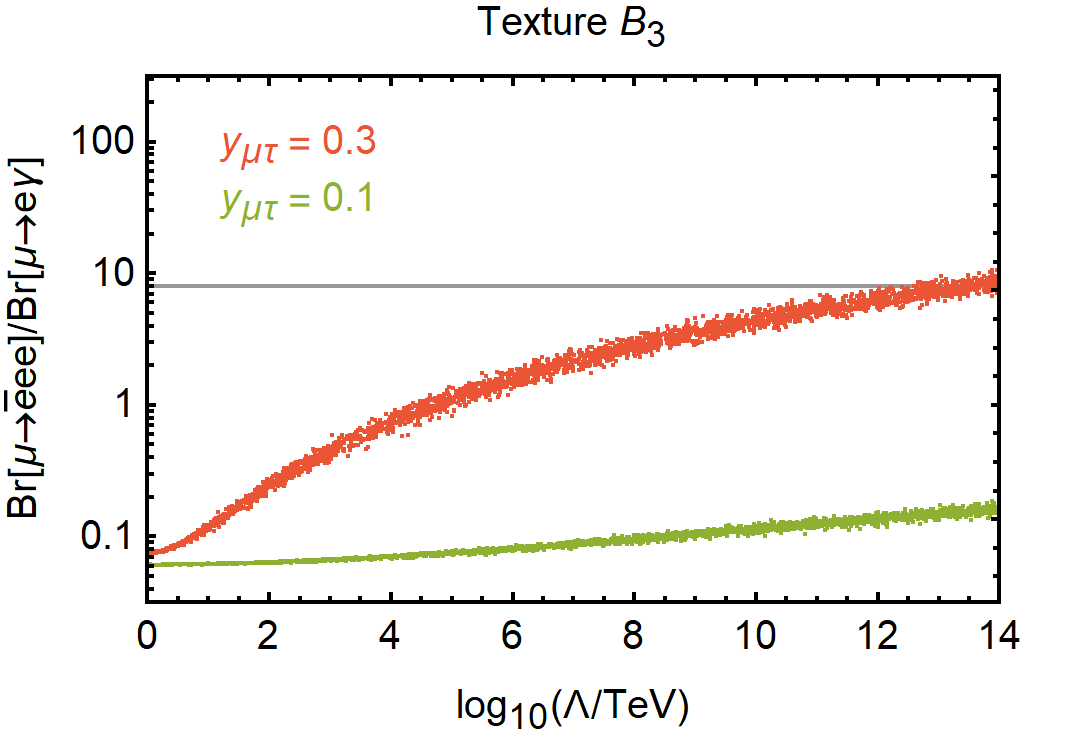}\quad
  \includegraphics[width=0.47\textwidth]{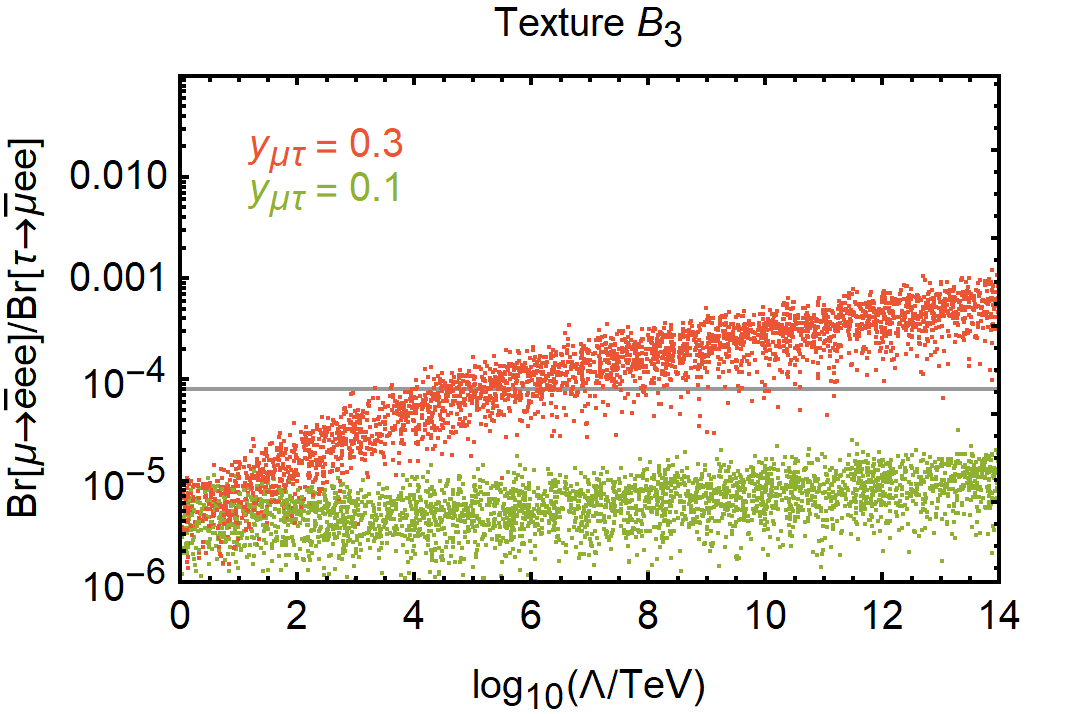}\\
    \caption[Impact of RG evolution on the ratios of the CLFV BRs.]{Ratios of CLFV BRs as functions of the scale $\Lambda_\text{UV}$ where a given texture is assumed to be exact, including the RG evolution of $Y_\Delta$ above $m_{\Delta}$. Green points correspond to $Y_{\Delta\mu\tau}=0.1$, red points to $Y_{\Delta\mu\tau}=0.3$. The gray solid lines show the ratios of the current experimental limits.}
   \label{RGE2}
\end{figure}

So far, we have not included any analysis on the robustness of the predicted CLFV patterns under the RG corrections. 
Without the protection of an explicit symmetry, 
the zero textures are not stable and the RG corrections to the `would-be' zero entries in $Y_{\Delta}$ 
must be taken into account for a complete study. 
For the physical observables involving on these entries, such corrections can become phenomenologically important even if they are small.
We define a given zero texture to be exact only at a high scale $\Lambda_\text{UV}$, with $Y_{\ell}(\Lambda_\text{UV})=\text{diag}\{y_e, y_{\mu}, y_{\tau}\}$, and show how the ratios of the CLFV BRs (observed at low energies) change with $\Lambda_{\text{UV}}$ in
Figure~\ref{RGE2}.
The gray lines indicating the ratios of the present experimental limits. 
If a point lying below the line, the process in the denominator sets the stronger constraint for the model, rendering that the RG corrections do not destabilize the CLFV patterns and the qualitative result shown in \fref{B1B4} and \fref{A1A2C} remain valid. 
For each point in the plots, we choose $Y_{\Delta\mu\tau}(m_{\Delta})=0.1$ (green) and 0.3 (red) as benchmark values for the absolute size of the Yukawa couplings.
We refer Section 5 of Ref.~\cite{Calibbi:2025ded} for more technical details.

The top-left and top-right panels of Figure~\ref{RGE2} confirm our analysis in Section~\ref{modelsection}, that $Y_{\Delta ee}=0$ in texture $\mathbf{A_1}$ and $\mathbf{A_2}$ is RG stable. 
Consequently, the process $\mu\to \bar e ee$, which vanishes at tree level, does not receive large addictive corrections from RG running, either.
The prediction for the ratio $\text{BR}(\mu\to \bar e ee)/\text{BR}(\mu\to e\gamma)$ 
is approximately scale-independent and thus valid without invoking flavour symmetries (and UV completions) to stabilize the zero entries. 
In contrast, the middle-left and middle-right panels show that, for textures $\mathbf{B_2}$ and $\mathbf{B_3}$, the ratio $\text{BR}(\mu\to \bar e ee)/\text{BR}(\mu\to e\gamma)$ can be enhanced by more than two orders of magnitude if there is a large hierarchy between $m_{\Delta}$ and $\Lambda_\text{UV}$ and $Y_{\Delta\mu\tau}$ is sizable. 
On one hand, the RG corrections change the CLFV correlations and weaken the predictivity.
On the other hand, such effects are sensitive to $\Lambda_\text{UV}$ and open up the interesting possibility of obtaining information on an ultra-high scale $\Lambda_\text{UV}$, given $Y_{\Delta \mu\tau}$ is known from TeV-scale physics.

Plots in the right-hand side of \fref{RGE2} answer in our view the most critical question: does $\tau\to\bar \mu ee$ become no longer more constraining than $\mu\to \bar e ee$, when RG corrections are considered? 
If $Y_{\Delta\mu\tau}\approx 0.1$, $\tau\to\bar \mu ee$ is always more constraining irrespective to $\Lambda_\text{UV}$. Nevertheless, approximate $U(3)_{\ell}$ flavour symmetry is restored in such a case. 
To demonstrate the predictivity of theories in the swampland of flavour symmetries, we must consider the scenario in which $Y_{\Delta\mu\tau}$ is sizable. In case $Y_{\Delta\mu\tau}\approx0.3$, $\tau\to\bar \mu ee$ may provide a weaker constraint indeed, which informs us that the zero-texture suppression of $\mu\to \bar e ee$ remains no longer effective when sizable RG effects are taken into account. 
However, this does not happen when the separation between $\Lambda_\text{UV}$ and $m_{\Delta}$ is less than about four orders of magnitudes. 
Such a large hierarchy is sufficient to ensure the $\Lambda_\text{UV}$-suppressed dimension-six operators remain negligible. 
In other words, although the $\mathbf{B_2}$, $\mathbf{B_3}$ textures require some kind of UV completion for stability, the corresponding UV physics can remain decoupling. 
The theories in the swampland of flavour symmetries can therefore remain predictive despite the RG corrections are taken into account.

At the end of this section, we remark that since the scalar triplet of type~II seesaw can lie at TeV scale when $Y_{\Delta\mu\tau}$ is sizable, it can lead to distinctive signals at colliders.
If $m_\Delta\approx 1$ TeV, LHC can directly produce the double-charged $\Delta^{\pm\pm}$ pairs through the electroweak Drell-Yan process --- see e.g.~Ref.~\cite{Cai:2017mow, Ashanujjaman:2021txz}
\begin{equation}
    \label{DrellYan}
    pp~\to ~\Delta^{++}\Delta^{--}~\to ~ \ell^+\ell^+ \ell^-\ell^-.
\end{equation}
Here, sizable $Y_{\Delta\mu\tau}$ requires $\Delta^{\pm\pm}$ to dominantly into the two same-sign charged leptons. 
This leads to spectacular 4-lepton signatures, which is much cleaner than the decay modes to the $W$ bosons\footnote{furthermore, the decay BR to $W^{\pm}W^{\pm}$ is negligible in the $\mu_{\Delta}\to 0$ limit.}. 
Assuming equal BRs for the $\Delta^{\pm\pm}$ decays, the current lower bound for $m_{\Delta}$ is 1080~GeV~\cite{ATLAS:2022pbd}. 
Directly producing $\Delta^{\pm\pm}$ pairs heavier than the TeV scale is possible at a future high-energy muon collider~\cite{InternationalMuonCollider:2025sys}, see Ref.~\cite{Li:2023ksw, Maharathy:2023dtp} for dedicated studies within the minimal type~II seesaw model.
In particular, $\Delta^{++}$ can also be singly produced at a $\mu^+\mu^-$ collider~\cite{Jia:2024wqi}, or at a $\mu^+\mu^+$ collider~\cite{Dev:2023nha, Das:2024kyk} such as the $\mu$TRISTAN proposal~\cite{Hamada:2022mua}
\begin{equation}
    \mu^+\mu^-~\to~ \mu^-\ell^-\Delta^{++}, \qquad
    \mu^+\mu^+~\to~ \Delta^{++}(\gamma)
\end{equation}
Compared to pair production, the single production modes require a lower energy threshold so that producing $2-3$~TeV $\Delta^{++}$ could be feasible for near-term technology.
Furthermore, even if $\Delta^{++}$ is too heavy to be produced on shell, it can leave an impact on high-$p_T$ observables, such as $\sigma(\mu^+\mu^+\to \ell^+_i\ell^+_j)$.
Interestingly, the cross sections predicted in the renormalizable model differ from those obtained from the dimension-six EFT (whose WCs can be inferred from the CLFV observables) by a factor of $\mathcal{O}(s/m_{\Delta}^2)$~\cite{Hamada:2022uyn, Fridell:2023gjx, Lichtenstein:2023iut}. At $\mu$TRISTAN with $\sqrt{s}=2$~TeV, such deviation is sizable even for $m_{\Delta}\simeq 4-5$ TeV. 
As a consequence, via combing the high-$p_T$ data with the low-energy observations such as $\text{BR}(\mu\to e\gamma)$, one can directly extract information on $m_\Delta$, even without producing it at a resonance peak.

Some the collider signals for $\Delta^{++}$ involve charged leptons with different flavours and can complement the low-energy CLFV data. 
Using the generated $10^4$ possible configurations of $Y_{\Delta}$,
we calculate the tree-level BRs of $\Delta^{++} \to \ell^+_i \ell^+_j$ modes under each of the seven possible two-zero textures.
The results are shown in \tref{branchingRatios}, in which the modes vanishing at tree level are indicated with the dashes. 
We also display the $3\sigma$ uncertainties, corresponding to the range that 99.7 \% of the output numbers lie within. 
Each texture features a distinctive pattern of the daughter $\ell_i^+\ell_j^+$ pairs, with relative uncertainties under control. 
Such patterns are to be compared with the prediction from the generic textures aligned with $M^{\nu}$, which is subject to large uncertainties from the unknown absolute mass scale and Majorana phases~\cite{Garayoa:2007fw, FileviezPerez:2008jbu, Cai:2017mow},
As highlighted in red, the dominant decay mode is $\Delta^{++}\to \mu^+\tau^+$ for all textures, distinct from the third-generation specific benchmark models for the double-charged scalars such as the $U(2)$ symmetric ones, which requires the $\tau^+\tau^+$ final state dominant. 
Moreover, the shown BR patterns are also different from those under the flavour anarchic textures, in which the six BRs are nearly equal. 
Hence, the (lucky) observation of $\Delta^{++}$ at colliders allows distinguishing the two-zero textures with many other flavour structures.

\begin{table}[t!]
\renewcommand\arraystretch{1.5}
\centering
\resizebox{1.\textwidth}{!}{
$\begin{array}{c ccccccc}
\toprule
\text{decay mode} & \bf{A_1}& \bf{A_2}& \bf{B_1} & \bf{B_2}& \bf{B_3}& \bf{B_4} & \bf{C} \\
 \midrule
\Delta^{++}\to e^+\mu^+ & \text{---} & 8.0_{-1.0}^{+1.7}~\% & 0.25_{-0.23}^{+0.50}~\% & \text{---} & \text{---} & 0.16_{-0.15}^{+0.26}~\% & 12_{-12}^{+10}~\% \\
\Delta^{++}\to e^+\tau^+ & 8.3_{-1.0}^{+1.8}~\%  & \text{---} & \text{---} & 0.25_{-0.24}^{+0.50}~\%  & 0.16_{-0.14}^{+0.26}~\%  & \text{---} & 9_{-9}^{+12}~\% 
   \\
\Delta^{++}\to \mu^+\tau^+  & \textcolor{red}{39_{-5}^{+8}~\%}  & \textcolor{red}{41_{-6}^{+9}~\%}  & \textcolor{red}{69.8_{-1.5}^{+0.3}~\%}  & \textcolor{red}{69.8_{-1.7}^{+0.3}~\%}  & \textcolor{red}{69.8_{-1.7}^{+0.4}~\%}  &
   \textcolor{red}{69.8_{-1.9}^{+0.4}~\%}  & \textcolor{red}{49_{-24}^{+23}~\%}  \\
\Delta^{++}\to e^+e^+ & \text{---} & \text{---} & 28_{-4}^{+4}~\%  & 28_{-4}^{+4}~\% & 28_{-4}^{+4}~\% & 28_{-4}^{+4}~\% & 30_{-20}^{+12}~\% \\
\Delta^{++}\to \mu^+\mu^+  & 27_{-8}^{+10}~\% & 27_{-7}^{+5}~\% & \text{---} & 2.5_{-2.3}^{+4.0}~\% & \text{---} & 2.1_{-1.9}^{+3.5}~\% & \text{---} \\
\Delta^{++}\to \tau^+\tau^+  & 25_{-7}^{+5}~\% & 24_{-8}^{+11}~\% & 2.5_{-2.2}^{+4.0}~\% & \text{---} & 2.0_{-1.8}^{+3.5}~\% & \text{---} & \text{---}  \\
\bottomrule
\end{array}$}
    \caption[The branching ratios of $\Delta^{++} \to \ell_i^+\ell_j^{+}$ decays.]{Central values and $3\sigma$ uncertainties for the branching ratios of $\Delta^{++} \to \ell_i^+\ell_j^{+}$ decays under the two-zero textures.} 
    \label{branchingRatios}
\end{table}

\section{Discussion}
\label{conclu}

Working in the context of the minimal type~II seesaw, our result demonstrate that the two-zero textures, originally imposed on the Majorana neutrino mass matrix $M^{\nu}$, can be consistently extended to new physics at a scale as low as $5-10$ TeV, leading to distinctive correlations among several CLFV transition rates. 
%
Compared with the flavor patterns protected by symmetries such as $U(2)_{\ell}, U(1)_e$, and $A_4$, some of the two-zero textures are equally predictive with respect to the following key aspects:
\begin{enumerate}[label={(\roman*)}]
    \item The dangerous $\mu\to e$ transitions can be suppressed, allowing a relatively low effective cutoff scale. TeV-scale new physics can thus remain compatible with sizable Yukawa coupling $Y_{\Delta\mu\tau}$.
    \item Several CLFV processes can simultaneously lie within the reach of future experiments. The ratios among their BRs are predicted, providing a possible method to distinguish some of the two-zero textures from each other and the other flavor structures.
    \item Although not all two-zero textures are stable under RG evolution, $\mu\to e$ transitions could remain suppressed even after the RG effects are included. 
\end{enumerate}
It is worth remarking that the less restrictive one-zero textures, such as $\mathcal A$ and $\mathcal B$ in Eq.~\eqref{onezero}, are not predictive within the present knowledge of neutrino parameters. For instance, one can see that textures $\mathbf{A_1}$ and $\mathbf{B_2}$ both satisfy the one-zero structure $\mathcal B$, but they give distinctive predictions for $\tau\to\overline{\mu}ee$.%

The physical implications of our result extend far beyond the seesaw mechanisms and zero textures.
We demonstrate that the TeV scale new physics interacting with the $\ell_L$ sector do not need a symmetry based protection mechanism to evade these stringent constraints.
The two-zero textures complement the patterns originate from explicit symmetries which have been studied in depth. 
Simple and predictive theories can emerge even when the underlying flavor symmetries (if any exist) remain implicit. 
Generalizing the framework to the full leptonic sector is straight-forward. 
One can add a double charged singlet scalar $\Delta_R^{++}$, whose interaction with $E_R$ can in general be sizable and explicitly break the $SU(3)_{E_R}$ symmetries. 
The $\mu_R\to e_R$ transition can remain suppressed in analog to the $\mu_L\to e_L$. 
Nevertheless, the universal $U(1)_{E_R}$ symmetry, which can be identified as combination of the gauged $U(1)_Y$ and $U(1)_L$, becomes relevant and remain unbroken. 
To ensure the full theory lies in the swampland of all leptonic flavour symmetries, one should exclude the lepton number. 
Such a definition for flavour symmetries is also motivated in practice, because $L$ can be interpreted as a known and nearly exact symmetry of nature, due to the tiny neutrino masses. 
On the other hand, generalizing to the quark sector is not trivial. 
Several features of quarks make suppressing the $s\to d$ transition without explicit symmetries more challenging. 
Different from $\mu\to e$, the SM quark FCNC are always non-vanishing. 
Due to the interference effects, the relevant observables involving quarks scale as $1/\Lambda^2$, instead of $1/\Lambda^4$ for those containing only leptons, where $\Lambda$ is the cut-off scale. 
Moreover, the magnitude of the QCD coupling strength is much larger than that from QED, relaxing the QCD induced loop suppression. 
Nevertheless, it is also interesting to explore maximal symmetry breaking terms which preserves the small quark FCNC. 
Constructing a realistic theory in which (most of) the quark flavour symmetries are explicitly broken can further demonstrate the rich and unexplored structures for the swampland of flavor symmetries from a more general perspective.

\chapter{Emerging Flavour Textures}
\label{calculating}
In this chapter, we discuss another question on the swampland of flavour symmetries: How can we make the flavour parameters calculable without explicit flavour symmetries?

Actually, the original Froggatt-Nielsen paper~\cite{Froggatt:1978nt} introduced two different ideas to explain the hierarchical SM fermion mass pattern: 
\begin{enumerate}[label={(\roman*)}]
    \item Interpreting the infrared values of SM flavour parameters as stable fixed points of RG equations. 
    \item Imposing approximate selection rules from soft-broken flavour symmetries. 
\end{enumerate}
The idea (i) is further developed in Ref.~\cite{Pendleton:1980as, Hill:1980sq}. 
If the approach based on fixed points works, the SM flavour parameters only depend on the $G_{\text{SM}}$ group of the low energy theory, irrespective of the flavour symmetries at UV. 
Comparing with the idea (ii), the fixed point approach is more promising, because the emergence of complexity from simple structures at small distance scales, as we expect from the RG evolution, is a universal phenomenon and not limited to particle physics. 
Well-known examples can be found in condensed matter and many other complex systems, see e.g.~\cite{Wilson:1974mb, Wolfram:2020jjc}.

Explaining the SM flavour structures as an emergent phenomena is not limited to interpreting them as infrared fixed points. 
In fact, as noted in Ref.~\cite{Hill:1980sq}, the fixed point for the top quark Yukawa coupling discussed in Ref.~\cite{Pendleton:1980as} is never significantly approached at the electroweak scale and is reached only when the QCD coupling starts to diverge. 
The mathematically exact fixed points, strictly speaking, can only be reached in the infinitely deep infrared. 
In practice, the running flavour parameters are always approaching the fixed points, but typically remain at intermediate values around the electroweak scale.  
In this case, we demonstrate that certain SM flavour parameters stay predictable and can agree with their measured values, even if they largely deviate from the patterns restricted by the underlying symmetries. 
Moreover, the perturbative RG evolution \typo{can not}{cannot} describe all emergent phenomena. 
Phase transitions due to non-perturbative effects can also erase the UV information and lead to the realistic flavour patterns at low energies.

In this chapter, we start with exact UV symmetries and show that even if these symmetries are strongly broken at low energies, the relevant flavour parameters can remain calculable. 
We introduce two simple frameworks as examples: one is a perturbative theory --- the minimal $SO(10)$, while the other involves non-perturbative effects of gravity. 
In both frameworks, the observed SM flavour parameters emerge due to dynamics \textit{irrelevant to} the flavour symmetries in the UV. 
Section~\ref{fixpoints} is mostly based on our work Ref.~\cite{Gao:2025gyo}. 
In subsection~\ref{SO10subsection}, we introduce the most minimal $SO(10)$ theory and its flavour implications. Then, in subsection~\ref{reshapeIR}, we go beyond the widely discussed `desert picture' and populate the TeV-scale spectrum with scalar leptoquarks (LQs) motivated by the long-standing anomalies in flavour-changing $B$ meson decays.
In subsection \ref{btausection} and \ref{flavourmixing}, we demonstrate that the TeV-scale LQs can lead to successful $b-\tau$ unification within the most minimal theory, by modifying the low-energy behavior of the Yukawa couplings. 
During the RG running, the flavour parameters evolve approaching their infrared fixed points and differ significantly from their values dictated by the UV symmetries. 
Comparing with Ref.~\cite{Gao:2025gyo}, we include more discussions on the threshold corrections to the $\tau$ lepton mass from the $R_2$ LQ, and how an accidentally enhanced PQ symmetry forbids proton decay induced by the $S_3, \overline{S}_1,$ or $\widetilde{S}_1$ LQ.
Section~\ref{phaseTransition} is based on Ref.~\cite{Dvali:2016uhn, Dvali:2017mpy} and our recent developments in Ref.~\cite{Gao:2026bvv}. 
In subsection~\ref{anomalymatching}, we introduce the non-perturbative gravity induced chiral symmetry breaking mechanism established in Ref.~\cite{Dvali:2017mpy} and demonstrate that assuming a non-vanishing vacuum topological susceptibility, the $U(48)$ symmetry in the gaugeless limit of the SM must be dynamically broken during a phase transition. 
Then, in subsection~\ref{numasssection}, we revisit the implication of this effect on neutrinos noticed in Ref.~\cite{Dvali:2016uhn} and argue that it explains the flavour puzzle on neutrinos \textit{fully} through the non-perturbative dynamics at low-energies. 
In particular, we identify an overlooked modification to the $U(1)_{B-L}$ gauge symmetry which ensures that the Dirac-type neutrino masses are calculable in this framework.
The phenomenological implications of the corresponding $B-L$ gauge bosons are discussed in subsection~\ref{BLsection}.

\section{Perturbative Theories: Fixed Points}
\label{fixpoints}

\subsection{SO(10) as a Flavour Symmetry}
\label{SO10subsection}
Considering a single generation of SM fermions shown by $\psi_{16}^i$ of Eq.~\eqref{16fermions}, and taking the gaugeless limit in which all SM gauge coupling strengths vanish, we have
\begin{equation}
    N_f~=~16.
\end{equation}
The maximal flavour symmetry is then $U(16)$, under which $\psi_{16}^3$ transform as a fundamental representation. 
$\psi_{16}^3$ is also an irreducible representation of $SO(10)$, a maximal special subgroup of $SU(16)$.
Considering the Euclidean space with $D=10$,
the Clifford Algebra is constructed by ten $32\times32$ Gamma matrices  
\begin{equation}
    \{\Gamma_{\mu}, \Gamma_{\nu}\}~=~2\eta_{\mu\nu},
\end{equation}
where $\mu,\nu=1,2,...,10$. $\eta_{\mu\nu}=1$ if $\mu=\nu$, otherwise, $\eta_{\mu\nu}=0$. 
One can then construct $D(D-1)/2=45$ generators represented by the $32\times 32$ matrices
\begin{equation}
    \Sigma_{\mu\nu}~=~\frac{1}{4 i} [\Gamma_{\mu}, \Gamma_{\nu}]. 
\end{equation}
It is straight forward to check that $\Sigma_{\mu\nu}$ satisfies the commutation relations for the special orthogonal groups in 10 dimension
\begin{equation}
    [\Sigma_{\mu\nu}, \Sigma_{\rho\sigma}]~=~i(\eta_{\mu\rho}\Sigma_{\nu\sigma}-\eta_{\mu\sigma}\Sigma_{\nu\rho}-\eta_{\nu\rho}\Sigma_{\mu\sigma}+\eta_{\nu\sigma}\Sigma_{\mu\rho}).
\end{equation}
In other words, the group represented by $32\times 32$ matrices $\exp(-i\theta_{\mu\nu}\Sigma_{\mu\nu})$ is locally isomorphic to the group of $10\times 10$ real orthogonal matrices with unit determinant. The corresponding $32$-dimensional representation $\Psi$ is called spinor representation of $SO(10)$, it transforms as
\begin{equation}
    \Psi_{\alpha}~\to~\Psi_{\alpha}~=~(e^{-i\theta_{\mu\nu}\Sigma_{\mu\nu}})_{\alpha\beta}\Psi_{\beta}, \quad \alpha, \beta ~=~1,2,...,32.
\end{equation}
However, $\Psi$ is reducible, because there exists a nontrivial matrix, the highest rank element of the $D=10$ Clifford Algebra, which commutes with all of the generators
\begin{equation}
     [\Gamma_*, \Sigma_{\mu\nu}]~=~0, \quad \text{where} \quad \Gamma_*~\equiv~i\Gamma_1\Gamma_2...\Gamma_{10}. 
\end{equation}
$\Gamma_*$ is in analog to the Dirac $\gamma_5$ matrix in $(3+1)$ dimension. 
Consequently, $\Psi$ can be projected onto two irreducible representations, 
\begin{equation}
    \psi^{\pm}~=~\frac12(1\pm\Gamma_*) \Psi.
\end{equation}
$\psi^{+}$ or $\psi^-$ is the Weyl spinors in $D=10$ and after equipped with the Lorentz structure, can be identified as $\psi_{16}^i$.  
In other words, one can always define the transformation under which the 16 free Weyl fermions `rotate' as spinors in the 10d Euclidean space.

The kinetic terms of the SM Lagrangian are automatically invariant under the spinorial transformation of $SO(10)$.
Ideally, the SM Yukawa and scalar sector should be restricted, so that if neglecting the SM gauge interactions, $SO(10)$ --- the maximal anomaly-free symmetry acting on $\psi_{16}^i$ --- is extended to a symmetry of the full theory. 
In particular, the $SO(10)$ symmetry acting on $\psi_{16}^i$ is compatible with the known $G_{\text{SM}}$.
Decomposing $\psi_{16}^i$ into representations of the $SU(3)\times SU(2)\times U(1)_Y$ subalgebra of $SO(10)$ gives
\begin{equation}
    \psi_{16}^i~=~(\textbf{3},\textbf{2},\frac{1}{6})+(\textbf{1},\textbf{2},-\frac12)+
    (\overline{\textbf{3}}, \textbf{1}, - \frac23)+(\overline{\textbf{3}}, \textbf{1}, \frac{1}{3})
    +(\textbf{1}, \textbf{1},1)+(\textbf{1},\textbf{1},0),
\end{equation}
which are exactly the $SU(3)_c, SU(2)_L$ representations and $U(1)_Y$ charges for the particles shown in \tref{SMfermions}.
As a consequence, each generation of SM fermions, including a right-handed neutrino, can compose neatly into $\psi_{16}^i$. 
Recently, Ref.~\cite{Herms:2024krp} pointed out that such `unifiability' is a rare feature among all possible anomaly-free assignments of $G_{\text{SM}}$ representation and strongly suggests the existence of the $SO(10)$ symmetry in Nature.

The global $SO(10)$ symmetry is anomaly-free so it can be gauged and serve as a fundamental principle.
Promoting a semi-simple group in which $G_{\text{SM}}$ is embedded, such as $SO(10)$, to be local, has a profound physical motivation on explaining the charge quantization puzzle. 
The quantization of electric charge in a way that neutron and neutrinos are electrically neutral has been tested to extremely high precision; for instance, the neutron charge is constrained to be about $10^{-21}$ times smaller than the electron charge~\cite{Bressi:2011yfa}. 
However, no fundamental principles within the SM can forbid deviations from this pattern \cite{Foot:1992ui,Babu:1989tq}. 
The group structure of $SO(10)$ requires the hypercharge group of SM to be quantized in units of the weak isospin of $SU(2)_L$ and predicts that the electric charges of all SM particles are integer multiples of the down quark charge. 
In addition, comparing with the global $SO(10)$ symmetry which only manifests itself in the gaugeless limit, the gauged $SO(10)$ is defined exact above a certain scale, commonly \typo{referred as}{referred to as} $M_{\text{GUT}}$.
At $M_{\text{GUT}}$, the three SM gauge coupling strengths (with proper normalization) merge. 
New gauge bosons represented by the $SO(10)$ generators arise and restore the $SO(10)$ invariance. 
This appealing framework is commonly referred to Grand Unified Theories (GUTs)~\cite{Georgi:1974sy, Fritzsch:1974nn}.
Various fits to gauge coupling unification for SO(10) suggest that $M_{\text{GUT}}$ lies around $10^{16}$ GeV~\cite{Deshpande:1992au, Deshpande:1992em, Bertolini:2009qj, Ohlsson:2019sja}. 
Although this scale is far above the limit that the current or future colliders can directly reach, the new vector bosons can induce baryon number violation processes such as proton decay.
Future experiments measuring the proton lifetimes~\cite{Hyper-Kamiokande:2018ofw, Dev:2022jbf} can make the GUT idea partly testable.
Due to the reasons above, the $SO(10)$ symmetries are more commonly assumed to be gauged in the literature. 
However, the local and global symmetries do not differ in the Yukawa sector, because it does not contain derivatives. 
Throughout this chapter, we focus on the role of $SO(10)$ on flavour parameters and do not specify whether it is gauged or not.

We not start to revisit the generic structure of the Yukawa sector constrained by the $SO(10)$ symmetry, which contains at most three different $SO(10)$ invariant terms~\cite{Mohapatra:1979nn}
\begin{equation}
\label{16Yukawa}
    -\mathcal{L}_Y~=~\overline{\psi_{16}^i}(Y_{10}^{ij}\phi_{10}+Y_{120}^{ij}\phi_{120}+Y_{126}^{ij}\phi_{126})(\psi^{j}_{16})^c.
\end{equation}
Here, $\phi_{10}, \phi_{120},$ and $\phi_{126}$ are respectively the vector, the rank-3 tensor, and the rank-5 (self-dual) tensor in the space spanned by the Gamma matrices of $SO(10)$
\begin{equation}
\label{16fields}
\phi_{10}~=~\Gamma_{\mu}\phi_{10}^{\mu},
\quad \phi_{120}~=~\Gamma_{\mu}\Gamma_{\nu}\Gamma_{\rho}\phi_{120}^{\mu\nu\rho},
\quad \phi_{126}~=~\Gamma_{\mu}\Gamma_{\nu}\Gamma_{\rho}\Gamma_{\eta}\Gamma_{\zeta}\phi_{126}^{\mu\nu\rho\eta\zeta},
\end{equation}
where $\mu,\nu,\rho,\eta,\zeta=1,2,...,10$.
$Y_{10}$ and $Y_{126}$ are both $3\times3$ symmetric complex matrices and $Y_{120}$ is anti-symmetric~\cite{Mohapatra:1979nn}. 
The simplest Yukawa sector only contains a vector $\phi_{10}$, in which case
Eq.~\eqref{16Yukawa} reduces to Eq.~\eqref{16Yukawa10} which we introduced in Section~\ref{Fpuzzlesection}. 
To illustrate the structure of $\phi_{10}$, we decompose it into the representations of $G_{\text{PS}}=SU(4)_C\times SU(2)_L\times SU(2)_R$~\cite{Pati:1974yy} subalgebra
\begin{equation}
    \phi_{10}~=~(\textbf{1}_C,\textbf{2}_L,\textbf{2}_R)+(\textbf{6}_C, \textbf{1}_L,\textbf{1}_R).\\
\end{equation}
$(\textbf{1}_C,\textbf{2}_L,\textbf{2}_R)$ contains one SM-like Higgs $H$, which is an $SU(4)_C$ singlet. Under the $SU(4)_C$ group, the leptons are treated as carrying a `forth color' and are contained in the fundamental representation together with the quarks.
As a consequence, $H$ interacts universally with quarks and leptons, so that the masses for top quark, bottom quark, charged $\tau$ lepton, and Dirac-type $\tau$ neutrino are degenerate at $M_{\text{GUT}}$
\begin{equation}
    m_t~=~m_b~=~m_{\tau}~=~m_{\nu_{\tau }}.
\end{equation}
For simplicity, we limit our analysis to the third generation fermions $\psi_{16}^3$, and neglect the small Yukawa couplings for the first-two generation fermions. 
From the known value of $m_t$, the other third generation fermions' masses become fully calculable due to the $SO(10)$ symmetry. Unfortunately, the degenerate mass spectrum is clearly inconsistent with the measured values, rendering the minimal theory containing only one $\phi_{10}$ unrealistic. 
One may alternatively try to replace the vector $\phi_{10}$ with a rank-three tensor $\phi_{120}$.
However, $Y_{120}$ shown in Eq.~\eqref{16Yukawa} is anti-symmetric in flavor space so its $(3,3)$ entry vanishes with one single generation. 
Even if more generations are considered, $\phi_{120}$ cannot generate heavy masses for the third-generation fermions if those in the first-two generations remain light.


To predict the realistic third-generation fermion mass spectrum, a representation with more structural complexity is needed. Indeed, the last remaining choice, the rank-five (self-dual) tensor $\phi_{126}$, improves the situation much. 
The decomposition of $\phi_{126}$ under $G_{\text{PS}}$ is
\begin{equation}
    \phi_{126}~=~(\textbf{6}_C,\textbf{1}_L,\textbf{1}_R)~+~(\textbf{10}_C, \textbf{3}_L, \textbf{1}_R)~+~(\overline{\textbf{10}_C}, \textbf{1}_L,\textbf{3}_R)~+~(\textbf{15}_C,\textbf{2}_L,\textbf{2}_R). 
\end{equation}
The following two features of $\phi_{126}$ make the fermion mass spectrum closer to reality. 
Firstly, $\phi_{126}$ is tailor-made for tiny neutrino masses. One of the $(\textbf{10}_C, \textbf{1}_L, \textbf{3}_R)$ components is singlet under $G_{\text{SM}}$. 
If it takes a high-scale VEV $\langle\Delta_R\rangle$, the right-handed neutrino $\nu_R$ can acquire a large mass and decouple from the low energy spectrum\footnote{This applies to all three generations of $\nu_R$, because even if the Yukawa couplings for the first-two generations are small, the hierarchy between $\langle\Delta_R\rangle$ is in general larger.}. 
Upon integrating out $\nu_R$, the neutrinos participating in electroweak interactions become all massless at leading order, and the dimension-5 Weinberg operator~\cite{Weinberg:1979sa} can generate a tiny mass term for $\nu_{\tau L}$~\cite{Minkowski:1977sc, Mohapatra:1979ia}. 
In addition, $\phi_{126}$ permits more freedom modifying the charged fermion masses. 
$(\textbf{15}_C,\textbf{2}_L,\textbf{2}_R)$ contains two Higgs doublets $H_u$ and $H_d$, which carry opposite $SU(2)_R$ isospins. 
The fermion masses thus do not only depend on the single Yukawa coupling strength, but also the ratio between the two Higgs VEVs, $\langle H_u\rangle/\langle H_d\rangle$, commonly \typo{referred as}{referred to as} $\tan\beta$. 
On the other hand, the Yukawa interaction involving one $\phi_{126}$ remains predictive, because it is accidentally invariant under the following global PQ-type $U(1)$ transformation
\begin{equation}
\label{U1pq}
    \psi_{16}^3\rightarrow \psi_{16}^3 e^{i\theta}, \quad \phi_{126}\rightarrow \phi_{126}e^{2i\theta}. 
\end{equation}
This symmetry is universal to all scalars in $\phi_{126}$, including the two Higgs doublets.  
The quark bilinears carrying different PQ charges cannot couple to both of the Higgs doublets, so the Yukawa interaction for charged fermion masses is restricted to
\begin{equation}
\label{yukawaDefHiggs}
    -\mathcal{L}_Y~=~y_t \overline{Q_L^3}t_R H_u  + y_b \overline{Q_L^3}b_R H_d+y_{\tau}  \overline{L_L^3}\tau_R H_d+\text{h.c.}
\end{equation}
Interaction terms such as $\overline{Q_L^3}b_R\widetilde{H}_u$ are forbidden by the PQ charge.
Since $H_u$ and $H_d$ live in the same $SU(2)_R$ doublet, we have
\begin{equation}
    y_t~=~y_b, \quad m_t~=~m_b\tan\beta. 
\end{equation}
$\tan\beta$ can generate the large hierarchy between the top and bottom quark mass.
Meanwhile, non-trivial relationships between the quark and lepton masses are also predicted. 
Contrary to $(\textbf{1}_C,\textbf{2}_L,\textbf{2}_R)$, the representation $(\textbf{15}_C,\textbf{2}_L,\textbf{2}_R)$ is not a singlet under $SU(4)_C$. $H_u$ and $H_d$ live in the direction of the $B-L$ generator of $SU(4)_C$
\begin{equation}
    T_{B-L}~=~\left(
    \begin{matrix}
        1 & 0 & 0 & 0\\
        0 & 1 & 0 & 0\\
        0 & 0 & 1 & 0\\
        0 & 0 & 0 & -3\\
    \end{matrix}
    \right),
\end{equation}
which is an $SU(3)_c$ singlet and carries unit $B-L$ charge. 
In other words, $y_b$ and $y_{\tau}$ are proportional to the $B-L$ charges for quarks and leptons, respectively. After eliminating the unphysical minus sign, we arrive at a quantitative prediction at $M_{\text{GUT}}$
\begin{equation}
    y_{\tau}~=~3 y_b, \quad m_{\tau}~=~3 m_{b}. 
\end{equation}
Comparing with the numbers shown in Eq.~\eqref{ynumbers} and Eq.~\eqref{ynumbers2}, 
this $b-\tau$ mass relation agrees with the measured values within the order of magnitude.

The predictions of the simplest $SO(10)$ invariant Yukawa sector, which only contains $\phi_{126}$ and $\psi_{16}^3$, can be summarized as
\begin{enumerate}[label={(\roman*)}]
    \item All three $\nu_R$ decouple from the low energy spectrum.
    \item The active neutrinos are approximately massless.
    \item The top and bottom quark mass ratio is equal to $\tan\beta$. 
    \item The $\tau$ lepton is three times heavier than the bottom quark.
    \item The charged fermions in the first-two generations are light. 
    \item All flavour mixings vanish. 
\end{enumerate}
Strictly speaking, (iii) implies as a prediction only if both $H_u$ and $H_d$ are relatively light so that $\tan\beta$ is experimentally measurable. 
Nevertheless, the fermion mass spectrum and mixing patterns predicted in (i)-(iv) more or less reflect the reality. We refer this setup as the \textit{most minimal} $SO(10)$. 
The question is: even if these predictions hold at a high scale, say $M_{\text{GUT}}$, above which $SO(10)$ is (approximately) exact, do they remain valid at the low energies?

In addition to $\psi_{16}^i$ and the vacuum state $\langle\phi_{126}\rangle$, the complete low energy theory also contains the $G_{\text{SM}}$ gauge bosons and the Higgs boson $H$. 
If we promote $\langle\phi_{126}\rangle$ to a spurion,
the $SO(10)$ symmetry is restored only when we take the gaugeless limit and (unphysically) require that the Higgs boson decouples. 
Therefore, although the $SO(10)$ symmetry can be exact above $M_{\text{GUT}}$, at low energies it is always broken by the SM gauge and Higgs bosons, and the constrains on the some flavour parameters are subject to sizable corrections.  
In other words, the SM does not lie in the landscape of $SO(10)$, but only in its swampland. 
The $SO(10)$ breaking effects are irrelevant to the qualitative predictions (i), (ii), and (v), and its correction to (iii) can be eliminated by choosing an approximate renormalization scheme for $\tan\beta$. 
Prediction (vi) remains unchanged because one can always take $Y_{126}$ diagonal so that the theory contains no source of flavour violation, irrespective of whether $SO(10)$ is broken or not. 
Only the prediction (iv) is modified by the $SO(10)$ breaking effects. Given $m_{\tau}=3m_b$ at a high scale of about $10^{16}$ GeV, and considering the loop corrections induced by SM gauge and Higgs bosons, we find $m_{\tau}\approx1.3~m_b$ at 1 TeV, a relation close to --- but still not reasonably agreeing with --- the measured quantities $m_{\tau}\approx0.7~m_b$ which can be inferred from Eq.~\eqref{ynumbers} and Eq.~\eqref{ynumbers2}.

Therefore, although the SM is in the swampland of $SO(10)$, it inherits all predictions of the most minimal $SO(10)$ theory except for (iv), which is subject to an $\mathcal{O}(1)$ correction. 
If one further includes the $SO(10)$ breaking effects from SM gauge and Higgs bosons, the prediction (i)-(iv) all qualitatively agree with experiments. However, quantitatively, (iv) and (vi) are inconsistent with the measurements and lead to two challenges
\begin{enumerate}[label={(\roman*)}]
    \item How to solve the $b-\tau$ mass discrepancy? 
    \item How to generate flavour mixings?
\end{enumerate}
For a long time, people believed that the mentioned failure comes from the minimal model
construction and next-to minimal model building is needed for a consistent $SO(10)$ theory in UV, as well as all realistic unification theories in general. 
More than one representation is required to achieve realistic fermion masses and mixings --- either heavier than $M_{\text{GUT}}$ (entering as effective operators~\cite{Ellis:1979fg, Preda:2022izo, Preda:2024vas}) or lighter (such as additional scalars containing Higgs doublets~\cite{Georgi:1979df, Bajc:2002iw, Babu:2016bmy, Preda:2025afo} or vector-like fermions~\cite{Dorsner:2014wva, Babu:2016cri}). The predictions of the most minimal GUT are replaced by fits within `minimal realistic' models which contain additional free parameters; see Refs.~\cite{Bajc:2005zf, Joshipura:2011nn, Dueck:2013gca, Babu:2016bmy, Ohlsson:2019sja, Mummidi:2021anm, Chen:2021zwn, Patel:2022xxu, Haba:2023dvo, Chen:2024cht} for examples of non-supersymmetric (SUSY) GUTs. 
Meanwhile, generating realistic flavour patterns becomes easier in the supersymmetric GUTs, see Refs~\cite{Deppisch:2018flu, Babu:2018tfi, Antusch:2019avd, Fu:2023mdu, Saad:2025cfb} for recent representative studies. 
The threshold correction to the matching relation between SM and minimal supersymmetric standard model (MSSM) Yukawa couplings can be enhanced when $\tan\beta$ is large~\cite{Hall:1993gn, Carena:1993ag, Carena:1999py, Diaz-Cruz:2000nvf, 
Girrbach:2011an, Deppisch:2018flu}, which corrects the $b-\tau$ mass relationship. 
Meanwhile, the flavour mixing effects from the soft-SUSY breaking terms can propagate into the charged fermion mass matrices at the low energies~\cite{Barbieri:1995tw, Hisano:1995nq,Calibbi:2006ne}. 
Comparing with the most minimal $SO(10)$ we introduced, a common feature of the proposed minimal realistic SUSY and non-SUSY models is, that the predictions on the flavour parameters are lost.

However, it would be premature to claim that the most minimal $SO(10)$ we introduced is unrealistic: The loophole is the assumption of a particle desert between the electroweak scale and $M_{\text{GUT}}$, usually deduced from the naturalness criteria~\cite{Georgi:1979md}. 
Recently, the authors of Ref.~\cite{Patel:2023gwt,Shukla:2024bwf} proposed a new idea, that the wrong $b-\tau$ mass relation in minimal $SU(5)$ can be resolved by introducing a large mass hierarchy among the particles within the same scalar multiplet. 
This hierarchy requires that some particles lie far below $M_{\text{GUT}}$, and suggests the desert picture together with the naturalness criterion should be reconsidered. 
In the view of RG evolution, these particles modify the infrared structure of the most minimal GUTs, and drive the UV parameters towards their infrared fixed-points.
Therefore, the wrong $b-\tau$ mass ratio does not necessarily falsify the most minimal $SO(10)$; rather, the low-scale particles\footnote{Including particles from the GUT representations in the light spectrum, in some cases, also leads to successful gauge coupling unification; see, for instance, Ref.~\cite{Dorsner:2005fq, Bajc:2006ia, Preda:2022izo, Goto:2023qch, Preda:2024vas}.} can further change the values of $y_b$ and $y_{\tau}$, rendering the $SO(10)$ symmetry (if exists) more implicit at the low energies.

Assuming this strategy makes the most minimal $SO(10)$ theory realistic, we then arrive at a successful theory for flavour, in which many flavour parameters are calculable. In particular, the $SO(10)$ symmetry exact above $M_{\text{GUT}}$, becomes implicit at low energies but remains predictive on restricting the SM flavour patterns, as the unbroken flavour symmetries can do. 
We believe this setup should also be regarded as a simple and concrete example showing that certain flavour structures can mostly originate from dynamics irrelevant to the underlying flavour symmetries.

\subsection{Reshaping the Infrared}
\label{reshapeIR}

To reshape the infrared structure of the most minimal $SO(10)$ theory, 
the needed low-scale particles can originate from $\phi_{126}$. 
In case the scalar sector is invariant under certain global symmetries $G_S$, light pNGBs can arise once $G_S$ is spontaneously broken into one of its subgroups~\cite{Anselm:1986um, Barbieri:1992yy, Barbieri:1993wz, Barbieri:1994kw}. 
However, $G_S$ is in general explicitly broken by the gauge and Yukawa interactions. The pNGB masses are thus subject to large additive loop corrections~\cite{Graf:2016znk}, so they still cannot stay at the low scales without fine-tuning. 
Although a further reduction of the pNGB masses is possible in supersymmetric theories, in which the additive corrections are forbidden and the pNGBs can lie at the SUSY-breaking scale, 
the existence of low-scale particles generally lead to conceptual puzzles without SUSY. The tree-level masses of these particles must be fined-tuned to cancel the loop corrections.

From a different perspective, requiring more states in $\phi_{126}$ to lie at low-scales is motivated by the experimental data on flavor-changing $B$ meson decays. 
For more than a decade several observables related to $b\to c$ and/or 
$b\to s$ decays have been found to deviate form their SM predictions. The current status of the `flavor anomalies' is as follows: $b\to c \tau\nu$ is probed through the ratios of branching ratios
\begin{equation}
    R(D^{(*)})~=~\frac{\text{BR}(B\to D^{(*)} \tau \nu)}{\text{BR}(B\to D^{(*)} \ell \nu)}, \quad \ell=e,\mu, 
\end{equation}
and polarization data. 
The SM prediction to $R(D^{(*)})$ is robust, because in the limit $m_{\tau}=0$, the non-perturbative effects cancel due to lepton flavour universality. 
In other words, the only non-perturbative quantity involved in the theory prediction of $R(D^{(*)})$ is a ratio of form factors multiplying a term suppressed by the mass ratio $m_\tau^2/m_B^2$.  
An analysis exploiting experimental information on form factor shapes finds the combined $b\to c\tau\nu$ data deviating from the SM predictions by 4.4$\sigma$ \cite{Iguro:2024hyk}. 
Moreover, if one tried to change the form factor ratio in $R(D^{*})$ to a level that the data are reproduced, predictions of measured polarization data in $B\to D^* \ell\nu$ decays with light leptons $\ell=e,\mu$ would instead severely deviate from their SM predictions \cite{Fedele:2023ewe}.
BaBar, Belle, Belle II and LHCb contribute to the $b\to c\tau\nu$ anomaly with mutually consistent measurements \cite{BaBar:2012obs,BaBar:2013mob,Belle:2019rba,LHCb:2023uiv,Belle:2015qfa,Belle:2016dyj,Belle:2017ilt,Belle-II:2025yjp} (combined in  Ref.~\cite{HFLAV:2022esi}). 
On the other hand, the $b\to s \ell^+\ell^-$ anomaly is supported by measurements of various decay branching ratios and angular distributions of $b$-flavored hadrons by LHCb~\cite{LHCb:2015ycz,LHCb:2015wdu,ATLAS:2018cur,CMS:2018qih, LHCb:2018jna, CMS:2019bbr,BELLE:2019xld,CMS:2020oqb,LHCb:2020dof,LHCb:2020gog, LHCb:2021vsc,LHCb:2021trn,LHCb:2021xxq,LHCb:2021lvy, LHCb:2021zwz, LHCb:2022qnv,LHCb:2022vje}
and, more recently, also by CMS \cite{CMS:2024atz}. 
In particular, the observed $B^+\to K^+\ell^+\ell^-$ decay at low $q^2$ region, defined as
\begin{equation}
    \Gamma(B^+\to K^+ \ell^+\ell^-)[1.1, 6.0]=\int_{(1.1~\text{GeV})^2}^{(6.0~\text{GeV})^2}\frac{d\Gamma(B^+\to K^+ \ell^+\ell^-)}{dq^2}dq^2, \quad q^2=(p_{\ell^+}+p_{\ell^-})^2, 
\end{equation}
are in tensions of $\mathcal{O}(4\sigma)$ with the SM predictions\footnote{All data are compatible with lepton-flavor universality in the first two generations, i.e.~equal decay width $B\to K^{(*)} e^+ e^-$ and $B\to K^{(*)} \mu^+ \mu^-$~\cite{LHCb:2022qnv}.}.
The combination of all data prefers beyond-SM scenarios with a significance above 5$\sigma$ \cite{Capdevila:2023yhq}, if the SM prediction of Ref.~\cite{Khodjamirian:2010vf} is used. The 
latter has been challenged by several alternative calculational methods \cite{Bobeth:2017vxj,Gubernari:2023puw,Bordone:2024hui,Isidori:2024lng,Isidori:2025dkp,Hurth:2025vfx}. While more conservative estimates of hadronic uncertainties reduce the significance of beyond SM physics, there is no convincing way to bring the all $b\to s \ell^+\ell^-$ data into good agreement with the SM predictions. 
Recently, Belle II reported an excess in $B\to K \overline{\nu}\nu$ over the SM prediction~\cite{Belle-II:2023esi}. The result obtained using the more recent inclusive tagging analysis (ITA)
method shows a $2.9\sigma$ departure. A combination of both conventional hadronic
tagging analysis (HTA) method and ITA method yields a $2.7\sigma$ discrepancy with the SM prediction. 
Statistically, it is unlikely that all these anomalies will disappear in the future~\cite{Crivellin:2023zui}, and their BSM explanation requires particles not far above the TeV scale.

\begin{table}[t!]
    \centering
    \renewcommand\arraystretch{1.5}
    \begin{tabular}{c c c c }
    \toprule
    ~ & $S_3$ & $S_1$ & $\widetilde{S}_1$  \\
    \midrule
        $b\rightarrow c \tau \nu$ & $(\overline{c}_L\gamma^{\mu}b_L)(\overline{\tau}_L\gamma_{\mu}\nu_L)$  &\renewcommand\arraystretch{1.5}
            \begin{tabular}{c}
                 $(\overline{c}_R \sigma^{\mu\nu} b_L)(\overline{\tau}_R\sigma_{\mu\nu} \nu_L)$  \\
                 $(\overline{c}_L\gamma^{\mu}b_L)(\overline{\tau}_L\gamma_{\mu}\nu_L)$\\
                 $(\overline{c_R}b_L)(\overline{\tau}_R \nu_L)$ 
            \end{tabular}  & $-$   \\
            \rule{0pt}{4ex}  
       $ b\rightarrow s \tau\tau$ & $(\overline{s}_L\gamma^{\mu}b_L)(\overline{\tau}_L\gamma_{\mu}\tau_L)$ &$-$&
       $(\overline{s}_R\gamma^{\mu}b_R)(\overline{\tau}_R\gamma_{\mu}\tau_R)$ \\
       
       $ b\rightarrow s  \nu\nu$  & $(\overline{s}_L\gamma^{\mu}b_L)(\overline{\nu}_L\gamma_{\mu}\nu_L)$ & $(\overline{s}_L\gamma^{\mu}b_L)(\overline{\nu}_L\gamma_{\mu}\nu_L)$ & $-$ \\
       \bottomrule
       \toprule
    ~ & $R_2$ & $\widetilde{R}_2$ & $\bar{S}_1$\\
    \midrule
        $b\rightarrow c \tau \nu$ & \renewcommand\arraystretch{1.5}
            \begin{tabular}{c}
                 $(\overline{c}_R \sigma^{\mu\nu} b_L)(\overline{\tau}_R\sigma_{\mu\nu} \nu_L)$  \\
                 $(\overline{c_R}b_L)(\overline{\tau}_R \nu_L)$ 
            \end{tabular}   & $-$  & $-$
                   \\
                   \rule{0pt}{4ex}  
       $ b\rightarrow s \tau\tau$ &  $(\overline{s}_L\gamma^{\mu}b_L)(\overline{\tau}_R\gamma_{\mu}\tau_R)$ & $(\overline{s}_R\gamma^{\mu}b_R)(\overline{\tau}_L\gamma_{\mu}\tau_L)$   & $-$\\
       \rule{0pt}{4ex}  
       $ b\rightarrow s  \nu\nu$    & $-$ & $(\overline{s}_R\gamma^{\mu}b_R)(\overline{\nu}_L\gamma_{\mu}\nu_L)$   & $-$\\
       \bottomrule
    \end{tabular}
    \caption[LQ induced effective operators relevant to the $b\to s, b\to c$ transitions.]{Effective operators relevant to the $b\to s, b\to c$ transitions, induced by the exchange of the six possible LQs with generic interacting patterns. The top quarks and right-handed neutrinos are omitted. Except for charm and strange quarks, we do not include the fermions in the third generation.} 
    \label{eftS}
\end{table}

In particular, LQs with masses between 1 TeV and 50 TeV are well-suited to remedy the flavor anomalies without harming predictions of observables which are in agreement with their SM predictions \cite{Sakaki:2013bfa,Dorsner:2016wpm,Dumont:2016xpj,Li:2016vvp,Bhattacharya:2016mcc,Chen:2017hir,Crivellin:2017zlb,Cai:2017wry,Jung:2018lfu,Aydemir:2019ynb,Popov:2019tyc,Crivellin:2019dwb,Bigaran:2019bqv,Bansal:2018nwp,Iguro:2020keo,Ciuchini:2022wbq,Dev:2024tto,Fedele:2023rxb,Bigaran:2024vnl}. 
The LQs are color triplets that can couple to a quark and a lepton. 
In total, there are six different types of scalar LQs, following the notation of Ref.~\cite{Dorsner:2016wpm} up to charge conjugation, the representations under $SU(3)_c\times SU(2)_L$ and hypercharges of each type of LQs are
\begin{equation}
\label{LQtypes}
    \begin{aligned}
        S_3~&\sim~(\textbf{3},\textbf{3},-1/3), \quad
&&S_1~\sim~ (\textbf{3},\textbf{1},-1/3), \quad
&&\widetilde{S}_1~\sim~(\overline{\textbf{3}},\textbf{1},4/3),\\
R_2~&\sim~(\overline{\textbf{3}},\textbf{2},-7/6), 
&&\widetilde{R}_2~\sim~(\overline{\textbf{3}},\textbf{2},-1/6),
&&\bar{S}_1~\sim~(\overline{\textbf{3}},\textbf{1},-2/3).
    \end{aligned}
\end{equation}
In the energy regime far below 1 TeV, the LQs can all be integrated out, generating effective operators suppressed by their masses. 
For each type, we collect the effective operators that can arise from tree-level matching given the most general Yukawa couplings~\cite{Dorsner:2016wpm}, and show the ones relevant for $b\to s, b\to c$ transitions in \tref{eftS}. 
For simplicity, we do not include any fermions of the first-two generations, except for the involved charm and strange quarks.
It is worth noting that the left-right symmetry is not manifest, because $\nu_R$ and the top quark are not included. 
As indicated in \tref{eftS}, among the six types of possible scalar LQs, five of them --- $S_3, S_1, \widetilde{S}_1, \widetilde{R}_2, R_2$ --- can address the long-standing anomalies observed in semi-leptonic B decays.
The $S_3, S_1,$ and $R_2$ LQs contribute to operators relevant to $b\rightarrow c$~\cite{Angelescu:2018tyl, Angelescu:2021lln}.
As the $SU(2)_L$ invariance restricts that $S_3$ induces the same $b\to c$ and $b\to s$ transition amplitudes, fully explaining the large $R(D^{(*)})$ discrepancy with merely one $S_3$ conflicts with the current bounds on $b\rightarrow s\overline{\nu}\nu$ transitions. A common way out of this problem is to combine $S_3$ with $S_1$, so their contributions to $b\rightarrow s\overline{\nu}\nu$ can partly cancel~\cite{Crivellin:2017zlb, Crivellin:2019dwb}\footnote{Assuming a cancellation of about $60\%$, and including the $b\to c$ operator induced by $S_1$, $R(D^{(*)})$ can be consistently explained at $1\sigma$  level~\cite{Crivellin:2025qsq}.}.
In addition to the $S_3$ and $S_1$ LQs, $\widetilde{R}_2$ can also induce $b\rightarrow s\overline{\nu}\nu$. In particular, if the coefficients for operators containing the left and right handed quarks cancel, $S_3$ (or $S_1$) and $\widetilde{R}_2$ can generate sizable $B\rightarrow K\overline{\nu}\nu$ decay rate without endangering $B\rightarrow K^*\overline{\nu}\nu$~\cite{Bause:2023mfe, He:2023bnk}, which addresses the recent Belle II excess.
The $S_3, \widetilde{S}_1, R_2,$ and $\widetilde{R}_2$ LQs are relevant for $b\rightarrow s\overline{\tau}\tau$ transitions. 
If closing the $\tau$ loop and attaching it to an off-shell photon, the penguin diagram can induce $b\rightarrow s\ell\ell$ ($\ell=e,\mu$) and satisfy the lepton flavor universality constraints~\cite{Crivellin:2018yvo}.

The state-of-the art is to postulate the required light LQs ad-hoc, which remains unsatisfactory until these particles are embedded into a meaningful 
theory addressing fundamental puzzles of the SM. 
The LQ explanation for $B$ anomalies become more convincing in the partial unification models, such as the Pati-Salam (PS) theories \cite{Pati:1974yy}.
In particular, the LQ masses are required light if PS unification is realized at the multi-TeV scale. Even so, additional vector-like fermions and/or extended gauge groups are typically needed to achieve TeV-scale partial unification~\cite{DiLuzio:2017vat, Bordone:2017bld, Greljo:2018tuh,Calibbi:2017qbu, Fuentes-Martin:2022xnb, Davighi:2022bqf}, in order to evade the bounds from processes involving light flavors, such as $K_L\rightarrow \mu e$. 
The flavour implications for the scalar LQs are also analyzed in the GUT frameworks, see Ref.~\cite{Becirevic:2018afm, Fajfer:2018hbq, Aydemir:2019ynb, Aydemir:2022lrq, Goto:2023qch} for examples. 
Here, we note that all six possible scalar LQs are contained in $\phi_{126}$ of the most minimal $SO(10)$ theory, with the embedding chains shown in \tref{LQembedding}. 
Based on this reasoning, the existence of scalar LQs becomes no longer a prior assumption, but a requirement for restoring $SO(10)$.

\begin{table}[t!]
\renewcommand\arraystretch{1.5}
    \centering
    \begin{tabular}{c c c}
    \toprule
       $SO(10)$  & $SU(4)_C\times SU(2)_L\times SU(2)_R$ & $SU(3)_c\times SU(2)_L\times U(1)_{Y}$ \\
       \midrule
       $\phi_{126}$  &  $(\textbf{6}_C,\textbf{1}_L,\textbf{1}_R)$ & $S_1(\textbf{3},\textbf{1},-1/3)$\\
       & & $S_1'(\overline{\textbf{3}},\textbf{1},1/3)$\\
       \rule{0pt}{4ex}  
       & $(\textbf{10}_C, \textbf{3}_L, \textbf{1}_R)$ & $S_3(\textbf{3},\textbf{3},-1/3)$\\
       \rule{0pt}{4ex} 
       & $(\overline{\textbf{10}_C}, \textbf{1}_L,\textbf{3}_R)$ & 
       $\bar{S}_1(\overline{\textbf{3}},\textbf{1},-2/3)$\\
       & & $S_1''(\overline{\textbf{3}},\textbf{1},1/3)$\\
       & & $\widetilde{S}_1(\overline{\textbf{3}},\textbf{1},4/3)$\\
       \rule{0pt}{4ex} 
       & $(\textbf{15}_C,\textbf{2}_L,\textbf{2}_R)$ & $R_2(\overline{\textbf{3}},\textbf{2},-7/6)$\\
       & & $\widetilde{R}_2(\overline{\textbf{3}},\textbf{2},-1/6)$\\
       & & $R_2'(\textbf{3},\textbf{2},7/6)$\\
       & & $\widetilde{R}_2'(\textbf{3},\textbf{2},1/6)$\\
       \bottomrule
    \end{tabular}
    \caption[Embedding of the six possible LQs]{Embedding of the six possible LQs in the Pati-Salam and the minimal $SO(10)$ framework.}
    \label{LQembedding}
\end{table}

If the LQs are all embedded in $\phi_{126}$, the operator structures shown in \tref{eftS}, ignoring flavour, do not receive additional constraints.
To demonstrate this, we firstly rewrite the $SO(10)$ invariant Yukawa interaction into a less compact form and drop the terms irrelevant to $b\to c$ or $b\to s$ transition
\begin{equation}
\label{chirality}
    \begin{aligned}
        -\mathcal{L}_{\text{LQ}}~=~&Y_3^{LL}\overline{Q_L}S_3L_L^c+
        Y_3^{RR}\overline{Q_R^c}
        \widehat{S}_3
        L_R+
        Y_1^{LL}\overline{Q_L}S_1L_L^c+Y_1^{RR}\overline{Q_R^c}\widehat{S}_1L_R\\
        &+Y_2^{LR}\overline{Q_L}\widehat{R}'L_R
        +Y_2^{RL}\overline{Q_R^c}\widehat{R}L_L^c+\text{h.c.}
    \end{aligned}
\end{equation}
Here, we keep the $SU(2)_R$ indices implicit:
$\widehat{S}_3$ represents the $SU(2)_R$ triplet composed by $\bar{S}_1$, $S_1'',$ and $\widetilde{S}_1$; $\widehat{S}_1$ is an $SU(2)_R$ singlet;
$\widehat{R}$ and $\widehat{R}'$ both contain two distinct $SU(2)_R$ doublets, but carry opposite $B-L$ charges.
Their explicit forms are
\begin{equation}
\label{LQSU2}
\begin{aligned}
    \widehat{S}_3~&=~
    \left(
        \begin{array}{cc}
           \overline{S}_1  & S''_1/\sqrt{2} \\
            S''_1/\sqrt{2} & \widetilde{S}_1
        \end{array}
        \right),
    \quad
    &&\widehat{S}_1~=~
    \left(
        \begin{array}{cc}
           0  & S'_1/\sqrt{2} \\
            -S'_1/{\sqrt{2}} & 0
        \end{array}
        \right),\\
    \widehat{R}~&=~
    \left(R_2~\widetilde{R}_2
    \right), \quad &&\widehat{R}'~=~\left(R_2'~\widetilde{R}_2'
    \right). 
\end{aligned}
\end{equation}
Eq.~(\ref{chirality}) and \eqref{LQSU2} agree with the gauge and Lorentz operator structures shown in \tref{eftS}.
As indicated by the Yukawa couplings, the $R$-type LQs couple to quarks and leptons with opposite chiralities, while the $S$-type LQs --- $S_3, S_1$ for left-handed fields and $\widehat{S}_3, \widehat{S}_1$ for right-handed fields --- are chirality-specific. 
Integrating out the LQs in Eq.~(\ref{chirality}) and \eqref{LQSU2} generates the scalar-type effective operators with the form $(\overline{q}\ell)(\overline{\ell'}q')$ at tree-level.
After Fierz transformation, these operators can be transformed into the ones shown in \tref{eftS}. 
On the other hand, the most minimal $SO(10)$ restricts the flavour structure of the operators generated by its embedded LQs. 
At $M_{\text{GUT}}$, the $3\times3$ Yukawa couplings matrices $Y_{1,2,3}$, irrespective of their chiral indices $^L$ or $^R$, are all aligned to the quark mass matrices and therefore inherit the third-generation specific pattern. 
In the limit that only $\psi_{16}^3$ interact with $\phi_{126}$
\begin{equation}
\label{FlavorUV}
    Y_{1,2,3}~\propto~Y_{126}~\sim~ \begin{pmatrix} ~0~ & ~0~ & ~0~ \cr ~0~ & ~0~ & ~0~ \cr ~0~ & ~0~ & ~1~ \end{pmatrix}. 
\end{equation}
This structure, if approximately preserved at low energies, yields the needed approximate $U(2)$ symmetry which can suppress FCNC involving the first-two generations. 
Therefore, in additional to their existence, the third-generation specific flavour structure for scalar LQs is also predicted instead of assumed. 
Nevertheless, generating the flavour patterns of the operators in \tref{eftS} requires mixing between the $3-2$ generation quarks at low energies. We show how these mixing angles can arise in Section~\ref{flavourmixing}.

In addition to the Yukawa couplings shown in Eq.~\eqref{chirality}, the $S$-type LQs can also couple to a pair of quarks. 
From the low-energy perspective, such `di-quark' couplings do not lead to proton or neutron decay, because they can be forbidden by the baryon number symmetry, under which each leptoquarks carries one unit of charge. 
The baryon number then remains exact at low energies. 
However, considering the $SO(10)$ embedding, the baryon number symmetry is broken at $M_{\text{GUT}}$, and \typo{can not}{cannot} remain exact in the matched low-energy theory. 
In the most minimal $SO(10)$, the $S_1$ LQ always couples to a pair of quarks~\cite{Patel:2022wya}, and are required to be as heavy as $M_{\text{GUT}}$ in order to suppress proton decay. 
If $S_1$ lies at the TeV scale, its di-quark coupling strength must be fine-tuned small~\cite{Dvali:1992hc, Dvali:1995hp, Dorsner:2024seb, Dvali:2024jhs} and the model setup must be extended. 
On the other hand, the diquark couplings involving $S_3, \overline{S}_1, $ and $\widetilde{S}_1$ are absent at tree level~\cite{Patel:2022wya}, which can be understood as a consequence of the accidental $U(1)_{\text{PQ}}$ symmetry shown in Eq.~\eqref{U1pq}. 
Taking $S_3$ as an example, although the SM gauge symmetry allows it to couple to both $(\overline{Q}_LL_L^c)$ and $(\overline{Q_L^c}Q_L)$, these two fermion bilinears contain opposite PQ charge so that $S_3$ can only be combined with one of them, which is $(\overline{Q}_LL_L^c)$. 
It is worth remarking that the high-scale VEV $\langle\Delta_R\rangle$ in $\phi_{126}$ is not invariant under PQ symmetry and spontaneously breaks it. 
However, $\langle\Delta_R\rangle$ is not a singlet of the $SU(5)\times U(1)_Z$ subgroup of $SO(10)$, either. 
Here, the $U(1)_Z$ charge $Q_Z$ is composed by $T_{3R}$ of $SU(2)_R$ and $B-L$, and we refer to Ref.~\cite{Bertolini:2009es, Bertolini:2012im} for the $U(1)_Z$ charge of the SM particles and the scalars in $\phi_{126}$. 
Considering this, $\langle\Delta_R\rangle$ only spontaneously breaks a combination of $U(1)_{\text{PQ}}$ and $U(1)_Z$, and is invariant under the modified PQ symmetry $U(1)_{\text{PQ}'}$
\begin{equation}
    U(1)_{\text{PQ}}\times U(1)_Z~\xrightarrow{\langle\Delta_R\rangle}~U(1)_{\text{PQ}'}, \quad 
    Q_{\text{PQ}'}=Q_{\text{PQ}}-\frac{1}{5} Q_Z, 
\end{equation}
This modified PQ symmetry forbids the dangerous di-quark couplings below $\langle \Delta_R\rangle$. 
Therefore, as long as the scalar sector of the most minimal $SO(10)$ is restricted to be PQ invariant\footnote{The scalar sector is automatically invariant under $U(1)_X$ because it is a subgroup of $SO(10)$.}, the embedded $S_3, \overline{S}_1, \widetilde{S}_1$ never couple to a pair of quarks. 
In other words, the $U(1)_{\text{PQ}'}$ breaking scale must be far lower than $M_{\text{GUT}}$ to suppress proton decay if these $S$-type LQs are light.

The requirement of low scale $U(1)_{\text{PQ}'}$ symmetry further constrains the scalar spectrum. 
The effective mixing term between the two Higgs doublets, $\mu_{H}^2 H_uH_d+\text{h.c.}$, is needed to generate the non-zero VEV for $H_d$, since $\langle H_d\rangle\sim \mu_{H}^2\langle H_u\rangle/m_{H_d}^2$~\cite{Branco:2011iw}. 
Moreover, $\mu_{H}$ explicitly breaks $U(1)_{\text{PQ}'}$ and the induced diquark couplings are proportional to $(\mu_{H}/M_{\text{GUT}})^2$, yielding $S$-type LQ mediated baryon number violating amplitude
\begin{equation}
    \mathcal{A}^{\Delta B=1}~\sim~\left(\frac{\mu_{H}}{M_{\text{GUT}}}\right)^2 \times \frac{1}{m_{S}^2}~\sim~  \left(\frac{m_{H_d}}{M_{\text{GUT}}}\right)^2 \times \frac{1}{\tan\beta~ m_{S}^2}. 
\end{equation}
Since $\tan\beta$ must be finite to generate the bottom quark masses, the $M_{\text{GUT}}^2$ suppression of $\mathcal{A}^{\Delta B=1}$ requires $m_{H_d}$, the mass of the Higgs other than the SM-like one, \typo{can not}{cannot} be much larger than the $S$-type LQ masses. The low-energy spectrum is thus enriched with an extra Higgs doublet.
If sufficiently light, its charged component can contribute to at tree-level $b\rightarrow c\tau\nu$~\cite{Crivellin:2012ye, Crivellin:2013wna, Iguro:2022uzz, Blanke:2022pjy} and induce $b\to s$ transition at one loop~\cite{Crivellin:2013wna, Kumar:2022rcf, Athron:2024rir}.

Therefore, the long-standing B anomalies provide data-driven motivations to reshape the infrared structure of most minimal $SO(10)$ theory. 
Among the scalar components of $\phi_{126}$, (some of) the $S_3, \widetilde{S}_1, R_2, \widetilde{R}_2$ LQ masses, together with two Higgs doublets, should lie around the TeV scale.

\subsection{\texorpdfstring{$b-\tau$}{b-tau} Unification}
\label{btausection}

Taking the limit that the Yukawa sector contains only the third generation fermions, and excluding $S_1$ and $\bar{S}_1$, Eq.~\eqref{chirality} reduces to
\begin{equation}
\label{yukawaDef}
\begin{aligned}
    -\mathcal{L}_{\text{LQ}}~=~y_1 \overline{b_R^c}\tau_R \widetilde{S}_1+y_2 \overline{b_R^c} L_L^{3c} \widetilde{R}_2+ y_3 \overline{Q_L^{3c}}L_L^3 S_3+
    y_4 \overline{Q_L}R_2'\tau_R+y_5\overline{t_R^c}R_2L_L^c+\text{h.c.}
\end{aligned}
\end{equation}
At $M_{\text{GUT}}$, the absolute Yukawa coupling strengths are related to $y_t$, through the group structure of $126_H$
\begin{equation}
 y_1~=~y_2~=~y_3~=~y_4~=~y_5~=~2\sqrt{3} y_t, 
\end{equation}
The factor $2\sqrt{3}$ can be interpreted as a generalized Clebsch-Gordan coefficient, which we take from Ref.~\cite{Patel:2022wya}. Its enhances the LQ Yukawa coupling over the top quark one by more than a factor of about 3.5.

Similar to the two Higgs doublets, $R_2$ and $R_2'$ can mix via a low-scale $U(1)_{\text{PQ}'}$ symmetry breaking term
\begin{equation}
    ~\mathcal{L}_{\text{mix}}~=~\mu_{R}^2 R_2 R_2'+\text{h.c.}
\end{equation}
The mixing term $\mu_{R}^2$, together with the top quark mass, connects the $\tau$ leptons with opposite chiralities and induces an inhomogeneous threshold correction to $m_{\tau}$
\begin{equation}
\label{Tauthreshold}
\begin{aligned}
    \delta m_{\tau}~&=~m_t  N_c\frac{y_{4}y_5}{16\pi^2} \frac{\mu_R^2}{m_{R}^2} ~ C_0\left(1, \frac{m_{R}'^2}{m_R^2}, \frac{m_{t}^2}{m_R^2}\right)\\
    ~&\approx~(1.73~\text{GeV})\times \left( \frac{y_4}{0.8} \right)\left( \frac{y_5}{0.7} \right) \left(\frac{\mu_R^2}{m_{R}^2} \right)C_0\left(1, \frac{m_{R}'^2}{m_R^2}, \frac{m_{t}^2}{m_R^2}\right). 
\end{aligned}
\end{equation}
Here, $m_R$ and $m_R'$ are the physical masses of $R_2$ and $R_2'$. Without loss of generality, we take $m_R> m_R'$. $C_0$ is the scalar Passarino-Veltman function~\cite{Passarino:1978jh} evaluated in the zero external momentum limit, which reads
\begin{equation}
    C_0\left(1, \frac{m_{R}'^2}{m_R^2}, \frac{m_{t}^2}{m_R^2}\right)~=~\frac{m_{R}^2}{m_{R}^2-m_R'^2} \log{\frac{m_{R}'^2}{m_{R}^2}}+\mathcal{O}\left(\frac{m_{t}^2}{m_{R}'^2}\right).
\end{equation}
It is straight-forward to check $C_0\sim \mathcal{O}(1)$, in the absence of a large hierarchy between $m_R$ and $m_R'$. $\left(\mu_R^2/m_{R}^2 \right)$ can be interpreted as the mixing angle between $R_2$ and $R_2'$. 
To fully address the large $b\to c\tau \nu$ anomaly, $\left(\mu_R^2/m_{R}^2 \right)$ and $y_4$ need to take $\mathcal{O}(1)$ values\footnote{$y_5$ is not directly related to the $b\to c$ transition, because it is the coupling strength for interaction involving $t_R$ instead of $c_R$.}. Therefore, Eq.~\eqref{Tauthreshold} leads to a large additive correction to the $\tau$ lepton mass, unless $y_5$ takes value much lower than $0.6$. 
We note this mechanism is in analogy to the SUSY threshold correction to the bottom quark mass in the large $\tan\beta$ regime~\cite{Hall:1993gn,Carena:1993ag,Carena:1999py}.
In particular, the precise value of $\delta m_{\tau}$ only depends on dynamics below the scale of $m_R$, and is \textit{insensitive} to the values of the Yukawa couplings in the UV theory.
As a consequence, although the TeV-scale $R_2$ LQ can generate the correct $b-\tau$ mass relationship and removes a weakness of the most minimal $SO(10)$, 
$m_{\tau}$ becomes an arbitrary parameter at low energies as in the theories without $SO(10)$ symmetry.

Since we want $m_{\tau}$ to be calculable, we exclude the $R_2$ explanation to the $R(D^{(*)})$ anomalies, and consider the scenario that $R(D^{(*)})$ is (partly) addressed by $S_3$ and the additional Higgs doublet. 
Without the data driven motivation, $R_2$ is supposed to lie around $M_{\text{GUT}}$. 
The TeV-scale spectrum contains $S_3, \widetilde{R}_2$, an additional Higgs doublet, and possibly $\widetilde{S}_1$. 
Comparing with $R_2$, these scalars do not lead to inhomogeneous loop corrections to the LQ and Higgs Yukawa interactions, because the $U(1)_{\text{PQ}}$ symmetry ensures each scalar only couples to a single type of fermion bilinear. 
The RG equations for the third-generation specific LQ and Higgs coupling strengths can be reduced from the generic ones provided in Refs.~\cite{Fedele:2023rxb, Grzadkowski:1987wr, Branco:2011iw}
\begin{equation}
\label{runningEq}
    \begin{aligned}
       16\pi^2 \frac{d}{d \log{\mu}}y_t~=&~y_t\left( -\frac{17 g_1^2}{12}-\frac{9 g_2^2}{4}-8 g_3^2+\frac{9 y_t^2}{2}+\frac{y_b^2}{2}+\frac{3 y_3^2}{2}\right), \\
16\pi^2\frac{d}{d \log{\mu}}y_b~=&~y_b\left( -\frac{5 g_1^2}{12}-\frac{9 g_2^2}{4}-8 g_3^2+\frac{y_t^2}{2}+\frac{9 y_b^2}{2}+y_{\tau
   }^2+\frac{y_1^2}{2}+y_2^2+\frac{3 y_3^2}{2}\right), \\
16\pi^2\frac{d}{d \log{\mu}}y_{\tau}~=&~y_{\tau}\left( -\frac{15 g_1^2}{4}-\frac{9 g_2^2}{4}+\frac{5 y_{\tau }^2}{2}+3 y_b^2+\frac{3 y_1^2}{2}+\frac{3 y_2^2}{2}+\frac{9
   y_3^2}{2}\right), \\
16\pi^2\frac{d}{d \log{\mu}}y_1~=&~y_1\left(-2 g_1^2-4 g_3^2+ y_b^2+\frac{y_{\tau }^2}{2}+3 y_1^2+y_2^2\right), \\
16\pi^2\frac{d}{d \log{\mu}}y_2~=&~ y_2\left(-\frac{13 g_1^2}{20}-\frac{9 g_2^2}{4}-4 g_3^2+y_b^2+\frac{y_{\tau }^2}{2}+\frac{y_1^2}{2}+\frac{7
   y_2^2}{2}+\frac{9 y_3^2}{2}\right), \\
16\pi^2\frac{d}{d \log{\mu}}y_3~=&~y_3\left( -\frac{g_1^2}{2}-\frac{9 g_2^2}{2}-4 g_3^2+\frac{y_t^2}{2}+\frac{y_b^2}{2}+\frac{y_{\tau }^2}{2}+\frac{3
   y_2^2}{2}+8 y_3^2 \right).\\
    \end{aligned}
\end{equation}
Here, $g_1,g_2,$ and $g_3$ are the gauge coupling strengths of $U(1)_Y, SU(2)_L,$ and $SU(3)_c$, respectively.
All coupling strengths in Eq.~\eqref{runningEq} should be understood as functions of the renormalization scale $\mu$. 
To illustrate the generic behavior of the RG running, one can view $g_1, g_2, g_3$ as slowly varying functions above the TeV scale. 
Then, if the Yukawa couplings take large values at $M_{\text{GUT}}$, they dominate over the gauge coupling strengths in the RG equations. The corresponding beta functions are thus all positive and drive the Yukawa coupling strengths to decrease during the evolution towards low energies. 
Conversely, if the Yukawa coupling strengths are small at $M_{\text{GUT}}$, their beta functions are dominated by the gauge coupling strengths and become positive. The Yukawa couplings strengths increase during the running from UV to infrared. 
No matter decreasing or increasing, the evolution slows down when the Yukawa coupling strengths get closed to the gauged ones.
In the deep infrared limit, all beta functions vanish and the running Yukawa coupling strengths become constants, commonly \typo{referred as}{referred to as} `fixed points'. 
The fixed points can be found by solving the system of equations
\begin{equation}
    \frac{d}{d\log\mu}y_f~=~0, \quad f~=~t,b,\tau,1,2,3. 
\end{equation}
For simplicity, we neglect $g_1$ and $g_2$. The fixed points solution reads
\begin{equation}
\begin{aligned}
    y_t~&\to~ 1.24 g_3, \quad y_b~\to~1.17 g_3, \quad y_{\tau}~\to~0,\\
    \quad y_1~&\to~0.88g_3, \quad y_2~\to~0.55g_3, \quad y_3~\to~0.51 g_3. 
\end{aligned}
\end{equation}
Therefore, irrespective of their values at the GUT scale, $y_t, y_b$ must be as large as $g_3$ at low energies. 
Since the running equation of $y_{\tau}$ does not involve $g_3$, $y_{\tau}$ evolves towards zero until it feels the effect of $g_1$ and $g_2$. 
This fixed point behavior qualitatively explains how the ratio $(y_{\tau}/y_{b})$ evolves from 3 at $M_{\text{GUT}}$ to about $0.7$ at the TeV scale. 
Moreover, the low energy LQ-fermion couplings are also numerically close to $g_3$. 
Such sizable couplings further support the scalar LQ explanation of the $B$ anomalies. 
We consider $y_2\approx y_3$ as a critical prediction. As explained in subsection~\ref{reshapeIR}, the $S_3$ and $\widetilde{R}_2$ contributions to $B\rightarrow K^{*}\overline{\nu}\nu$ have to moderately cancel each other, which becomes much less ad-hoc given $y_2\approx y_3$.
Interestingly, the behaviors of $y_1, y_2,$ and $y_3$ are very similar to what was recently found in Ref.~\cite{Fedele:2023rxb}, that the LQ-fermions couplings approach the infrared fixed-points around $0.5\sim1.0$. Interestingly, the LQ-type considered here is different, suggesting that this feature is a rather general consequence of RG evolution.

\begin{figure}[t!]
    \centering
    \includegraphics[width=0.49\linewidth]{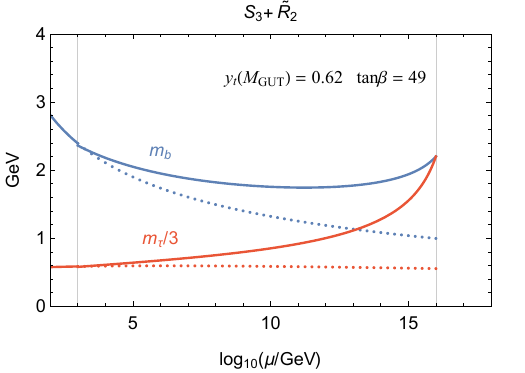}
    \includegraphics[width=0.50\linewidth]{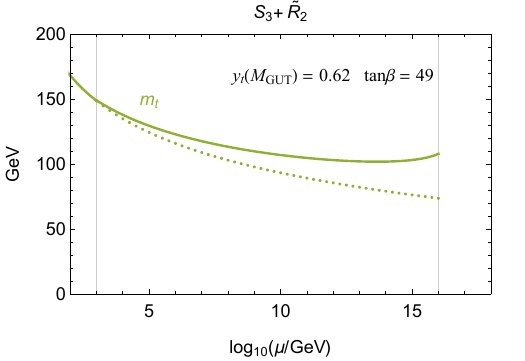}\\
    \vspace{10pt}
    \includegraphics[width=0.49\linewidth]{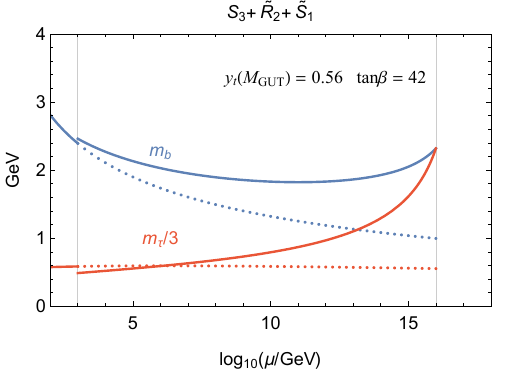}
    \includegraphics[width=0.5\linewidth]{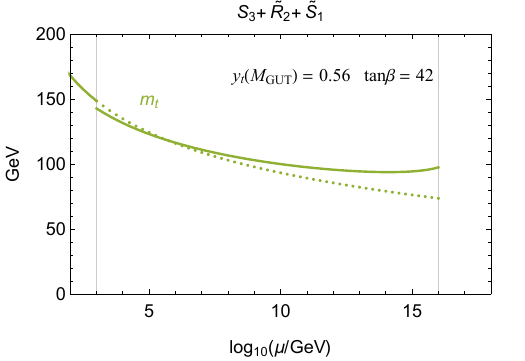}
    \caption[RG evolution of the third-generation charged fermions masses.]{RG evolution of the third-generation charged fermions masses from $10^2$ to $10^{16}$ GeV. The solid (dotted) lines indicate the scenario with (without) LQs at TeV scale. 
    We fix $y_t(M_{\text{GUT}})=0.56$ to get the correct value of $m_t$ and Yukawa unification implies $\tan\beta=42$. The gray vertical lines indicate $M_{\text{GUT}}$ and the light LQ threshold.}
    \label{S3R2}
\end{figure}

As mentioned at the beginning of this chapter, the fixed points can only be reached at very low energies, when the QCD coupling strength starts to diverge and should no longer be viewed as a slowly varying parameter. 
At the TeV-scale, the Yukawa couplings are only `approaching' the fixed points.
To get quantitative predictions, information from both UV and infrared theories are needed. 
There are two free input parameters in total: the GUT-scale value of $y_t$ and the Higgs VEV ratio $\cot\beta$. 
On the other hand, three low energy flavour parameters are to be predicted: the three masses of the third-generation charged fermions $m_t, m_b,$ and $m_{\tau}$. 
We find in case $S_3, \widetilde{R}_2$, and an additional Higgs doublet in $\phi_{126}$ are as light as 1 TeV, $y_t\approx 0.62$ at $M_{\text{GUT}}=10^{16}$ GeV, and $\tan\beta=49$, the resulting $b, \tau, t$ masses at 1 TeV agree well with the values extracted from the SM\footnote{Varying the masses of the LQs or $M_{\text{GUT}}$ around the values we choose do not significantly change the results, because the low-energy outputs only depend on the logarithm of these scales.}.
RG evolution of these three masses from $10^{16}$ GeV down to $10^2$ GeV are illustrated by the solid lines in the top-left and top-right panels of Figure~\ref{S3R2}. 
For the running above $1~\text{TeV}$, we use the RG equation shown in Eq.~(\ref{runningEq}), with $y_1$ fixed as zero.
As shown in \fref{S3R2}, $(m_{\tau}/3)$, represented by the red solid curve, decreases rapidly towards zero, while $m_t$ and $m_b$ only decrease when $\mu$ is large and then start to grow up, analogous to the QCD coupling strength $g_3$. 
At $1~\text{TeV}$, all three masses matches perfectly with the interpolated the SM running data, which we plot using the SMDR package~\cite{Martin:2019lqd}. 
For comparison, we also show the evolution without TeV-scale LQs, by extending the SM running curve to regions of $\mu>1$ TeV~\cite{Martin:2019lqd} with dots. 
In this scenario, $(m_{\tau}/3)$, represented by the red dotted curve, is small and remains nearly a constant. $m_b$, shown with the blue dotted curve, decreases as $\mu$ goes up mainly due to QCD effects, but its evolution is not fast enough to meet $(m_{\tau}/3)$ at $M_{\text{GUT}}$. 
Consequently, the red and blue dotted curves diverge significantly at $10^{16}$ GeV, indicating the relation $m_b=m_{\tau}/3$ cannot be reached if the theory below $M_{\text{GUT}}$ is SM alone.

We also explore scenarios with $\widetilde{S}_1$ also as light as the TeV scale. 
For this scenario, we find $y_t(M_\text{GUT})=0.56$ and $\tan\beta=42$ can also lead to reasonably good matching with the SM data at 1 TeV, which we show in the bottom-left and bottom-right panels of \fref{S3R2}. 
Although the solid lines are not exactly continuous at 1 TeV, yielding the matching is not as perfect as the scenario without $\widetilde{S}_1$, the gaps visible in the plots are not problematic. At leading-log level, TeV-scale LQs can reduce the $b-\tau$ mass tension from an $\mathcal{O}(1)$ discrepancy to a suppressed small mismatch, which is subject to other higher-order corrections we do not include here. 
Rather, all panels in \fref{S3R2} demonstrate improved $b-\tau$ unification with TeV-scale LQs as a universal feature. 
The outcome seems not accidental as a particular LQ choice, but rather a model-independent effect of the colored scalar fields. We understand the underlying reason as follows. 
As shown in Eq.~(\ref{runningEq}), the LQ contribution to the homogeneous terms in the $\beta$-function of $y_{\tau}$ are always three times larger than those for $y_{b}$. 
The one-loop LQ corrections to $y_{b}$ and $y_{\tau}$ originate from the self-energy diagrams for bottom quarks and $\tau$ leptons, which contain the same couplings strengths and integral structures but different color factors.
The LQ-quark loop for lepton self-energies involves a color factor $N_c=3$ while the LQ-lepton loop for quark self-energies does not, so we have
\begin{equation}
    \text{LQ-quark loop for leptons}~=~\text{LQ-lepton loop for quarks}\times N_c.
\end{equation}
This relation can be intuitively understood as three quarks with different colors run in the loop for the lepton self-energies, while only one lepton runs in the loop for the self-energies of a quark with one specific color index. 
Therefore, the TeV-scale LQs, no matter which type, always drive the lepton masses to decrease more rapidly than quarks, yielding $(m_{\tau}/m_b)$ to take smaller low-energy values than the $SO(10)$ prediction.

The discussion above shows the problem on the wrong $b-\tau$ mass relationship in the most minimal $SO(10)$ theory can be solved through modifying the infrared structure of the theory. 
The existence of TeV-scale LQs explicitly breaks the $SO(10)$ symmetry and leads to large but calculable corrections to the symmetry-protected prediction $m_{\tau}=3m_b$. 
This example demonstrates that although the theory of SM with TeV-scale LQs lies in the swampland of the $SO(10)$ symmetry, some of the low energy flavour parameters remain calculable in analogy to those restricted by explicit flavor symmetries. The predicted relationships can stay consistent with observed values. 

\subsection{Flavour Mixing}
\label{flavourmixing}

The implementation of flavour mixing still needs to be discussed. 
In SM, the $3-2$, $3-1$ quark mixing angles are small. 
The $2-1$ quark mixing angle (Cabibbo angle) is sizable as of $0.2$, but it becomes unphysical in the limit that the first-two generation quarks are massless. 
Similar arguments apply to the large neutrino mixing angles. 
The B anomalies involve $b\to s$ and $t\to c$ mixing beyond the SM. 
If the responsible LQs lie at around 1 TeV, the $b\to s$ mixing angles also stay small. 
The only exception is on the $b\to c$ transition, since the $R(D^{(*)})$ discrepancy to the SM is large, a sizable $t\to c$ mixing angle is needed. 
Despite this, the measured $b\to c\tau\nu$ transition rate remains subject to statical uncertainties, and future experimental data can also bring it closer to the SM predictions. 
Therefore, we view the third-generation specific Yukawa coupling patterns as the leading order approximation, and consider the flavour mixing to be the sub-leading corrections.

If one views the most minimal $SO(10)$ as an effective theory at $M_{\text{GUT}}$, 
it is then necessary to include flavor mixing corrections for a complete analysis.  
The EFT motivation is that the flavor-conserving limit is not protected by a fundamental principle such as gauge invariance, but comes out as a result of minimality, \textit{i.e.} because the theory contains only one unique Yukawa coupling matrix. 
Since $M_{\text{GUT}}$ is close to the the scale the EFT description for gravity becomes non-perturbative, new dynamics are believed to lie not far beyond $M_{\text{GUT}}$, which in general involve additional Yukawa coupling matrices.
Potential flavor violating corrections can arise from additional scalars that couple to $\psi_{16}^i$ via loops~\cite{Witten:1979nr,Bajc:2004hr, Bajc:2005aq}, or higher-dimensional operators~\cite{Ellis:1979fg} generated by vector-like fermions or other gravity related GUT multiplets. 

\begin{figure}
    \centering
    \includegraphics[width=0.45\linewidth]{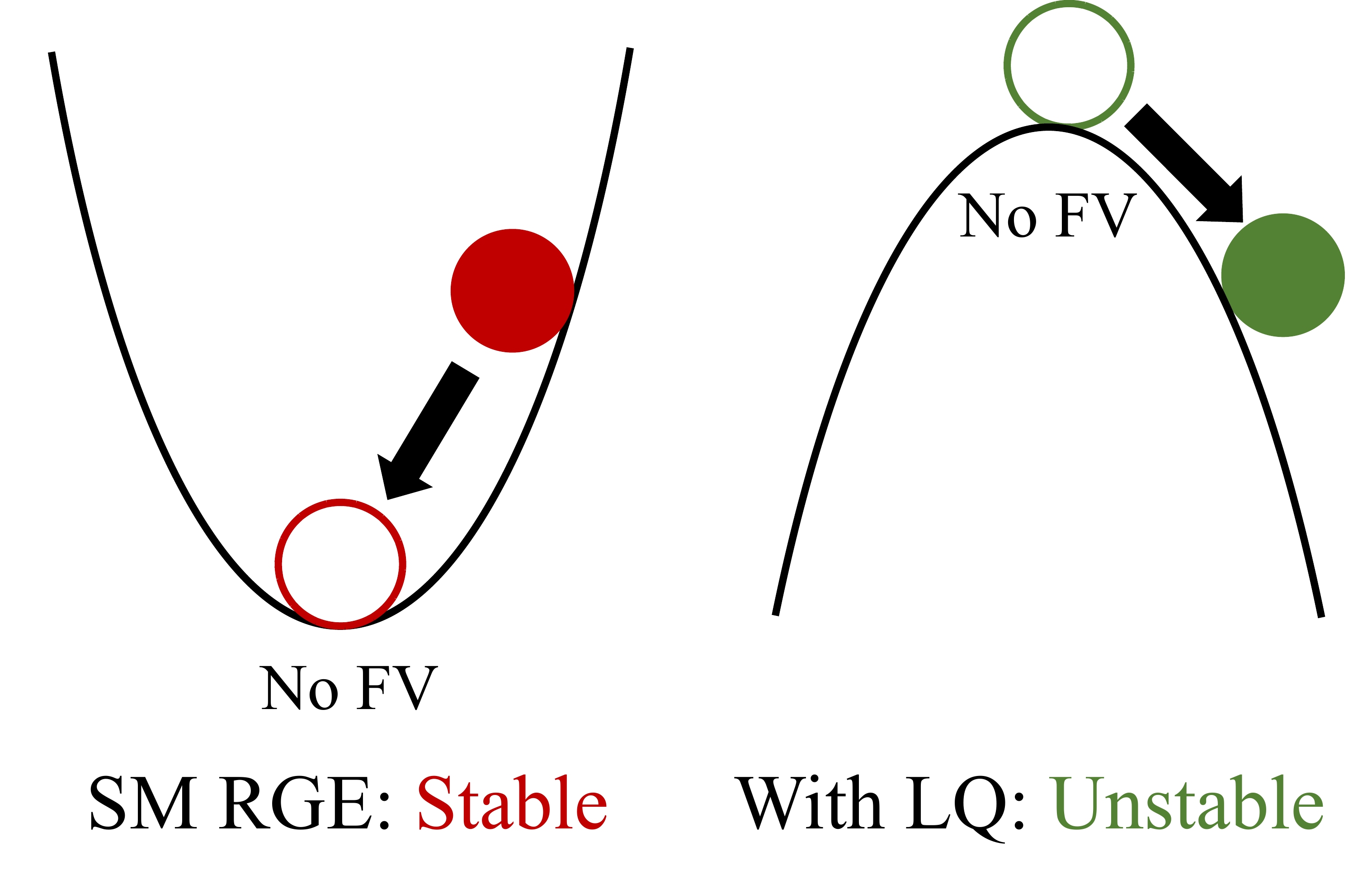}
    \caption[The low energy stability of the flavour conserving limit.]{A schematic illustration on at low energy stability of the flavour conserving limit in the SM (left panel) and the theory containing TeV scale LQs (right panel).}
    \label{stability}
\end{figure}

Our approach is to add the maximally allowed flavor-breaking term $\mathcal{L}_Y^{\epsilon}$ at low energies. 
This follows a well-established methodology of including the soft SUSY-breaking Lagrangian for the MSSM, which is also chosen maximal to cover all possible SUSY-breaking scenarios. 
Therefore, we do not specify the underlying mechanism of flavor breaking, but instead add 
the following four new interacting terms
\begin{equation}
\label{nondiag}
    \mathcal{L}_Y^{\epsilon}~=~ \epsilon^{ct} y_t \overline{Q_L^2}t_RH_u+ \epsilon^{sb} y_b \overline{Q_L^2}b_RH_d+\epsilon^{bs} y_b \overline{Q_L^3}s_RH_d+ \epsilon^{bs}_1y_1 \overline{s_R^c}\tau_R \widetilde{S}_1+\text{h.c.}
\end{equation}
This equation captures \emph{all} possible $3-2$ generation quark flavor violating effects which should be extended to the over-restrictive flavor conserving interactions. 
Here, the flavor basis is chosen without loss of generality, such that the $S_3$ and $\widetilde{R}_2$ couplings are aligned to $Q_L^3$ and $b_R$, respectively. 
The Higgs Yukawa coupling matrices are not diagonal. 
Once $H_u, H_d$ acquire VEVs, the diagonalisation of the quark mass matrix provides the needed flavor-changing couplings for $S_3$ and $\widetilde{R}_2$. 
The $t_R-c_R$ rotation remains unphysical, so we do not need to include the $\overline{Q_L^3}c_RH_u$ term. 
Mixings involving leptons or first-generation fermions are omitted, as they are irrelevant for  the $b\to s$ and $b\to c$ processes under consideration. 
We do not consider $R_2$ due to the reasons explained in subsection~\ref{btausection}, but if $R_2$ is also included in the TeV-scale spectrum, the relevant flavour violating terms can be incorporated accordingly. 
Therefore, no matter which theory lies behind Eq.~(\ref{nondiag}), it serves as the maximally possible extension and covers all possible models generating flavour mixing. 
We believe it is also important to remark here, that the flavor mixing terms in Eq.~(\ref{nondiag}) do not render the relation $m_{\tau}=3m_b$ arbitrary at $M_{\text{GUT}}$. 
This relation is predicted by the $T_{B-L}$ generator of the adjoint representation of $SU(4)_c$, which is the relevant symmetry to unify quarks and leptons. 
To change $m_{\tau}=3m_b$, one needs a $(\textbf{1}_C,\textbf{2}_L,\textbf{2}_R)$ Higgs field  or an effective operator containing $(\textbf{15}_C,\textbf{2}_L,\textbf{2}_R)\times(\textbf{15}_C,\textbf{2}_L,\textbf{2}_R)$. 
By contrast, any extensions to the minimal Yukawa sector can lead to flavor violation. 
For instance, two $(\textbf{15}_C,\textbf{2}_L,\textbf{2}_R)$ Higgs fields (which can originate from two $\phi_{126}$) incorporate flavor but do not change the $b-\tau$ mass ratio.

Nevertheless, simply adding the flavour violating interactions without considering their origin is ad-hoc and sacrifices the elegance of minimality. 
To make flavour mixing arise with more convincing motivations, we reflect on the statement that the most minimal $SO(10)$ theory must be exactly flavour conserving at low energies and reexamine whether it accounts for a no-go theorem for the most minimal $SO(10)$ theory. 
We note that the validity of a physical prediction is not only related to its logical rigor but also involves the robustness. 
Under the SM RG evolution it is well-known that the quark flavour mixing angles all decrease from the high scales to low scales~\cite{Ma:1979cw, Sasaki:1986jv, Grzadkowski:1987tf}.
Then, if flavour conservation is exact at $M_{\text{GUT}}$, the vanishing of all quark mixing angles at low energies yields an stable infrared limit. 
We find that this situation changes when TeV-scale LQs are introduced, which we schematize in \fref{stability}.  
If flavor conservation is not exact at $M_{\text{GUT}}$, the RG flow can amplify small deviations and generate larger flavor violating interactions at low energies. A similar example is the RG amplification of the neutrino mixing angles in the MSSM~\cite{Babu:1993qv, Tanimoto:1995bf, Balaji:2000gd, Balaji:2000ma, Hagedorn:2004ba}. Therefore, flavor mixing can manifest itself as an emergent phenomenon with small `seeds' above $M_{\text{GUT}}$. The complexity of the low-energy theory of the most minimal $SO(10)$ can arise dynamically through the RG evolution.

\begin{figure}[t!]
    \centering
    \includegraphics[width=0.95\linewidth]{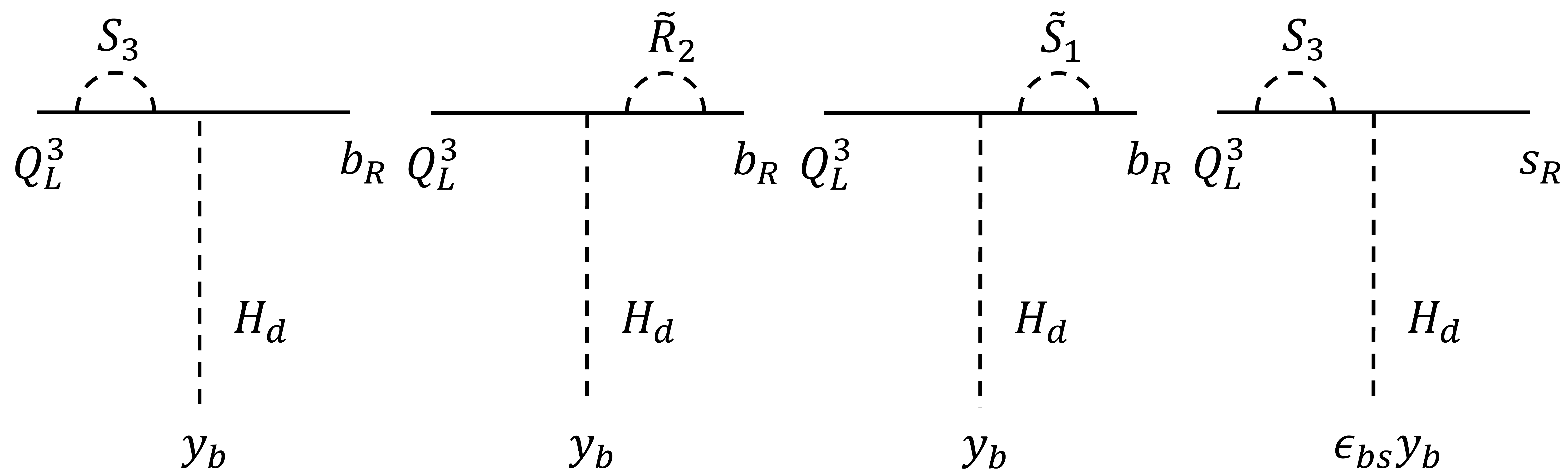}
    \caption[Feynman diagrams for the LQ contributions to $y_b$ and $(\epsilon^{bs}y_b)$.]{Feynman diagrams illustrating the LQ contributions to the running of $y_b$ and $(\epsilon^{bs}y_b)$. The self-energy correction to $Q_L^3$, arising from $S_3$, is universal  to both $y_b$ and $(\epsilon^{bs}y_b)$. $\widetilde{S}_1$ and $\widetilde{R}_2$ couple to $b_R$ only so only contribute to $y_b$. $s_R$ receives no self-energy corrections.}
    \label{feyndiag}
\end{figure}

We take the evolution of $\epsilon^{bs}$ as an example to explain how the flavour conserving limit becomes unstable in the infrared. 
We firstly include the generic flavour violating terms $\mathcal{L}_Y^{\epsilon}$ shown in Eq.~(\ref{nondiag}) at a scale $\Lambda_f$\footnote{It is worthy to remark that although Eq.~(\ref{nondiag}) is defined at the low energies when we introduce it, it is well defined at all energy scales, because it contains the general flavor mixing terms, with no new physical counter-terms from renormalization.}. Typically $\Lambda_f\gtrsim M_{\text{GUT}}$, and new dynamics correcting the most minimal $SO(10)$ ansatz arise above the $\Lambda_f$ threshold.
Below $M_{\text{GUT}}$, the RG equations relevant to the running of this flavour violating coupling is given by:
\begin{equation}
\label{runningeqnondiag}
    \begin{aligned}
    16\pi^2\frac{d}{d \log{\mu}}(\epsilon^{bs}y_b)~=&~\epsilon^{bs}y_b\left( -\frac{5 g_1^2}{12}-\frac{9 g_2^2}{4}-8 g_3^2+\frac{y_t^2}{2}+\frac{9 y_b^2}{2}+y_{\tau}^2+\frac{3 y_3^2}{2}\right). \\
    \end{aligned}
\end{equation}
Here, we neglect higher order flavor-violating terms. Comparing with the running equation of $y_b$ shown in Eq.~(\ref{runningEq}), the $\left(\frac{y_1^2}{2}+y_2^2\right)$ term is absent here. This is a generic consequence for the third-generation specific LQ couplings, which do not contribute to the self-energy corrections for $s_R$. 
For illustration, we compare the Feynman diagrams contributing to $y_b$ and $(\epsilon^{bs}y_b)$ in \fref{feyndiag}, which shows the $\widetilde{S}_1$ and $S_3$ contribution to $(\epsilon^{bs}y_b)$ is absent. 
As a consequence, during the evolution from UV to infrared, $(\epsilon^{bs}y_b)$ decreases more slowly than $y_b$, and the mixing parameter $\epsilon^{bs}$ effectively increases. 
This behavior is more explicit in the RG equation for $\epsilon^{bs}$ itself
\begin{equation}
\label{runningeqmixing}
    16\pi^2\frac{d}{d \log{\mu}}\epsilon^{bs}=-\epsilon^{bs}\left( \frac{y_1^2}{2}+y_2^2\right). \\
\end{equation}
The beta-function for $\epsilon^{bs}$ is negative. As a generic feature, the sign for the beta-function remains negative above $M_{\text{GUT}}$, which can be cross checked by the relevant RG equations and matching conditions shown in Ref.~\cite{Aulakh:2002zr, Meloni:2016rnt, Aydemir:2018cbb, Fang:2025hlc}. 
Therefore, if $\epsilon^{bs}$ vanish at $\Lambda_f$, it remains zero at all energy scales because $(d\epsilon^{bs}/d\log{\mu})$ is proportional to $\epsilon^{bs}$. 
On the other hand, if $\epsilon^{bs}\neq0$ at $\Lambda_f$, the RG evolution then drives it towards sizable values at the low energies.

\begin{figure}
    \centering
    \includegraphics[width=0.65\linewidth]{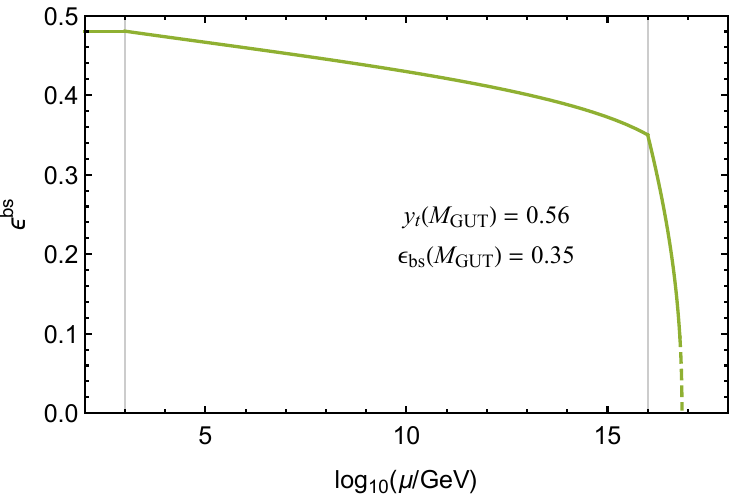}
    \caption[The RG evolution of $\epsilon^{bs}$.]{RG evolution of $\epsilon^{bs}$ with $y_t(M_{\text{GUT}})=0.58$. The dashed line lies in the region where the Yukawa couplings become non-perturbative and thus may not reflect the physical reality.}
    \label{mixbs}
\end{figure}

As a consequence, the $s_R-b_R$ mixing is enhanced at low energies with respect to its value at $\Lambda_f$. 
The evolution of $\epsilon^{bs}$ is illustrated with the solid line in Figure~\ref{mixbs}. 
We fix $y_t=0.58$ and $\epsilon^{bs}=0.35$ at $M_{\text{GUT}}$. 
$\epsilon^{bs}$ increases as $\mu$ decreases, so the preservation of flavors is clearly not a stable solution of the RG equations. 
It is worthy to remark that the evolution below $M_{\text{GUT}}$ is slow: Achieving $\epsilon^{bs}\approx 0.5$ at the TeV scale typically requires its GUT scale value be at least $0.35$, which one may consider as too large for a perturbation. 
On the other hand, we find $\epsilon^{bs}\sim0.35$ at $M_{\text{GUT}}$ can arise from tiny values at around $10 M_{\text{GUT}}$. 
This is partly due to a change of the RG equations.
Above the $M_{\text{GUT}}$ threshold, the $(\textbf{6}_C,\textbf{1}_L,\textbf{1}_R)$, $(\textbf{10}_C, \textbf{3}_L, \textbf{1}_R),$ and $(\textbf{15}_C,\textbf{2}_L,\textbf{2}_R)$ multiplets contained in $\phi_{126}$ all contribute to the $b_R$ self-energy. 
We include their effects using the results shown in Ref.~\cite{Aulakh:2002zr, Meloni:2016rnt, Aydemir:2018cbb}. 
Another reason is that the absolute value of the Yukawa coupling becomes large above $M_{\text{GUT}}$. 
Using to the one-loop running equations for the gauge and Yukawa couplings of $SO(10)$~\cite{Vaughn:1978st, Vaughn:1981qi, Fang:2025hlc}, we find $y_{126}$ increases rapidly and approaches the non-perturbative regime when $\mu\gtrsim \Lambda_f\sim 10^{17}$ GeV.
We show $\epsilon^{bs}$ lying in the non-perturbative region with the dashed line in \fref{mixbs}, indicating that extrapolating the perturbative calculation results may not reflect the physical reality near $\Lambda_f$. 
Therefore, if physics at the scale $\Lambda \gtrsim 10\, M_{\text{GUT}}$ gives a tiny correction to the most minimal $SO(10)$ theory, the RG evolution can yield sizable flavor mixings at the low energies. 
It is worth remarking that this enhancement originates from the self-energy correction to all components in $\psi_{16}^3$, which is universal to the quarks and leptons and therefore does not change the relation $m_{\tau}=3m_b$.

Although the $y_{126}$ become non-perturbative around $\Lambda_f$, asymptotic freedom of most minimal $SO(10)$ theory (excluding the scalar self-couplings) can be restored in the deep UV~\cite{Vaughn:1981qi}.
Above $M_{\text{GUT}}$, large $y_{126}$ can induce a sizable $SO(10)$ gauge coupling strength $g_{10}$ through two-loop effects, then the sizable $g_{10}$ in turn slows down the running of $y_{126}$ and drives it to decrease.
Nevertheless, comparing with invoking the dynamics between $M_{\text{GUT}}$ and $\Lambda_f$, a more convincing approach to generate large flavour mixings is further modifying the infrared structure of the most minimal $SO(10)$. 
This can be achieved by requiring more light scalar components in $\phi_{126}$, such as di-quarks~\cite{Arnold:2009ay, Giudice:2011ak, Patel:2022nek, Crivellin:2023saq}, which can also speed up the running of quark Yukawa couplings as the LQs do. 
$\epsilon^{bs}$ then grows more rapidly when $\mu$ evolves towards low energies. 
The existence of light di-quarks are also motivated by data, because they can also explain the $b\rightarrow s\ell\ell$ anomalies~\cite{Crivellin:2023saq} and furthermore are related to the interpretation of the observed CP asymmetry in $D_0\rightarrow\pi^+\pi^-$ and $D_0\rightarrow K^+K^-$ decays~\cite{LHCb:2019hro,LHCb:2022lry, Altmannshofer:2012ur, Iguro:2024uuw}.

The behaviors of left-handed quark mixings $\epsilon^{sb}$ and $\epsilon^{ct}$ are similar to that of $\epsilon^{bs}$. 
Both $\epsilon^{sb}=0$ and $\epsilon^{ct}=0$ are unstable infrared limits. 
Given small $\epsilon^{sb}$ and/or $\epsilon^{ct}$ at $\Lambda_f$, they firstly increase rapidly and the growing rates slow down below $M_{\text{GUT}}$. 
For the left-handed quarks, the $b_L-s_L$ mixing should be well aligned with $t_L-c_L$ mixing at low energies, to preserve the small $V_{cb}\sim -V_{ts}\sim 0.04$. 
Since $SU(2)_L$ gauge symmetry stays unbroken above the electroweak scale, the RG evolution from $\Lambda_f$ down to the TeV scale is identical for $\epsilon^{sb}$ and $\epsilon^{ct}$. 
The only requirement is that $\epsilon^{sb}=\epsilon^{ct}$ must hold up to a good precision at $\Lambda_f$ and this relation can be protected by the $SU(2)_R$ symmetry.

\section{Non-perturbative Theories: Bootstrap}
\label{phaseTransition}

\subsection{Gravitational Anomaly Matching}
\label{anomalymatching}
We now consider the maximal possible flavour symmetry --- $U(48)$ symmetry --- introduced at the beginning of Section~\ref{Fpuzzlesection}.
We would like to ask: if gravity is included in the free field theory shown in Eq.~\eqref{gravity}, can the $U(48)$ symmetry remain explicit at low-energies? 
In other words, for the theory constructed by 48 Weyl fermions and gravity, does it lie in the landscape or swampland of $U(48)$?

The global or weakly gauged symmetries are widely viewed as incompatible with gravity, see Ref.~\cite{Arkani-Hamed:2006emk, Banks:2010zn, Harlow:2018tng} for reviews. 
If this conjecture is true, the $U(48)$ symmetry must be broken at the low energies. 
To restore consistency in the effective theories, one can include the so-called `biased terms', which explicitly break the would-be global symmetries and allow the 48 Weyl fermions to acquire masses and mix with each other.
Although one can stay agnostic on the origin of such biased terms, it would be more convincing to precisely understand how gravity responds to global symmetries.
Without loosing generality, the origin of such biased terms can be 
classified by
\begin{enumerate}[label=(\roman*)]
    \item the UV completions for the standard gravity, for instance, from string theories, or 
    \item non-perturbative effects at the low energies. 
\end{enumerate}
Option (i) implies that a theory containing merely gravitationally interacting fermions is inconsistent at all scales and must be embedded into the extended frameworks. 
By contrast, option (ii) indicates that the apparent inconsistency only manifests itself in the perturbative frameworks, and can be explained via non-perturbative effects. This dynamical symmetry breaking mechanism implies that the induced flavour parameters are, in principle, calculable, although not necessarily within the perturbative frameworks. 
Similar non-perturbative effects are well-known in non-abelian gauge theories~\cite{tHooft:1973alw, tHooft:1976rip, tHooft:1976snw, Witten:1979vv, Veneziano:1979ec}.

According to Ref.~\cite{Dvali:2017mpy}, if the topological susceptibility of the vacuum in pure gravity is non-zero, the $U(N_f)$ chiral flavour symmetry of a theory containing $N_f$ massless Weyl fermions and gravity must dynamically break down to one of its maximal anomaly-free subgroups. This mechanism is similar to and shares the same topological origin as the chiral symmetry breaking in QCD, while the difference is that the gravity induced fermion condensate do not necessarily lead to confinement. 
Here, to be self-contained, we introduce the gravity induced condensation for the 48 massless Weyl fermions $\psi$, using the gravitational anomaly matching technique established in Ref.~\cite{Dvali:2005an, Dvali:2005ws, Dvali:2013cpa, Dvali:2016uhn, Dvali:2017mpy, Dvali:2022fdv}.  
In the standard (Einstein-Hilbert) theory for pure gravity, one can include a unique Chern-Simons term in the Lagrangian
\begin{equation}
\label{EinsteinGravity}
    \mathcal{L}_{\text{Gravity}}~=~M_P^2\mathcal{R}+ \theta_G R\widetilde{R}. 
\end{equation}
Here, $\mathcal{R}$ is the Ricci scalar, $R$ is the Riemann tensor, and $\widetilde{R}~\equiv~^*R$ represents the Hodge dual. 
The topological susceptibility of the vacuum is defined as
\begin{equation}
\label{corre}
    \langle R\widetilde{R}, R\widetilde{R}\rangle_{p\to 0}~\equiv~\lim_{p\to 0}
    \int d^4x e^{ipx}\langle T[R\widetilde{R}(x), R\widetilde{R}(0)]\rangle. 
\end{equation}
It is reasonable to expect this correlator to be non-zero, because it receives contributions from the microscopic black holes states 
\begin{equation}
    \langle R\widetilde{R}, R\widetilde{R}\rangle_{p\to 0}~=~
    \sum_k \langle R\widetilde{R}|\text{black hole}_k\rangle\langle\text{black hole}_k | R\widetilde{R}\rangle~+ ~...
\end{equation}
The challenge is how to calculate the $\langle R\widetilde{R}|\text{black hole}_k\rangle$ correlator without more detailed control of quantum gravity. Here, we take a non-zero $\langle R\widetilde{R}, R\widetilde{R}\rangle_{p\to 0}$ as a prior but physically motivated assumption.

Non-zero $\langle R\widetilde{R}, R\widetilde{R}\rangle_{p\to 0}$ indicates that the vacuum of the theory shown in Eq.~\eqref{EinsteinGravity} breaks CP-symmetry, which is equivalent to the statement that $\theta_G$ is physical. 
It is worthy noting that the $\theta_G$ term in Eq.~\eqref{EinsteinGravity} is a UV parameter while $\langle R\widetilde{R}, R\widetilde{R}\rangle_{p\to 0}$ is defined in the zero momentum limit and represents the vacuum structure relevant to the non-perturbative phenomena\footnote{Intuitively, one may understand it as that the strongly interacting gravitons break the CP invariance.}. 
To study this low energy behavior, expanding $R$ with graviton fields $h_{\mu\nu}$ is not a convenient framework. 
Instead, one can take the Chern-Simons three-from and its field strength
\begin{equation}
    C~=~\Gamma\wedge d\Gamma+\frac{2}{3} \Gamma\wedge \Gamma\wedge \Gamma, \quad E=~^*dC,
\end{equation}
where $\Gamma$ is the Christoffel connection, and it is straight forward to check $E=R\widetilde{R}$. 
Under the general coordinate transformations, $C$ shifts by an exterior derivative of a two-form field $w$:
\begin{equation}
\label{redundancy}
    C~\to~C+dw. 
\end{equation}
All components in $C$ can be eliminated by this gauge redundancy, so $C$ does not carry any propagating degrees of freedom. 
Nevertheless, $C$ can mediate long range interactions, because the non-vanishing $\langle R\widetilde{R}, R\widetilde{R}\rangle_{p\to 0}$ (or equivalently $\langle E, E\rangle_{p\to 0}$) implies the propagator of the $C$ field has a pole at $p^2=0$
\begin{equation}
\label{pole}
    \langle C, C\rangle_{p\to 0}~\sim~\frac{1}{p^2}. 
\end{equation}
In other words, the non-zero vacuum topological susceptibility implies a massless (non-propagating) three-form field in the low energy theory. The $C$-field description for the vacuum is more transparent than that with the gravitons. 
Considering the Lagrangian of an effective theory constructed by the $C$-field, the vacuum solution requires that its variation with respect to $C$ must vanish. 
It is then straight-forward to check that the field strength satisfying
\begin{equation}
\label{VacuumSol}
    \langle E\rangle~=~\text{const},
\end{equation}
is a vacuum solution.
This constant field strength is in analogy to the ordinary electric field created by a charged capacitor.
As $E$ is CP-odd, Eq.~\eqref{VacuumSol} yields the vacuum of gravity --- represented by a
distance-independent static force --- can break the CP symmetry.

\begin{figure}[t!]
    \centering
    \includegraphics[width=0.45\linewidth]{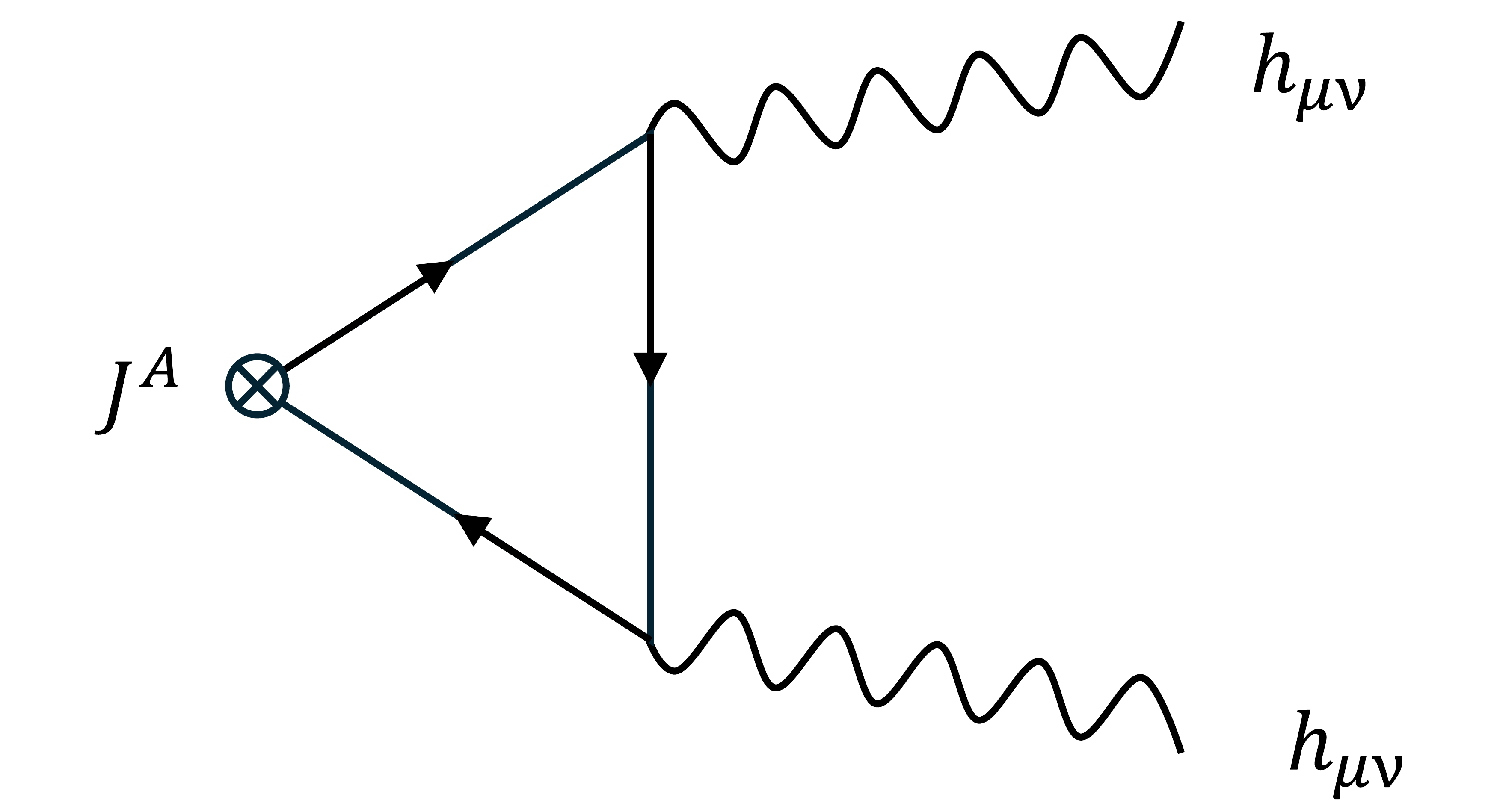}
    \caption{The triangle diagram for the gravitational anomaly.}
    \label{anomaly}
\end{figure}

Now we include the 48 Weyl fermions to Eq.~\eqref{EinsteinGravity}. The theory contains is $U(48)$ global symmetry, which can be decomposed by
\begin{equation}
    U(48)~=~SU(48)_f\times U(1)_A. 
\end{equation}
Under the $U(1)_A$ symmetry, all Weyl fermions in $\psi$ transform by an universal phase rotation
\begin{equation}
    \psi_\alpha~\to~\psi_{\alpha}~=~e^{i \theta'_G} \psi_\alpha, \quad \alpha=1,2,...,48. 
\end{equation}
Due to the so-called Adler-Bell-Jackiw chiral anomaly~\cite{Adler:1969gk, Bell:1969ts}, the $\theta_G$ term shown in Eq.~\eqref{EinsteinGravity} shift under this $U(1)_A$ transformation as
\begin{equation}
    \theta_G~\to~\theta_G+\kappa \theta'_G, \quad \kappa\neq0.
\end{equation}
This shift can be understood as originating from 
the one-loop diagram shown in \fref{anomaly}. 
The $\theta_G$ term is therefore unphysical, because it can be eliminated by a $U(1)_A$ transformation satisfying $\theta_G'=-\kappa\theta_G$. 
The unphysical $\theta_G$ implies that the CP symmetry is unbroken from the view of an UV observer.
Correspondingly, the low energy CP-violating observable $\langle R\widetilde{R}, R\widetilde{R}\rangle_{p\to 0}$ must also vanish. 
This vanishing correlator implies that the $p^2=0$ pole of the two-point function of the three-form field $C$ is removed. 
Eq.~\eqref{pole} becomes
\begin{equation}
    \langle C, C\rangle_{p\to 0}~\sim~\frac{1}{p^2-m_{C}^2}. 
\end{equation}
The massive three-form field contains one propagating degree of freedom\footnote{Conversely, the massive $C$ field explains how the correlator $\langle R\widetilde{R}, R\widetilde{R}\rangle_{p\to 0}$ vanishes without explicitly calculating the contribution from gravitons and fermions. 
With a mass term, $C$ can no longer mediate long-range interactions. 
Even if its field strength $E$ is non-zero locally, it decays exponentially fast~\cite{Dvali:2005an}, which one can intuitively understand as `screened' by the fermion-antifermion pairs.}. 
Therefore, we demonstrated that the standard perturbative theory of gravitationally interacting massless fermions fails in the low energy limit, and the corresponding effective theory must contain a mass gap. 
We note that the precise scale of this mass gap \typo{can not}{cannot} be directly calculated within the perturbative framework.
Rather, it is `bootstrapped' by the requirement that the UV and low energy theory must describe the same physics.


We now study the origin of the $C$ mass. 
The direct mass term \typo{can not}{cannot} exist because it explicitly breaks the gauge redundancy shown in Eq.~\eqref{redundancy}. Instead, $C$ must be Higgsed and become massive by `eating up' a two-form field $B$
\begin{equation}
\label{twoformmass}
    \mathcal{L}~=~\frac{1}{2}(dC)^2+m_C^2\left(C-d B \right)^2. 
\end{equation}
Here, $B$ plays the role of the Stueckelberg field~\cite{Stueckelberg:1938hvi}, and the gauge invariance is restored given $B$ transforms as
\begin{equation}
    B~\to~B+w.
\end{equation}
The massless two-form field $B$ contains one \typo{pesudo}{pseudo}-scalar degree of freedom, which is eaten-up by $C$ and becomes its longitudinal component. 
One can further formally treat $P\equiv dB$ as a fundamental three-form, and rewrite Eq.~\eqref{twoformmass} as
\begin{equation}
\label{LagrangeMultiplier}
        \mathcal{L}~=~\frac{1}{2}(dC)^2+m_C^2\left(C-P \right)^2+\eta_G'~^*dP. 
\end{equation}
Here, $^*dP=0$ obeys the Bianchi identity so $\eta_G'$ appears as a Lagrange multiplier field. 
Upon integrating out $P$, Eq.~\eqref{twoformmass} turns into its dual form in which the two-form $B$ field is replaced by a \typo{pesudo}{pseudo}-scalar field $\eta_G'$
\begin{equation}
\label{dualsol}
    \mathcal{L}~=~\frac{1}{2}(dC)^2+\frac{1}{2}(d \eta_G')^2-\frac{\eta_G'}{\Lambda_G}~^*dC. 
\end{equation}
Here, we normalize the $\eta_G'$ field such that its kinetic term takes the canonical form. The equation of motion for $C$ requires
\begin{equation}
    ~^*dC~=~\frac{\eta_G'}{\Lambda_G}+\text{const}. 
\end{equation}
Using this relationship, the $C$ field can be eliminated in Eq.~\eqref{dualsol}. We then get the Lagrangian for a massive \typo{pesudo}{pseudo}-scalar field $\eta_G'$
\begin{equation}
    \mathcal{L}~=~\frac{1}{2}(d\eta_G')^2-m_{\eta_G'}^2 (\eta_G')^2.
\end{equation}
Here, $\eta_G'$ is in analogy to the $\eta'$ particle in QCD, and its mass $m_{\eta_G'}$ originates from a bootstrapped mass gap in the low energy theory, which \typo{can not}{cannot} be understood within the perturbative framework.
Intuitively, $\eta_G'$ can be interpreted as a composite particle containing the strongly interacting gravitons and massless Weyl fermions.

Except for Higgsing with the $B$ field, there are no other way to generate $C$ mass without breaking the  gauge redundancy shown in Eq.~\eqref{redundancy}. 
As a consequence, eliminating $\theta_G$ in Eq.~\eqref{EinsteinGravity} with $\psi_{\alpha}$ is exactly compensated by the $B$ field or --- as its dual description --- by $\eta_G'$ at the low energies. In other words, the $U(1)_A$-gravity-gravity triangle anomaly is \textit{fully} matched to $\eta_G'$. 
Interestingly, this conclusion leads to more indications.
Assuming the deep infrared theory contains $N_f'$ generations of massless fermions $\lambda_{\alpha}$, which can be either a sub-set of $\psi$ or their massless composite states. There then exists a chiral symmetry
\begin{equation}
\lambda_{\alpha}~\to~\lambda_{\alpha}~=~e^{i\theta_{\lambda}}\lambda_{\alpha}, \quad \alpha=1,2,..., N_f', 
\end{equation}
which is anomalous with respect to gravity. 
However, from the viewpoint of an UV observer, $U(1)_A$ is the \textit{only} symmetry anomalous with respect to gravity and the relevant anomaly is fully matched to $\eta_G'$. 
In addition to $U(1)_A$, the other subgroups of $U(48)_f$ involves no mixed anomalies with gravity, because the all $SU(N)$ generators are traceless. 
Therefore, massless $\lambda_{\alpha}$ creates a contradiction, that it leads to the wrong anomaly structures in the infrared theory, which \typo{can not}{cannot} be compensated in UV. 
Therefore, the arguments above demonstrates that assuming non-zero topological susceptibility of the gravitational vacuum, no massless fermions can exist, no matter fundamental or composite.
It is worthy to remark that this conclusion replies on an important feature of gravity, that the gravitational interaction is universal and there are no `gravity-singlets' in nature. 
By contrast, within a non-abelian Yang-Mills theory, such as QCD, we can only prove that massless colored fermions cannot exist. The massless color singlets cannot lead to the mixed $U(1)_A-SU(3)_c-SU(3)_c$ anomaly.

The absence of massless fermions furthermore indicates the breaking of the $SU(48)_f$ symmetry at low energies. 
Since $SU(48)_f$ is defined chiral, there also exists $[SU(48)]^3$ triangle anomalies in the UV theory, which must also manifest in the low energy limit\footnote{Technically, one should no longer introduce the spectator fermions for $[SU(48)]^3$ anomaly matching, because these spectator fermions also interact with the gravitons and \typo{can not}{cannot} be distinguished with $\psi_{\alpha}$. Following the technique established in~\cite{Dvali:2017mpy}, one can take the Wess-Zumino-Green-Schwarz-type axions as spectators.}. 
The $[SU(48)]^3$ anomaly \typo{can not}{cannot} be matched to the massless fermions which do not exist. 
Rather, it could only be compensated by a set of massless bosons $\pi_{\alpha'}$. 
The anomalous $SU(48)$ transformation among $\psi_{\alpha}$ must be matched to shifts of $\pi_{\alpha'}$. Then, if $SU(48)$ is weakly gauged, the gauge bosons related to the $[SU(48)]^3$ anomaly are all Higgsed by $\pi_{\alpha'}$ and decouple from the massless spectrum, so that the gauged theory is anomaly-free in the deep infrared. 
$\pi_{\alpha'}$ is therefore represented by the generators of $SU(48)/G_{\text{free}}$. Here, $G_{\text{free}}$ is one of the anomaly-free subgroups of $SU(48)$ such that the $[G_{\text{free}}]^3$ triangle anomaly vanish, e.g. $SO(48)$ or $SO(10)\times SO(10)\times SO(10)$. 
In other words, the $SU(48)$ symmetry must be spontaneously broken into one of its anomaly-free subgroup in the low energy limit.

In summary, using the technique of gravitational anomaly matching and assuming non-zero vacuum topological susceptibility, we have bootstrapped the conclusion that the $U(48)$ symmetry of 48 gravitationally interacting Weyl fermions must be spontaneously broken to one of its anomaly-free subgroups\begin{equation}
    U(48)~\xrightarrow{f_G}~G_{\text{free}}.
\end{equation}
The physical degrees of freedom in the UV and low energy theories are different
\begin{equation}
\label{spectrum}
    \mu\gg f_G:~ \psi_{\alpha}+\text{gravitons}; \quad 
    \mu\lesssim f_G:~ \pi_{\alpha'}+\eta_G'+...
\end{equation}
Clearly, there exist at least a phase transition at some scale which can be indicated by $f_G$. 
The low energy effective theory can no longer be described by the massless fermions and gravitons in the perturbative framework. Instead, it contains multiple composite states: massive Goldstone bosons $\pi_{\alpha'}$ and a massive \typo{pesudo}{pseudo}-scalar $\eta_G'$, in analog to those in the chiral perturbation theory matched from QCD. 
Notably, confinement, which appears irrelevant to gravity, is not required at the low energies, so the fundamental fermions $\psi_{\alpha}$ can in principle coexist with the composite states, as long as they are massive. 
The mass terms for $\psi_{\alpha}$ can originate from the $U(48)$ breaking order parameters
\begin{equation}
    -\mathcal{L}~=~m^{\alpha\beta}\overline{\psi_{\alpha}}\psi_{\beta}^c, \quad m^{\alpha\beta}~\equiv~\frac{1}{f_G^2}\langle \psi_{\alpha}\overline{\psi_{\beta}^c}\rangle. 
\end{equation}
$m^{\alpha\beta}$ is a symmetric $48\times 48$ matrix, and can be chosen diagonal without loss of generality. 
It is worth remarking that elements of $m^{\alpha\beta}$ or $\langle \psi_{\alpha}\overline{\psi_{\beta}^c}\rangle^{1/3}$ are not necessarily far smaller than $f_G$. 
In such a case, the $U(48)$ symmetry is \textit{strongly} broken below $f_G$ so that the generation of $m^{\alpha\beta}$ is distinguished from the Froggatt-Nielsen mechanism introduced in subsection~\ref{FNsection}.

\subsection{Effective Neutrino Masses}
\label{numasssection}
The non-perturbative gravity induced $U(48)$ breaking has important phenomenological implications, even if the SM gauge interactions are turned on. 
In the low energy limit, the quarks, charged leptons, and $W,Z$ bosons are massive due to electroweak symmetry breaking, while light quarks and gluons are confined into the bound states. 
Then, the light spectrum can only contain
\begin{equation}
    \text{Photons};~\text{Gravitons};~3\times \nu_L,~3\times \nu_R. 
\end{equation}
The photons and gravitons are strictly massless due to gauge invariance.
On the other hand, the neutrinos are defined as gauge singlets below the electroweak scale and can acquire masses via
\begin{equation}
\label{numass}
-\mathcal{L}~=~m_L\overline{\nu_L^c}\nu_L+m_R\overline{\nu_R^c}\nu_R+m_D\overline{\nu_L}\nu_R+\text{h.c.}.
\end{equation}
In the limit $m_R\to \infty$, $\nu_R$ decouple and $\nu_L$ can be massive only when $m_L\neq 0$. 
However, $m_L$ violates the lepton number symmetry $L$, which is preserved by all SM interactions if excluding $\nu_R$. 
As a consequence, the \textit{minimally defined} (so $\nu_R$ are excluded) SM requires three generations of exactly massless neutrinos.
Since the observation of neutrino oscillations~\cite{Super-Kamiokande:1998kpq, SNO:2001kpb, SNO:2002tuh} implies that the neutrinos are massive, the minimally defined SM are widely considered as an incomplete theory for phenomenology. 
This statement is revisited in Ref.~\cite{Dvali:2016uhn}:
although the neutrinos are predicted massless in UV, they can acquire masses in the deep infrared limit due to the non-perturbative effects of gravity. 
We interpret the loop-hole for the $m_L=0$ prediction as related to the global $L$ symmetry. 
As introduced in subsection~\ref{anomalymatching}, assuming non-zero gravitational vacuum topological susceptibility, only the anomaly free subgroup of the would-be $U(48)$ maximal flavour symmetry can remain unbroken in the low energy limit. Since $L$ is chiral and thus not anomaly free, it \typo{can not}{cannot} co-exist with gravity and must be spontaneously broken below some scale $f_G$. 
In other words, although $L$ is conserved in the minimally defined SM, it \typo{can not}{cannot} be applied to predict the observables in the low energy limit, such as the vanishing of neutrino masses, when gravity is taken into consideration.

The mechanism generating $m_L$ through the low-scale non-perturbative gravitational effects are commonly \typo{referred as}{referred to as} the neutrino condensate models~\cite{Dvali:2016uhn, Addazi:2016oob, Dvali:2017mpy, deLima:2022dht} (see Refs.~\cite{Barenboim:2008ds, Azam:2010kw, Barenboim:2010db, Barenboim:2019fmj} for discussions on the high scale condensation models). 
In this framework, the active neutrino mass terms can be effectively described as
\begin{equation}
    \label{Mneutrinomass}
    m^{ij}_{L}~=~\frac{1}{f_G^2}\langle \nu_{iL}^c\overline{\nu}_{jL}\rangle, \quad 4\pi f_G~\gtrsim~0.1~\text{eV}. 
\end{equation}
Chiral $U(3)$ --- the global symmetry accidentally enhanced for the three massless neutrinos in UV --- is spontaneously broken by the neutrino condensate order parameter $\langle \nu_{iL}^c\overline{\nu}_{jL}\rangle$. 
The scale of $U(3)$ breaking is characterized by $f_G$, which lies around the eV or sub-eV scales if $\langle \nu_{iL}^c\overline{\nu}_{jL}\rangle^{1/3}\sim f_G$. 
The physical picture is then very similar to the chiral symmetry breaking of QCD, while it is worthy remarking that the light quarks are all confined into the bound states QCD but neutrinos can stay fundamental in the gravity-induced condensate models. 
$m_{L}$ is a $3\times 3$ symmetric matrix. 
It must be full rank to break the $U(1)_L$ symmetry, so no neutrinos can stay massless. 
$m_L$ is not necessarily aligned with the Yukawa coupling matrix for charged leptons $Y_e$. In the basis that $Y_e$ is diagonal, $m_L$ is in general not diagonal and can be diagonalized by a bi-unitary transformation, so the neutrino mixing angles are also generated.

It is crucial to notice that the gravity-induced neutrino masses conceptually differ from those arising from the UV models, e.g. the type~II seesaw model we introduced in Eq.~\eqref{massmatrix}. 
Within the minimal type~II seesaw model, the neutrino masses and mixing patterns depend on the UV parameters $v_{\Delta}$ and $Y_{\Delta}$, which are not calculable unless one embeds the minimal model into some other frameworks. 
By contrast, $m_L$ originates from $\langle \nu_{iL}^c\overline{\nu}_{jL}\rangle$ and $f_G$ in the neutrino condensate model, and these values are fixed as long as one works within the theory in which the neutrinos are massless in UV. 
In other words, $m_L$ do not involve free UV parameter, so all entries in $m_L$ must be understood as calculable, although not within the well-established perturbative framework. 
Furthermore, the see-saw models explicitly break the $U(3)$ flavour symmetry at all scales. In the neutrino condensate models, $U(3)$ remains invariant in the UV theory, and is only spontaneously broken at low energies. Given $f_G\gtrsim\langle \nu_{iL}^c\overline{\nu}_{jL}\rangle^{1/3}$, the $U(3)$ symmetry becomes implicit below $f_G$ and \typo{can not}{cannot} be distinguished with strongly broken symmetries, so that the effective theory below $f_G$ as lying in the swampland of the $U(3)$ symmetry.\footnote{The $U(3)$ symmetry can remain approximately exact below $f_G$, given the hierarchy $\langle \nu_{iL}^c\overline{\nu}_{jL}\rangle^{1/3}\ll f_G$. The effective theory then remains lying in the landscape of $U(3)$.}
The information on $U(3)$ is fully lost at low energies, due to the gravity-induced phase-transition around $f_G$.

The mass term $m_L$ generated in Eq.~\eqref{Mneutrinomass} is Majorana type.
If $\nu_R$ do not decouple in the low energy limit (which requires $m_R\to 0$), the neutrino condensate model can also generate the Dirac type neutrino masses $m_D$. 
However, unlike $m_L$, the entries in $m_D$ do not automatically vanish in the bare Lagrangian, and strictly speaking, their values are unpredicted. 
In other words, gravity only leads to a shift of the free parameter $m_D$
\begin{equation}
    \label{Dneutrinomass}
    m^{ij}_{D}~\to~\widetilde{m}_D^{ij}~=~m_D^{ij}+\frac{1}{f_G^2}\langle \nu_{iL}\overline{\nu}_{jR}\rangle. 
\end{equation}
Since $m_D^{ij}$ is arbitrary, the shifted mass term $\widetilde{m}_D^{ij}$ \typo{can not}{cannot} be accounted as a prediction. 
The physical effects of $\langle \nu_{iL}\overline{\nu}_{jR}\rangle$ can therefore be eliminated by refining the renormalization schemes. 
To ensure the Dirac neutrino masses to be calculable, $m_D$ must be restricted to be zero in the UV theory, by, for instance, flavour symmetries acting differently on the left-hand and right-hand neutrinos.
Such chiral symmetries are widely studied for explaining the large mass hierarchies between neutrinos and charged fermions, which span $7-13$ orders of magnitudes, see Ref.~\cite{Ma:2015raa, Ma:2015mjd, Bonilla:2016zef, CentellesChulia:2016rms,  Bonilla:2016diq, Ma:2016mwh, CentellesChulia:2018gwr, Ma:2026tyk} for representative examples.
In this framework, the tiny Dirac neutrino masses can arise naturally from the sub-eV scale VEVs.
Yet, to induce these VEVs, further extensions to the SM, such as additional fundamental Higgs fields, are commonly believed necessary. 
To compare, within the neutrino condensate models, the sub-eV scale VEVs can directly emerge from the non-perturbative gravitational effects so no additional new physics is need, as we noted in Ref.~\cite{Gao:2026bvv}.

\begin{table}[t]
    \centering
\renewcommand\arraystretch{1.5}
\resizebox{0.95\textwidth}{!}{
    \begin{tabular}{c  c c c  c  c  c  c}
    \toprule
    & $U(1)_{B-L}$ & $U(1)_{L_{\mu}-L_{\tau}}$ & $U(1)_{L_e-L_{\mu}}$ & $U(1)_{L_{\mu}-L_{\tau}}^{\ell}$  & $U(1)_{L_{\mu}-L_{\tau}}^{\nu}$ & $U(1)_{B-L}'$ & $U(1)_{B-L}''$\\
    \midrule
    $(Q_L^{\alpha i}, u_R^{\alpha i}, d_R^{\alpha i})$  & $+\frac{1}{3}$ & 0 & 0 &  0 &  0 & $+\frac{1}{3}$ & $+\frac{1}{3}$\\
    $(e_L, \nu_{eL}, e_R)  $ & $-1$ & 0 & $+1$ & 0 & 0 & $-1$ & $-1$\\
    $(\mu_L, \nu_{\mu L}, \mu_R)  $ & $-1$ & $+1$ & $-1$ & $+1$ & 0 & $-1$ & $-1$\\
    $(\tau_L, \nu_{\tau L}, \tau_R)  $ & $-1$ & $-1$ & 0 & $-1$ & 0 & $-1$ & $-1$\\
    $\nu_{eR}  $ & $-1$ & 0 & $+1$ & 0  & 0 &  $+5$ & $-3+\mathcal{O}(\epsilon^2)$\\
    $ \nu_{\mu R}  $ & $-1$ & $+1$ & $-1$ & 0 & $+1$ & $-4$ & $+\frac{1}{\epsilon}+\mathcal{O}(\epsilon)$\\
    $\nu_{\tau R}  $ & $-1$ & $-1$ & 0 & 0 & $-1$ & $-4$ & $-\frac{1}{\epsilon}+\mathcal{O}(\epsilon)$\\
    \midrule
    $m_L$ & 0 & $*$ & $*$ &  $*$ &  $**$ &  0 & 0\\
    $m_R$ & 0 & $*$ & $*$ &  $**$ &  $*$ &  0 & 0\\
    $m_D$ & $**$ & $*$ & $*$ &  $*$ &  $*$ &  0 & 0\\
   \bottomrule
\end{tabular}}
    \caption[Certain anomaly-free charge assignments and their implication.]{Row $1-7$: Certain anomaly-free charge assignments to SM fermions with three right-handed neutrinos. $\alpha$ and $i$ take values among $1-3$, representing three different colors and flavors respectively. 
    Row $8-10$: The symmetry constrains on the different neutrino mass matrices.
    The symbol $*~(**)$ indicate some (all) of the entries in $m_L, m_R, m_D$ can take non-zero values, while $0$ denote all possible mass terms are forbidden.
    }
    \label{charge1}
\end{table}

Ideally, the flavor symmetry forbidding the neutrino masses should be gaugeable, so that the $m_D=0$ becomes a consequence of fundamental principles. 
We therefore explore the anomaly-free abelian symmetries in the SM (including $\nu_R$). 
In addition to the hypercharge group, the maximal one is~\cite{Foot:1992ui, Araki:2012ip}
\begin{equation}
\label{max}
    G_{\text{max}}~=~U(1)_{B-L}\times U(1)_{L_e-L_{\mu}}\times U(1)_{L_{\mu}-L_{\tau}}.
\end{equation}
The corresponding charge assignments are shown in columns $1-3$ of Table~\ref{charge1}. 
$G_{\text{max}}$ is a vectorial group and acts universally on the left and right handed fermions. 
As a consequence, it \typo{can not}{cannot} forbid the Dirac neutrino mass matrix $m_D$, although both $m_L$ and $m_R$ vanish if $B-L$ is conserved.
The SM can also accommodate the anomaly-free chiral symmetries, for instance
\begin{equation}
\label{chiralfree}
    G_{\chi}~=~U(1)_{L_{\mu}-L_{\tau}}^{\ell} \times U(1)_{L_{\mu}-L_{\tau}}^{\nu},
\end{equation}
whose charge assignments are shown in the $4-5$ columns of Table~\ref{charge1}. Here, $U(1)_{L_{\mu}-L_{\tau}}^{\ell}$, which can be understood as the $T_3$ generator of the anomaly-free $SU(2)^{\ell}$ horizontal symmetry group, is the maximal anomaly free abelian extension to the SM when the right-handed neutrinos are excluded~\cite{Foot:1992ui, He:1990pn, He:1991qd}. 
$U(1)_{L_{\mu}-L_{\tau}}^{\nu}$ acts on two of the three right-handed neutrinos. 
The vectorial subgroup of $G_{\chi}$, under which the left and right hand neutrinos are charged equally, is same as the $U(1)_{L_{\mu}-L_{\tau}}$ symmetry contained in $G_{\text{max}}$. The remaining axial part of $G_{\chi}$ can also be made equivalent to $U(1)_{L_{\mu}-L_{\tau}}$ by exchanging $\nu_{\mu R}$ and $\nu_{\tau R}$, since they \typo{can not}{cannot} be distinguished in the SM.
Nevertheless, $G_{\chi}$ is distinguished with $G_{\text{max}}$ because its axial and vectorial subgroups \typo{can not}{cannot} be simultaneously equivalent to $U(1)_{L_{\mu}-L_{\tau}}$.   
The Dirac masses for $\nu_{\mu}$ and $\nu_{\tau}$ are forbidden due the chiral nature of $G_{\chi}$, but $\nu_e$ can remain massive because $G_{\chi}$ does not act on it. 
Ref.~\cite{Montero:2007cd} pointed out an chiral and anomaly-free symmetry which acts on all three generations of neutrinos. 
This symmetry can be identified as modified $B-L$, because the quarks and charged leptons carries the canonical $B-L$ charges. 
This modified $B-L$ charge assignments are shown in the sixth column of Table~\ref{charge1}.
The left-handed and right-handed neutrinos carry different modified $B-L$ charges and the Dirac neutrino mass terms are all forbidden.

In Ref.~\cite{Gao:2026bvv}, we generalize the $B-L$ charge assignment to the real domain, which, to the best of our understanding, represents a previously unexplored regime. 
The $B-L$ symmetry is anomaly-free provided that the $B-L$ charge for the three generations of $\nu_R$ satisfies~\cite{Montero:2007cd}:
\begin{equation}
\label{neutrinocharge}
\begin{aligned}
    Q_{\nu_{eR}}+Q_{\nu_{\mu R}}+Q_{\nu_{\tau R}}~&=~-3, \\
    Q_{\nu_{eR}}^3+Q_{\nu_{\mu R}}^3+Q_{\nu_{\tau R}}^3~&=~-3.\\     
\end{aligned}
\end{equation}
Note that the indices $(e,\mu,\tau)$ are used solely to label the three right-handed neutrinos $\nu_R$ and must not be confused with the lepton flavor indices associated with electroweak interactions.
Up to permutations among $\nu_{eR}, \nu_{\mu R},$ and $\nu_{\tau R}$, the solution to Eq.~\eqref{neutrinocharge} in the integer domain is
\begin{equation}
\left(Q_{\nu_{eR}}, Q_{\nu_{\mu R}}, Q_{\nu_{\tau R}}\right)~=~(-1, -1,-1)\quad \text{or}\quad (+5, -4,-4). 
\end{equation}
It is widely assumed that the gaugeable charges must be quantized. 
The irrational charges inevitably lead to global $U(1)$ symmetries, under which only the fields with irrational charges rotate.
These global symmetries as remanent of gauge invariance seemingly contradict the `no-global symmetry' conjecture when gravity is taken into consideration~\cite{Banks:2010zn}. 
However, we note that this conjecture can only select the consistent effective theories and should not be interpreted as constraints for UV model building. 
As pointed out in Ref.~\cite{Dvali:2017mpy}, the gravity response to the global symmetries at low energies is dynamically explained in the neutrino condensation framework: the global symmetries are broken spontaneously due to the non-perturbative effects. 
Therefore, the assigned $B-L$ charges are not necessarily quantized and in general can take real values.
After relaxing the charge quantization assumption, we note that Eq~(\ref{neutrinocharge}) admits infinitely many solutions. 
We identify a novel solution which, to our knowledge, has not been explored previously.
By studying the limit $Q_{\nu_{e R}}\to -3$ and  $Q_{\nu_{\mu R}}\to -Q_{\nu_{\tau R}}$, we find that the following $B-L$ charges also satisfy Eq.~\eqref{neutrinocharge}
\begin{equation}
    \label{solReal}
    \begin{aligned}
        Q_{\nu_{eR}}~&=~-3-8\epsilon^2, \\
        Q_{\nu_{\mu R}}~&=~ +\frac{\sqrt{\epsilon ^2+1} }{\epsilon }\left(4 \epsilon ^2+1\right)+4 \epsilon^2 ~\approx~+\left(\frac{1}{\epsilon}+\frac{9}{2}\epsilon+4\epsilon^2+\mathcal{O}(\epsilon^3)\right), \\
        Q_{\nu_{\tau R}}~&=~-\frac{\sqrt{\epsilon ^2+1} }{\epsilon }\left(4 \epsilon ^2+1\right)+4 \epsilon ^2
        ~\approx~-\left(\frac{1}{\epsilon}+\frac{9}{2}\epsilon-4\epsilon^2+\mathcal{O}(\epsilon^3)\right). \\  
    \end{aligned}
\end{equation}
We show the corresponding charge assignments in the last column of Table~\ref{charge1}, in the limit $\epsilon\to0$.

Importantly, $Q_{B-L}$ for two of the three generations of $\nu_R$ can be enhanced by arbitrarily large factors, so we refer to this setup as the `enhanced $B-L$ symmetry'.
It is worth remarking that if this enhanced $B-L$ symmetry is gauged, $Q_{B-L}\gg1$ does not imply a violation of perturbative unitarity, as long as the corresponding gauge coupling strength $g_{B-L}$ is sufficiently small.
The $1/\epsilon$ factor can be absorbed by redefining $g_{B-L}$ as
\begin{equation}
    g_{B-L}~\to~g_{\text{eff}}^{\nu}~=~\frac{1}{\epsilon}g_{B-L}. 
\end{equation}
When $\epsilon\to0$, $g_{\text{eff}}^{\nu}$ can stay finite while the $B-L$ gauge interaction vanishes for quarks, charged leptons, and $\nu_{eR}$. 
Then, $g_{B-L}=0$ if $\epsilon=0$ and the enhanced $B-L$ symmetry reduces to the $U(1)_{L_{\mu}-L_{\tau}}^{\nu}$ symmetry we introduced in Eq.~\eqref{chiralfree}. 
This neutrino-philic $L_{\mu}-L_{\tau}$ symmetry is non-chiral and therefore automatically anomaly-free, providing a consistency check of our solution. 
However, it is distinct from the enhanced $B-L$ symmetry: for $\epsilon = 0$, both Majorana and Dirac neutrino mass terms are allowed, whereas for $0 < \epsilon \ll 1$, these mass terms are forbidden.
We summarize the implications on the neutrino mass terms altogether in the last three rows of \tref{charge1}.

We now promote the enhanced $B-L$ symmetry to be gauged. 
Then, $m_D$ is forbidden by the principle of gauge invariance and cannot arise as a bare parameter in the UV theory. 
Due to the effects from non-perturbative gravity, the Dirac neutrino masses do not vanish in the low-energy theory, but are proportional to the order parameter $\langle \nu_{iL}\overline{\nu}_{jR}\rangle$. Its tiny values are therefore dynamically explained. 
$\langle \nu_{iL}\overline{\nu}_{jR}\rangle$ break the $U(3)_L^{\nu}\times U(3)_R^{\nu}$ chiral flavor symmetry of three generations of massless $\nu_L$ and $\nu_R$ down to one of its anomaly-free subgroups.
Generating three different neutrino masses requires that the symmetry breaking is maximal
\begin{equation}
    U(3)_L^{\nu}\times U(3)_R^{\nu}~\xrightarrow{f_G}~U(1)_V^3~=~U(1)_V^{\nu_e}\times U(1)_V^{\nu_{\mu}}\times U(1)_V^{\nu_{\tau}}. 
\end{equation}
The $U(1)_V^{\nu_i}$ symmetry is vectorial and acts on $\nu_i$ only. 
The enhanced $B-L$ symmetry is different from the unbroken $U(1)_V^3$, 
so it must be spontaneously broken. 
The corresponding $B-L$ gauge boson $A'$ acquires a mass $m_{A'}$, satisfying
\begin{equation}
    m_{A'}~=~\frac{1}{\epsilon}g_{B-L} f_G. 
\end{equation}
Within the minimal scenario, $m_{A'}$ is bounded from above by $4\pi f_G$. Heavier $m_{A'}$ requires extensions, such as including the Stueckelberg fields~\cite{Stueckelberg:1938hvi} or other additional symmetry breaking order parameters.

\subsection{Enhanced \texorpdfstring{$B-L$}{B-L} Gauge Symmetry}
\label{BLsection}
We discuss about the phenomenology of the enhanced $B-L$ gauge symmetry in this subsection.
Given $f_G\sim \langle \nu_{iL}\overline{\nu}_{jR}\rangle^{1/3}\sim \mathcal{O}(0.1)$ eV, $A'$ stays light and manifests itself as a long-range force for the neutrons in the same manner as the canonical $B-L$ gauge bosons, often called `the fifth force'. 
Effectively, such $B-L$ interaction leads to an inhomogeneous correction to the Newtonian potential
\begin{equation}
    V(r)~=~-G_N\frac{M_1M_2}{r}\left(1+ \alpha e^{-m_{A'}r}\right), \qquad \alpha~=~\eta_1\eta_2\frac{g_{B-L}^2}{4\pi G_N m_{n}^2}. 
\end{equation}
Here, $G_N$ is the Newtonian gravitational constant, $m_n$ represents the neutron mass, and $\eta_1 (\eta_2)$ indicates the ratio of neutron number over the total baryon number contained in the object with mass $M_1 (M_2)$. 
When $g_{B-L}\gtrsim G_N^{1/2} m_n\approx 10^{-19}$, we have $\alpha\gtrsim 1$ and the $B-L$ force becomes comparable or dominant over gravity at the distance smaller than $m_{A'}^{-1}$,
which can leave imprints in the form of violations of the weak equivalence principle~\cite{Schlamminger:2007ht, ADELBERGER2009102, MICROSCOPE:2022doy} and/or deviations from the inverse-square law of gravity~\cite{Kapner:2006si, Sushkov:2011md, Yang:2012zzb, Chen:2014oda, Tan:2020vpf}.
Therefore, $g_{B-L}$ is constrained to be tiny and the $B-L$ gauge interaction is widely believed to be feeble and to decouple from the other sectors.

We note that even if $g_{B-L}$ is tightly constrained, the $B-L$ gauge interaction can be enhanced by sufficiently small $\epsilon$ and become sizable for neutrinos. 
In particular, we emphasize that $\epsilon$ should not be regarded as a conventional free parameter, such as a mixing angle. Rather, its value is fixed once the principle of gauge invariance is specified. 
The arbitrariness instead reflects that the chiral structure of the known quarks and leptons permits an uncountably infinite number of anomaly-free $U(1)$ symmetries, with $\epsilon$ serving to label the distinct theories within that moduli space. 
Furthermore, the weak-gravity conjecture~\cite{Arkani-Hamed:2006emk}, which roughly states that gravity should be the weakest force, requires only the theories with (fairly) small $\epsilon$ are consistent in the limit of vanishing $g_{B-L}$. 
If the $g_{B-L}$ is far smaller than $G_N^{1/2}m_\nu$, tiny $\epsilon$ ensures that the enhanced $B-L$ gauge interaction with neutrinos can stay stronger than gravity.
In this scenario, we have
\begin{equation}
    g_{B-L} ~\ll~ G_N^{1/2}m_\nu~\approx~ 10^{-30}~\lesssim~ \frac{1}{\epsilon}g_{B-L}.
\end{equation}
Therefore, we argue that the enhanced $B-L$ gauge coupling strength $\epsilon^{-1}g_{B-L}$ should be interpreted as a fundamental parameter of nature.  
Probing its magnitude is essential for a complete understanding of the SM and we believe an immediate experimental priority is to determine whether $\epsilon^{-1}g_{B-L}$ is smaller than $\mathcal{O}(1)$. 
Currently, the upper bound of $\epsilon^{-1}g_{B-L}$ remains largely unexplored, even in the presence of existing phenomenological constraints.

Assuming $\epsilon^{-1}g_{B-L}\sim \mathcal{O}(1)$, $A'$ dominantly couples to two of the three $\nu_R$. 
In the limit $g_{B-L}\to 0$, the interactions among $A'$ and the other fermions are possible only through the Dirac neutrino mass portal which connects $\nu_R$ to $\nu_L$. 
The gravity induced neutrino mass $m_{D}$ is temperature dependent.
In general, it vanishes above the critical temperature of a phase transition $T_c\sim f_G$ and becomes non-zero only below $T_c$. If such phase transition is supercooled, $T_c$ can be lower than $f_G$ and these two scales are independent parameters~\cite{Lorenz:2018fzb}.
In case $T_c$ is lower than the temperature of the photon decoupling epoch $0.26~\text{eV}$ (or equivalently around $3000~\text{K}$), the early universe observables reduces to those of the standard cosmology with massless $\nu_L$~\cite{Dvali:2016uhn}.
As a consequence, the early universe constraints on the absolute neutrino masses are invalidated~\cite{Dvali:2016uhn, Koksbang:2017rux, Lorenz:2018fzb}. 
As pointed out in Ref.~\cite{Lorenz:2018fzb}, if the neutrinos acquire masses via a supercooled phase transition in late universe ($a_s\sim 1$, where $a_s$ is the scale factor), the cosmological constraint to the absolute neutrino masses can be relaxed to $\sum m_{\nu_i}<4.8~\text{eV}$, at 95\% confidence level, which becomes much weaker than terrestrial bound~\cite{KATRIN:2024cdt}. In early universe, the physical effects of $A'$ decouple together with $\nu_R$.
Only when $T<T_c$, $A'$ is relevant to phenomenology, i.e., the terrestrial and/or astrophysical neutrino experiments performed in the (our) late universe.

Since $A'$ is ultra-light, its physical effects are only suppressed by the characteristic energy scale $E$.
Comparing with the high-energy colliders such LHC, the experimental implications of $A'$ are enhanced in the low-energy frontier experiments.,
This feature is to be distinguished from a broad class of neutral gauge boson models discussed in the literature, see, e.g., Refs.~\cite{He:1990pn, He:1991qd, Foot:1994vd, Ma:1997nq, Araki:2012ip, Heeck:2014zfa,Batell:2016zod, Farzan:2016wym, Ballett:2019xoj, Herbermann:2025uqz}. In our framework, neutrino decay and low-energy neutrino--electron ($\nu$--$e$) elastic scattering as well as coherent elastic neutrino--nucleus scattering (CE$\nu$NS), provide the most stringent constraints.
Consequently, currently operating and next-generation neutrino observatories and telescopes—such as DUNE~\cite{DUNE:2020lwj, DUNE:2015lol, LBNE:2013dhi, Ghoshal:2020hyo}, JUNO~\cite{JUNO:2015zny, JUNO:2021vlw, Abrahao:2015rba}, Hyper-Kamiokande~\cite{Hyper-KamiokandeProto-:2015xww}, IceCube-Gen2~\cite{IceCube-Gen2:2020qha}, and ESS$\nu$SB~\cite{Chakraborty:2020cfu}—as well as dark matter direct-detection experiments, including liquid noble-gas detectors such as PandaX-4T~\cite{PandaX:2024qfu, PandaX:2024cic}, XENONnT~\cite{XENON:2022ltv, XENON:2023cxc}, LUX-ZEPLIN~\cite{LZ:2025igz, LZ:2024zvo}, DarkSide-50~\cite{DarkSide-50:2022qzh}, and DARWIN~\cite{DARWIN:2020bnc}, as well as cryogenic solid-state detectors such as SENSEI~\cite{SENSEI:2023zdf} and DAMIC-M~\cite{DAMIC-M:2025luv, DAMIC-M:2025ltz}, are especially important to $A'$.

Firstly, we identify the neutrino decay $\nu_i\to\nu_j A'$ as a golden channel for probing $A'$, if $A'$ is lighter than the heaviest neutrino. 
For this process, $E\simeq m_{\nu_i}$ so the chirality flipping suppression is eliminated.
Naively, one may expect that the neutrino decay depends on unknown mixing angles. 
However, following the same logic discussed above on $\epsilon$, 
we note that the information of lepton flavor for $\nu_{R}$ is contained within the definition of gauge invariance and the non-perturbative gravity.
Through this work, we \textit{define} $(\nu_{eR}, \nu_{\mu R}, \nu_{\tau R})$ for the enhanced $B-L$ symmetry such that in the basis where $m^{ij}_{D}$ --- fixed by non-perturbative gravity --- is diagonal, the $A'$ interaction takes the form
\begin{equation}
\label{massbasis}
    \mathcal{L}_{A'}~=~\frac{g_{B-L}}{\sqrt{2}\epsilon}A_{\mu}'
    \left(\overline{\nu}_{3}\gamma^{\mu}P_R \nu_{2}+\overline{\nu}_{2}\gamma^{\mu}P_R \nu_{1}\right)+\text{h.c.}.
\end{equation}
The enhanced $B-L$ symmetry can also be gauged in other directions of the flavour space, which leads to different physical effects.
Nevertheless, the neutrino life and decay branching ratios stay non-ambiguous as long as the definition of $B-L$ is specified, and our setup in Eq.~\eqref{massbasis} reflects the very generic features\footnote{The only exception is the (tuned) flavor alignment limit, in which the flavor and mass eigenstates of $\nu_R$ are approximately the same and neutrino decay rates are suppressed.}.
Starting with Eq.~\eqref{massbasis} and using the analogy of the top quark decay $t_L\to b_L W^+$~\cite{Bigi:1986jk}, the $\nu_i\to \nu_j A'$ decay width reads
\begin{equation}
\label{decayR2}
\begin{aligned}
    \Gamma_{\nu_i}~=~&\epsilon^{-2}g_{B-L}^2\frac{m_{\nu_i}}{64\pi^2 } \frac{\lambda^{1/2}(m_{\nu_i}^2, m_{\nu_j}^2,m_{A'}^2)}{m_{A'}^2}\\
    &\times\left( \left(1-\frac{m_{\nu_j}^2}{m_{\nu_i}^2}\right)^2
    +\left(1+\frac{m_{\nu_j}^2}{m_{\nu_i}^2}\right)\frac{m_{A'}^2}{m_{\nu_i}^2}-2\frac{m_{A'}^4}{m_{\nu_i}^4}\right),
\end{aligned}
\end{equation}
where $\lambda(\alpha,\beta,\gamma)=\alpha^2+\beta^2+\gamma^2-2\alpha\beta-2\alpha\gamma-2\beta\gamma.$ 
In the limit $m_{\nu_i}\gg m_{\nu_j}$, Eq~(\ref{decayR2}) reduces to 
\begin{equation}
\label{decayR}
    \Gamma_{\nu_i}~=~\epsilon^{-2}g_{B-L}^2\frac{m_{\nu_i}}{64\pi^2}\left(2+\frac{m_{\nu_i}^2}{m_{A'}^2}\right) \left(1-\frac{m_{A'}^2}{m_{\nu_i}^2}\right)^2. 
\end{equation}
We note that a non-negligible $m_{\nu_j}$ \typo{can not}{cannot} change $\Gamma_{\nu_i}$ by orders of magnitude, unless the neutrino masses are in the degenerate limit, in which case $\Gamma_{\nu_i}$ is kinematically suppressed.
Since the KATRIN experiment constrains the absolute neutrino masses~\cite{KATRIN:2024cdt}, this suppression \typo{can not}{cannot} be arbitrarily large.

As pointed out in Ref.~\cite{Dvali:2016uhn}, the neutrinos decay widths can be large in the neutrino condensate models, even without gauging $B-L$. 
In the gaugeless theory, the neutrinos can decay into the Majoron $\phi_M$, which corresponds to combination of composite Goldstone bosons $\pi_{\alpha}$ shown in Eq.~\eqref{spectrum}. 
Here, we compare the neutrino decay widths to the gauge boson $A'$ and the Majoron $\phi_M$. 
In the gauged theory, the neutrino dominantly decays to the longitudinally polarized $A'$ when $m_{A'}\ll m_{\nu_i}$, similar to the top quark decays. 
We note $m_{A'}\ll m_{\nu_i}$ corresponds to the gaugeless limit, in which the $\nu_i\to \nu_j A'$ decay width is equal to that of $\nu_i\to \nu_j \phi_M$. 
Considering neutrino decay is kinetically forbidden in the $m_{A'}\geq m_{\nu_i}$ regime, gauging $B-L$ \typo{can not}{cannot} significantly enhance the neutrino decay rates comparing with the gaugeless theory. 
Nevertheless, as long as $m_{A'}< m_{\nu_i}$, the neutrino decay always provides a robust constraint on the enhanced gauge coupling strength $\epsilon^{-1}g_{B-L}$.

There exists another portal for $A'$ to couple with the charged particles, that $A'$ can directly mix with the visible photon $A$ by
\begin{equation}
    \label{mixing}
    \mathcal{L}_{\text{mix}}~=~\frac{1}{2}\chi F'_{\mu\nu}F^{\mu\nu}. 
\end{equation}
However, astrophysics puts stringent constraints on the magnitude of $\chi$; see, e.g. Refs.~\cite{Davidson:2000hf, Hardy:2016kme, Caputo:2021eaa, Li:2023vpv}. 
For $\epsilon^{-1}g_{B-L} \simeq 1$, the $A'$ bremsstrahlung and $A'$ mediated right-handed neutrino production can carry energy away during stellar evolution~\cite{Harnik:2012ni, Heeck:2014zfa, Jeong:2015bbi}. Constraints from solar and red-giants cooling then require $\chi\lesssim 10^{-14}$~\cite{Harnik:2012ni}. Hence, neglecting the bare kinetic mixing term is a good approximation. 
In Ref.~\cite{Gao:2026bvv}, we notice the stringent astrophysics bounds on $\chi$ can, in principle, be evaded similar to the cosmological constraint on $m_{\nu}$. The kinetic mixing $\chi$ can also originate from an order parameter which vanishes above a certain critical temperature. 
This order parameter may also emerge from non-perturbative gravity as both $A'$ and $A$ are contained in the low-energy spectrum and couple directly to gravity. 
We emphasize the distinction: the origin of the temperature dependent $\chi$  should be understood as an additional hypothesis.

With the non-vanishing kinetic mixing at zero temperature, $A'$ can effectively couple to the quarks and charged leptons and mediates $\nu-e$ elastic scattering and CE$\nu$NS. In the $m_{A'}\to 0$ limit, the differential cross section of $\nu-e$ elastic scattering on an atom carrying effective electric charge $Z_{\text{eff}}$ reads~\cite{Harnik:2012ni, Bilmis:2015lja, Cerdeno:2016sfi, DeRomeri:2022twg}:
\begin{equation}
\label{scattering}
\begin{aligned}
    \frac{d \sigma_{\nu e} }{d E_r}~=~Z_{\text{eff}}\epsilon^{-2}g_{B-L}^2\alpha_{EM} \frac{\chi^2 P_{L\to R}}{m_e E_r^2} \left( 2-\frac{2E_r}{E_{\nu}}+\frac{E_r^2}{E_{\nu}^2}\right),
\end{aligned}
\end{equation}
where $E_{\nu}$ is the neutrino energy, and $E_{r}$ is the electron recoil energy. 
The out-going neutrinos from the $A'$ mediated scattering are right-handed polarized, so Eq~(\ref{scattering}) contains no interference terms with electroweak interactions.
We use $P_{L\to R}$ to represent the possibility of observing $\nu_{\mu R}$ or $\nu_{\tau R}$ in the incoming neutrino flux.
$P_{L\to R}$ can directly arise from non-vanishing neutrino masses, in which case it scales as $(m_{\nu}/E_{\nu})^2$ and leads to a suppression. 
In certain reasonably extended scenarios, enhancements are possible.
For instance, if the incoming neutrinos are from the Sun\footnote{Because of the larger flux~\cite{Vitagliano:2019yzm}, the solar neutrino experiments lead to better sensitivities, however, the short-baseline reactor and accelerator experiments can also be relevant due their large statistics at short distances between the source and detector.}, some of them can lose their chiral origin due to solar magnetic fields. 
We refer Ref.~\cite{Barranco:2014cda, Barranco:2017zeq, Joshi:2019dcj} for analyses on neutrino chirality flipping through their magnetic moments in the Sun.
Moreover, if $m_R\neq0$ due to neutrino condensate, $\nu_R$ become sterile neutrinos which maximally mix with $\nu_L$ while differing in masses, modifying the standard oscillation picture. 
For solar neutrinos, $m_R\simeq 10^{-5}~\text{eV}$ is sufficient to convert a significant fraction of solar neutrinos into $\nu_R$~\cite{Harnik:2012ni}.

The ultra-light $A'$ mediated CE$\nu$NS cross section on an nuclei carrying $Z$ protons reads~\cite{DeRomeri:2022twg}
\begin{equation}
\label{Nscattering}
    \frac{d \sigma_{\nu N} }{d E_r'}~=~Z^2\epsilon^{-2}g_{B-L}^2\alpha_{EM} \frac{\chi^2 P_{L\to R}}{m_N E_r'^2} \left( 2-\frac{2E_r'}{E_{\nu}}+\frac{E_r'^2}{E_{\nu}^2}\right) F_W(|\vv{q}|^2),
\end{equation}
Here, $E_r'$ represents the nuclei recoil energy and  $F_W(|\vv{q}|^2)\simeq 1$ is the weak nuclear form factors for protons. 
We note that although the differential cross section in Eq.~\eqref{Nscattering} is enhanced by an additional factor of $Z$ due to the coherent effects, it is suppressed by the heavy nuclei mass $m_N$, which is different from the models with heavy mediators.
As a consequence, CE$\nu$NS currently provides a weaker constraint on the $A'$ coupling strength compared to the $\nu-e$ elastic scattering.
Nevertheless, as shown Ref.~\cite{Dent:2025drd}, the future low-threshold detectors can yield comparable bounds for both processes. 
If both of them are discovered in future, comparing the two cross sections allows one to distinguish our scenario and the other generic models.

The differential cross sections in Eq.~\eqref{scattering} and Eq.~\eqref{Nscattering} are very different from what derived from the effective four-fermion operators. 
The critical behavior for the demonstrated expressions is that they increase as the recoil energy $E_r$ decreases. 
This behavior changes only when $E_r\lesssim 0.1$ eV so that the neutrino masses effect matters. 
Therefore, experiments with lower recoil energy thresholds are tailor-made for this process. 
In particular, the proposed superfluid $^4\text{He}$~\cite{vonKrosigk:2022vnf, SPICE:2023tru, DELight:2024bgv} or superconductor~\cite{Hochberg:2015pha, Hochberg:2015fth} based detectors feature the thresholds at eV or lower scales, and are ideal for the ultra-light $A'$, see Figure 11 of Ref.~\cite{Alexander:2016aln} for comparison. 
In general, ultra-light new particles are subject to tight cosmological and astrophysical constraints, yielding their effects on the terrestrial experiments are suppressed and stay far below the reach of current sensitivities. 
In our set-up, the gravity induced interactions for the ultra-light particle $A'$ only arise in the low temperature limit, which automatically evade the tight cosmological and astrophysical constraints. 
Here, we refer such emergent interactions as the \textit{non-perturbative gravity portal}, which can serve as a very simple benchmark for experiments on ultralight new physics.

\begin{figure}[t!]
    \centering
    \includegraphics[width=0.9\linewidth]{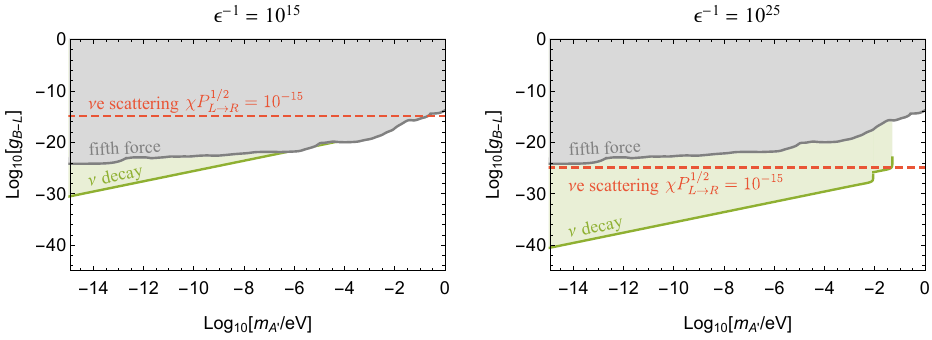}
    \caption[Constraints on the $B-L$ gauge coupling strength.]{Constraints on the $B-L$ gauge coupling strength $g_{B-L}$, enhanced by $\epsilon^{-1}=10^{15}$ (left panel) and $\epsilon^{-1}=10^{30}$ (right panel), as a function of $m_{A'}$. The gray regions are excluded by the fifth-force tests. The green region denotes bounds from the neutrino lifetime. The red dashed line shows the current limit from $\nu-e$ elastic scattering, evaluated with the benchmark value $\chi P_{L\to R}^{1/2}=10^{-15}$.}
    \label{BLbound}
\end{figure}

In \fref{BLbound}, We illustrate the $A'$ implications on neutrino decay, $\nu-e$ scattering, and the fifth force search. The exclusion limits on the gauge coupling strength $g_{B-L}$ are shown as a function of $m_{A'}$. We compare $\epsilon^{-1}=10^{15}$ and $\epsilon^{-1}=10^{30}$ to demonstrate the $\epsilon^{-1}$ enhancement. 
The experimental inputs and analysis we take to generate the plots are summarized as follows. 
\begin{itemize}[label={}, leftmargin=*]
    \item \textit{Neutrino decay}:~We follow Ref.~\cite{Funcke:2019grs} and take the lifetime bound for Dirac neutrinos with normal non-degenerate mass ordering:
\begin{equation}
\label{lifetime2}
    \frac{\tau_2}{m_{\nu 2}}~>~1\times 10^{-3}~\frac{\text{s}}{\text{eV}}, \quad \frac{\tau_3}{m_{\nu_3}}~>~2\times 10^{-10}~\frac{\text{s}}{\text{eV}},
\end{equation}
which comes from the analysis on long-baseline and solar neutrino data, respectively. 
We note that many other astrophysical and terrestrial experiments can also constrain the neutrino lifetime; for instance, see Refs~\cite{Ando:2003ie, Fogli:2004gy, Bustamante:2016ciw, Denton:2018aml, Abdullahi:2020rge,  Martinez-Mirave:2024hfd, Valera:2024buc} and Refs~\cite{Gonzalez-Garcia:2008mgl, Gomes:2014yua, Pagliaroli:2016zab, Gago:2017zzy, Choubey:2018cfz, Huang:2018nxj, Porto-Silva:2020gma, Picoreti:2021yct, Ivanez-Ballesteros:2023lqa, Dey:2024nzm} respectively. 
Given the associated uncertainties, we do not take the constraints from long-traveling astrophysical neutrinos, such as those observed by IceCube~\cite{IceCube:2015rro, Abbasi:2025fjc} and those from SN1987A~\cite{Kamiokande-II:1987idp, Bionta:1987qt, Alekseev:1988gp}. 
Nevertheless, if these uncertainties are brought under control, measurement of astrophysical neutrino flavor ratios at IceCube can yield dramatically improved bounds of $\tau/m_{\nu_i}\gtrsim \mathcal{O}(10)~(\text{s}/\text{eV})$ for all neutrino mass eigenstates~\cite{Bustamante:2016ciw, Valera:2024buc}. 
In~\fref{BLbound}, the regions excluded by the neutrino lifetime, derived in our work, are shown in green. 
To generate these plots, the neutrino masses hierarchy is assumed maximal. We have checked that the the generic non-degenerate mass spectra leads to only minor changes in this logarithmic plot. 
\item \textit{$\nu-e$ scattering}:~
Many neutrino and dark matter detection experiments can measure $\nu-e$ elastic scattering and CE$\nu$NS; for instance, see analysis in Ref.~\cite{Lindner:2018kjo, Khan:2022bel, Khan:2019cvi,  Khan:2020csx, Dev:2021xzd, CONUS:2021dwh, Chakraborty:2021apc, A:2022acy, DeRomeri:2022twg, DeRomeri:2024iaw, Demirci:2025qdp}.
We extract from the neutrino scattering bound to $g_{B-L}$ from the universal tensor model analyzed in Ref.~\cite{Demirci:2025qdp}, by comparing Eq.~(2.9) of that work with our Eq.~(\ref{scattering}). 
We use the tensor limits rather than the vector ones because the tensor mode contribution does not lead to interference the with electroweak interactions, which is consistent with our $\nu_R$ scattering scenario. 
The extracted bound for $g_{B-L}$ is shown with the red dashed line in \fref{BLbound}. We make this line dashed to indicate its dependence on $\chi P_{L\to R}^{1/2}$, which we fix as $10^{-15}$ for a benchmark. If $\chi P_{L\to R}^{1/2}$ takes other values, the constraint to $g_{B-L}$ can be rescaled accordingly. 
\item \textit{Fifth force search}:
The regions excluded by the high-precision tests on gravity are shown in gray. 
The bounds summarized in Ref.~\cite{Heeck:2014zfa} remain mostly unchanged and we adopt most of them directly. 
Yet, we update the constraints in the mass ranges $m_{A'}\lesssim 10^{-13}$ eV and $m_{A'}\gtrsim 10^{-2}$ eV, which is improved by roughly one order of magnitude by the updated results from the MICROSCOPE satellite~\cite{MICROSCOPE:2019jix, MICROSCOPE:2022doy} (reinterpreted by Ref.~\cite{Amaral:2024tjg}) and the IUPUI group~\cite{Chen:2014oda}, respectively. 
We also consider the effect that $A'$ can also couple to the neutrons through mixing with the $Z$ boson~\cite{Chauhan:2020mgv, Chauhan:2022iuh}. The magnitude of the induced coupling strength is around $\epsilon^{-1}g_{B-L}G_Fm_{D}^2/(16\pi^2)\simeq \epsilon^{-1}g_{B-L}\times 10^{-28}$, which is far smaller than $g_{B-L}$ for the values of $\epsilon$ we take. This mixing contribution is therefore irrelevant for \fref{BLbound}, but we remark that if $\epsilon\lesssim 10^{-28}$, the region excluded by the fifth force search must be extended.
\end{itemize}
Comparing the left and right panels of \fref{BLbound}, it is obvious that when $\epsilon$ decreases, the neutrino decay and scattering bound to $g_{B-L}$ is tightened, while that from the fifth force test stays unchanged. 
As a result, the neutrino experiments can serve as important complimentary probes for the ultra-light $B-L$ gauge boson. 
Furthermore, we note that for $m_{A'}\simeq 0.01-0.1$ eV, $\epsilon^{-1}g_{B-L}$ are allowed to be as large as $\mathcal{O}(1)$, which can be understood as a \textit{leading-order} correction to the SM.

\section{Discussion}
Comparing with the phenomenology-driven study in Chapter~\ref{twozerochapter}, we demonstrate here that studying the swampland of flavour symmetries is also motivated from the theory perspective. 
Even without explicit flavour symmetries, the low energy flavour parameters could stay calculable: 
on one hand, if the flavour symmetries become implicit due to the perturbative effects, the low-energy behavior of the flavour parameters are dominated by the infrared physics and always evolve towards the  fixed points; on the other hand, in certain scenarios, the consistency requirements between the UV and infrared theory allows bootstrapping the observed low-energy flavour parameters, even if the underlying dynamics can only be described in the non-perturbative frameworks. 
In other words
\begin{equation}
    \text{No Symmetries}~\neq~\text{No Predictions}.
\end{equation}
Therefore, the swampland of flavour symmetries does not veto a predictive flavour theory. Rather, it provides a path different from the conventional flavour model building, that the observed flavour structure may fully originate from the low energy dynamics. 
We note such emerging behaviors do not conflict with the canonical paradigm of reductionism, because the fundamental theory in UV remains (maximally) symmetric. 
In this view, the SM flavour pattern is similar to the properties of chemical elements --- both exhibit complicated behaviors at low energies but incorporate simple UV origins. 
The difference is that atoms can be divided into smaller parts while the quarks and leptons in SM are fundamental. The low-energy complexities arise only in the Yukawa couplings.

We note that such dynamical approaches to the flavour puzzle need non-decoupling physics. 
That is to say, the low-energy instead of UV theories deviate from the `standard expectations', which is of particular interest for phenomenology. 
Interpreting the values of the known flavour parameters as infrared fixed points indicates that the needed new particles \typo{can not}{cannot} lie at an ultra-high scale.
The lightness of (some of) these particles are even already hinted at by the long-standing flavour anomalies. 
Moreover, the non-perturbative gravity-induced spontaneous flavour symmetry breaking mechanisms lead to phase transitions and massless goldstone bosons and sometimes require ultra-light gauge bosons. 
The enriched infrared spectrum can provide testable implications in neutrino and dark matter direct detection experiments.

\chapter{Generalization to Flavoured Axions}
\label{flavouredaxions}
In Chapter~\ref{calculating}, we have demonstrated that the SM fermion masses and mixing angles can originate from low-energy dynamics insensitive to the underlying flavour symmetries in the UV. 
Nevertheless, to argue this approach leads to an ultimate theory of flavour, one must consider that the SM itself may not be complete and non-trivial flavour structures may also arise in the extended theories. 
In particular, the fundamental particle spectrum of nature is motivated to contain a \typo{pesudo}{pseudo}-scalar boson $a$ --- commonly \typo{referred as}{referred to as} `axion'~\cite{WW2} --- to address the so-called strong CP problem in QCD.  
In analog to $\theta_G$ term of gravity shown in Eq.~\eqref{EinsteinGravity}, the Lagrangian of QCD can also include an unique Chern-Simons term
\begin{equation}
    \mathcal{L}_{\text{QCD}}~=~\frac{1}{2}FF+\theta_c F\widetilde{F}.
\end{equation}
Here, $F$ is the field strength tensor of gluons.
$\theta_c$ is a free parameter of QCD whose value \typo{can not}{cannot} be predicted within the SM. $\theta_c$ shifts by a factor proportional to $\arg\det Y_uY_d$ under the chiral $U(1)$ transformation for the quark fields. 
Without loss of generality, we take the basis for which $\arg\det Y_uY_d=0$, so that $\theta_c$ is physical.
Then, the value of $\theta_c$ can be constrained by experiments. The measurement of the neutron electric dipole moment~\cite{Abel:2020pzs} implies that~\cite{Pospelov:1999mv, DiLuzio:2020wdo}
\begin{equation}
    |\theta_c|~\lesssim~10^{-10}
\end{equation}
The smallness of $\theta_c$ leads to a conceptual puzzle. The axions couple to gluons in the form of $aF\widetilde{F}$ and shift by a constant $a\to a+f_a$ under certain global symmetries, rendering the $\theta_c$ term in QCD unphysical\footnote{
We note that in the framework established in Refs.~\cite{Dvali:2013eja, Dvali:2014gua, Dvali:2017eba, Dvali:2020etd, Dvali:2021kxt, Berezhiani:2021zst, Dvali:2022fdv}, the existence of a light \typo{pesudo}{pseudo}-scalar degree of freedom, which differs with $a$ only by the UV completions, becomes a consistency requirement. 
The $S$-matrix formulation of gravity excludes the de Sitter vacua and yields that the other vacuua which do not asymptote to the Minkowski vacuum are equally problematic.
On the other hand, physical $\theta_c$ leads to an infinite number of super-selection sectors of non-Minkowski vacua, which leads to a contradiction.}. 
However, $a$ leads to more free flavour parameters through its interactions with the SM fermions~\cite{Georgi:1986df, Bauer:2017ris, Bjorkeroth:2018dzu, Calibbi:2020jvd, Bauer:2020jbp, Bauer:2021mvw, MartinCamalich:2025srw, Ziegler:2026kis}, which can be represented as
\begin{equation}
\label{axionEFT}
    \mathcal{L}~=~\frac{\partial^{\mu}a}{f}\sum_{\psi}\overline{\psi_L}\gamma_{\mu}X_{\psi}\psi_L, \quad \psi_L~=~Q_L, \ell_L, U_R^c, D_R^c, E_R^c, \nu_R^c. 
\end{equation}
$X_{\psi}$ are $3\times 3$ hermitian matrices in the generation space. 
In other words, the number of arbitrary flavour parameters increase dramatically when $a$ is included in the SM. 
A commonly adopted framework is restricting $X_{\psi}$ with the symmetries in UV~\cite{Choi:2017gpf, MartinCamalich:2020dfe, Chala:2020wvs, Bauer:2020jbp, Bauer:2021mvw}. 
Under these symmetries, $X_{\psi}$ can exhibit certain flavour structures at leading order, i.e., the diagonal ones. Then, the SM quark mixing effects contribute as next-to leading order corrections to $X_{\psi}$. 
Our question is: if these UV symmetries do not exist or exist but become implicit the low energies, could the patterns of $X_{\psi}$ stay under control?

Ideally, if $X_{\psi}$ are observed in future and appear to be flavour anarchic or quasi-anarchic, they can remain calculable under reasonable assumptions. 
We note that the ideas introduced in Chapter~\ref{calculating} can be applied to $X_{\psi}$ accordingly. 
In particular, $a$ is contained in $\phi_{126}$ of the most minimal $SO(10)$ theory we introduced in Section~\ref{fixpoints} and its shift symmetry is part of the accidental PQ symmetry shown in Eq.~\eqref{U1pq}. 
The axion needed to solve the strong CP problem is thus automatically embedded without additional model building. The $a$-fermion couplings in UV are predicted by the $SO(10)$ symmetry and an their low-energy values are outputs of the RG evolution. 
The TeV-scale LQs also contribute to the RG evolution and their effects are more sizable than those from the SM quark mixing. 
Therefore, 
we expect that the light spectrum of the most minimal $SO(10)$ also contains an axion, whose low energy flavour structures are calculable but deviate significantly from the symmetry protected patterns in UV.

Despite so, we emphasis that in our view, an important preliminary work --- fully understanding the landscape of these symmetries --- is incomplete yet. 
Therefore, in the rest of this chapter, we will \textit{not} delve into the swampland of the flavour symmetries for decoupling axions. 
Rather, we will revisit one of the minimal UV complete axion models, the DFSZ model~\cite{DFSZ1,DFSZ2}, and thoroughly study its low-energy implications on the axion flavour-violating couplings, based on our recent work Ref.~\cite{Gao:2025ohi}.
By exploring this minimal benchmark model, we will discuss two critical questions on the connection between the UV and low-energy descriptions of the axion interactions:
\begin{enumerate}[label={(\roman*)}]
    \item Is one loop matching always a good approximation?
    \item Which low-energy basis is consistent with UV completions? 
\end{enumerate}
These questions will be addressed in Section~\ref{DFSZmodel}. 
Comparing with Ref.~\cite{Gao:2025ohi}, we will show more detailed analysis on the relevant phenomenological implications in Section~\ref{BtoKa}.

\section{\texorpdfstring{$b\to s a$}{b->sa} in the DFSZ Model}
\label{DFSZmodel}

We start by revisiting the DFSZ model.
The Yukawa and scalar sector of the model read~\cite{DFSZ1,DFSZ2,DiLuzio:2020wdo}
\begin{equation}
\begin{aligned}
\label{DFSZpotential}
    V_{\Phi}~=&~ \Tilde{V}_{\text{moduli}}(|\Phi_u|,|\Phi_d|,|\Phi_u \Phi_d|,|\Phi_s|)+\lambda \Phi_s^{2} \Phi_u^{~} \Phi_d^{\dagger} +\text{h.c.},\\
    -\mathcal{L}_{Y}~=&~ Y_u \overline{Q_L}u_R \Phi_u +Y_d \overline{Q_L}d_R \widetilde{\Phi}_d+\text{h.c.}
\end{aligned}
\end{equation}
Here, $\Tilde{V}_{\text{moduli}}$ contains all gauge invariant combination of $\Phi_u,\Phi_d,\Phi_s$ without phase dependence. $Y_u$ and $Y_d$ are $3\times3$ matrices and we do not show their generation indices explicitly. 
The $\Phi_u$ and $\Phi_d$ transform in the same manner as the SM Higgs doublet $H$ under $G_{\text{SM}}$, and $\phi_s$ is a gauge singlet. They can be represented by
\begin{equation}
    \label{Fields}
\begin{aligned}
    \Phi_s~=~f+\frac{r_0+i a_0}{\sqrt{2}}, \quad \Phi_{\alpha}~=~
    \left(
    \begin{array}{c}
        \phi_{\alpha}^-   \\
          v_{\alpha}+(\rho_{\alpha}+i \eta_{\alpha})/\sqrt{2}
    \end{array}\right), \quad  \alpha~=~u,d. \\
\end{aligned}
\end{equation}
Here, $v_u^2+v^2_d=(246~\text{GeV})^2$, while $f$ is unconstrained and can take arbitrarily large values.

We firstly take $\lambda=0$ for insights. 
In this limit, the DFSZ model reduces to the so-called PQWW model~\cite{PQ1,PQ2,WW1,WW2} plus an irrelevant complex singlet scalar. 
Then, the interaction of Eq.~(\ref{DFSZpotential}) admits three $U(1)$ symmetries, corresponding to three independent phase rotations for $\Phi_u, \Phi_d,$ and $\Phi_s$. All these $U(1)$ symmetries are spontaneously broken when the scalar fields take their VEVs, so three massless Goldstone modes arise. One of them is gauged by the hypercharge $Y$, and the other two are physical. They can be identified as the $U(1)_{\text{PQWW}}$ axion $A_{0}$ and the $U(1)_s$ axion $a_0$ as a massless radial mode of $\Phi_s$. 
Since the $U(1)_{\text{PQWW}}$ symmetry is broken together with the electroweak symmetries, $A_0$ \typo{can not}{cannot} be decoupled from the SM. 
Consequently, as analyzed in~\cite{Wise:1980ux, Hall:1981bc, Frere:1981cc}, the $b\rightarrow s A_0$ amplitude \typo{can not}{cannot} be made arbitrarily small. 
On the other hand, $b\rightarrow s a_0$ is clearly zero due to the $a_0\rightarrow -a_0 $ (or $\Phi_s\rightarrow\Phi_s^{*})$ symmetry. 
Given a tiny $a_0-A_0$ mass matrix that only softly breaks the $U(1)$ symmetries, the $a_0$ and $A_0$ states can mix with a physical mixing angle $\theta$. If indicating the light physical state by $a$ and take the limit $\theta\to 0$, the $b\to s a$ amplitude reads 
\begin{equation}
    \label{LigetiResult}
    \mathcal{A}(b\rightarrow s a)_{\text{DFSZ}}~=~-\theta\times \mathcal{A}(b\rightarrow s A_0)_{\text{PQWW}}.
\end{equation}
This agrees with the argument of Ref.~\cite{Freytsis:2009ct}, that $b\rightarrow s a $ in the DFSZ model is induced by mixing.

\begin{figure}
    \centering
    \includegraphics[width=0.7\textwidth]{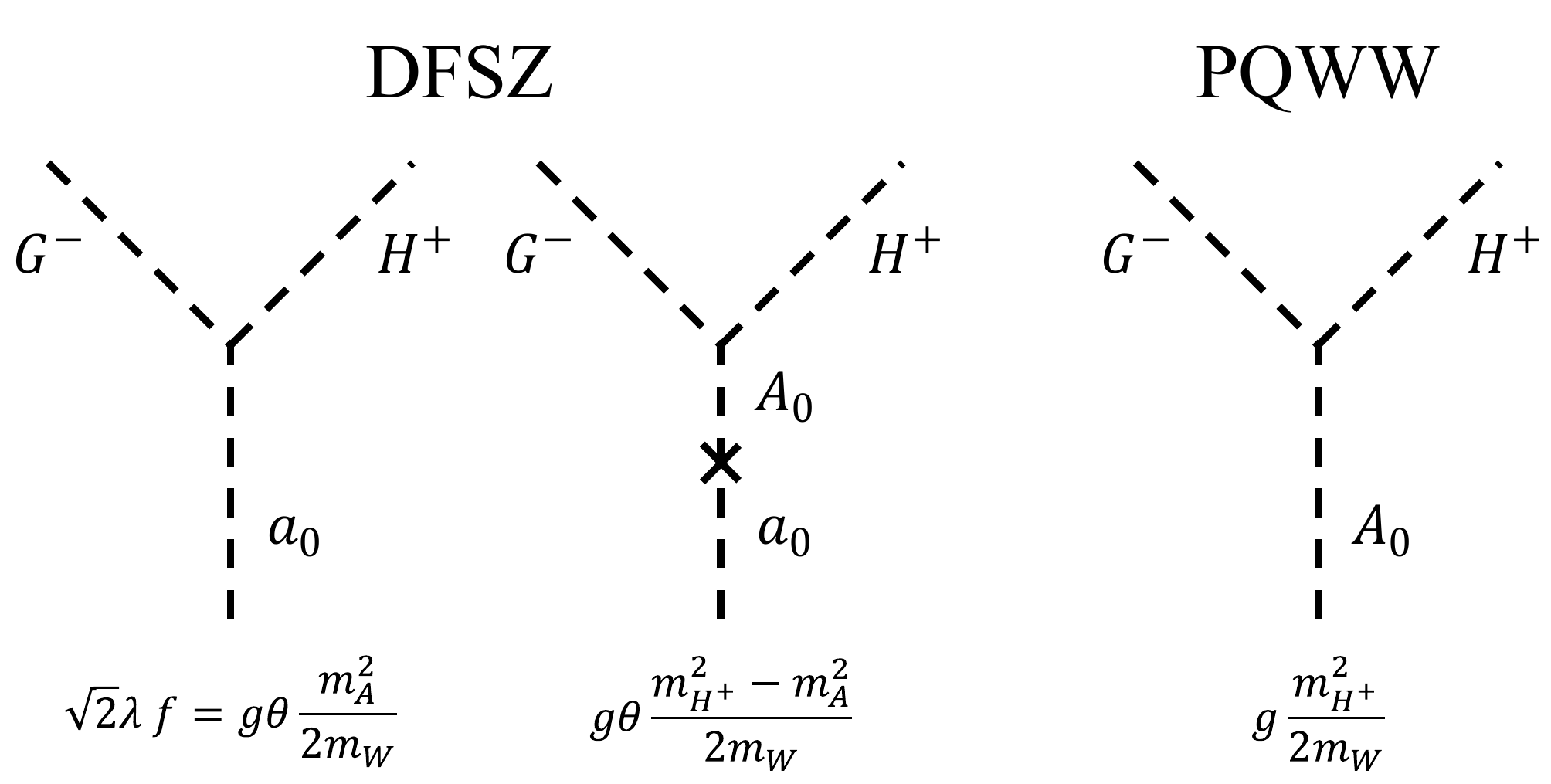}
    \caption[The Feynman rules and diagrams for the $G^-H^+a$ interaction.]{Feynman rules and diagrams for $G^-H^+-$ALP interaction in the DFSZ (left) and the PQWW (right) models.}
    \label{feynrules1}
\end{figure}

However, Eq.~(\ref{LigetiResult}) is derived in the limit $\lambda=0$. 
When $\lambda$ does not vanish, the $\lambda \Phi_s^{2} \Phi_u^{~} \Phi_d^{\dagger}$ term explicitly breaks the $U(1)_{\text{PQWW}} \times U(1)_{s}$ group into its diagonal subgroup, which we denote as $U(1)_{\text{PQ}}$. 
Then, one of the would-be massless Goldstone boson, the $a_0-A_0$ mixing state $A$, becomes massive. In case $m_A>m_b$, the $b\rightarrow s A_0$ amplitude  turns to be off shell. In addition, some Feynman rules change: as shown in Fig.~\ref{feynrules1}, a new $G^-H^+a_0$ vertex appears and the $G^-H^+A_0$ vertex rule is modified by a $m_{A}^2$ term. Therefore, we conclude: 
\begin{equation}
\begin{aligned}
    \mathcal{A}(b\rightarrow s a)_{\text{DFSZ}}~&\neq~-\theta\times\mathcal{A}(b\rightarrow s A_0)_{\text{DFSZ}}, \\
    \mathcal{A}(b\rightarrow s A_0)_{\text{DFSZ}}~&\neq~\mathcal{A}(b\rightarrow s A_0)_{\text{PQWW}}.\\
\end{aligned}
\end{equation}
Strictly speaking, the off-shell amplitude $\mathcal{A}(b\rightarrow s A_0)_{\text{DFSZ}}$ is unphysical because its gauge dependent. Despite this, we find that the Feynman rules of the effective $G^-H^+a_0$ vertex in the DFSZ model and the $G^-H^+A_0$ vertex of the PQWW model only differ by a factor of $\theta$. As illustrated in \fref{feynrules1}, the $m_A^2$ term for the direct and mixing contributions to the $G^-H^+a_0$ vertex cancels. 
Therefore, Eq.~(\ref{LigetiResult}), the explicit expression shown in Ref.~\cite{Freytsis:2009ct}, is correct even if $\lambda$ takes generic values. 
We understand the reason for the agreement as that the $\lambda \Phi_s^{2} \Phi_u^{~} \Phi_d^{\dagger}$ term does not contain the physical state $a$, when parameterized exponentially. In the non-linear basis, the $G^-H^+a$ term becomes $G^-\overset{\leftrightarrow}{\partial^{\mu}}H^+ \partial_{\mu}a$, and is manifestly independent of $m_A$.

Nevertheless, the non-vanishing $\lambda$ indicates that $a$ can directly interact with the $G^+H^-$ and contribute to $b\to s a$. 
Then, the reason why $\lambda$ does not appear in the $b\to s a$ amplitude still needs to be explained. 
We find this can be understood by analyzing the symmetries in the broken phase. 
In the limit $v_d\ll v_u\ll f$, the mass and interaction basis for all scalar fields are aligned. 
So we rewrite Eq.~\eqref{Fields} as
\begin{equation}
\label{notionZ2}
\begin{aligned}
    \Phi_s\simeq f+\frac{r+i a}{\sqrt{2}}, \quad 
    \Phi_u\simeq H_u= 
    \left(\begin{array}{c}
        G^-   \\
          v+\frac{h+i G^0}{\sqrt{2}}
    \end{array}\right),\quad 
    \Phi_d\simeq H_d=\left(\begin{array}{c}
        H^-   \\
          \frac{H+i A}{\sqrt{2}}
    \end{array}\right).\\
\end{aligned}
\end{equation}
If in addition $\lambda\ll1$, the DFSZ model are invariant under the following $Z_2$ transformation 
\begin{equation}
    \label{Z2sym}
    d_R\rightarrow-d_R, \quad H_d\rightarrow-H_d. 
\end{equation}
If the limits discussed above are relaxed, the $Z_2$ symmetry of the model can be formally promoted as exact when interpreting
\begin{equation}
\label{alignmentLimit}
\begin{aligned}
    \theta~=~\frac{2 v_d}{f}, \quad \cot{\beta}~=~\frac{v_d}{v_u}, \quad  \lambda
\end{aligned}
\end{equation}
as $Z_2$ spurions.
We note the $b\to sa$ amplitude does not violate the $Z_2$ symmetry, but it must at least contain the $Z_2$ spurion $\theta$ so that it vanishes in the $\theta\to 0$ limit. 
Consequently, the $Z_2$ invariant effective Hamiltonian for $b\rightarrow s a$ decay must contain at least two $Z_2$ spurions. Its form is then fixed up to the loop factors by
\begin{equation}
\label{Hamilton}
\begin{aligned}
    \mathcal{H}^{bs}~=&~ \theta \frac{g^3 V_{ts}^*V_{tb}}{128\pi^2} \frac{m_t^2}{m_W^3}
    \left(
    X_1\frac{1}{\tan\beta}+X_2\frac{1}{\tan^3\beta}+X_3\frac{\lambda}{16\pi^2}
    \right)\overline{s}\gamma^{\mu}P_Lb~\partial_{\mu} a.
\end{aligned}
\end{equation}
Here, $X_1,X_2,$ and $X_3$ denote the loop factors which \typo{can not}{cannot} be extracted by symmetry arguments. 
$X_1$ and $X_2$ are given in Ref.~\cite{Freytsis:2009ct}, while we note their contribution correlated to the spurion $\cot\beta$ needed to restore the $Z_2$ invariance. 
In case $\tan\beta$ is large, their effects become irrelevant. 
Interestingly, we find $\lambda$ enters the expression of $\mathcal{H}^{bs}$ because $\lambda$ is also a $Z_2$ spurion like $\cot\beta$. 
The difference is that $\lambda$ is a coupling constant and carries Planck Units $\hbar$. It arises only at the next order of the perturbative expansion and its effect is suppressed at least by an additional loop factor $1/16\pi^2$. 
As a consequence, its contribution is absent in the one-loop expressions shown given in Ref.~\cite{Freytsis:2009ct}. 
We notice that the $\lambda$ contribution can dominant if the one-loop result is suppressed by sizable $\tan\beta$.

\begin{figure}[t!]
    \centering
    \includegraphics[width=0.7\textwidth]{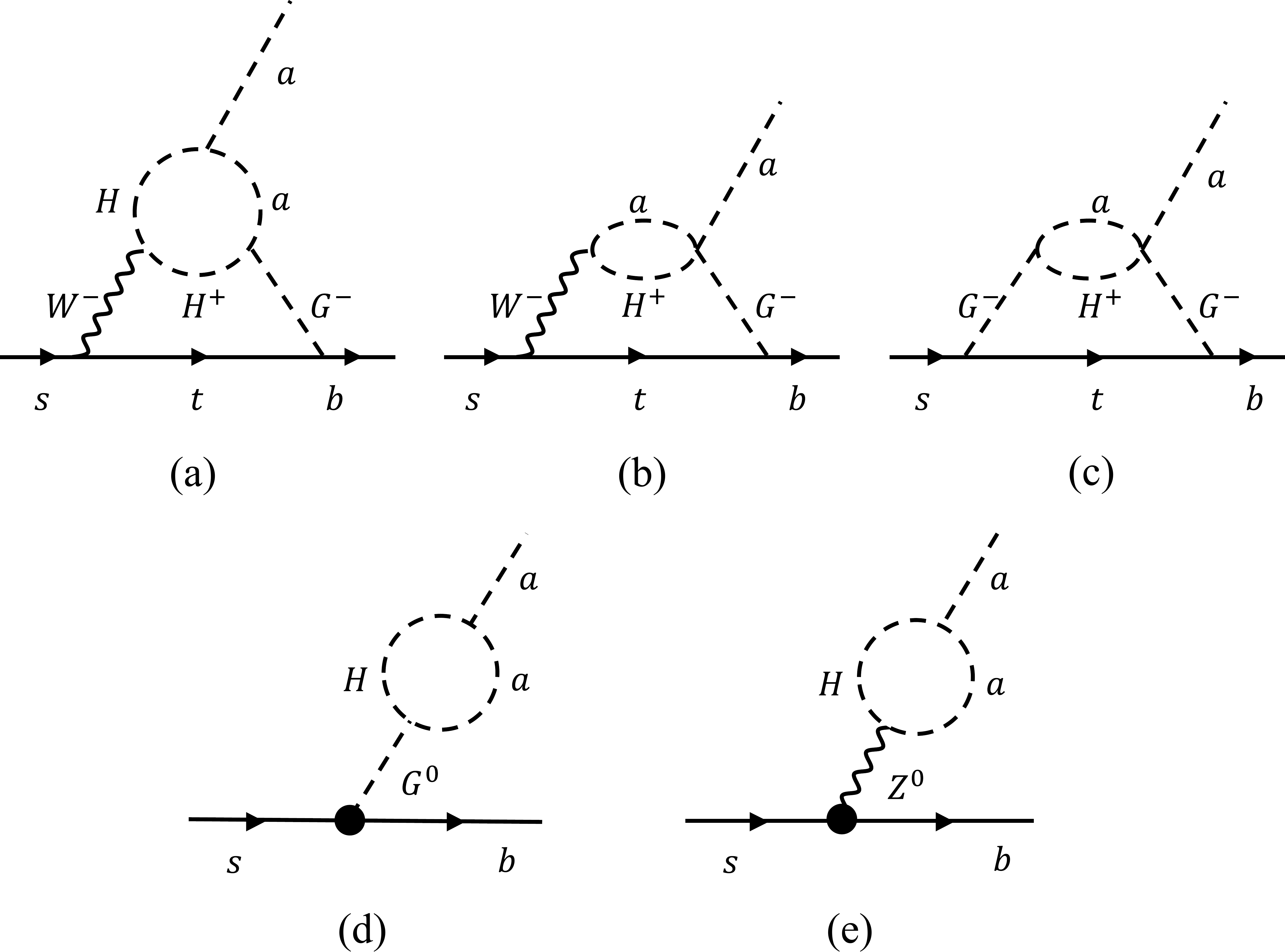}
    \caption[Two-loop Feynman diagrams for $b\to sa$.]{Illustration of the two-loop Feynman diagrams contributing to $b\to sa$ at the given order. The black dots in (d) and (e) indicate one-loop vertices in SM.
    }
    \label{feynrules2}
\end{figure}

We define $m_H\equiv m_{H^+}$ and for simplicity take the limit $m_H^2\simeq m_A^2=\lambda f^2\tan\beta \gg m_W^2$. The loop factors in Eq.~(\ref{Hamilton}) are calculated to be
\begin{equation}
\begin{aligned}
\label{X1X2X3}
    X_1~=&~-\log{\frac{m_H^2}{m_{t}^2}}+\frac{3 m_W^4}{(m_t^2-m_W^2)^2}\log{\frac{m_t^2}{m_W^2}}
    +\frac{3(m_t^2-2m_W^2)}{m_t^2-m_W^2},\\
    X_2~=&~0,\\
    X_3~=&~\log{\frac{m_{H}^2}{m_t^2}}+\frac{6m_W^2}{m_t^2-m_W^2}\log{\frac{m_t^2}{m_W^2}}+\frac{1}{2}.\\
\end{aligned}
\end{equation}
Our results for $X_1$ and $X_2$ agree with Ref.~\cite{Freytsis:2009ct}. 
The expression for $X_3$ is newly identified in our recent work, Ref.~\cite{Gao:2025ohi}. 
To calculate it, we firstly use the packages FeynRules~\cite{Christensen:2008py,Alloul:2013bka} to derive the scalar interacting vertexes contained in $\lambda \Phi_s^{2} \Phi_u^{~} \Phi_d^{\dagger}$, add them to the THDM model file of the FeynArts package~\cite{Hahn:2000kx}, and then generate the $b\to sa$ Feynman diagrams using the updated model. 
In total, there are thousands of diagrams contributing to $b\rightarrow s a$ at two-loop level. Fortunately, only very few of them are relevant for calculating $X_3$, which we show in \fref{feynrules2}. We note that the diagrams from exchanging $b\leftrightarrow s$, and the ones replacing the internal $a,H$ with $r,A$\footnote{$m_r\ll m_A$ in the large $\tan\beta$ limit.} are not shown explicitly. 
Next, we extract the scattering amplitudes in \fref{feynrules2} and Taylor expand them in the small external momentum limit $p_i^{\mu}\sim m_b$~\cite{Fleischer:1994ef}: 
\begin{equation}
    \label{Tayler}
    \mathcal{A}~=~\mathcal{A}|_{p_i^{\mu}=0}+p_i^{\mu}\left.\frac{\partial \mathcal{A}}{\partial p_i^{\mu}}\right|_{p_i^{\mu}=0}+\mathcal{O}(p_i^2).
\end{equation}
We find the $\mathcal{A}|_{p_i^{\mu}=0}$ term do not contribute to $X_3$. In the large $\tan\beta$ limit, its effect is canceled after renormalizing the quark-mixing matrix, which arise from the same diagrams by replacing the external $a$ with the vacuum tadpole $f$. Only the $\mathcal{O}(p_i^{\mu})$ term is relevant, and we evaluate it using FeynCalc~\cite{Shtabovenko:2016sxi,Shtabovenko:2020gxv, Mertig:1990an}, FIRE~\cite{Smirnov:2019qkx}, and FeynHelpers~\cite{Shtabovenko:2016whf}. The new functions for multi-loop tensor reduction and topological identification in Feyncalc 10~\cite{Shtabovenko:2023idz} are applied. The master integrals are reduced to the vacuum bubble ones, whose analytical expressions are given in~\cite{Davydychev:1992mt,Nierste:1995fr}.
We note that although these master integrals contain the transcendental-weight-two functions, only the logarithm functions appear after expanding in $1/m_H^2$.

To cross check our results, we work within the $R_{\xi}$ gauge and demonstrate $X_3$ is independent of $\xi$.
We note the cancellation of the gauge parameter $\xi_W$ is similar to the case of $b\rightarrow s \mu^+\mu^-$ in SM~\cite{Inami:1980fz}. Intuitively, the $\xi_W$ dependent contribution from the box-like diagrams (a-c) shown in \fref{feynrules2} cancel those from the penguin-like diagrams (d-e). Thus, the loop-induced kinetic-mixing term $\partial_{\mu}G^0\partial^{\mu}a$ 
(and $Z_{\mu}\partial^{\mu}a$, which vanishes in Landau gauge $\xi_Z=0$)
plays an important role. To perform $a-G^0$ wave-function renormalization, we apply the standard method of subtracting the Goldstone boson self-energies at zero momentum~\cite{Santos:1996vt, Grinstein:2015rtl}. This eliminates the tadpole contributions and simplifies the calculation.

\begin{figure}[t!]
    \centering
    \includegraphics[width=0.45\linewidth]{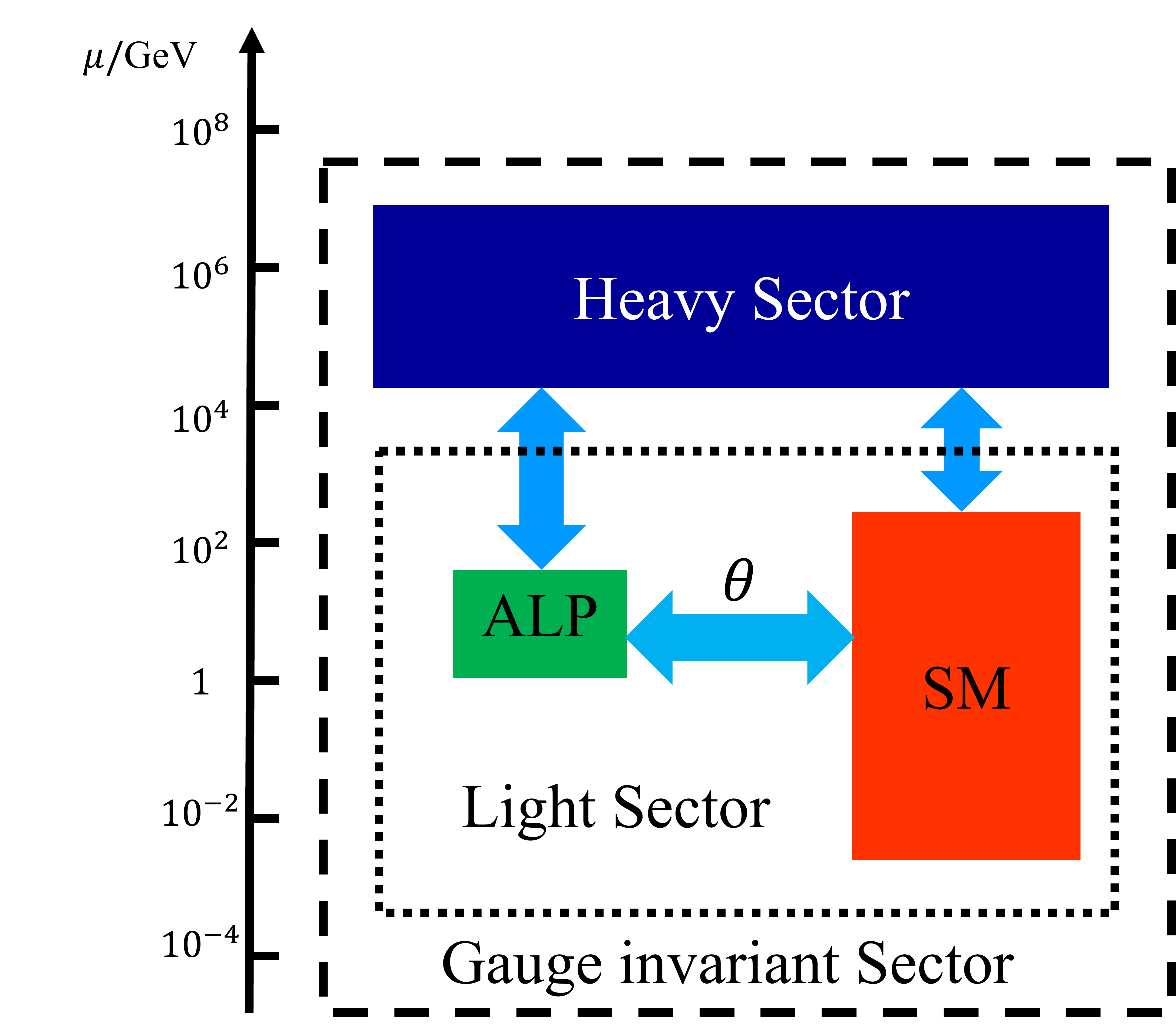}
    \includegraphics[width=0.45\linewidth]{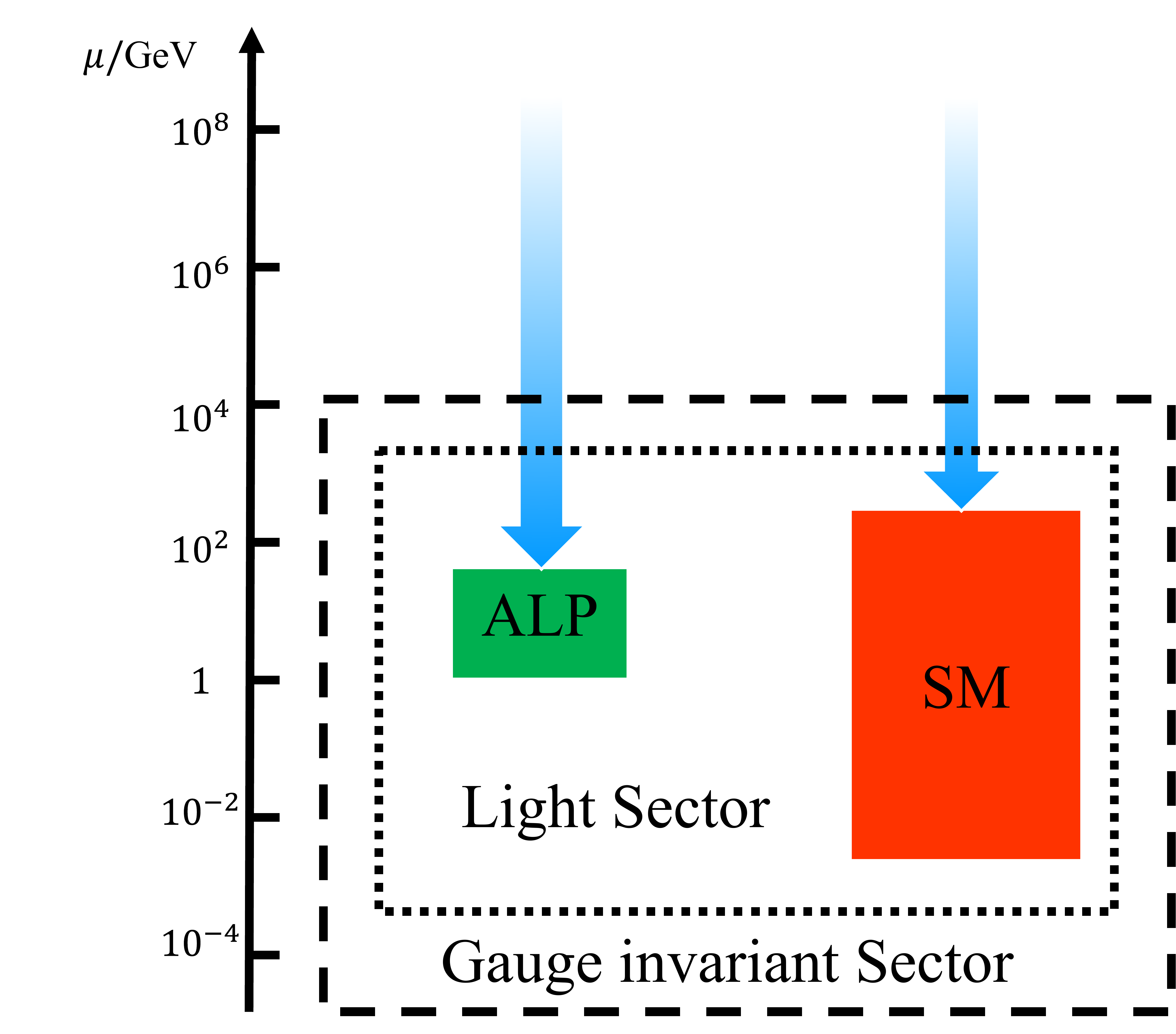}
    \caption[The apparent non-decoupling heavy sector.]{A schematic illustration on the apparent non-decoupling feature heavy particles in a theory containing SM and an ALP. Effects of the heavy sector vanishes only when $\theta=0$.}
    \label{decoupleFig}
\end{figure}

We note $X_3$ does not vanish in the limit $m_H\to \infty$ but diverges logarithmically. 
This is not a special feature of the two-loop integrals because the same behavior is also reported on the $X_1$ term~\cite{Freytsis:2009ct}. 
Rather, such `apparent non-decoupling behavior' is intrinsic for the model. 
Decoupling is hidden in the mixing angle 
\begin{equation}
\label{Nondecoupling}
    \theta~\sim~\frac{1}{f} ~\sim~ \frac{1}{m_H}. 
\end{equation}
The full $b\to s a$ amplitude is proportional to $\theta$ and vanishes in the $m_H\to \infty$ limit. 
This behavior is uncommon but not unique. 
For instance, as pointed out in Ref.~\cite{Chang:1993kw}, the $\mu\rightarrow e  \gamma$ decay amplitude in a 2HDM is not directly suppressed by the heavy mass either. 
Instead, it is proportional to the misalignment parameter which scales as $c_{\alpha\beta}\sim m_H^{-2}$~\cite{Gunion:2002zf}.
Apparently, the behaviors introduced in the two examples above challenges the Wilsonian decoupling picture~\cite{Wilson:1974mb}, because the dimensionless couplings inherited with $\theta$ or $c_{\alpha\beta}$ cannot take the `naively expected' $\mathcal{O}(1)$ values. 
Based on the argument on the necessary condition for decoupling provided in Ref~\cite{Senjanovic:1979yq}, we understand as the origin of this apparent non-decoupling feature that the light sub-sector is not invariant under $SU(2)_L\times U(1)_Y$ without the heavy degrees of freedom. 
As illustrated in \fref{decoupleFig}, the light sector composed by SM particles and the axion $a$ \typo{can not}{cannot} be made gauge invariant if $\theta\neq 0$. An extended sector is required for model building and the particles contained in this sector \typo{can not}{cannot} lie at arbitrary heavy scales. 
Notably, the contributions of these new particles to the low energy processes are not necessarily suppressed by their masses. 
If such extended sector decouples, the portal $\theta$ must vanish. The light sector is gauge invariant because it only contains SM and a free particle $a$.

\begin{figure}[t!]
    \centering
    \includegraphics[width=0.7\linewidth]{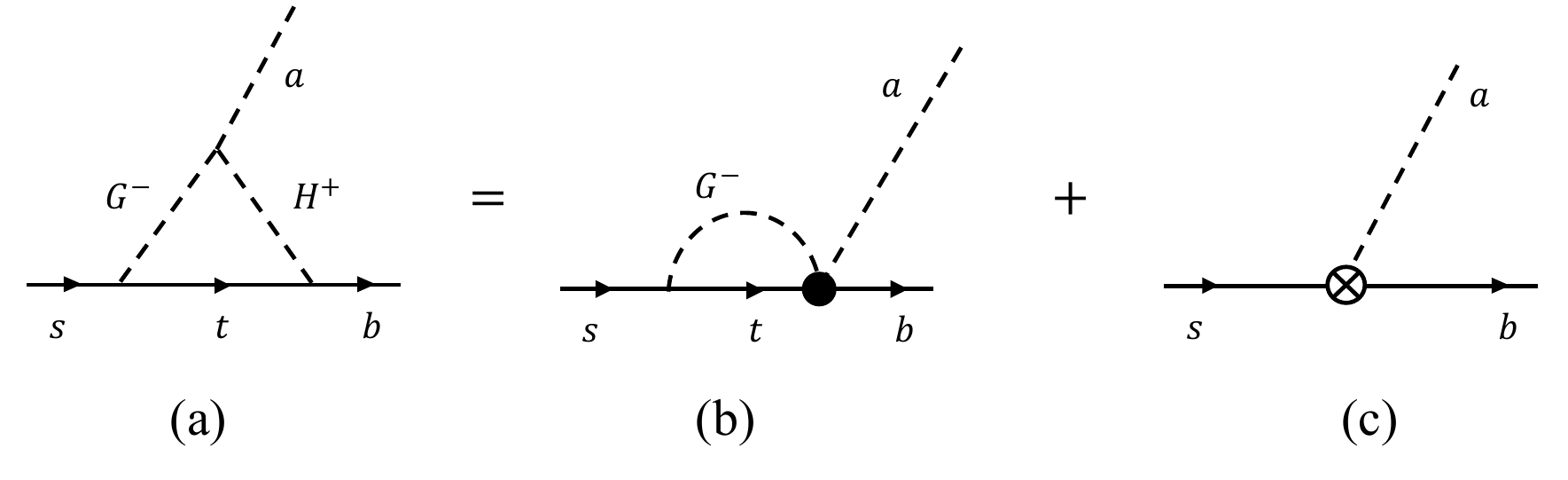}
    \caption{The Feynman diagrams relevant to the apparent non-decoupling effects.}
    \label{nondecoupFeynmanDiags}
\end{figure}

This apparent non-decoupling feature indicates that the $H^+$ and $W^+$ contributions to $b\to s a$ are equally important in the DFSZ model. 
Taking the one-loop contributions as an example, we find the unsuppressed loop factors all originate from the diagram (a) shown in~\fref{nondecoupFeynmanDiags}. 
Specifically, the Feynman rule of the $G^-H^+a$ vertex is proportional to $\theta m_H^2$ and leads to the apparent non-decoupling behavior. 
In the limit $m_H\to \infty$, the large coupling strength then cancels the $m_H^2$ term contained in the denominator of the Feynman integral, so the full amplitude is suppressed by $\theta$ only. 
If one isolates the heavy $H^+$ interactions, the light theory, although seemingly renormalizable, cannot reproduce the full $b\to sa$ amplitude. To construct a consistent effective theory at low energies, one must \textit{integrate out} $H^+$ and include the matched effective operators which contain both $a$ and $G^-$.

The arguments above motivate us to check the consistency of the basis describing the axion interactions. 
We start with the renormalizable Lagrangian of Eq.~(\ref{DFSZpotential}) and isolate all heavy particles
\begin{equation}
    \label{lightsubtheory}
    -\mathcal{L}~=~  i a \sum_{q=t,b} c_{q}~\overline{q} \gamma_5 q. 
\end{equation}
This is a seemingly reasonable basis: the $a$-fermion interactions can be renormalizable, because the low-energy effective field theory (LEFT)~\cite{Jenkins:2017jig,Jenkins:2017dyc} only respects the QED$\times$QCD symmetry, under which the quarks and leptons are vector-like. 
We match $c_{q}$ from the DFSZ model and calculate the $b\rightarrow s a$ amplitude, which is UV divergent.
We then replace the $1/\epsilon$ pole of the amplitude with the logarithm function containing a UV cut-off $\Lambda_{\text{UV}}$, and get
\begin{equation}
\label{leadinglogtermR}
    \mathcal{A}'(b\rightarrow s a)~\sim~\log {(\Lambda_{\text{UV}}^2/m_t^2)}.
\end{equation}
If the basis shown in Eq.~\eqref{lightsubtheory} is consistent, $\mathcal{A}'(b\rightarrow s a)$ must agree with the leading-log term of the one-loop $b\to s a$ amplitude calculate in the DFSZ model, after replacing $\Lambda_{\text{UV}}$ with $m_H$\footnote{As pointed out in \cite{Alonso-Alvarez:2021ett}, $\Lambda_{\text{UV}}$ is not necessarily equal to the axion decay constant $f$.}. However, we find the calculated coefficient of Eq.~(\ref{leadinglogtermR}) disagrees with the expressions shown in Eq.~\eqref{Hamilton} and Eq.~(\ref{X1X2X3}), which leads to a clear contradiction. 
We find the mismatch originates from the diagram (a) of \fref{feynrules2}, which is not suppressed by $1/m_H^2$ although it contains a heavy particle. 
Eq.~\eqref{lightsubtheory} yields a basis which \typo{can not}{cannot} capture the non-decoupling $H^+$ contribution.

The LEFT basis shown in Eq.~\eqref{lightsubtheory} is not invariant under the $SU(2)_L\times U(1)_Y$ gauge symmetry. In a consistent low-energy, the gauge symmetry must be respected. 
Unlike the QED$\times$QCD, the $SU(2)_L\times U(1)_Y$ symmetry is chiral and requires the SM Higgs boson $H$ for chirality flipping. 
$H$ leads to dimension five operators and renormalizability \typo{can not}{cannot} be preserved. The gauge invariant Lagrangian then reads
\begin{equation}
\begin{aligned}
    \label{EFTS2}
    -\mathcal{L}
     ~=&~i \frac{a}{v}\left( c_b \overline{Q}_L  b_R \widetilde{H}_u+c_t ~\overline{Q}_Lt_RH_u+\text{h.c.}  \right) \\
    =&~i a \sum_{q=t,b} c_{q}\overline{q} \gamma_5 q
    + i \frac{a}{v}\left[ c_b V_{tb}^{}  \overline{t}_L  b_R G^++c_t \left(V_{tb}^{*}\overline{b}_L t_R G^-+ V_{ts}^{*}\overline{s}_L t_R G^-\right)+\text{h.c.}  \right]+...
\end{aligned}
\end{equation}
The key difference between Eq.~\eqref{lightsubtheory} and Eq.~\eqref{EFTS2}  is the appearance of non-renormalizable operators containing unphysical Goldstone modes $G^{\pm}$. 
We illustrate the effects of such operators in diagram (b) of~\fref{feynrules2}. If splitting the propagator of $H^-$ into two   pieces~\cite{Bilenky:1993bt}, one gets
\begin{equation}
\label{splitting}
    \frac{1}{k^2-m_H^2}~=~-\frac{1}{m_H^2}+\frac{1}{m_H^2}\frac{k^2}{k^2-m_H^2}. 
\end{equation}
The $-1/m_H^2$ term plays the role of the of the non-renormalizable operator containing $G^{\pm}$, 
This effective operator, indicated with a solid circle in diagram (b) of \fref{feynrules2}, leads to a divergent amplitude, and the $m_H^2$ suppression is compensated by the $G^-H^+a$ vertex proportional to $m_H^2$. 
We have checked that this diagram exactly reproduces the leading-log term missing in Eq.~(\ref{leadinglogtermR}). 
The basis shown in Eq.~\eqref{EFTS2} sacrifices renormalizability but restores the $SU(2)_L\times U(1)_Y$ gauge invariance and the consistent decoupling behavior. 
We refer the reader to Ref.~\cite{Altmannshofer:2020shb} for details about a closely related example on $\mu\rightarrow e\gamma$ in 2HDM, in which the contributions from certain effective operators containing Goldstone bosons are not suppressed. 
Furthermore, Eq.~(\ref{EFTS2}) reduces to Eq.~(\ref{lightsubtheory}) under the unitary gauge, as the gauge fixing condition sets $G^+(x^{\mu})= 0$. We have checked that in this framework, the missing leading-log term of Eq.~(\ref{leadinglogtermR}) originates from the the longitudinal part of $W$ boson propagator $k_{\mu}k_{\nu}/m_W^2$. Decoupling works because the gauge symmetry is strictly speaking still preserved although hidden by gauge fixing. Again, the cost is losing renormalizability, known as a consequence of the unitary (non-renormalizable) gauge.

There are other equivalent basis in addition to the one shown in Eq.~\eqref{EFTS2}. 
By applying the equations of motions, Eq.~(\ref{EFTS2}) reduces to the derivative basis for $a$-fermion interactions shown Eq.~\eqref{axionEFT}, where the $a\to a+f_a$ shift symmetry is manifest\footnote{The anomalous terms such as $a\tilde{W}_{\mu\nu}W^{\mu\nu}$~\cite{Bauer:2020jbp} are higher order for flavor violating processes~\cite{Izaguirre:2016dfi}, so we do not show them explicitly in Eq.~\eqref{axionEFT}.}~\cite{Georgi:1986df, Choi:2017gpf, Bauer:2020jbp}. 
The derivative basis is non-renormalizable but gauge invariant, under which the $b\rightarrow s a$ amplitude can be correctly reproduced at the leading-log level.
We remark that the authors of Ref.~\cite{Dolan:2014ska} had commented on this discrepancy between the derivative basis and the \typo{pesudo}{pseudo}-scalar basis in Eq.~(\ref{lightsubtheory}) with a footnote and correctly connected the underlying reason to the missing dimension-5 operators. 
Here, we emphasize that only the derivative basis and its equivalent forms are consistent and the \typo{pesudo}{pseudo}-scalar basis (without gauge fixing) leads to the non-decoupling behaviors. 
Such a conclusion supports the argument that gauge invariance is more fundamental, while renormalizability should not be regarded as a basic principle of quantum field theories~\cite{Weinberg:1995mt}.

Before finishing this section, we want to emphasize the low-energy effective theory can only reveal the leading-log term contained in the $b\to sa$ amplitude, which dominants only when $\log{(\Lambda_{\text{UV}}^2/m_t^2)}$ is sufficiently large. 
Without specifying UV physics, the rational functions are not available. 
As one can refer from the third term of Eq.~(\ref{splitting}), part of the $H^+$ contribution is hidden in diagram (c) of \fref{feynrules2} as a counter term. 
This agrees with the fact that the generic axion effective theory allows tree-level flavour violating couplings. 
Therefore, the $b\rightarrow s a$ amplitude itself is strictly speaking, a \textit{definition} of the renormalization scheme, on how \fref{feynrules2}(c) cancels the divergence of \fref{feynrules2}(b), instead of a prediction of the low-energy theory.

\section{Implications on the Belle II Excess}
\label{BtoKa}

The $b\to s a$ transition in the DFSZ model is relevant to the $B\to K\overline{\nu}\nu$ excess recently reported at Belle II, as we introduced in subsection~\ref{reshapeIR}. 
Since the outgoing neutrinos \typo{can not}{cannot} be directly detected at Belle II, the excess (if confirmed) does not necessarily indicate that the $b\to s\overline{\nu}\nu$ transition rate deviates from the SM prediction. Rather, 
the inconsistency with the SM includes the transition of $b\to s$ and any invisible states. 
As pointed out in Ref.~\cite{Altmannshofer:2023hkn}, the excess is rather localized in the visible kaon energy and a fit under the assumption of a two-body decay $B^+ \to K^+ X$ with invisible $X$ also gives an excellent fit to the data. 
Using Belle II data the authors of Ref.~\cite{Altmannshofer:2023hkn} obtain a significance of $3.6 \sigma$ for $m_X \approx 2~\text{GeV}$ and ${\rm BR} (B^+ \to K^+ X) = (8.8 \pm 2.5) \times 10^{-6}$. 
Recently, the Bayesian analysis performed in Ref.~\cite{Gartner:2026clx} also indicate a very strong preference for the SM
plus resonance over the SM-only hypothesis.
However, the lightness of $X$ remains puzzling. A common interpretation is that $X$ is the \typo{pesudo}{pseudo}-Goldstone boson which shifts as of $X\to X+f_X$ under a spontaneously broken global symmetry; see for example~Refs.~\cite{Batell:2009jf, Izaguirre:2016dfi, Aloni:2018vki,Chakraborty:2021wda, Gavela:2019wzg, Bauer:2021mvw, Calibbi:2025rpx, MartinCamalich:2025srw}.
The mass of $X$ can the arise naturally light, because it vanished when the shift symmetry is exact.

We note that $X$ is distinguished with the axion $a$ and by itself \typo{can not}{cannot} address the strong CP problem. 
The explicit mass term for $a$ must vanish and the $a$ could only dynamically acquire an effective mass from QCD condensate, which is always much lower than 2 GeV.
Nevertheless, an explicit mass term for $a$ does not change the analysis on $b\to s a$ transition performed in Subsection~\ref{DFSZmodel}.
Based on this fact, we assume that the nature of $X$ is a \typo{pesudo}{pseudo} Goldstone boson similar to the axion --- commonly \typo{referred as}{referred to as} axion-like particles --- and keep denoting it with $a$. 
A self-consistent explanation on the Belle II excess requires $m_a~\simeq~2~\text{GeV}$ and 
\begin{equation}
    \begin{aligned}        
    \text{Br}(B\rightarrow K a)\cdot \mathcal{P}_a^{\text{inv}}~&\simeq~9\times 10^{-6}. \\
    \text{Br}(B\rightarrow K a)\cdot (1-\mathcal{P}_a^{\text{inv}})~&\lesssim~\text{exclusive search limits}.
    \end{aligned}
\end{equation}
Here, we follow Ref.~\cite{Gartner:2026clx} and use $\mathcal{P}_a^{\text{inv}}$ to indicate the probability that $a$ and/or its  decay products are undetected.

Analysis based on the generic axion effective theories cannot unambiguously connect $b\rightarrow s a$ with the other physical processes, such as the rates of $a$ decaying to SM particles. 
Therefore, we choose the DFSZ model as a minimal benchmark and follow the analysis performed in Section~\ref{DFSZmodel}.
Br$(B\rightarrow K a)$ can be derived from the effective Hamilton $\mathcal{H}^{bs}$ shown in Eq~\eqref{Hamilton}, whose explicit expression reads~\cite{Freytsis:2009ct}
\begin{equation}
\label{BtoKaBR}
\begin{aligned}
    \text{Br}(B\rightarrow K a)~=&~\frac{1}{\Gamma_B}~ \theta^2 \frac{G_F^3 \left|V_{ts}V_{tb}^*\right|}{\sqrt{2}2^{12}\pi^5}~ m_t^4 m_B^3\left(\frac{X_1}{\tan\beta}+\frac{X_2}{\tan\beta^3}+\frac{\lambda X_3}{16\pi^2}  \right)^2 [f_0(m_a^2)]^2\\
    &\times \frac{\sqrt{(m_B^2-m_a^2-m_K^2)^2-4 m_a^2 m_K^2} }{m_B^2} \left(1-\frac{m_K^2}{m_B^2}\right)^2.
\end{aligned}
\end{equation}
Here, the form factor $f_0$ is defined as
\begin{equation}
    \langle K(p-q)| \bar s \slashed{q}  P_L b | B(p)\rangle ~=~ \frac12 (m_B^2 - m_K^2)\, f_0(q^2)
\end{equation}
Approximately, $f_0(m_a^2)\approx f_0(0)\approx 0.33$~\cite{Parrott:2022rgu}.

In the DFSZ model, $a$ mostly decays to hadrons. 
In the framework of quark-hadron duality~\cite{Braaten:1988hc,Braaten:1991qm}, the inclusive hadronic decay rate is equal to the sum of $\Gamma(a\rightarrow \overline{s}s)$ and $\Gamma(a\rightarrow g g)$~\cite{Bauer:2017ris}. 
After including next-to-next-to leading order (NNLO) QCD corrections~\cite{Sakai:1980fa,Gorishnii:1990zu,Chetyrkin:1998mw}, we find
\begin{equation}
\label{decay}
    \begin{aligned}
        \Gamma(a\rightarrow \text{hadrons})~=~&\frac{3 m_a}{ 4 \pi} \frac{m_s^2}{f^2}\sqrt{1-\frac{4 m_s^2}{m_a^2}}\left(1+\frac{17}{3}\left(\frac{\alpha_s}{\pi}\right)+ c_2\left(\frac{\alpha_s}{\pi}\right)^2\right)\\
        &+~\frac{\alpha_s^2}{16\pi^3}\frac{m_a^3}{f^2}\left(1+\frac{83}{4}\left(\frac{\alpha_s}{\pi}\right)+c_2'\left(\frac{\alpha_s}{\pi}\right)^2\right),
\end{aligned}
\end{equation}
where the two-loop coefficients read
\begin{equation}
    c_2~=~\frac{9631}{144}-\frac{17}{2}\zeta(2)-\frac{35}{2}\zeta(3), \quad c_2'~=~\frac{35935}{96}-\frac{243}{8}\zeta(2)-\frac{465}{8}\zeta(3).
\end{equation}
Here, $\zeta$ is the Riemann zeta function. 
We extract the NNLO QCD corrections from the SM Higgs boson decay with $N_f=5$ replaced by $N_f=3$\footnote{We note that although $4m_c^2\gg m_a^2$ is not a good approximation when for $m_a\simeq 2$ GeV, the finite charm mass contribution is small comparing with the full decay width.}.
On the other hand, the partial decay rate for each exclusive channel \typo{can not}{cannot} be calculated within the perturbative framework. 
Despite this, information on these partial rates are available. 
Since $a$ is a CP-odd scalar and both $agg$ and $a\overline{s}s$ interactions conserve CP, $a$ \typo{can not}{cannot} decay into two \typo{pesudo}{pseudo}-scalar mesons. 
The other hadronic decay modes are all multi-body ones and are suppressed by phase factors. 
Comparing with the unpressed total hadronic decay rate shown in Eq.~\eqref{decay}, we refer that there are many relevant hadronic decay channels but none of them dominant over the total decay width~\cite{Aloni:2018vki, Chakraborty:2021wda, BaBar:2013lkw}. 
For illustration, we display here some typical hadronic decay modes extracted from Ref.~\cite{BaBar:2013lkw}
\begin{equation}
\begin{aligned}
          a~\to~&\pi^+ \pi^-\pi^0, ~\pi^+ \pi^-2\pi^0, ~2\pi^+ 2\pi^-,~ 2\pi^+ 2\pi^-\pi^0,~\pi^+ \pi^-\eta,~2\pi^+ 2\pi^-2\pi^0,~3\pi^+ 3\pi^-,\\
         &2\pi^+ 2\pi^-\eta,~3\pi^+ 3\pi^-2\pi^0,~4\pi^+ 4\pi^-,~K^+ K^-\pi^0,~K^{\pm}K_s^0\pi^{\mp},~K^+ K^-2\pi^0, \\
        &K^+ K^-\pi^+\pi^-,~ K^+ K^-\pi^+ \pi^-\pi^0, ~K^{\pm}K_s^0\pi^{\mp}\pi^+\pi^-,~ K^+ K^-\eta,~...
        \end{aligned}
\end{equation}
These modes are sorted by the total mass of the decay products. We emphasize that the large total hadronic decay rate emerges only after summing over the multiple suppressed partial ones. 
This feature makes detection for the visible the $a\to \text{hadrons}$ decay modes challenging. 

In reasonable extensions to the DFSZ model, $\mathcal{P}_a^{\text{inv}}$ can be as large as $1$. 
The simplest one is to add a light gauge singlet Majorana fermion $\chi$ which couples to $a$ via
\begin{equation}
\begin{aligned}
    \mathcal{L}~=~iy_{\chi}\Phi_s^{\dagger}\overline{\chi}\chi+\text{h.c.}
    ~=~ \sqrt{2}y_{\chi} a \overline{\chi}\chi. 
\end{aligned}
\end{equation}
Here, $y_{\chi}$ (chosen real) does not necessarily vanish in the limit $f\to \infty$. Even if $y_{\chi}$ breaks the $a\to a+f_a$ shift symmetry, its induced inhomogeneous loop contribution to $m_a$ is proportional to $m_{\chi}$
and vanishes in the limit $m_{\chi}\to 0$.
Therefore, as long as $\chi$ exist, the invisible $a\to \overline{\chi}\chi$ decay rate is motivated to dominate over the $f$ suppressed visible ones. 
This setup can be straightforwardly generalized to the $a$ portal light dark sector models. 
We do not consider the dark matter models in the work, but take $\mathcal{P}_a^{\text{inv}}\simeq 1$ in the following analysis. 
The constraints on the model parameters can be extracted accordingly if $\mathcal{P}_a^{\text{inv}}$ takes the other values.

The experimental constraints for the visible $a$ decay modes vanish in the limit $\mathcal{P}_a^{\text{inv}}\to 1$. 
If $\mathcal{P}_a^{\text{inv}}$ takes generic values, the non-hadronic decay branching ratios remain suppressed or can be suppressed. 
For instance, the $a\rightarrow  \gamma\gamma$ decay rate reads~\cite{Bauer:2017ris}
\begin{equation}
    \Gamma(a\rightarrow \gamma\gamma)~=~\frac{\alpha_{\text{EM}}^2}{18\pi^3}\frac{m_a^3}{f^2}
\end{equation}
Comparing this with the inclusive hadronic decay width shown in Eq.~\eqref{decay}, the branching ratio for $a\rightarrow  \gamma\gamma$ must be smaller than at least $(\alpha_{\text{EM}}/\alpha_s)^2\sim 10^{-4}$. 
As a result, although the photonic decay modes can provide much cleaner signature than the hadronic ones~\cite{BaBar:2021ich}, the corresponding statistics are significantly lower. 
Moreover, either $H_d$ (DFSZ-I) or $\widetilde{H}_u$ (DFSZ-II) must couple to $\overline{\ell_L}E_R$ to generate the charged lepton masses, which leads to the $a\to \ell^+\ell^-$ signals. 
The decay width read~\cite{Bauer:2017ris}
\begin{equation}
     \Gamma(a\rightarrow \ell^+\ell^-)~=~ t_{\ell}^4\frac{m_a}{4 \pi} \frac{m_{\ell}^2}{f^2}\sqrt{1- \frac{4 m_\ell^2}{m_a^2}}, \quad \ell=e, \mu. 
\end{equation}
Here, $t_\ell$ is equal $\sin\beta$ and $\cos\beta$ in the DFSZ-I and DFSZ-II models, respectively. 
Since $y_t\sin\beta= \sqrt2m_t/v$, small $\beta$ leads to non-perturbative Yukawa couplings~\cite{DiLuzio:2020wdo}. Consequently, $\sin\beta$ \typo{can not}{cannot} take values far smaller than unity and the $a\to \ell^+\ell^-$ decay width is comparable with the inclusive hadronic one in the DFSZ-I model. In DFSZ-II model, the $a\to \ell^+\ell^-$ decay can be suppressed when $\beta\simeq \pi/2$.
The leptonic $a$ decay leads to a resonance in the di-lepton spectrum of $B\to K\mu^+\mu^-$, which peaks at $q^2\simeq (2~\text{GeV})^2$. However, the lepton flavour universality is violated in the minimal model: since $m_\mu\gg m_e$, the $a\to\mu^+\mu^-$ decay width is much larger than that of $a\to e^+e^-$. 

\begin{figure}[t!]
    \centering
    \includegraphics[width=0.7\linewidth]{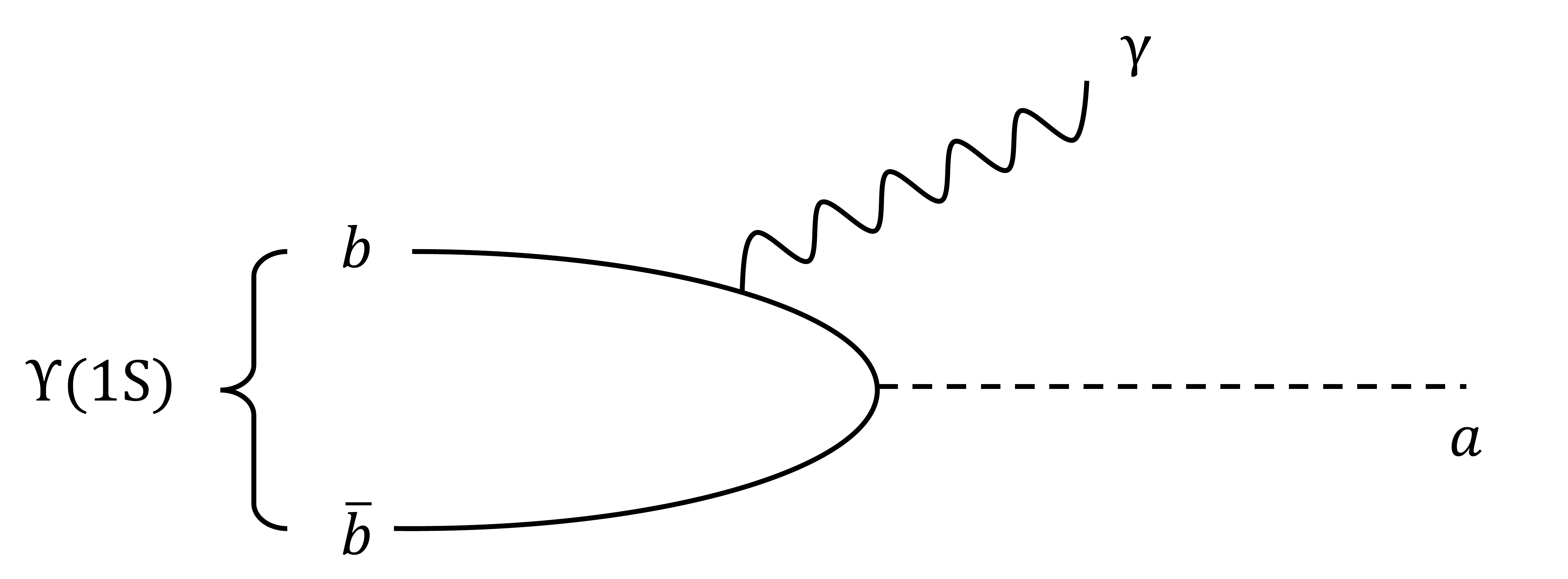}
    \caption{The Feynman diagram for the single photon decays of $\Upsilon(1S)$.}
    \label{upsilondecay}
\end{figure}

In addition to the $B\to K a$ transition, the flavour conserving $a\overline{b}\gamma_5b$ interaction leads to $\Upsilon(1S)\to \gamma a$ decay. 
This process is induced at tree level, and we show the related Feynman diagram in~\fref{upsilondecay}. 
Comparing with $B\to K a$, $\Upsilon(1S)\to \gamma a$ transition is not suppressed by the loop factor or small quark mixing angles, however, it is a rare process because the $\Upsilon(1S)$ meson is about $10^8$ times shorter living than the $B$ mesons. 
The $\Upsilon(1S)$ meson mostly decay into the hadrons via the unsuppressed QCD interactions. 
As a consequence, the $\Upsilon(1S)\to \gamma a$ branching ratio stays small and does not endanger the $a$ explanation to the $B\to K \overline{\nu}\nu$ excess. 
The $\Upsilon(nS)\to \gamma a$ decay rate, normalized by the leptonic $\Upsilon(nS)$ decay rate, reads~\cite{Nason:1986tr}
\begin{equation}
    \frac{\Gamma(\Upsilon(nS)\rightarrow\gamma a)}{\Gamma(\Upsilon(nS)\rightarrow\ell^+\ell^-)}~=~\frac{G_F^2 m_b^2}{\sqrt{2}\pi \alpha} \sqrt{1-\frac{m_a^2}{m_{\Upsilon}^2}}(\theta \tan\beta)^2 \left(1-\left(\frac{\pi^2}{6}+\frac{8}{3}\log{2}\right)\left(\frac{\alpha_s}{\pi}\right)\right).
\end{equation}
Here, we include the NLO QCD corrections. Since $\alpha_s(m_{\Upsilon})\approx 0.18$, the effects of QCD corrections are much smaller than those to $a\to \text{hadrons}$.
We note that comparing with $\text{Br}(B\to Ka)$ shown in Eq.~\eqref{BtoKaBR}, the $\Upsilon(nS)\rightarrow\gamma a$ decay branching ratio scales as $1/f^2\sim (\theta\tan\beta)^2$, which is enhanced by a factor of $\tan^2\beta$\footnote{The enhancement becomes $\tan^4\beta$ in the limit $\lambda \to 0$.}.
Therefore, the $\Upsilon(1S)$ decay can serve as an important complementary probe in the large $\tan\beta$ regime.

\begin{figure}[t!]
  \centering
  \includegraphics[width=0.95\textwidth]{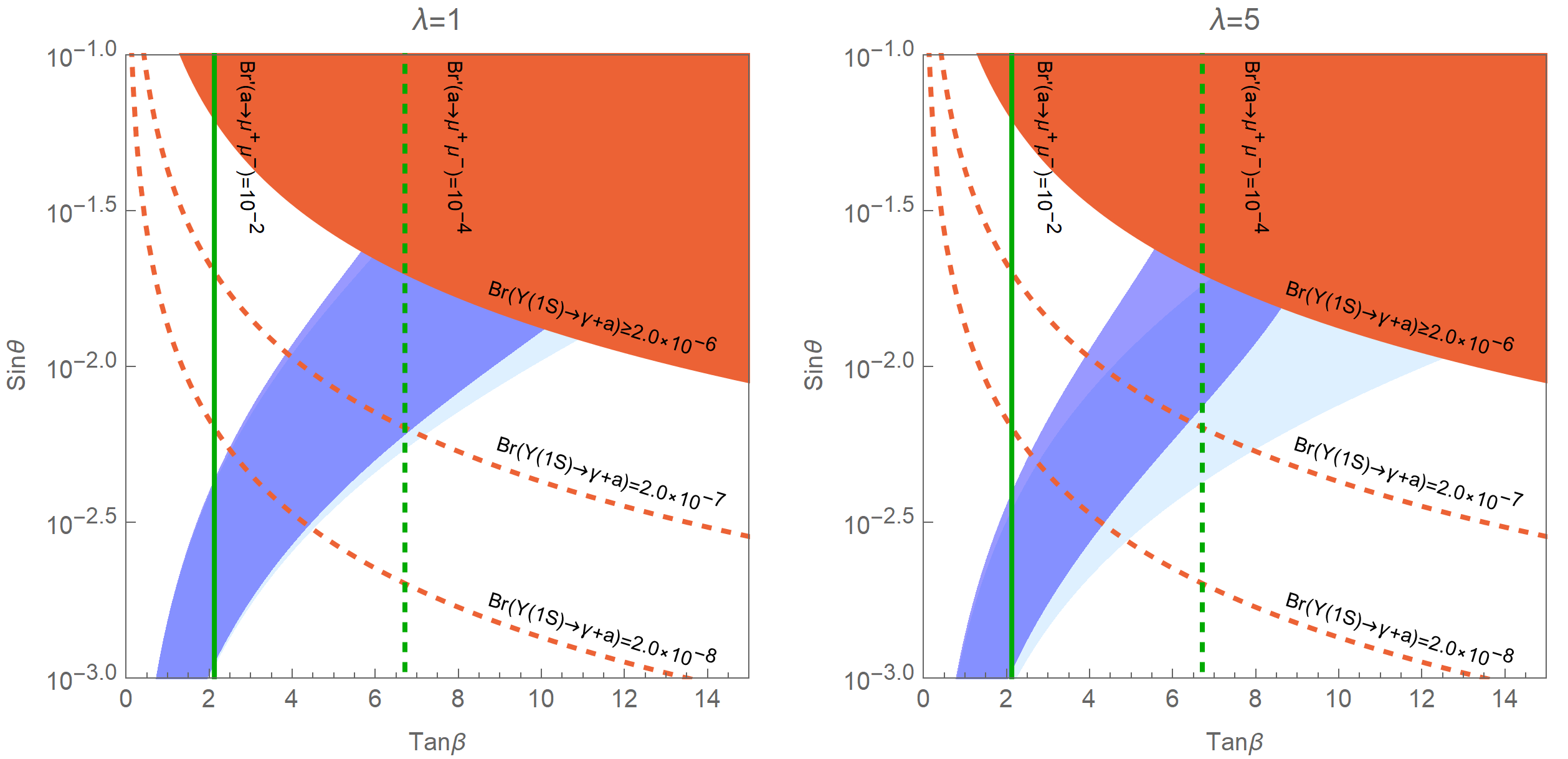}
    \caption[Parameter space explaining the Belle II excess.]{Parameter space explaining the Belle II excess in the DFSZ model, for $\lambda=1$ (left) and $\lambda=5$ (right). The dark (light) blue region gives Br$(B\rightarrow K a)=(1-9)\times 10^{-6}$, with complete (one loop only) calculation. The red region is excluded by the search for $\Upsilon(nS)\rightarrow \gamma a$. The green contours indicate the visible branching ratio $\text{Br}'(a\rightarrow \mu^+\mu^-)=\text{Br}(a\rightarrow \mu^+\mu^-)/\text{Br}(a\rightarrow \text{visible})$.}
   \label{DFSZII}
\end{figure}

In~\fref{DFSZII}, we show preferred and excluded regions in the $\sin\theta-\tan\beta$ plane, taking $\lambda=1$ and $\lambda=5$. 
The regions that can explain the $B\to K\overline{\nu}\nu$ excess are indicated with dark blue. We calculate them by requiring
\begin{equation}
    \text{Br}(B\to Ka)~=~(1-9)\times 10^{-6}, \quad \mathcal{P}_a^{\text{inv}}~=~ 1.
\end{equation}
To demonstrate that the newly founded two-loop effects are relevant, we also show the corresponding regions calculated using the one-loop expressions alone in light blue. 
When $\lambda=1$, the dark blue regions are mostly overlap with the blue ones, implying that ignoring the two loop corrections is a good approximation. 
On the other hand, when $\lambda=5$ and $\tan\beta\gtrsim 5$, the dark and light blue regions deviate from each other, which indicates the two-loop factor $X_3$ we newly computed becomes important. 
Notably, the two-loop and one-loop amplitudes contain opposite signs and partly cancel, so a larger value for $\theta$ is favored when the two-loop corrections are considered. 
$\tan\beta\gtrsim 10$ are excluded by the experimental constraint that~\cite{BaBar:2010eww}
\begin{equation}
    \text{Br}(\Upsilon(1S)\to \gamma+\text{invisible})~\lesssim~2.0\times 10^{-6}, 
\end{equation}
This bound assumes the (invariant) mass of the invisible state is around 2 GeV. 
We display the regions excluded by this process in red. 
The red dashed counters yields $\text{Br}(\Upsilon(1S)\to \gamma+\text{invisible})=2\times10^{-7}$ and $2\times 10^{-8}$, which illustrate the future sensitivities.

If the probability of $a$ decaying into visible states is large, the di-lepton signals from $a\to \mu^+\mu^-$ decay is also relevant.
We define the visible branching ratio for $a$ decaying into the muons as
\begin{equation}
\label{visibleBR}
    \text{Br}'(a\rightarrow \mu^+\mu^-)~=~\frac{\text{Br}(a\rightarrow \mu^+\mu^-)}{1-\mathcal{P}_a^{\text{inv}}}~\simeq~\frac{\text{Br}(a\rightarrow \mu^+\mu^-)}{\text{Br}(a\to \text{visible})}. 
\end{equation}
When the detection efficiency for the visible states is equal to 1, $1-\mathcal{P}_a^{\text{inv}}$ is equal to $\text{Br}(a\to \text{visible})$. 
$\text{Br}'(a\rightarrow \mu^+\mu^-)$ is independent of $\theta$ because all visible $a$ decay rates are equally $\theta$ suppressed. 
On the other hand, $\text{Br}'(a\rightarrow \mu^+\mu^-)$ depends on $\tan\beta$ and in case of DFSZ-II, it is inversely proportional to $\tan\beta$. 
Within this scenario, the values of $\tan\beta$ which leads to $\text{Br}'(a\rightarrow \mu^+\mu^-)=10^{-2}$ and $10^{-4}$ are indicated with solid and dashed green lines of \fref{DFSZII}, respectively. 
Assuming a sizable visible branching ratio so that $1-\mathcal{P}_a^{\text{inv}}\simeq0.1$, 
the small $\tan\beta$ regions can then be excluded by measurements on the $B\to K \mu^+\mu^-$ decay branching ratios. 
Then, only the regions with moderate $\tan\beta$ are unconstrained and the $\sin\theta-\tan\beta$ parameter space favored by the $B\to K\overline{\nu}\nu$ excess becomes compact.
However, if the total visible branching ratio takes generic values, merely small $\tan\beta$ \typo{can not}{cannot} indicate the physical observable $\text{Br}(a\rightarrow \mu^+\mu^-)$ is sizable.
According to Eq.~\eqref{visibleBR}, large $\text{Br}'(a\rightarrow \mu^+\mu^-)$ can always be compensated by reducing the magnitude of $(1-\mathcal{P}_a^{\text{inv}})$, so that $\text{Br}(a\rightarrow \mu^+\mu^-)$ can stay small.

The discussions above demonstrates that the DFSZ model with a moderately coupled dark sector can consistently explain the $B\to K\overline{\nu}\nu$ excess at Belle II. 
It is worthy to remark that the data prefers $\theta\sim 0.001-0.03$, which indicates that $f$ is not an ultra-high scale and may lie as low as TeV. 
The rare $B\rightarrow K a$ decay rate is mostly suppressed by a loop factor $1/(16\pi^2)$, small quark mixing angles, and possibly large $\tan\beta$, rather than only by a high scale. 
As a consequence, the charged Higgs needed for UV completion may not be super-heavy. 
Considering
\begin{equation}
    m_H~=~(\lambda\tan \beta)^{\frac12} \cdot f~=~(\lambda\tan\beta^{-1})^{\frac12}\cdot \frac{2v}{\theta}. 
\end{equation}
Perturbative unitarity requires that $m_H$ is bounded from above. 
If $\lambda$ is small, it can even lie below or around the TeV scale. 
In this scenario, the additional Higgs doublet contained in the DFSZ model can be directly produced at the high-energy frontier experiments such as the high-luminosity LHC. 
Even if not directly produced, the collider search can constrain the $m_H$ and $\tan\beta$ (since large $\tan\beta$ enhances the production cross section~\cite{Branco:2011iw}), so that $\lambda$ can be bounded from below\footnote{We note that if small $\lambda$ is excluded in future, our two-loop calculation will become always important.}. 
The additional Higgs doublet may also leads to other phenomenologically relevant processes, such as neutral $B$ meson mixing.

\section{Discussion}
In this chapter, we revisit the $B\rightarrow K a$ transition rate in the DFSZ model, which is a minimal UV-complete benchmark model for the axions or axion-like particles. 
From a practical side of view, we remark that the operator $a\overline{s}\gamma_5 b$ alone is sufficient to explain the Belle II excess. 
However, a ``mixing-only'' framework, in which $a$ inherit its coupling from $A$ with a mixing angle $\theta$, is widely used for analyzing the experimental data and showing the exclusion plots.
In our opinion, the physical picture of this framework is not very clear, because the `calculated' $b\rightarrow s a$ amplitude is divergent and gauge-dependent. 
We argue that the `mixing-only' framework brings no more information than the EFT with $a\overline{s}\gamma_5 b$ operator alone, but introduces unnecessary ambiguities when correlating $b\rightarrow s a$ to the other processes involving $a$.

On the other hand, the $U(1)$ and $Z_2$ symmetries of the DFSZ model make the physical picture of $b\to sa$ decay more transparent.
By studying approximate symmetries suppressing the known one-loop amplitudes, we determine new unsuppressed two-loop contributions. In particle, when $\tan\beta$ is sizable, our result becomes essential for a complete analysis. Moreover, while it is possible to capture the key features of DFSZ model with a bottom-up approach, the choice of the low energy basis is subtle. We find only the gauge invariant EFT yields the correct leading-log term, although at the cost of loosing renormalizability. 
Moreover, the rare $B\rightarrow K a$ decay rate is suppressed by flavour symmetries in this framework, which allows the needed UV completion for axion interactions to lie around or even below the TeV scale. Ideally, some other beyond-SM processes should not be far away from detection. We expect detecting $\Upsilon\rightarrow \gamma +a$ and $B\rightarrow K a\rightarrow K \mu^+\mu^-$ signals in future experiments, as well as an additional Higgs doublet. 
In this view, the DFSZ model represents a wide class of axion models satisfying the MFV criteria, and perhaps more generally, all axion models lying in the landscape of flavour symmetries. 
Thoroughly studying the DFSZ model is therefore important to fully understand the generic features of the landscape of flavour symmetries, which is a critical preliminary work for the future studies on the flavored axion models in the swampland.

At the end of this chapter, we encourage the readers to reflect on the following two lessons illustrated by this specific example, which is relevant to both light and heavy new physics:
\begin{enumerate}[label={(\roman*)}]
    \item In models more complicated than the SM, it is common that the rate of a physical process can be suppressed in some limits. In such cases, the higher-loop diagrams are not always subdominant and can sometimes become relevant for revealing the full processes. 
    \item Decoupling heavy particles is conditional. Modifying the low-energy structure of the SM ad-hoc, such as changing coupling strengths or adding new light states without considering a complete theory, may lead to incomplete predictions. 
\end{enumerate}
Although these statements are not new to the quantum field theory, they are sometimes overlooked in practice. 
Here, we remark that they are related to all theories no matter they lie in the landscape or swampland of the flavour symmetries, and moreover, phenomenologically revelant for Belle~II physics.

\clearpage

\chapter{Conclusion and Outlook}
\label{concluchapter}

In this thesis, we have demonstrated the feasibility of constructing a predictive and testable theory of flavour without explicit flavour symmetries. 
Our work paves the way for future studies on what we call the swampland of flavour symmetries.

We revisited the flavour puzzle in the SM and its extensions, and reviewed a number of popular symmetry-based flavour models, in particular those applying the idea of flavour deconstruction. 
Although the deconstructed flavour non-universal gauge interactions lead to $N_f=1$ and eliminate the replication of chiral fermions, generating the needed soft symmetry breaking terms at low energies nevertheless requires many new heavy particles and involves free model parameters. 
To unveil a simple flavour theory (assuming it exists), we argued to go one step further than the idea of flavour deconstruction and explore the generic models beyond the symmetry-based frameworks.
We found that apart form certain extra dimensional models, such a possibility has been highly overlooked so far.

The most urgent question on the swampland of flavour symmetries is the testability.
To demonstrate that verifying or falsifying a model without symmetry protection is possible, we revisited a very simple extension to the SM, the minimal type~II seesaw model, which involves only an $SU(2)_L$ triplet scalar field. 
We found that even if the triplet scalar lies at the TeV-scale, the model can stay consistent with all CLFV constraints, including the stringent bounds from $\mu\to e$ transitions, as long as the relevant Yukawa interactions obey certain two-zero textures. 
Interestingly, the needed two-zero textures do not lead to enhanced flavour symmetries acting on the three generations of $\ell_L$. 
In other words, suppressing the $\mu\to e$ transitions to not require any flavour symmetries in the relevant sub-sector of the theory. 
Our conclusion holds even if the two-zero textures are not exact. The deviations can be induced by the RG running across different scales.

The next challenge is on how to render the low-energy flavour parameters calculable when the underlying flavour symmetries do not exist or stay implicit at the scales that the experiments can reach. 
In this context, we argued that the low-energy flavour parameters should appear dynamically and referred to them as the emerging flavour textures. 
We firstly limited our study to perturbative frameworks. 
For simplicity, we started with only one single generation of chiral fermions, in which case $N_f=16$ in the gaugeless limit and the maximal anomaly-free flavour symmetry is $SO(10)$. 
It is evident that the SM lies in the swampland of $SO(10)$ due to its non-trivial gauge and Yukawa interacting structures. 
Yet taking the $SO(10)$ constraints as a high-scale initial condition for the RG evolution, certain flavour textures remain predictable at low energies, although they turn out to largely deviate from the commonly inferred $SO(10)$ constraints found for the usual ``grand desert'' scenario. 
In particular, in the presence of the TeV-scale leptoquarks motivated by the long-standing flavour anomalies, we found the low-energy mass ratio between the bottom quark and the $\tau$ lepton can agree well with the experimental observations. 
The TeV-scale leptoquarks drive the flavour parameters towards the infrared fixed points, indicating their low-energy patterns are mostly controlled by the SM gauge group instead of the $SO(10)$ boundary condition. 
Furthermore, after including the first two generation fermions, we found that the TeV-scale leptoquarks render the flavour conserving limit to be unstable at low energies. To some degree, this effect explains why the flavour mixing angles in the SM and the leptoquark models motivated by the flavour anomalies appear to be non-zero.

We also explored a non-perturbative framework in which all neutrino masses are bootstrapped to be non-zero. 
We revisited the $N_f=48$ free field theory and re-examined the fate of the $U(48)$ maximal flavour symmetry when gravity is taken into consideration, motivated by the fact that gravity \typo{can not}{cannot} be shielded. 
According to the framework established in Ref.~\cite{Dvali:2005an, Dvali:2005ws, Dvali:2013cpa, Dvali:2016uhn, Dvali:2017mpy, Dvali:2022fdv}, as long as the topological susceptibility of the vacuum in pure gravity is non-vanishing, no massless fermions can exist in the infrared limit and the $U(48)$ group must be broken into one of its anomaly-free subgroups at the low energies. 
The high and low energy theories correspond to different phases, so a phase transition, which cannot be described within the perturbative framework, must appear. 
We noted that such non-perturbative effects are not postulated ad-hoc, but bootstrapped by requiring the anomaly structures in different phases of the same theory to match. 
Importantly, although the dynamically generated fermion masses \typo{can not}{cannot} be directly calculated without more detailed control of the infrared behavior of gravity, they should all be viewed as calculable parameters, because the UV theory (assumed consistent) contains only one relevant free parameter, the gravitational constant. 
This non-perturbative framework fully addresses the problem of how to generate the Majorana neutrino masses. 
Yet, if the known neutrinos are all Dirac fermions, a gaugeable chiral symmetry acting on the neutrinos is needed to forbid bare neutrino mass terms and render them calculable parameters. We identified such a symmetry as a variation of the known $B-L$ symmetry, \typo{referred as}{referred to as} the enhanced $B-L$ symmetry, under which the two out of the three right-handed neutrinos can carry enhanced $B-L$ charges. 
We therefore concluded that the neutrino masses, irrespective of their Majorana or Dirac nature, can arise as calculable dynamical parameters, even if the underlying flavour symmetries are strongly broken by the non-perturbative effects.

The swampland of flavour symmetries should also involve light new physics, in particular the axions motivated by solving the strong CP problem. 
We noted that the flavourful axions can be naturally embedded into the swampland of $SO(10)$, together with the explanation on the known SM flavour parameters. In such a case, the flavour structure of the axion interactions depends mostly on the leptoquark Yukawa couplings and is quite different from the symmetry protected patterns.
However, instead of going deeper into such a model, we focused more on a complete understanding about the axion models in the landscape of flavour symmetries, because we found many important preliminary work is still missing. 
We revisited the axion flavour violating interactions in one of the minimal renormalizable axion models, known as the DFSZ model.
We clarified the interpretation based on the mixing picture and identified an overlooked two-loop contribution, which is enhanced if $\tan\beta$ is sizable. 
Furthermore, we pointed out a rare feature which we refer to as apparent non-decoupling and identified the consistent low-energy basis in which the effect of ultra-heavy particles \textit{truly} decouples. 
We also analyzed the parameter space addressing the excess of the $B\to K+\text{invisible}$ events at Belle II, and showed its indications on other relevant processes, including the single photon decays of $\Upsilon(1S)$ and a possible excess in $B\to K \mu^+\mu^-$.

In our view, the most promising path towards a simple flavour theory is the above-mentioned $SO(10)$ framework. 
In future, identifying a concrete benchmark model which generates not only the SM flavour parameters but also those needed for explaining the known $B$ anomalies will make this approach more convincing.
For this purpose, one can start with the minimal Pati-Salam model, whose Yukawa sector only involves $(\textbf{15}_C,\textbf{2}_L,\textbf{2}_R)$ and $(\overline{\textbf{10}}_C, \textbf{1}_L, \textbf{3}_R)$, responsible for reproducing the SM charged-fermions masses and decoupling the right-handed neutrinos, respectively. 
Similar to minimal SO(10), an incorrect $b-\tau$ mass relation arises in the UV due to quark-lepton unification, but it can be modified by the RG evolution induced by TeV-scale leptoquarks contained in the minimal Yukawa sector. 
The new feature is that the theory involves more than one independent Yukawa couplings, which \typo{can not}{cannot} always be simultaneously diagonalized. 
As a result, sizable flavor mixing angles can be generated. 
Such a model can be extended with more scalar representations, which all together compose the $\phi_{126}$ multiplet of $SO(10)$. 
We expect the data to prefer a particularly interesting limit of this extended model in which all Yukawa couplings become quasi-aligned, so that the theory can be approximated by the minimal $SO(10)$ model in the UV. 
In such a case, the misalignment terms can further be replaced by $SO(10)$ invariant higher-dimensional operators, which can be UV-completed with next-to minimal extensions. 
We note the simplest one could contain a singlet fermion $\psi_{1}$ and a scalar field $\phi_{16}$ which transforms as an $SO(10)$ spinor. 
The induced one-loop threshold corrections can modify the Yukawa interactions in the most minimal $SO(10)$. 
An appealing feature of such an extension is that the $\psi_{1}$ and $\phi_{16}$ fields both carry matter parity opposite to the SM chiral fermion fields in $\psi_{16}$. As a result, their neutral components are stable even if being light and can naturally serve as the stable dark matter candidates. 
Due to these appealing features, we expect that this model will yield a concrete unified theory addressing the SM flavor puzzles, the long-standing $B$ anomalies, and hopefully dark matter, with neither the Pati-Salam group nor $SO(10)$ being explicit at the relevant low energy scales. 
Examining this model only requires the information on TeV scale new physics with sizable interaction strengths. In principle, the next-generation colliders can directly verify or falsify it in near future.

In addition to the flavour mixing angles, the origin of the CP violating phase can also be explained by non-decoupling new physics. 
The two Higgs doublet model can contain two degenerate electroweak vacua with opposite CP violating phases and allows the so-called spontaneous CP violation~\cite{Lee:1973iz}, given the physical scalar states contained in the model are all lighter than about $500$ GeV~\cite{Nebot:2018nqn, Nierste:2019fbx, Miro:2024zka}.
In the limit that all model parameters are real, the CP symmetry is only spontaneously broken below the electroweak scale and restores above it, so that CP invariance remains as a fundamental principle of nature. 
From this perspective, another physical parameter of the SM --- the CP violating phase in the CKM matrix --- then arises as the vacuum solution and its origin gets dynamically explained by non-decoupling new physics. 
Embedding this mechanism into a minimal realistic $SO(10)$ model without light LQs has been explored in our earlier work~\cite{Gao:2024xte}. 
Here, we note that in principle, the model we discussed in Ref.~\cite{Gao:2024xte} could  be further simplified if the light LQs are taken into consideration.

The framework based on non-perturbative gravity can be applied to the ultra-light new physics in general. 
It is a common practice to postulate the feeble interactions between the SM and ultra-light new physics ad-hoc. 
Although such a set-up provides benchmark models useful for analyzing the experimental data, it nevertheless leads to many free model parameters in most cases. 
We think it is more convincing to promote these coupling strengths as dynamical observables generated during a gravitationally induced phase transition at a low scale $f_G$, so that there are no additional parameters in the UV theory. 
We refer to this framework as the non-perturbative gravity portal to ultra-light new physics. 
In this thesis, we have discussed about interpreting the mixing angle between the $B-L$ gauge boson and the visible photon as a dynamical parameter, while the non-perturbative gravity portal can in principle be generalized to all interactions for the ultra-light new particles. 
It is worth to note that the SM is only recovered in the limit $f_G\to 0$. In other words, signals for ultra-light new physics, as deviations to the SM predictions, are always proportional to powers of $f_G/E$, where $E$ is the relevant energy scale squared. 
Comparing with the collider experiments in which $E$ is around the $\mathcal{\text{TeV}}$ scale, the low-energy frontier experiments are more feasible for testing such portals because they enforce $E\to0$. Notably, the dark matter direct detection experiments with ultra-low thresholds are particularly motivated. 
Furthermore, we highlight that the phase transitions change the standard cosmological model, which we notice leads to two important indications. Firstly, the existing constraints from cosmology exclude the possibility of directly detecting many ultra-light particles in near-future terrestrial experiments. Yet the phase transition predicts time varying interaction strengths, which can escape the stringent cosmological bounds from early universe so that the direct measurements on earth become possible again. 
Secondly, the phase transition, although typically occurring at the low temperatures, can still leave observable imprints on the late universe observables, such as the matter power spectrum. 
We note that analyzing the impact of these effects on the cosmological data remains \typo{somehow}{widely} overlooked in the literatures.

At the very end of this thesis, we emphasize that our approach has broader importance beyond particle physics. 
The ultimate origin of complexity in nature, e.g. the mechanism leading to turbulence, has been debated for a long time. 
On one hand, it is commonly believed that the randomness in most of the physical systems can be traced back to external noise. In this view, the anarchic flavour structures are more likely to stem from randomness in other unknown sectors, so that models with such structures can never manifest themselves as an ultimate theory for flavour. 
On the other hand, people have also argued that complexity does not always need random inputs but can emerge dynamically from the non-linear evolution of the system~\cite{Wolfram:1985qr}. 
Then, an anarchic flavour structure with no underlying flavour symmetries, if uncovered by future experiments and showing consistency with certain very simple patterns in the UV,
will indicate that even the fundamental laws for elementary particles are governed by emergence, supporting the hypothesis that all complexity ultimately emerges from simple structures. 
In this thesis we have pointed out two examples for the second avenue, namely the non-trivial $m_b/m_{\tau}$ ratio and the emergence of neutrino masses from a chirally-symmetric theory.


\newpage

\phantomsection
\addcontentsline{toc}{chapter}{Bibliography}
\bibliographystyle{JHEP} 
\bibliography{Dissertation.bib}

\end{document}